\documentclass[12pt,a4paper]{book}

\usepackage{mathptmx}	%% palatino
\newcommand{\mathbold}{\bm}
\DeclareMathVersion{bold}

\usepackage{mypackages}
\usepackage{mymacros}
\usepackage{mylayout}

\begin{document}
\selectlanguage{english}

\thispagestyle{empty}

\begin{center}

{\fontsize{28}{40}\usefont{OT1}{\rmdefault}{b}{n}\selectfont 
Probing
%} \\[6pt]
%{\fontsize{35}{42}\usefont{OT1}{\rmdefault}{b}{n}\selectfont 
the transverse spin  of quarks} \\[6pt]	
{\fontsize{28}{40}\usefont{OT1}{\rmdefault}{b}{n}\selectfont  in deep
inelastic scattering}

%{\fontsize{28}{40}\usefont{OT1}{\rmdefault}{b}{sc}\selectfont 
%Probing
%%} \\[6pt]
%%\fontsize{35}{42}\usefont{OT1}{\rmdefault}{b}{n}\selectfont 
%the transverse spin  of quarks} \\[6pt]	
%{\fontsize{28}{40}\usefont{OT1}{\rmdefault}{b}{sc}\selectfont  in deep inelastic scattering}

%{\Huge \bf Probing the transverse spin of quarks}\\
%{\Huge \bf in deep inelastic scattering}

\vfill

\includegraphics[width=12cm]{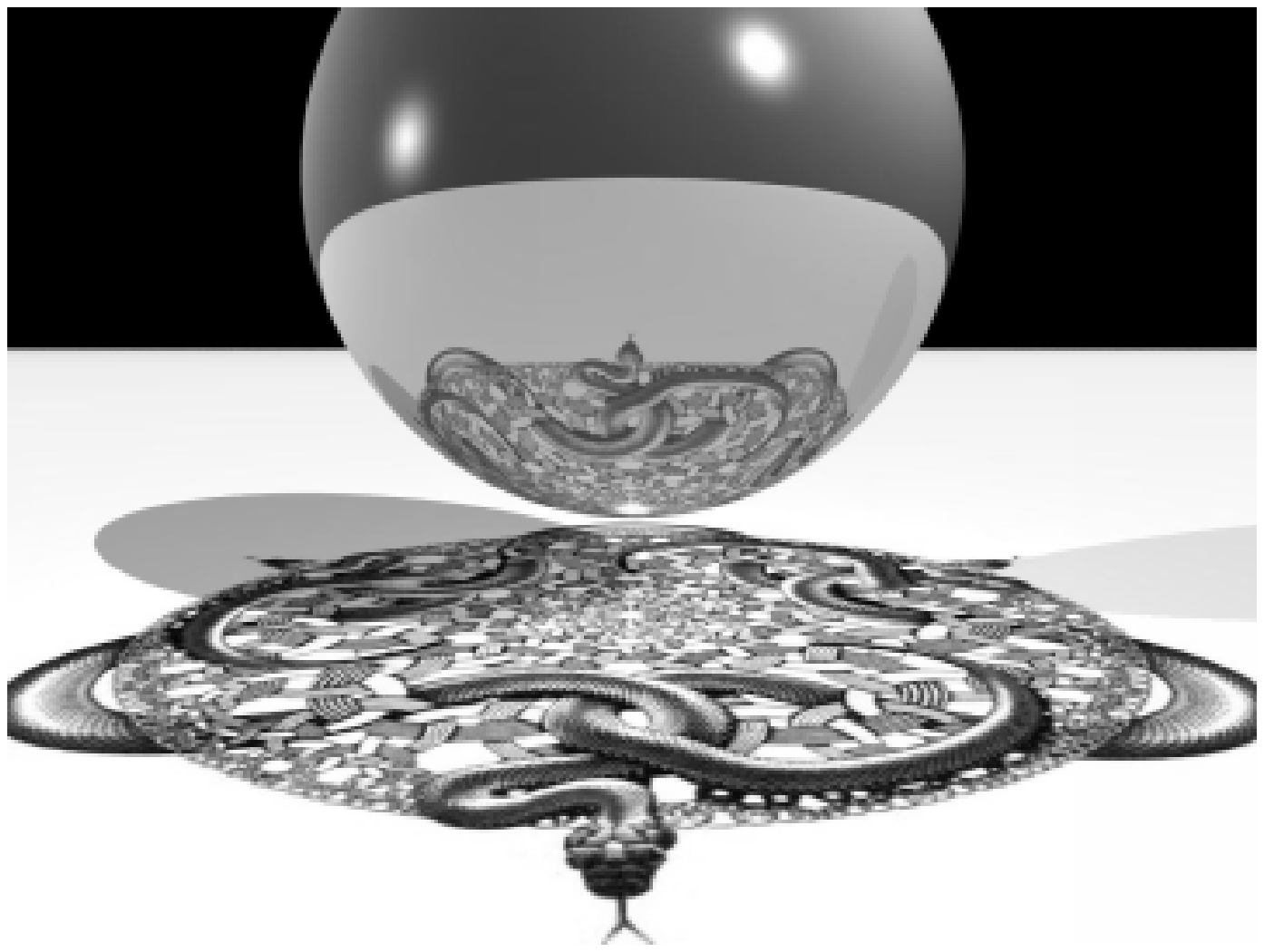}

\vfill

{\fontsize{22}{40}\usefont{OT1}{\rmdefault}{b}{n}\selectfont  
 Alessandro Bacchetta}
\end{center}

\clearpage

%%%%%%%%%%%%%%%%%%%%%%%%%%%%%%%%%%%%%%%%%%%%%%%%%%%%%%%%%%%%

\newpage

\thispagestyle{empty}
{\noindent Front cover: {\it The Hermetic
Truth of Hadrons} by Anders Sandberg, inspired by M.C. Escher's last drawing,
{\it Snakes} (1969). Printed with the permission of the author.}

\vfill
\begin{center}
\fbox{\parbox{\textwidth}{{\bf
        {\large This is a version of my PhD thesis (defended on Oct
4th, 2002) prepared for
submission to the arXiv. It is almost identical to the
official copy. However, to comply with the arXiv requirements some changes
had to be done, resulting in  a few differences and layout
mistakes. A more faithful version of the thesis can be found
at the
web address http://www.nat.vu.nl/~bacchett/research/thesis.pdf.\par}}}}
\end{center}

\vfill

\noindent
\includegraphics[width=2cm]{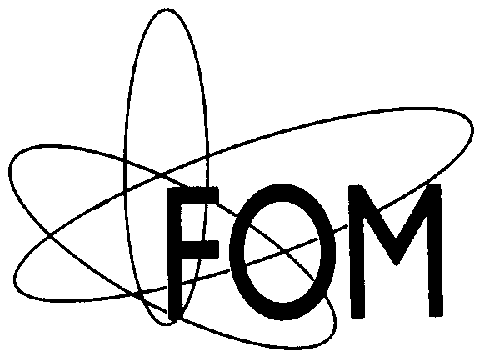}\\
The work described in this thesis is part of the research programme of
the {\em Stichting voor Fundamenteel Onderzoek der Materie} (FOM), which is
financially supported by the {\em Nederlandse Or\-ga\-ni\-sa\-tie voor
Wetenschappelijk Onderzoek} (NWO).

\newpage

\thispagestyle{empty}
\begin{center}
        {VRIJE UNIVERSITEIT}
        \\
        \vspace*{2.5cm} {\bf
        {\LARGE Probing the transverse spin of quarks\\ in deep inelastic
scattering\par}}
        \vspace*{2.5cm}
        {ACADEMISCH PROEFSCHRIFT}
        \\
        \vspace*{1.5cm}
%       \parbox{8.2cm}{
        ter verkrijging van de graad van doctor aan\\
        de Vrije Universiteit Amsterdam,\\
        op gezag van de rector magnificus\\
        prof.dr.\ T.~Sminia,\\
        in het openbaar te verdedigen \\
	ten overstaan van de promotiecommissie \\
        van de faculteit der Exacte Wetenschappen\\
        op vrijdag 4~oktober~2002 om 15.45 uur\\
        in het auditorium van de universiteit, \\
        De Boelelaan 1105 %}
        \vspace*{1.5cm}

        door \\
        \vspace*{1.5cm}

        {\large \bf Alessandro Bacchetta} \\
        \vspace*{1cm}
        geboren te Borgosesia, Itali\"e
\end{center}

\newpage

\thispagestyle{empty}
\noindent promotor: prof.dr.\ P.J.G. Mulders

\newpage

\setcounter{page}{1}
\pagenumbering{roman}

\tableofcontents

\renewcommand{\quot}{%
\parbox{7cm}{All cultural products contain a mixture of two elements:
conventions and inventions.} \\ J. G. Cawelti
}

%%%%%%%%%%%%%%%%%%%%%%%%%%%%%%%%%%%%%%%%%%%%%%%%%%%%%%%%%%%%
\chapter*{Notations and conventions}
\markboth{Notations and conventions}{Notations and conventions}
\addcontentsline{toc}{chapter}{Notations and conventions}

The conventions will mainly follow the book of Peskin and Schroeder~\cite{Peskin:1995ev}. We use the metric
tensor 
\renewcommand{\arraystretch}{1}
\[
g^{\mspace{2mu}\mu \nu} = \begin{pmatrix}
	1 & 0 & 0 & 0 \\
	0 & -1 & 0 & 0 \\
	0 & 0 & -1 & 0 \\
	0 & 0 & 0 & -1 \end{pmatrix},
\]
with Greek indices running over 0,1,2,3.
Repeated indices are summed in all cases. Light italic roman type will be used 
for four-vectors, while boldface italic will be used for three-vectors.

\section*{Light-cone vectors}

Light-cone vectors will be indicated as
\[
a^{\mspace{2mu}\mu} = \lf[a^-,\; a^+ ,\; \mb{a}_T\rg] = 
\lf[\frac{a^0 - a^3}{\sqrt{2}},\; \frac{a^0 + a^3}{\sqrt{2}},\; a^1 ,\; a^2\rg].
\]
The dot-product in light-cone components is
\[ \begin{split}
a \cdot b &= a^+ b^- + b^- a^+ - \mb{a}_T \cdot \mb{b}_T \\
 &= a^+ b^- + b^- a^+ - a^i b^i \\
 &= a^+ b^- + b^- a^+ - a_x b_x - a_y b_y
\end{split}
\]
The two-dimensional transverse parts of the vectors will be written 
in boldface with an index $T$ and Latin indices will be used to denote the two 
transverse components only. Note that
\begin{align*}
\mb{a}_T &= (a_{x},\; a_{y}), & a_T^{\mspace{2mu}\mu} &= [0,\; 0,\; \mb{a}_T], &  a_{T \mu} &= [0,\; 0,\; -\mb{a}_T].
\end{align*} 
We introduce the projector on the transverse subspace
\[
g_T^{\mspace{2mu}\mu \nu} =  \begin{pmatrix}
	0 & 0 & 0 & 0 \\
	0 & -1 & 0 & 0 \\
	0 & 0 & -1 & 0 \\
	0 & 0 & 0 & 0 \end{pmatrix},
\]
We define the antisymmetric tensor so that
\begin{align*} 
\eps^{0123} &= +1, & \eps_{\mspace{2mu}0123} &= -1. 
\end{align*} 
and we define the transverse part of the antisymmetric tensor as
\[
\eps_T^{\mspace{2mu}\mu \nu} = \eps^{- + \mu \nu} = \eps^{0 3 \mu \nu}.
\]

\section*{Dirac matrices}

Dirac matrices will be often expressed in the chiral or Weyl representation, i.e.\
\begin{align*}
\g^0 &= \begin{pmatrix}
	0 & \mb{1}	\\
	\mb{1} & 0 
	\end{pmatrix},&
\g^i &= \begin{pmatrix}
	0 & -\bm{\sigma}^i \\	
	\mb{\sigma}^i & 0 
	\end{pmatrix},&
\g_5 &= \begin{pmatrix}
	\mb{1} & 0 \\	
	0 & -\mb{1} 
	\end{pmatrix}, 
\end{align*} 
and we will make use of the Dirac structure 
\[ 
\sig^{\mspace{2mu}\mu \nu} \equiv \frac{\ii}{2} \lf[\g^{\mspace{2mu}\mu},\g^{\nu}\rg].
\]
\renewcommand{\arraystretch}{1.5}

\setcounter{page}{1}
\pagenumbering{arabic}

\renewcommand{\quot}{%
\parbox{6cm}{The most difficult part of a trip is to cross the doorway.}\\[2mm]
P. Terentius Varro 
}
%\renewcommand{\quot}{%
%\parbox{6cm}{If you cannot -- in the long run -- tell everyone what
%you have been doing, your doing has been worthless.}\\[2mm]
% E. Schr\"odinger  
%}

%%%%%%%%%%%%%%%%%%%%%%%%%%%%%%%%%%%%%%%%%%%%%%%%%%%%%%%%%%%%%%%%%%%%%%%%%%%%
\chapter{Introduction}

In this thesis I will discuss three different ways to observe the
transverse spin of quarks inside the nucleons.
Before embarking on such an undertaking, I would like to spend a few pages on
explaining what makes this problem so interesting to justify
investing years of research on it. 
This introduction is meant especially for nonexperts, 
since I will review notions well known 
to the experts in the field.

%%%%%%%%%%%%%%%%%%%%%%%%%%%%%%%%%%%%%%%%%%%%%%%%%%%%%%%%%%%%
\section{The structure of matter}

When we talk about {\em quarks} inside {\em nucleons} we are referring to the
best paradigm we currently have to describe the {\em elementary} structure of
matter. 
The comprehension of this elementary structure is a question that has
allured philosophers and scientists since the historical 
origins of philosophical thought. It is striking to observe that 
as early as six hundred years BC, 
Greek philosophers were already wondering: are there fundamental
elements in nature, what are they and how do they interact? 
Today, after more than two millennia, we learned a lot about the
structure of matter, but some of the most important questions still elude our
comprehension. We are still engaged in one of the oldest quests of human mind.

Since 1803, when Dalton suggested his atomic hypothesis~\cite{Dalton:1803}, 
we have gradually realized that almost all 
matter on earth is made up of atoms. Atoms
contain electrons -- identified for the first time by J.~J. Thomson in
1897~\cite{Thomson:1897,Thomson:1899} -- and nuclei -- 
introduced for the first time by
E.~Rutherford in 1911~\cite{Rutherford:1911,Geiger:1909}. 
The efforts to explain precisely 
the structure of atoms and the electromagnetic interaction binding together
electrons and nuclei lead to two of the major achievements of
physics in the last century: Quantum Mechanics and Quantum Electrodynamics
(QED).

In the meantime, more investigations were carried out to grasp the structure
of the nucleus inside atoms. The smallest known nucleus was identified with a
single particle~\cite{Rutherford:1920}, the {\em proton}, while a second
constituent of heavier nuclei, the {\em neutron}, 
was eventually observed by J. Chadwick~\cite{Chadwick:1933}. Since they are
the constituents of the nucleus, protons and
neutrons are referred to as {\em nucleons}. They are kept together by the
nuclear force, of which at the moment we have only an incomplete
understanding.  

Although the electrons are responsible for the chemical properties of 
atoms, they account for a very small fraction of the mass of the
atom. The mass of an electron is about 0.511 MeV, while the mass of a
proton is about 938 MeV\@. 
%Since in any atom there is
%at least the same number of protons and electrons, it is easy arithmetic
%to conclude that 
Therefore, nucleons make up for more than 99.9\% of ordinary atomic matter.
If we want to understand matter, we cannot set aside the problem of explaining 
the structure of nucleons.
Nucleons belong to the more
general class of {\em hadrons}, of which they are the most abundant specimen. 
At first, hadrons were classified as {\em elementary
particles}, i.e.\ without any internal substructure. Very soon this
appeared to be an unsatisfactory hypothesis, in particular 
since 
there are so many of them (several dozens). 
Nowadays, the study of the structure 
of hadrons represents a 
field of research on its own, often designated with
the name of {\em hadronic physics}. An up-to-date review of the field can be
found in Refs.~\citen{Bjorken:2000ni} and~\citen{Capstick:2000dk}.

%%%%%%%%%%%%%%%%%%%%%%%%%%%%%%%%%%%%%%%%%%%%%%%%%%%%%%%%%%%%
\section{Hadrons and deep inelastic scattering}

To interpret the information available on the properties of hadrons
in 1964, M.~Gell-Mann~\cite{Gell-Mann:1964nj} and G.~Zweig~\cite{Zweig:1964} 
independently suggested that
hadrons are
composed of smaller constituents, the {\em
quarks}, having spin $1/2$, a fractional
electric charge and a new degree of freedom, called flavor. This model is 
often referred to as {\em constituent quark model}.
Gell-Mann himself seemed not to believe in the existence of
quarks as real entities, but rather regarded them as convenient concepts~\cite{Gell-Mann:1964nj}.
One of the reasons to be skeptical about the real existence of quarks was
that they have 
a charge that is just a fraction of the electron charge, while the charge of 
all
other elementary particles is an integer multiple 
of that.

The quark model aimed at describing the mass, charge and spin of the hadrons.
For instance, the proton has a mass of about 1 GeV, a charge $+e$ (the same as 
the electron, but with opposite sign) and spin equal to $1/2$.
According to the model, a proton with its spin, for instance,
in the up direction 
is made of two quarks with flavor {\em up} and charge $2e/3$ 
plus one quark with flavor {\em down} and charge $-e/3$.
Two of the quarks have spin $1/2$ in the up direction 
and one has spin $1/2$ in the down direction. Each of the three quarks carries
 about one-third of the mass of the proton.

Looking at the ``extrinsic'' properties of hadrons -- like their mass, charge
and  spin 
-- was not enough to unravel
the details of their structure. 
%To follow an analogy, looking at the body of a 
%car is not sufficient to understand how it works. To 
%investigate further, it turns out to be convenient to\ldots destroy the car. 
%In
%fact, this is what is usually done in 
To glance at the inside of hadrons, physicists 
resorted to 
{\em deep inelastic scattering} (DIS)
experiments, as in the pioneering experiments led by Friedman,
Kendall and Taylor at the Stanford Linear
Accelerator Center (SLAC)~\cite{Bloom:1969kc,Breidenbach:1969kd}.
In scattering experiments, a focused beam of particles
 is dispersed
by the interaction with a target. The way this dispersion takes place yields
information on the structure of the target. 
%When we are looking at an
%object, we see it because photons hit it and they are subsequently 
%detected by our eyes. Our brain is constantly processing information from
%scattering experiments. 
For instance,  
the existence of the nucleus was suggested by Rutherford as an explanation to
the scattering experiments of Geiger and Marsden~\cite{Geiger:1909}. 
%About
%those experiments, Rutherford himself said: 
%``It was as if you had fired a fifteen inch
%shell at a piece of tissue paper and it came back and hit you\@!''
%In reality, because of the existence of the nucleus, those early experiments
%corresponded more or less to shooting soccer balls at a car. 
%They did tell us a lot about the
%existence of a hard object, but they could not say much about its structure.
%In comparison, deep inelastic scattering experiments correspond to shooting
%at a car with a machine gun, penetrating its hood and recording the way
%bullets bounce on its internal components -- engine, gear box and all. 
%%That's a way to glance at the internal structure of the car.

Basically,
particle accelerators as the one at SLAC 
are exploited as microscopes of extremely high
resolution. The experiments at SLAC scattered electrons off hydrogen. The interaction 
proceeds via
the exchange of a virtual photon with high energy and momentum. A measure of
the resolution of the experiment is given by the four-momentum squared of
the virtual photon, $Q^2$, or rather by the associated wavelength $\hslash
/Q$. The SLAC experiments reached a  maximum $Q^2$ of
7.4 GeV$^2$, corresponding to a resolution of the order of $1/10$ of the
proton size.
 
The results of the SLAC experiments indicated that the scattering data did not
exhibit a (strong) dependence on $Q^2$. They
depended rather on the variable that was later to be named $x$-Bjorken,
$\xbj$, in honor of J.~Bjorken.\footnote{We will properly define $\xbj$ in Chap.~\ref{c:trans}.}
 This property, called {\em scaling}, was
predicted by Bjorken himself~\cite{Bjorken:1969dy} and 
explained by 
R.~Feynman~\cite{Feynman:1969ej,Bjorken:1969ja}, 
who introduced the {\em parton model}:
the proton was pictured to be a collection of almost 
{\em free} point-like constituents off which the electrons
scatter incoherently. The constituents of the proton were initially 
called {\em partons}, but it soon became clear that they had a lot in common
with the quarks of Gell-Mann and Zweig.

Feynman's partons have spin $1/2$, fractional electric charge and flavor, 
but they have a very small mass compared to Gell-Mann's quarks, a few
MeV against about 300 MeV. Consequently, we call them {\em
current} quarks, to distinguish them from the {\em constituent} quarks of the
quark model.
But this is not the only difference between the two models.
In the constituent quark model, the proton is made up just of three quarks,
while in the parton model it turns out that there is a huge number 
 of quark-antiquark pairs, together with a huge number of
electrically neutral particles, later to be identified as {\em gluons}. 

Deep inelastic scattering experiments are performed in some of the 
world's largest experimental 
facilities for high energy physics, such as CERN, SLAC, DESY, BNL.
They usually employ beams of
electrons, positrons, muons or, more rarely, neutrinos. 
They scatter off different kinds of fixed
targets or off a beam of protons,
and they operate at different energies and kinematic
coverage.

%%%%%%%%%%%%%%%%%%%%%%%%%%%%%%%%%%%%%%%%%%%%%%%%%%%%%%%%%%%%
\section{Quantum Chromodynamics and confinement}

The parton model raised a profound question. We experience that matter, at
least in the normal conditions on earth, is 
composed of hadrons -- it is the so-called {\em hadronic matter}. 
If quarks are the hypothetical constituents of hadrons, they must be bound 
extremely tight
to explain why 
we have never directly observed a single isolated quark, nor a different state 
of matter other than hadronic. This essential feature of quark dynamics 
is known as {\em confinement}. 
Yet, deep inelastic scattering suggests that in
the interaction with a high-$Q^2$ virtual photon, quarks behave as if they
were essentially free. This property is known as {\em asymptotic freedom}. The 
question is then: how is
it possible to devise a theory to reconcile these two opposing
properties, confinement and asymptotic freedom?

A first attempt to implement confinement was done by
postulating that quarks have
a color charge and that all detectable 
objects have to be colorless (cf.\ Refs.~\citen{Nambu:1966,Nambu:1974pb} and 
references
therein).  By virtue of
this assumption, it is impossible to see an isolated, colored quark.
On the other hand, such a point of view is not suited to describe
asymptotic freedom.
There was the need of a theory that could describe the binding of colored 
quarks as a dynamical mechanism.

\begin{figure}
	\centering
	\begin{tabular}{c}
	\rput(-0.3,3.1){\rotatebox{90}
	{{$\alpha$ }} }
	\rput(4.7,-0.5){$\log(Q^2/\rm{GeV}^2)$}
	\rput(0.25,0.48){\scriptsize \boldmath $/$}
	\includegraphics[width=8.5cm]{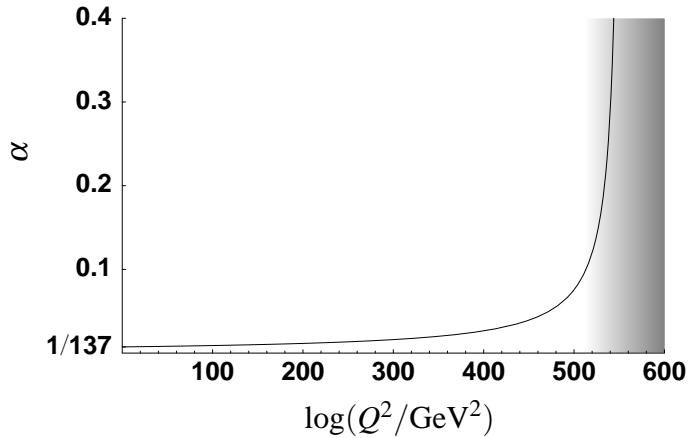}
	\\
	\\
	\end{tabular}
	\caption{The running of the electromagnetic 
	coupling constant. The gray area roughly 
	indicates where perturbation theory
	is not trustworthy anymore. 
	Note the {\em huge}
	scales necessary to achieve a appreciable increase of the coupling.}
	\label{f:runningalpha}
        \end{figure}

To shape a new theory of color interactions, it seemed natural to follow the
example of Quantum Electrodynamics, the quantum field theory
of electromagnetic
interactions. QED is in essence a {\em perturbative} theory, which works
because electromagnetic interactions are weak. In fact, electrons can be
separated quite easily from atoms and observed as free
particles. A measure of the strength of the
electromagnetic interaction is given by the value of the electromagnetic
coupling 
constant $\alpha \approx 1/137$. From renormalization of QED, it is known
that in reality $\alpha$ is not constant, but it has to increase as the 
momentum exchange of the interaction increases, or equivalently as the
interaction takes place over shorter distances. However, the increase of the
coupling is so weak (e.g.\ $\alpha \approx 1/135$ at $Q^2 =1000$ GeV$^2$) 
that perturbative QED works brilliantly for any
electrodynamics experiment we might do. Fig.\ \vref{f:runningalpha} shows
approximately the way 
the electromagnetic coupling constant changes with $Q^2$.

A breakthrough in the comprehension of quark interactions came in 1973, when
D.~Gross and F.~Wilczek~\cite{Gross:1973id,Gross:1973ju} and
D.~Politzer~\cite{Politzer:1973fx} showed that {\em non-Abelian} quantum field 
theories can display the crucial properties of {asymptotic
freedom}, i.e.\ the interaction they describe is weak at high momentum
transfer (or long distances).
This discovery prompted the birth of Quantum Chromodynamics (QCD), a
non-Abelian field theory of color interactions.

The difference between QED and QCD can be likened to the difference between
the attraction forces of two opposite magnetic poles and of two
ends of a spring. In the first case, we know that increasing the distance
between the magnets, the attraction diminishes, while, in the second case,
separating the two ends the force will increase more and more.

In field theory language, the electromagnetic coupling 
constant is reduced at large distances due to the
effect of vacuum polarization, which is responsible for a {\em screening} of
the bare electric charge. On the contrary, the color coupling constant is
reduced at short distances because the vacuum polarization induces an {\em
antiscreening} of the charge, or equivalently 
an enhancement of the charge at large
distances. The reason for this different behavior is that {\em gluons}, the
mediator of the color interaction, carry color charge themselves, while
photons, the mediator of the electromagnetic interaction, are chargeless. 
Ref.~\citen{Peskin:1995ev} (p.~541) and Ref.~\citen{Field} (p.~5) 
present enlightening discussions on antiscreening and asymptotic freedom.

	\begin{figure}
	\centering
	\begin{tabular}{c}
	\rput(-0.3,3.1){\rotatebox{90}
	{{$\alpha_s$ }} }
	\rput(4.7,-0.5){$\log(Q^2/\rm{GeV}^2)$}	
	\includegraphics[width=8.5cm]{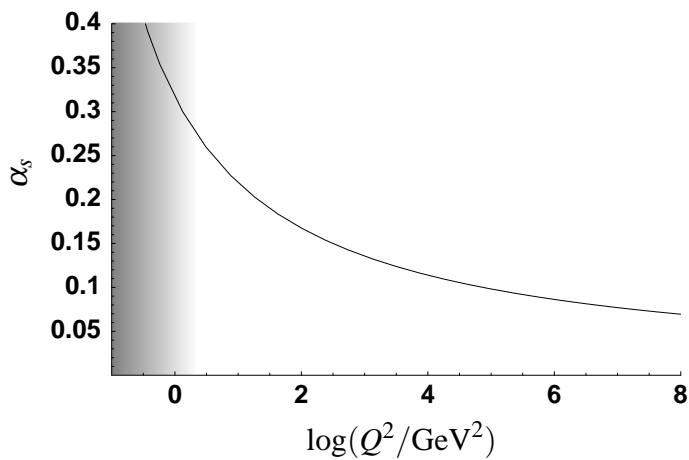}\\
	\\
	\end{tabular}
	\caption{The running of the strong 
	coupling constant. 
	The gray area roughly indicates where perturbation theory
	is not trustworthy anymore. }
	\label{f:runningalphas}
        \end{figure}

QCD is able to justify asymptotic freedom, but what about confinement?
At the moment, we know that QCD is not in contradiction with
confinement and 
might in fact explain it, but we are not able to demonstrate 
this statement. As in the case of QED, 
the strength of color interaction is measured by the strong  
coupling constant $\alpha_s$,
which has a value of about 0.117 at $Q^2 = 8.3 \times 10^3$
GeV$^2$. But at lower energy scales, e.g.\ $Q^2 \approx 1$
GeV$^2$, the coupling constant grows and becomes of the order of 1. The
running of the strong coupling constant is illustrated in
Fig.\ \vref{f:runningalphas}. 
We might deduce that 
the increase of the coupling constant
 is a sign of the onset of confinement. In reality,
we can only conclude that at low energies we enter a regime where perturbation
theory cannot be trusted. Therefore, even if QCD is in principle a consistent
theory at any energy scale, we cannot use standard techniques
to draw conclusions about its
behavior in the nonperturbative regime. To a certain extent, 
we cannot be sure that QCD  is the correct 
theory in this regime: maybe it is simply an asymptotic approximation of a
more profound theory. 

\begin{figure}
	\centering
	\includegraphics[width=11.5cm]{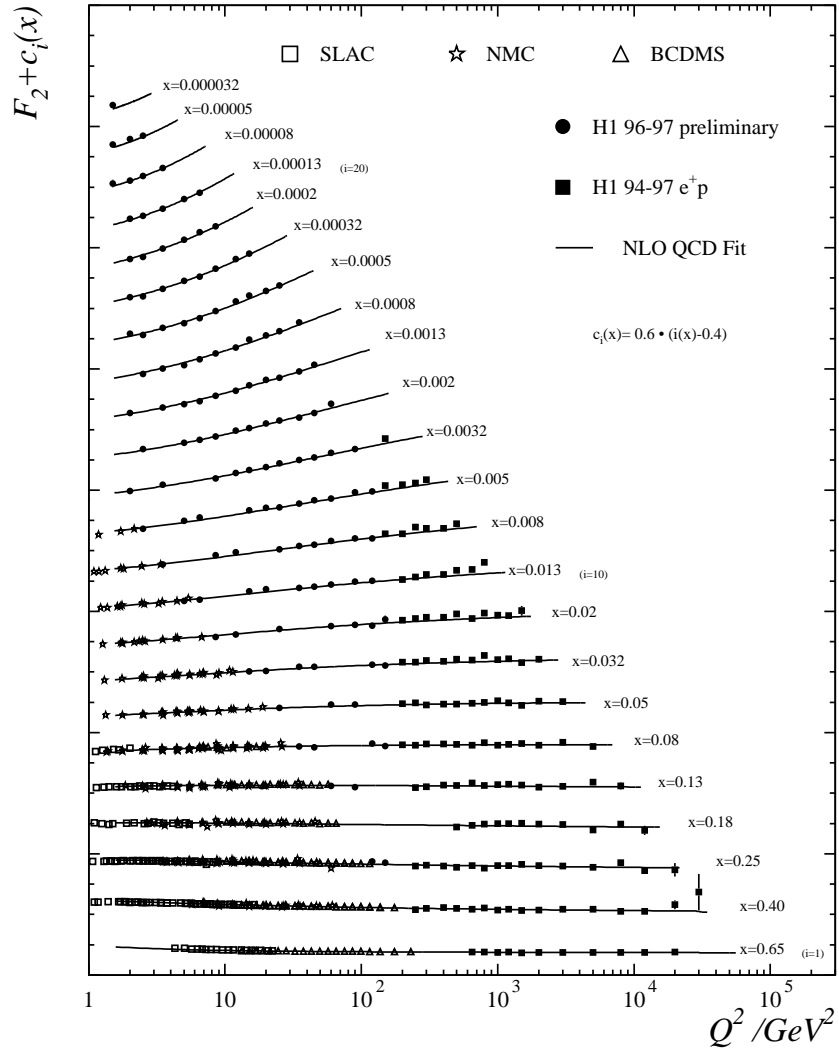}
	\caption{One of the greatest achievements of perturbative QCD:
	world data on the $Q^2$ dependence of the unpolarized 
	structure function $F_2$, compared to theoretical expectations (solid
	curves)~\cite{Makins:2000}. }
	\label{f:evolution}
        \end{figure}

In practice, we have to make a distinction between
two major branches of QCD: perturbative 
and nonperturbative, or short-distance and long-distance.
Perturbative QCD is relatively well understood. It 
is essentially similar to QED, it is based on Feynman-diagram 
approach, although it often requires larger sets of diagrams to attain the
desired accuracy. The theory contains pointlike and almost massless fermions
(the so-called {\em current} quarks) and massless bosons to carry their
interactions (the gluons). Probably one of the most important achievements of
perturbative QCD is the study of the way deep inelastic scattering data
change with $Q^2$. The striking agreement between
theory and experiments is shown in Fig.\ \vref{f:evolution}.
On the other hand, nonperturbative QCD is poorly understood and it is a
challenging playground for fundamental physics. At the moment, our
understanding of this theory relies on lattice calculations, effective chiral
field theories, and phenomenological models. 

We know that nonperturbative QCD should display confinement as a fundamental
property, at least under normal conditions. Lattice calculations
already provide strong evidence
that the quark-quark interaction potential increases linearly, 
and is therefore a confining potential~\cite{Bali:2000gf}. It is essential to
understand from first principles why this occurs, and it is
desirable 
to
explore how it is possible to achieve a deconfined phase of QCD, maybe
under extreme conditions (e.g. neutron stars).
We also 
know that in QCD chiral symmetry is approximately valid. On the other hand, 
the existence of pions, which are nearly massless, 
suggests that chiral symmetry should be spontaneously broken, with pions being 
Goldstone bosons. Nonperturbative QCD should be able to explain this feature.
Nonperturbative QCD  should also explain the transition
between massless current quarks and constituent quarks. 

Finally and more generally, 
nonperturbative QCD should lead to a reliable quantitative 
description of the
structure of hadrons and of hadronic phenomena.
The question at the heart of hadronic physics is: what is the structure of
hadrons in terms of their quark and gluon constituents? Therefore, we might 
define 
hadronic physics as the branch of physics that deals with understanding QCD,
and in particular nonperturbative QCD.

%%%%%%%%%%%%%%%%%%%%%%%%%%%%%%%%%%%%%%%%%%%%%%%%%%%%%%%%%%%%
\section{Spin physics and the transversity distribution}

\begin{figure}
	\centering
	\includegraphics[height=14cm, angle=-90]{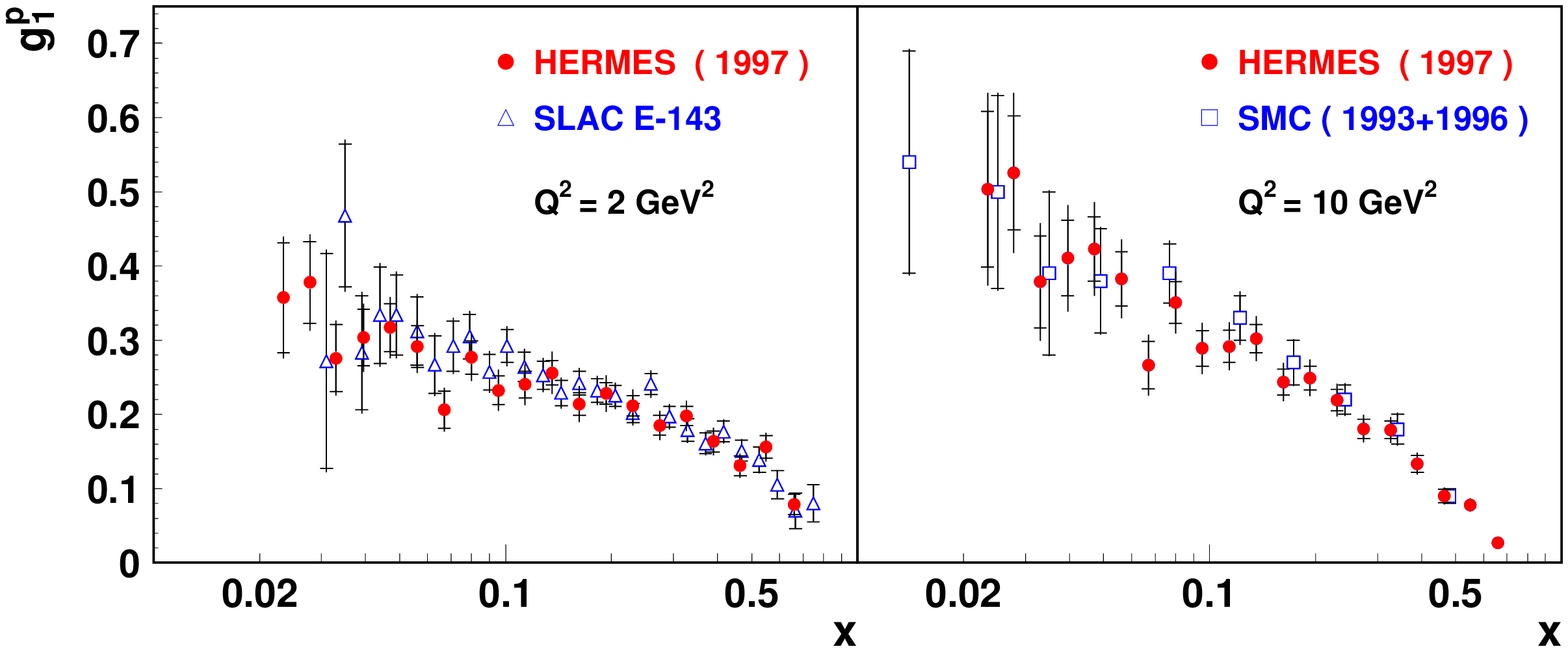}
	\caption{The helicity distribution of the proton, $g_1$ as a function
        of the fractional momentum $x$ carried by
	quarks~\cite{Airapetian:1998wi}.} 
	\label{f:g1}
        \end{figure}

One of the key questions in understanding the structure of hadrons is: where
does the spin of the nucleons come from?
In the constituent quark model, the spin of the quarks adds up to
yield the total spin of the proton. Deep inelastic scattering experiments,
 however, show the importance of other contributions, such as the spin of the
gluons and the orbital angular momenta of quarks and gluons.
A measure of the quark spin contribution is given by the distribution function
$g_1$, often denoted as $\Delta q$ and usually called the {\em helicity 
distribution}. 
In a frame of reference where the hadron is moving with a very large speed 
({\em
infinite momentum frame})
and if the direction of its spin is {\em longitudinal} to its motion, 
the helicity distribution 
describes the number of quarks with their spin aligned with that of the hadron 
minus the number of quarks with opposite spin, it is therefore a measure of
the {\em longitudinal} spin of the quarks in the hadron.
The quark helicity distribution has been measured with a good precision,
as shown in Fig.\ \vref{f:g1}.
Naively, if the spin of the hadron is entirely due to the
quark spin as in the constituent quark model, 
we expect to have a net balance of one 
quark spinning in the direction of the proton and thus accounting for the
whole proton spin.
In reality, it turns out that (the integral of) the helicity distribution
accounts for only about 30\% of the proton spin!
We expect thus that the missing spin is provided by the gluon spin and by the
orbital angular momentum of quarks and gluons. 
These two quantities have not
been measured yet. Even worse, we don't know if it is possible to measure 
the orbital angular momentum
directly~\cite{Ji:1997ek,Harindranath:1998ve,Bashinsky:1998if}. 
Tab.~\vref{tab:exps} shows a list of all polarized 
deep inelastic
scattering experiments, together with their typical energies and 
the characteristics of their beams and
targets. The kinematic coverage of each experiment
is indicated in the table by its average $Q^2$ (GeV$^2$) and
$\xbj$ range. Columns $P_B$ and $P_T$ give
the average or typical beam and
target polarizations as quoted by each experimental group. 
The column 
labeled $\mathcal{L}$ is an estimate of the total nucleon luminosity
(\# of nucleons/cm$^2$ times \# of beam particles/s) in units of
$10^{32}$ nucleons/cm$^2$/s for each experiment. 
\begin{table}
\caption{Summary of polarized deep inelastic measurements~\cite{Filippone:2001ux}.
}
\vspace{0.01cm}
\begin{center}
\setlength{\tabcolsep}{4pt}
\begin{tabular}{|>{\small}c|>{\small}c|>{\small}c|>{\small}c|>{\small}c|>{\small}c|>{\small}c|>{\small}c|>{\small}c|>{\small}c|}
\hline
 Lab & Exp. & Year & Beam & $\langle Q^2\rangle$   & $\xbj$ & $P_B$ & Target &$P_T$ &$\mathcal{L}$\\ 
%     &      &      &      & GeV$^2$ &     &       &        &      &  cm$^{-2}$-s \\
\hline
SLAC & E80   &   75 & 10-16 GeV $e^-$    & 2   &  0.1 -- 0.5  & 85\% & H-butanol & 50\% & 400 \\
     & E130  &   80 & 16-23 GeV $e^-$    & 5   &  0.1 -- 0.6  & 81\% & H-butanol & 58\% & 400 \\
     & E142  &   92 & 19-26 GeV $e^-$    & 2   & 0.03 -- 0.6  & 39\% & $^3$He    & 35\% & 2000 \\
     & E143  &   93 & 10-29 GeV $e^-$    & 3   & 0.03 -- 0.8  & 85\% & NH$_3$    & 70\% & 1000 \\ 
     &       &      &                    &     &              &      & ND$_3$    & 25\% & 1000 \\
     & E154  &   95 & 48 GeV $e^-$       & 5   & 0.01 -- 0.7  & 82\% & $^3$He    & 38\% & 3000 \\
     & E155  &   97 & 48 GeV $e^-$       & 5   & 0.01 -- 0.9  & 81\% & NH$_3$    & 90\% & 1000 \\
     &       &      &                    &     &              &      & LiD       & 22\% & 1000 \\
     & E155' &    99& 30 GeV $e^-$       & 3   & 0.02 -- 0.9  & 83\% & NH$_3$    & 75\% & 1000 \\
     &       &      &                    &     &              &      & LiD       & 22\% & 1000 \\
\hline
CERN & EMC  &    85 &100-200 GeV $\mu^+$ & 11  & 0.01 -- 0.7  & 79\% & NH$_3$    & 78\% & 0.3 \\
     & SMC  &    92 &100 GeV $\mu^+$     & 4.6 & 0.006 -- 0.6 & 82\% & D-butanol & 35\% & 0.3 \\
     &      &    93 &190 GeV $\mu^+$     & 10  & 0.003 -- 0.7 & 80\% & H-butanol & 86\% & 0.6 \\
     &      & 94-95 &                    &     &              & 81\% & D-butanol & 50\% & 0.6 \\
     &      &    96 &                    &     &              & 77\% & NH$_3$    & 89\% & 0.6 \\
\hline
DESY &HERMES&    95 & 28 GeV $e^+$       & 2.5 & 0.02 -- 0.6  & 55\% & $^3$He    & 46\% & 1 \\
     &      & 96-97 &                    &     &              & 55\% & H         & 88\% & 0.1 \\
     &      & 98    & 28 GeV $e^-$       &     &              & 55\% & D         & 85\% & 0.2 \\
     &      & 99-00 & 28 GeV $e^+$       &     &              & 55\% & D         & 85\% & 0.2 \\
     &      & 01-? & 28 GeV $e^-$       &     &              & 55\% & H
& 85\%$^*$ & 0.2 \\\hline
CERN &COMPASS&   01-? &190 GeV $\mu^+$     & 10  & 0.005 -- 0.6 & 80\% & NH$_3$    & 90\% & 3 \\
     &       &      &                    &     &              &      &  LiD      & 40\% & 3 \\
\hline
BNL  & RHIC &    02-? & 200 GeV $p-p$      & $\sim 100$  & 0.05 -- 0.6 & 70\% & Collider  & 70\% & 2 \\
\hline
DESY & ZEUS/H1 & 02-? & $28\times 800$ GeV $e-p$   & 22  &0.00006 -- 0.6& 50\%
& Collider  &   & 0.2 \\
\hline
\end{tabular}
\end{center}
$^*$ Transversely polarized target
\setlength{\tabcolsep}{5pt}
\label{tab:exps}
\end{table}

So far we talked about the longitudinal spin of the quarks inside the proton,
but what about the transverse spin?
The observable we have to take into consideration is the {\em transversity
distribution}. In the infinite momentum frame with the proton spin
{\em transverse} to
the direction of motion, 
the transversity distribution 
describes the number of quarks with their spin aligned with that of the hadron 
minus the number of quarks with opposite spin, it is therefore a measure of
the {\em transverse} spin of the quarks in the hadron.
The transversity distribution looks very similar to the helicity distribution, 
as at first sight they seem related by rotational symmetry. 
However, we cannot forget the fact
that the interpretation of helicity and transversity holds true only in the
infinite momentum frame, where the direction of motion of the hadron breaks
rotational symmetry. 

In the rest frame of the proton, there is a probability to find quark spins 
aligned with the proton's spin. This probability is obviously the same 
no
matter what the spin orientation is. 
If we now boost the proton to a large speed in the direction of its 
spin, the alignment probability
will correspond to the helicity distribution. If we boost the
proton in a direction transverse to its spin, the alignment probability
will correspond to the transversity 
distribution. In a nonrelativistic situation, Galilean boosts will not affect
the spin distribution and we would still expect helicity and transversity to
be equal to each other and to the spin distribution in the rest frame. But in
a relativistic context, Lorentz boosts can affect the spin distribution and
can cause helicity and transversity to be different from each other and from
the rest-frame distribution. The way this difference arises depends on the
inner structure of the nucleon.
%An artistic view of the effect of Lorentz boosts
%is illustrated in 
%(see Fig.~\vref{f:boost}).
%	\begin{figure}[b]
%	\centering
%	\includegraphics[width=8cm]{Figures/boost}
%	\caption{An artistic impression of the different effects of two
%	orthogonal boosts.}
%	\label{f:boost}
%        \end{figure}

The transverse spin of quarks is thus another missing piece in 
the proton spin puzzle. It can give new information on the dynamics of quarks
inside hadrons, complementary to the helicity distribution. 
In spite of this, the transversity distribution
escaped notice
until 1979, when it was introduced by J.~Ralston and D.~Soper~\cite{Ralston:1979ys}.
In the last decade, it has been evaluated in 
models~\cite{Jaffe:1992ra,Jakob:1997wg,Pobylitsa:1996rs,Barone:1997un}
and lattice computations~\cite{Aoki:1997pi}. At this point, an experimental 
measurement will be needed to
put all these calculations on test, but 
unfortunately the transversity distribution
is an elusive object to measure. 
%In fact, due to the difficulty of accessing it,
%experimental physicists did not take the transversity distribution into
%consideration 
%until a few years ago.
Today, 
looking for a practical way to observe transversity is still an open
problem.
Experimental collaborations are planning its
measurement at last~\cite{hermes,compass,Bunce:2000uv}, and some of them will
resort to the methods discussed in this thesis.

%%%%%%%%%%%%%%%%%%%%%%%%%%%%%%%%%%%%%%%%%%%%%%%%%%%%%%%%%%%%
\section{Outline of the thesis}

The goal of the thesis is to
discuss three different ways to observe the quark transversity
distribution in deep inelastic scattering.

To start with, 
in chapter~\ref{c:trans} I will review the formalism of deep inelastic
scattering. I will introduce the {\em parton distribution functions}
and I will devote a particular attention to the transversity distribution
function.  I will discuss totally inclusive DIS, where only the scattered
electron is detected, and I will show that it is not possible to measure the
transversity distribution in this kind of process.

In chapter~\ref{c:collins}, I will turn the attention to one-particle inclusive
DIS, where one of the
outgoing hadrons is detected in coincidence with the scattered electron. I will introduce {\em fragmentation
functions} to describe the production of hadronic fragments. 
In particular, I will demonstrate 
that the presence of the
transverse momentum of the outgoing hadron allows the introduction of the
Collins fragmentation function. In the cross section of one-particle inclusive
DIS,  I will
show the occurrence of the product of the transversity distribution and the
Collins function. Therefore, this suggests a first way to observe the
transversity distribution of the quarks.

In chapter~\ref{c:twohadron}, I will examine the more complex case of
two-particle inclusive DIS, where two of the outgoing hadrons are detected 
together with the scattered electron. I will discuss how the presence of the
relative transverse momentum between the two hadrons permits the definition of 
a new function, $H_1^{\newangle}$, 
to be connected to the transversity distribution. In the same
chapter, I will study what happens when we assume that
the two hadrons are produced only in $s$
and $p$ waves. In addition to the usual $s$ wave contributions, 
I will distinguish the pure $p$-wave contributions and the
$sp$ interference contributions. This will 
lead to the introduction of two new 
fragmentation functions that can be observed in connection
 with the transversity. They are two distinct components of the function 
$H_1^{\newangle}$
and they generate the second and third way I will 
consider to access the 
transversity distribution. 

In chapter~\ref{c:spinone}, I will analyze the formalism needed to deal with
spin-one hadrons in deep inelastic scattering. In the first part of the
chapter I will focus on spin-one targets, while in the second part I will
study the production of spin-one hadrons in the final states. This process has 
something in common with one-particle inclusive DIS (because the production of 
a single hadron is addressed), but also with two-particle inclusive DIS
(because the spin-one hadron has to decay into two hadrons to yield
information on its polarization). In particular, I will clarify 
 the connection between spin-one fragmentation functions and the pure $p$-wave
sector of the analysis of two-particle production.

The various options described to measure the transversity distribution
all involve the class of {\em T-odd} fragmentation functions.\footnote{In fact, 
in the thesis I will not deal with the well known case of spin-half
production, which involves a T-even fragmentation function.}
To attempt some quantitative assessments on the magnitude of T-odd
fragmentation functions, in chapter~\ref{c:model} I will present a model 
calculation of the Collins
function and of some of the measurable quantities in which it appears.

\renewcommand{\quot}{%
\parbox{7.5cm}{\begin{flushright}You have never given me a transverse look.\end{flushright}}\\ A. Chekhov  
}

%%%%%%%%%%%%%%%%%%%%%%%%%%%%%%%%%%%%%%%%%%%%%%%%%%%%%%%%%%%%%%%%%%%%%%%%%%%%
\chapter[Distribution functions and transversity]{Distribution functions \\ and transversity}
\label{c:trans}

In this chapter, we will introduce 
the concept of parton distribution functions. 
In order to do this, first of all we will review 
the general formalism of polarized deep inelastic scattering, 
starting from
the simplest case of inclusive processes. This subject is covered in detail
in books (e.g.\ Refs.~\citen{Roberts:1990ww,Peskin:1995ev,Elliot}),
PhD theses (e.g.\ Refs.~\citen{Levelt} and~\citen{Tangerman}) and
reports (e.g.\ Refs.~\citen{Anselmino:1995gn} and~\citen{Barone:2001sp}). 
Nevertheless, it is useful to examine the formalism from the point of view
we will adopt throughout the remaining chapters. 
In the analysis of distribution functions, we will include beam and target 
polarization and partonic transverse momentum. We will limit the analysis to
leading order in $1/Q$ and we will only briefly mention $\alpha_s$ corrections.

A particular attention in this chapter and in the rest of the 
thesis will be reserved to
the quark transversity distribution. 
The quark transversity distribution $h_1$~\cite{Jaffe:1991kp} -- also called
transverse spin distribution~\cite{Cortes:1992ja} --  was
first introduced by Ralston and Soper
\cite{Ralston:1979ys} and it is an essential component in the description of 
the nucleon spin. It is a chiral-odd object describing the difference of 
probabilities to find in a
transversely polarized hadron a quark with spin aligned or antialigned to the
spin of the hadron. 
The transversity distribution has been 
upstaged for many years by the helicity distribution, $g_1$, which is
easier to measure.  However, some experimental
collaborations are planning to  measure it
in the next years~\cite{hermes,compass,Bunce:2000uv}, 
possibly using one of the techniques we will
outline in the thesis.

%%%%%%%%%%%%%%%%%%%%%%%%%%%%%%%%%%%%%%%%%%%%%%%%%%%%%%%%%%%%
\section{Inclusive deep inelastic scattering}

\begin{figure}
	\centering
	\rput(2.51,5.51){$'$}
	\rput(7.0,1.4){\scriptsize ${\cal X}$}
	\includegraphics[width=7cm]{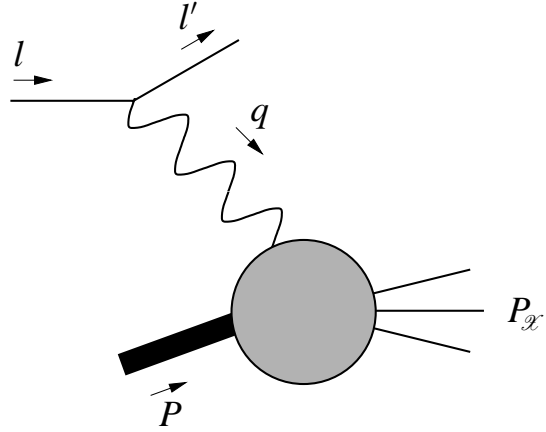}
	\caption{Inclusive deep inelastic scattering.}
	\label{f:0p}
        \end{figure}

In deep inelastic scattering, an electron scatters off a nucleon via a large
momentum transfer,
the nucleon is destroyed and many hadrons are formed as a consequence of the
collision.  
In {\em inclusive} events, only the scattered electron is detected, while
the hadronic final states are unobserved. A schematic view of the process is
provided by Fig.~\vref{f:0p}.

%%%%%%%%%%%%%%%%%%%%%%%%%%%%%%%%%%%%%%%%%%%%%%%%%%%%%%%%%%%%
\subsection{Kinematics}
\label{s:kin1}

In electron-nucleon scattering, an electron with momentum $l$
scatters off a nucleon with momentum $P$, mass $M$ and spin $S$, 
via the exchange of a
virtual photon with momentum $q$. The electron final momentum is $l'$.

We define the invariants
\begin{align} 
s &= (P+l)^2,	& W^2 &= (P+q)^2, &
Q^2 &= -q^2 = -(l-l')^2, 
\end{align} 
and we introduce the variables
\begin{align}
\xbj&=\frac{Q^2}{2 P \cdot q}, & y= \frac{P \cdot q}{P \cdot l}.
\label{e:xandy}
\end{align} 
In deep inelastic scattering it is required that $Q^2, P\cdot q \gg
M^2$. Usually, the {\em Bjorken limit} is assumed 
($Q^2, P\cdot q \rightarrow \infty$, $\xbj$ fixed).
In particular, $Q^2$ represents the
{\em hard scale} of the process. In this thesis, only the leading terms in an
expansion in $1/Q$ will be retained. In agreement with the working
redefinition of {\it twist} proposed by Jaffe,  we will very often identify the
expression ``leading twist'' with the expression ``leading order in $1/Q$''~\cite{Jaffe:1996zw}. 

When working in the Bjorken limit, the vectors $P$ and $q$ can be
conveniently parametrized (in light-cone coordinates) as
\begin{align} 
P^{\mu}& = \lf[\frac{\xbj M^2}{A Q \sqrt{2}},\; \frac{A Q}{\xbj \!\sqrt{2}} ,\;
\mb{0}\rg]  = \lf[\frac{M^2}{2 P^+},\; P^+ ,\; \mb{0}\rg], \\ 
q^{\mu}& = \lf[\frac{Q}{A \sqrt{2}},\; -\frac{A\, Q}{\sqrt{2}} ,\; \mb{0}\rg]
=\lf[\frac{Q^2}{2 \xbj P^+},\; - \xbj P^+ ,\; \mb{0}\rg] .
\end{align}
This parametrization holds in any frame of reference
 where the virtual photon direction 
is antiparallel to the $z$ axis. Any frame fulfilling this
requirement will be simply called
{\em collinear}. The parameter $A$ specifies uniquely a
specific collinear 
frame of reference. For instance, for $A=M \xbj /Q$ we select the nucleon rest
frame, where $P$ is purely timelike, while for $A=1$ we select the
so-called infinite momentum frame, where $q$ is purely spacelike. 

In a $1/Q$ expansion, it turns out that the plus component of $P$ plays a
dominant role. This statement holds regardless of the value of $A$, i.e.\
in any collinear frame. In the nucleon rest frame $P^+$ is of the order of 1, while
in the infinite momentum frame it is of the order of $Q$. However, if we take for instance 
the scalar combination $P \cdot q$, we see that the component $P^+ q^-$ is of
the order of $Q^2$, whereas $P^- q^+$ is of the order of 1, independently of the frame.
Therefore, we can say that the plus
component of the nucleon's momentum is the {\em relevant} or {\em dominant}
one, although only in the infinite momentum frame it is truly dominant. 

We are going to define a process as
{\em soft} if the relevant component of all momenta remains the same. 
In contrast, in a {\em hard} process, such as the interaction with the 
hard momentum $q$, the relevant component has to change. 
Similarly, when we describe a momentum as soft with respect 
to another, we mean that their relevant component is the same.

In the Bjorken limit, the electron and the proton 
can be considered to be massless,
 $2 P \cdot l \approx s$ and $Q^2 = s \,\xbj\, y$.
The lepton momenta can be parametrized as 
\begin{subequations}
\begin{align} 
l^{\mu}& = \lf[\frac{Q}{A y \sqrt{2}},\; \frac{A\,(1-y) Q}{y \sqrt{2}} ,\;
\frac{\sqrt{1-y}}{y},\; \mb{0}\rg], \\
l^{\prime \mu}& = \lf[\frac{(1-y) Q}{A y \sqrt{2}},\; \frac{A Q}{y
\sqrt{2}},\; \frac{\sqrt{1-y}}{y},\; \mb{0}\rg]. 
\end{align}
\end{subequations}
This parametrization implies that we chose the $y$ axis of our system as
pointing in the direction of the vector product $(- \mb{q} \times \mb{l}')$.
Normally, transverse vectors and azimuthal angles will be defined as lying on a
plane perpendicular to the direction of the virtual photon (see
Fig.~\vref{f:planes1}). 

%%%%%%%%%%%%%%%%%%%%%%%%%%%%%%%%%%%%%%%%%%%%%%%%%%%%%%%%%%%%
\subsection{The hadronic tensor}
\label{s:hadro1}

\begin{figure}
	\centering
	\includegraphics[width=13cm]{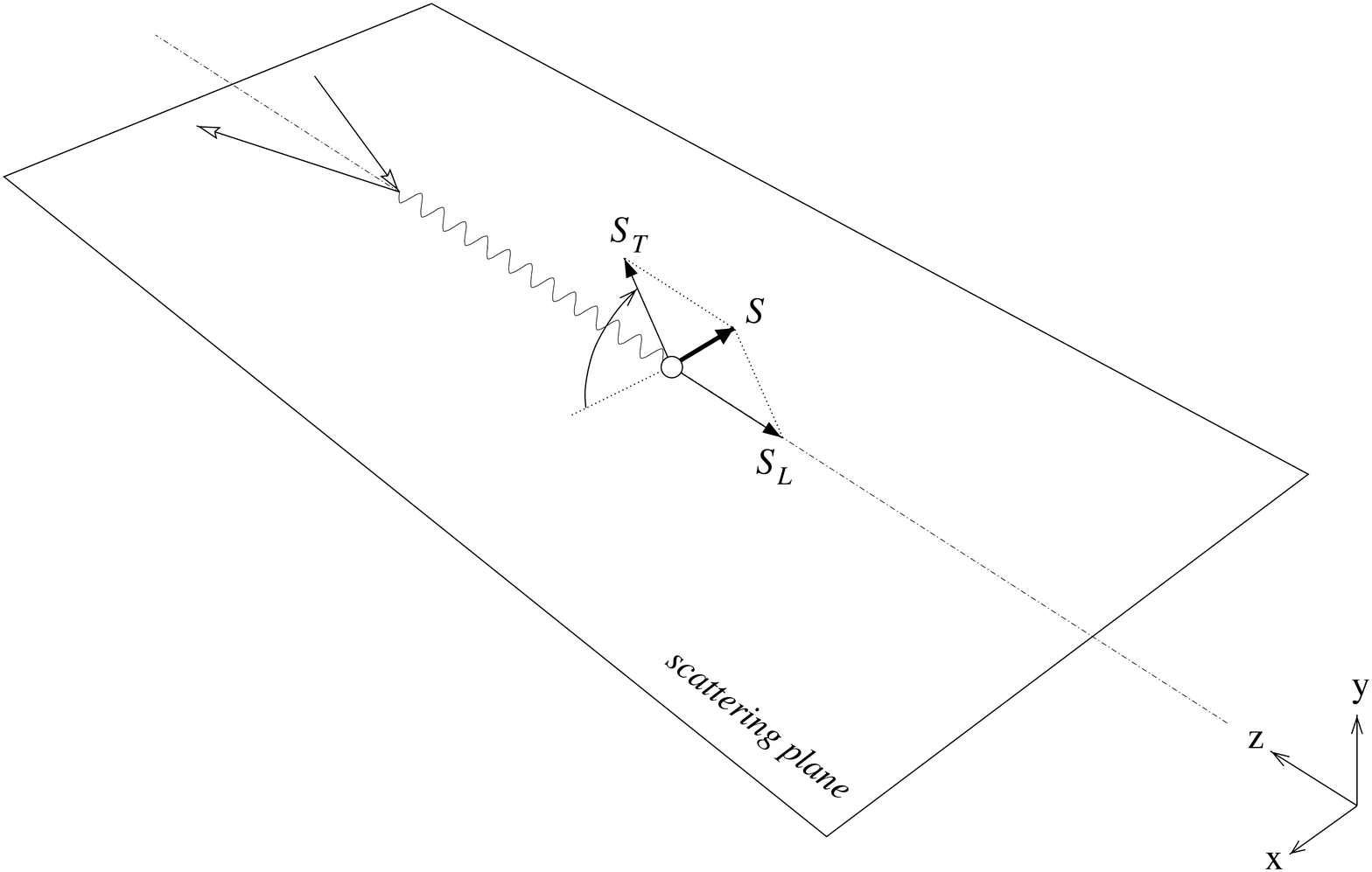}
	\rput(-9.7,7.5){$l$}
	\rput(-11.2,6.6){$l'$}
	\rput(-7.9,4.9){$\phi_S$}
	\caption{Description of the vectors involved in totally 
	inclusive deep inelastic scattering and of the azimuthal angle $\phi_S$.}
	\label{f:planes1}
        \end{figure}

The cross section for polarized electron-nucleon  scattering 
can be written in a
general way as the contraction between a leptonic and a hadronic tensor
\begin{equation} 
\frac{\de^3\!\sigma}{\de \xbj \de y \de \phi_S} = \frac{\alpha^2}{2\, s\,\xbj\,Q^2} \, 
	L_{\mu \nu}(l, l', \lambda_e)\; 2M W^{\mu \nu}(q, P, S),
\label{e:crossinc}
\end{equation} 
where the vector $S$ denotes the spin of the nucleon and
$\phi_S$ its azimuthal angle, $\lambda_e$ denotes the
helicity of the electrons and   $\alpha =e^2/4\pi$.
 Fig.\ \vref{f:planes1} illustrates the definition of the scattering plane,
the $z$ axis of our collinear frame and the azimuthal angle
$\phi_S$.

Considering the lepton to be longitudinally polarized, in the massless limit 
the leptonic tensor is given by~\cite{Mulders:1996dh}
\begin{equation} \begin{split} 
L_{\mu \nu} &= \sum_{\lambda_e'}
	\Bigl( \bbar{u}(l',\lambda_e') \,\g_{\mu}\, u(l,\lambda_e)\Bigr)^*\;
	\Bigl( \bbar{u}(l',\lambda_e') \,\g_{\nu}\, u(l,\lambda_e)\Bigr) \\
&=- Q^2 g_{\mu \nu}+2 \lf(l_{\mu} l'_{\nu} + l'_{\mu} l_{\nu} \rg) + 2 \ii\, \lambda_e\,
\eps_{\mu \nu \rho \sigma}\, l^{\rho} l^{\prime \sigma}.
\end{split} \end{equation}  
The leptonic tensor contains all the information on the leptonic probe, which
can be described by means of perturbative QED, while
the information on the hadronic target is contained in 
the hadronic tensor
\begin{align}  
2 M W^{\mu \nu}(q,P,S) &= \frac{1}{2 \pi} \sum_{\cal X} 
\int \frac{\de^3 \! \mb{P}_{{\cal X}}} {(2 \pi)^3\, 2
P_{{\cal X}}^0}\; (2 \pi)^4 \, \delta^{(4)}\Bigl(q+P-P_{{\cal X}}\Bigr)\, H^{\mu \nu}(P,S,P_{\cal X}), \\
H^{\mu \nu}(P,S,P_{\cal X})& =
\bra{P,S} J^{\mu} (0) \ket{\cal X} \bra{\cal X} J^{\nu} (0) \ket{P,S}.
\end{align} 
The state ${\cal X}$ symbolizes any final state, with total
 momentum $P_{\cal X}$. It is integrated over since in inclusive processes the 
final state goes undetected.
By  Fourier transforming the delta function and translating one of the current
operators, we can rewrite the hadronic tensor as
\begin{equation}
 2 M W^{\mu \nu}(q,P,S) =\frac{1}{2 \pi} \int \de^4\! \xi\; \e^{\ii q \cdot
\xi}\; 
	\bra{P,S} J^{\mu} (\xi)\,  J^{\nu} (0) \ket{P,S}.
\end{equation} 

In general, the structure of the hadronic tensor cannot be specified further, 
because this would require an understanding of its inner dynamics. At most, it 
can be parametrized in terms of {\em structure functions}.
However, the phenomenology of DIS taught us that at sufficiently high $Q^2$
we can  
assume that the scattering of the electron takes place off a quark of mass
$m$ inside
the nucleon. The final state ${\cal X}$ can be split in a quark with momentum
$k$ plus 
a state $X$ with momentum $P_X$. Considering the electron-quark interaction at tree level only, 
the hadronic tensor can be written as
\begin{equation} \begin{split} 
2 M W^{\mu \nu}(q,P,S) &= \frac{1}{2 \pi} \sum_q e_q^2 \sum_{X} 
\int \frac{\de^3 \! \mb{P}_{X}} {(2 \pi)^3\, 2 P_{X}^0} 
\int \frac{\de^3 \! \mb{k} }{(2 \pi)^3\, 2 k^0}\; (2 \pi)^4\,
\delta^{(4)}\Bigl(P+q-k-P_{X}\Bigr) \\
&\quad \times
	\Bigl(\bra{P,S} \bbar{\psi}_i (0) \ket{X} 
	\bra{X} \psi_j (0) \ket{P,S} \,
\g^{\mspace{2mu}\mu}_{ik} \lf(\kslash + m\rg)_{kl} \g^{\nu}_{lj}  \\
&\quad+	\bra{P,S} \psi_j (0) \ket{X} 
	\bra{X} \bbar{\psi}_i (0) \ket{P,S} 
\, \g^{\nu}_{ik} \lf(\kslash - m\rg)_{kl} \g^{\mspace{2mu}\mu}_{lj} \Bigr),
\end{split} \end{equation}  
where $k$ is the momentum of the struck quark, the index $q$ denotes the quark 
flavor and $e_q$ is the fractional
charge of the quark. Note that, for simplicity, 
 we omitted the flavor 
indices on the quark fields.
The integration over the phase space of the final-state quark can be
replaced by a four-dimensional integral with an on-shell condition,
%\begin{equation}
%\int \frac{\de^3 \! \mb{k} }{2 k^0} \longrightarrow \int \de^4 \! k \; 
%\delta\lf(k^2 -m^2\rg)\, \theta\lf(k^0 -m\rg),
%\end{equation} 
so that the hadronic tensor can be rewritten as
\begin{equation} \begin{split} 
2 M W^{\mu \nu}(q,P,S) &= \sum_q e_q^2 \sum_{X} 
\int \frac{\de^3 \! \mb{P}_{X}} {(2 \pi)^3\, 2 P_{X}^0} 
\int \de^4 \! k \; \delta\lf(k^2 -m^2\rg)\, \theta\lf(k^0 -m\rg)
  \\
&\quad \times\delta^{(4)}\Bigl(P+q-k-P_{X}\Bigr)\,
	\Bigl(\bra{P,S} \bbar{\psi}_i (0) \ket{X} 
	\bra{X} \psi_j (0) \ket{P,S} \,
\g^{\mspace{2mu}\mu}_{ik} \lf(\kslash + m\rg)_{kl} \g^{\nu}_{lj}  \\
&\quad+	\bra{P,S} \psi_j (0) \ket{X} 
	\bra{X} \bbar{\psi}_i (0) \ket{P,S} 
\, \g^{\nu}_{ik} \lf(\kslash - m\rg)_{kl} \g^{\mspace{2mu}\mu}_{lj} \Bigr).
\end{split} \end{equation} 
Next, we Fourier transform the Dirac delta function 
%according to
%\begin{equation}
% \delta^{(4)}\Bigl(P+q-k-P_{X}\Bigr) \longrightarrow
%	 \int \frac{\de^4 \! \xi} {(2 \pi)^4} \; \e^{\ii
%\,(P+q-k-P_{X}) \cdot \xi} 
%\end{equation} 
and we introduce the momentum $p = k-q$ to obtain
\begin{equation} \begin{split} 
2 M W^{\mu \nu}(q,P,S) &= \sum_q e_q^2 \sum_{X} 
\int \frac{\de^3 \! \mb{P}_{X}} {(2 \pi)^3\, 2 P_{X}^0} 
\int \de^4 \! p \; 
%\\ & \quad \times
\delta\lf(\bigl(p+q\bigr)^2 -m^2\rg)\, 
\\ & \quad \times
\theta\lf(p^0+q^0 -m\rg)
%\\ &\quad \times 
\int \frac{\de^4 \! \xi} {(2 \pi)^4} \,
%\nn \\
%&\quad \times  
 \e^{\ii\, (P-p-P_{X}) \cdot \xi }\;
\\ & \quad \times
	\Bigl(\bra{P,S} \bbar{\psi}_i (0) \ket{X} 
	\bra{X} \psi_j (0) \ket{P,S} \,
\g^{\mspace{2mu}\mu}_{ik} 
\lf(\pslash + \qslash + m\rg)_{kl} 
\g^{\nu}_{lj}  
\\ &\quad
+	\bra{P,S} \psi_j (0) \ket{X} 
	\bra{X} \bbar{\psi}_i (0) \ket{P,S} 
\, \g^{\nu}_{ik} \lf(\pslash + \qslash  - m\rg)_{kl} \g^{\mspace{2mu}\mu}_{lj}
\Bigr). 
%\nn
\end{split} \end{equation}  
Finally, we use part of the exponential to perform a translation of the
field operators and we use completeness to eliminate the unobserved $X$
states, so that
\begin{equation} \begin{split} 
2 M W^{\mu \nu}(q,P,S) 
&
= \sum_q e_q^2  
\int \de^4 \! p \;\delta\lf(\bigl(p+q\bigr)^2 -m^2\rg)\, 
\theta\lf(p^0+q^0 -m\rg)
\int \frac{\de^4 \! \xi} {(2 \pi)^4} \, \e^{-\ii p \cdot \xi }\\
&\quad 
\times
	\Bigl(\bra{P,S} \bbar{\psi}_i (\xi) 
	\; \psi_j (0) \ket{P,S} \,
\g^{\mspace{2mu}\mu}_{ik} \lf(\pslash + \qslash + m\rg)_{kl} \g^{\nu}_{lj}  \\
&\quad
+	\bra{P,S} \psi_j (\xi)\;\bbar{\psi}_i (0) \ket{P,S} 
\, \g^{\nu}_{ik} \lf(\pslash + \qslash  - m\rg)_{kl} \g^{\mspace{2mu}\mu}_{lj} \Bigr).
\end{split} \end{equation}

\begin{figure}
	\centering
	\includegraphics[width=8cm]{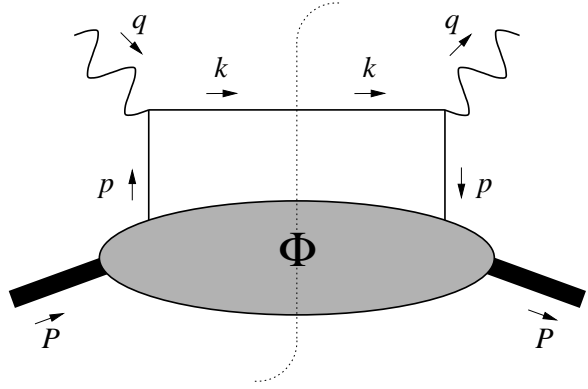}
	\caption{The handbag diagram, describing the hadronic tensor at tree
	level.} 
	\label{f:handbag}
        \end{figure}

The hadronic tensor can be written in a more compact way by introducing the
quark-quark correlation function $\Phi$ and the antiquark-antiquark
correlation function $\bbar{\Phi}$
\begin{equation} \begin{split} 
2 M W^{\mu \nu}(q,P,S)& = \sum_q e_q^2 \int \de^4 \! p \;\delta\lf(\bigl(p+q\bigr)^2 -m^2\rg)\, 
\theta\lf(p^0+q^0 -m\rg) \\
& \quad \times \tr \lf[ \Phi(p,P,S)
\g^{\mspace{2mu}\mu} \lf(\pslash + \qslash + m\rg) \g^{\nu} + \bbar{\Phi}(p,P,S)
\g^{\nu} \lf(\pslash + \qslash - m\rg) \g^{\mspace{2mu}\mu} \rg]
\label{e:hadrocorre}
\end{split} \end{equation}  
where
\begin{subequations} \begin{align} 
\begin{split}
\Phi_{ji}(p, P,S)&=\frac{1}{(2\pi)^4} \int \de^4 \! \xi\;\e^{-\ii p \cdot \xi }
\bra{P,S} \bbar{\psi}_i (\xi) 
	\; \psi_j (0) \ket{P,S} \\
&= \sum_{X} 
\int \frac{\de^3 \! \mb{P}_{X}} {(2 \pi)^3\, 2 P_{X}^0}\;\bra{P,S}\bbar{\psi}_i (0) \ket{X} 
	\bra{X} \psi_j (0) \ket{P,S} \;\delta^{(4)}\Bigl(P-p-P_{X}\Bigr),
\label{e:phi1}
\end{split}
\\
\begin{split}
\bbar{\Phi}_{ji}(p, P,S)
 &= \frac{1}{(2\pi)^4} \int \de^4 \! \xi\;\e^{-\ii p \cdot \xi }
\bra{P,S} \psi_j (\xi) 
	\; \bbar{\psi}_i (0) \ket{P,S} \\
&=\sum_{X} 
\int \frac{\de^3 \! \mb{P}_{X}} {(2 \pi)^3\, 2 P_{X}^0}\;\bra{P,S} \psi_j (0) \ket{X} 
	\bra{X} \bbar{\psi}_i (0) \ket{P,S} \;\delta^{(4)}\Bigl(P-p-P_{X}\Bigr). 
\end{split}
\end{align} \end{subequations} 
As the quark fields should carry a flavor index that we omitted, also the
correlation functions are flavor dependent and they should be indicated more
appropriately as $\Phi^q$ and $\bbar{\Phi}^q$.
For simplicity, $\bbar{\Phi}$ will be omitted henceforth. It can be
accounted for simply by extending the summation over quarks to a summation over
quarks and antiquarks. A
graphical representation of the hadronic tensor at tree level in the parton
model is given by the so-called {\em handbag diagram}, depicted in
Fig.~\vref{f:handbag}.

We parametrize the quark momentum $p$ in the following way
\begin{equation}
p^{\mspace{2mu}\mu} = \lf[\frac{p^2 + |\mb{p}_T|^2}{2 x P^+},\; x P^+ ,\;\mb{p}_T\rg].
\end{equation} 
In our approach, we assume that neither the virtuality of the quark, $p^2$, 
nor its
transverse momentum squared, $|\mb{p}_T|^2$,
can be large in comparison with the hard scale $Q^2$. 
Under these conditions, the quark momentum is soft with respect to the hadron
momentum and its relevant component is $x P^+$.
In Eq.~\eqref{e:hadrocorre}, neglecting terms which are $1/Q$
suppressed,  we can
use an approximate expression for the delta function and 
\begin{align} 
2 M W^{\mu \nu}(q,P,S)& \approx \sum_q e_q^2 \int \de^2 \!
\mb{p}_T \de p^- \de x \;
\frac{P^+}{2 P \cdot q}\;\delta\lf(x -\xbj \rg)\, 
\tr \lf[ \Phi(p,P,S)\,\g^{\mspace{2mu}\mu} \lf(\pslash + \qslash + m\rg) \g^{\nu}
\rg]  \nn \\
&= \sum_q e_q^2 \;
\frac{1}{2}\tr \lf[ \Phi(\xbj,S)\,\g^{\mspace{2mu}\mu} \frac{P^+}{P \cdot
q}\lf(\pslash + \qslash + m\rg) \g^{\nu}\rg]
\label{e:hadrohadro}
\end{align} 
where we introduced the integrated correlation function
\begin{equation} \begin{split} 
\Phi_{ji}(x,S) &= \int \de^2 \! \mb{p}_T  \de p^-  \;\Phi_{ji}(p,P,S)
\biggr\vert_{p^+ = x P^+} \\
&=\int \frac{\de \xi^-}{2 \pi} \; \e^{-\ii p \cdot \xi}  
\bra{P,S} \bbar{\psi}_i (\xi)	\; \psi_j (0) \ket{P,S}
	\biggr\vert_{\xi^+ = \bm{\xi}_T = 0}.
\end{split}
\label{e:phiint} \end{equation}  
Notice that there is a contradiction between the fact that we assumed the
transverse momentum of the quark to be small in comparison to the hard
scale, yet we are integrating over the entire space of $\mb{p}_T$. Indeed,
when dealing with transverse momentum of perturbative origin (i.e. arising
from the radiation of gluons, see next section), which typically
falls down as $1/|\mb{p}_T|^2$, we have to
impose a cut-off on the maximum value the transverse momentum can reach. This
cut-off depends on the scale $Q^2$. On the other hand, the transverse momentum 
of nonperturbative origin, usually called {\em intrinsic transverse
momentum}, is supposed to fall off very rapidly so that there is effectively
almost no intrinsic transverse momentum above a typical scale of 1 GeV$^2$.

Finally, from the outgoing quark
momentum, $p+q$, we can select only the minus component and obtain the final
form for the hadronic tensor at leading twist
\begin{equation} 
2 M W^{\mu \nu}(q,P,S)\approx \sum_q e_q^2\;
\frac{1}{2}\tr \lf[ \Phi(\xbj,S)\,\g^{\mspace{2mu}\mu} \g^+ \g^{\nu}\rg].
\label{e:hadrophi}
\end{equation} 
A few words to justify the last approximation are in order. The dominance of
 the minus component is most easily seen in the
infinite momentum frame, where  $p^-\!+q^-$ is of the order of $Q$, while
$p^+\!+q^+ =0$, and
$\mb{p}_T$ and $m$ are
of the order of 1. However, 
if we
perform a $1/Q$ expansion of the full expression, including 
the correlation function $\Phi$ [starting from Eq.~\eqref{e:decomphi1}], we
would be able to check that in any collinear frame
the dominant terms arise only from the
 combination of plus component in the correlation
function and minus components in the outgoing quark momentum.

%%%%%%%%%%%%%%%%%%%%%%%%%%%%%%%%%%%%%%%%%%%%%%%%%%%%%%%%%%%%
\subsection{One gluon additions}
\label{s:gluon}

Up to now, we took into consideration only quark-quark correlation functions
at tree level. 
The addition of a gluon can either lead to the introduction of
a quark-gluon-quark correlation function or can give rise to perturbative
corrections to the photon-quark scattering~\cite{Collins:1988gx}. 
In this thesis, we will not take quark-gluon-quark correlation
functions into account, since they start contributing only at the
twist-three level, and we will not examine perturbative corrections, since 
they give only origin to a logarithmic 
scale dependence of the quark-quark
functions. 
Nevertheless, for completeness we will now give 
a sketchy view of these very important issues.

	\begin{figure}
	\centering
	\includegraphics[width=7cm]{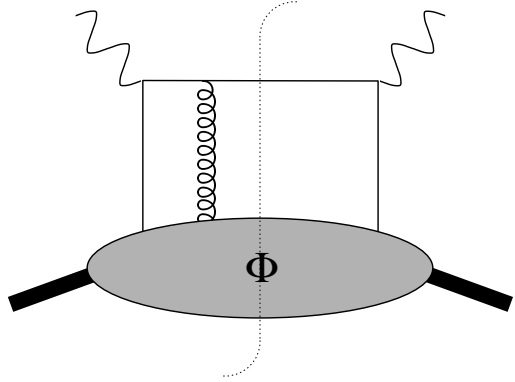}
	\caption{One gluon addition: gluon coming from the soft blob.}
	\label{f:qgq}
        \end{figure}
When the gluons come directly from the soft blob, as in the diagram of 
Fig.~\vref{f:qgq}, 
longitudinally polarized
gluons ($A^+$) are
the dominant ones, while transversely polarized gluons ($A_T$) are subject to
a $1/Q$ suppression. If we choose a physical gauge where $A^+=0$, then
this kind of diagrams contribute only at twist three and
higher and they require the introduction of quark-gluon-quark 
correlators~\cite{Ellis:1983cd}. 
On the other hand, in a different gauge the contributions of
longitudinal gluons is present and is not necessarily 
suppressed by any power of $1/Q$. 
Then, we have to sum all the contributions with an arbitrary
number of longitudinal gluons. The result of this summation can be cast in the 
form of a {\em gauge link} to be inserted in the definition of the quark-quark 
correlation function
\begin{equation} 
\Phi_{ji}(p, P,S)=\frac{1}{(2\pi)^4} \int \de^4 \! \xi\;\e^{-\ii p \cdot \xi }
\bra{P,S} \bbar{\psi}_i (\xi) 
	\,{\cal L}(\xi,0;\mbox{path})\, \psi_j (0) \ket{P,S} 
\label{e:phiwithgaugelink}
\end{equation} 
where the gauge link is a path-ordered exponential
\begin{equation}
{\cal L}(\xi,0;\mbox{path})= {\cal P} \exp \lf(-\ii g_s \int_0^\xi \de s^{\mu}
A_{\mu}(s) \rg).
\end{equation} 
with a straight path along the light-cone minus direction~\cite{Boer:1999si,Ellis:1983cd}.

In $A^+=0$ gauges the link is equal to unity (although some subleties have
been recently analyzed in Ref.~\citen{Brodsky:2002ue}). We will
henceforth neglect it and trade off manifest color gauge invariance for a
lighter notation. Finally, we mention that recently it has been suggested that 
the gauge link could
play an important role in the context of T-odd distribution
functions~\cite{Collins:2002kn} (see Sec.~\vref{s:inclusion}). 
In particular, much care should be taken when 
including partonic transverse momentum. In this case, the gauge link path
cannot simply run along the light-cone but has to have a transverse component
and it might not be reducible to unity anymore~\cite{Ji:2002aa,Collins:2002kn}.

	\begin{figure}
	\centering
	\includegraphics[width=5cm]{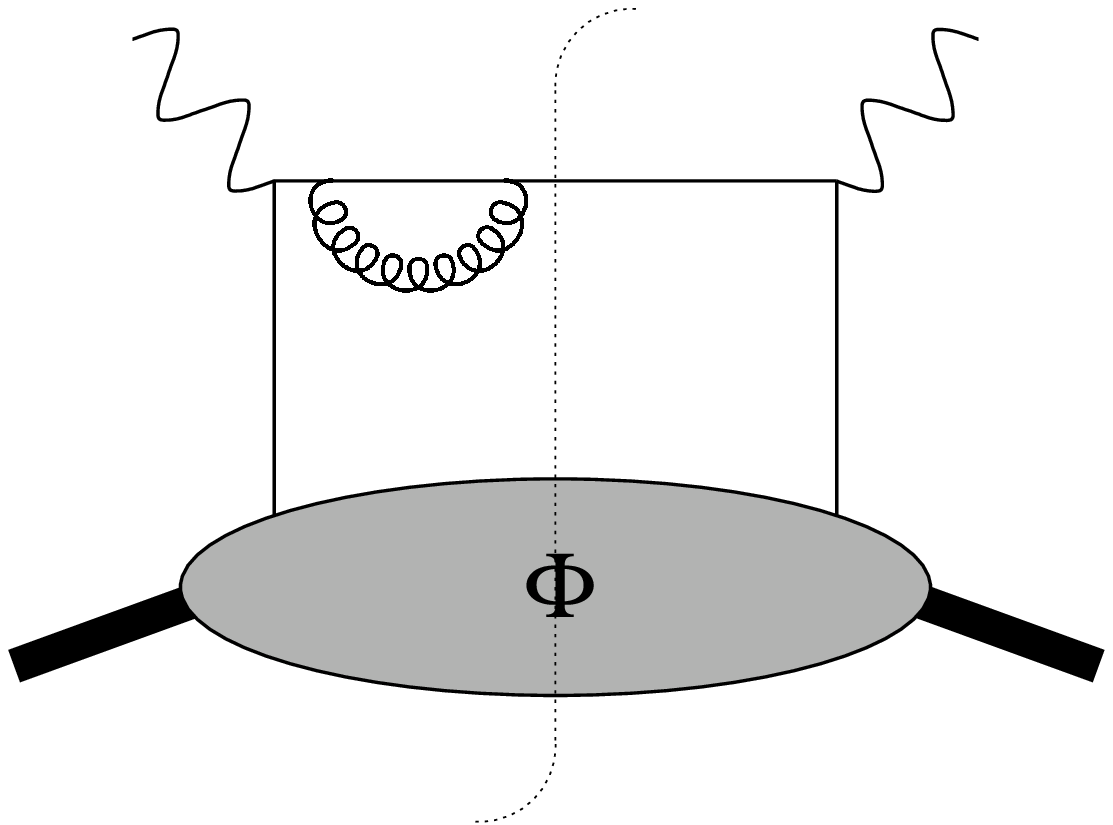}\hfil
	\includegraphics[width=5cm]{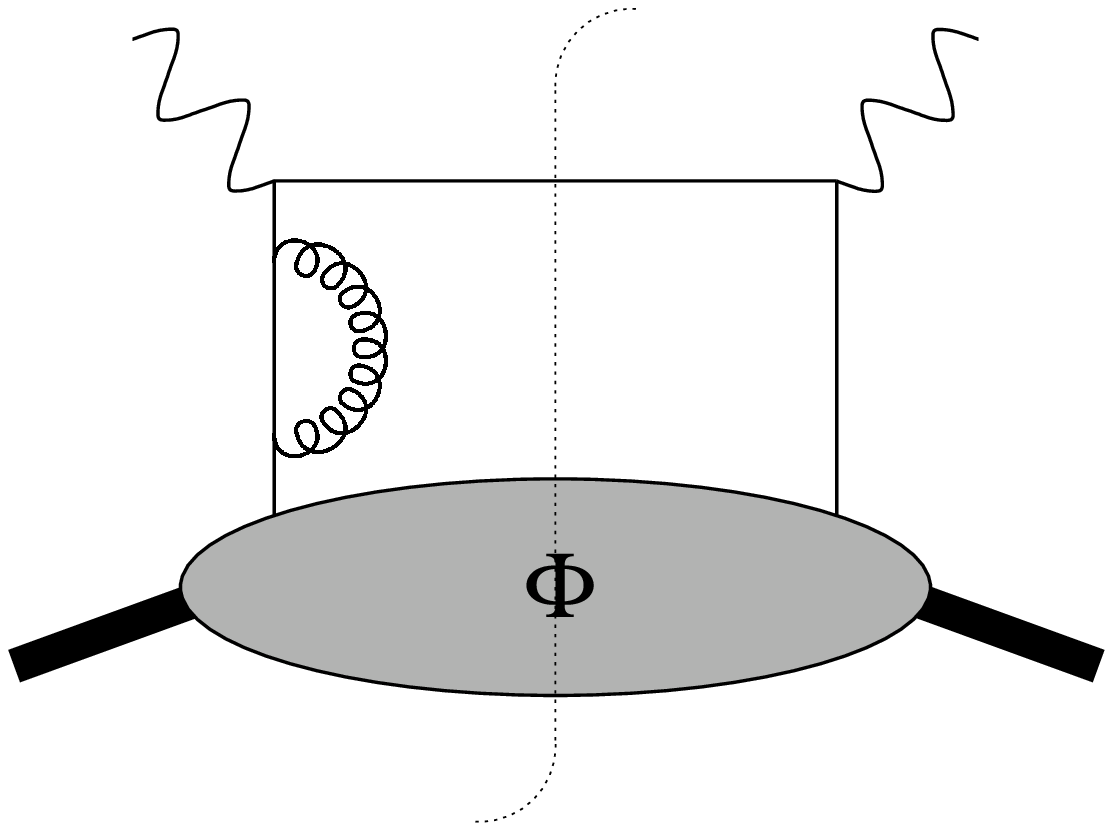}\hfil
	\includegraphics[width=5cm]{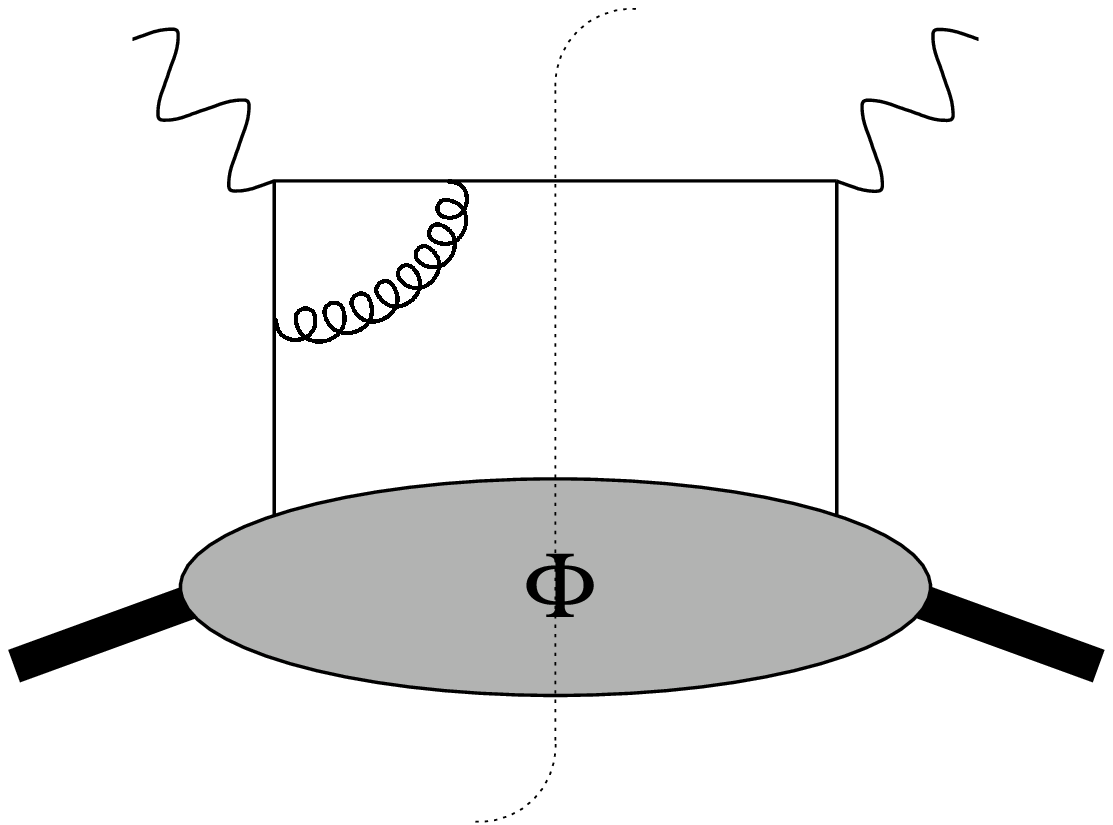}
	\caption{One gluon addition: virtual loop diagrams.}
	\label{f:virtualloop}
        \end{figure}
\begin{figure}
	\centering
	\includegraphics[width=5cm]{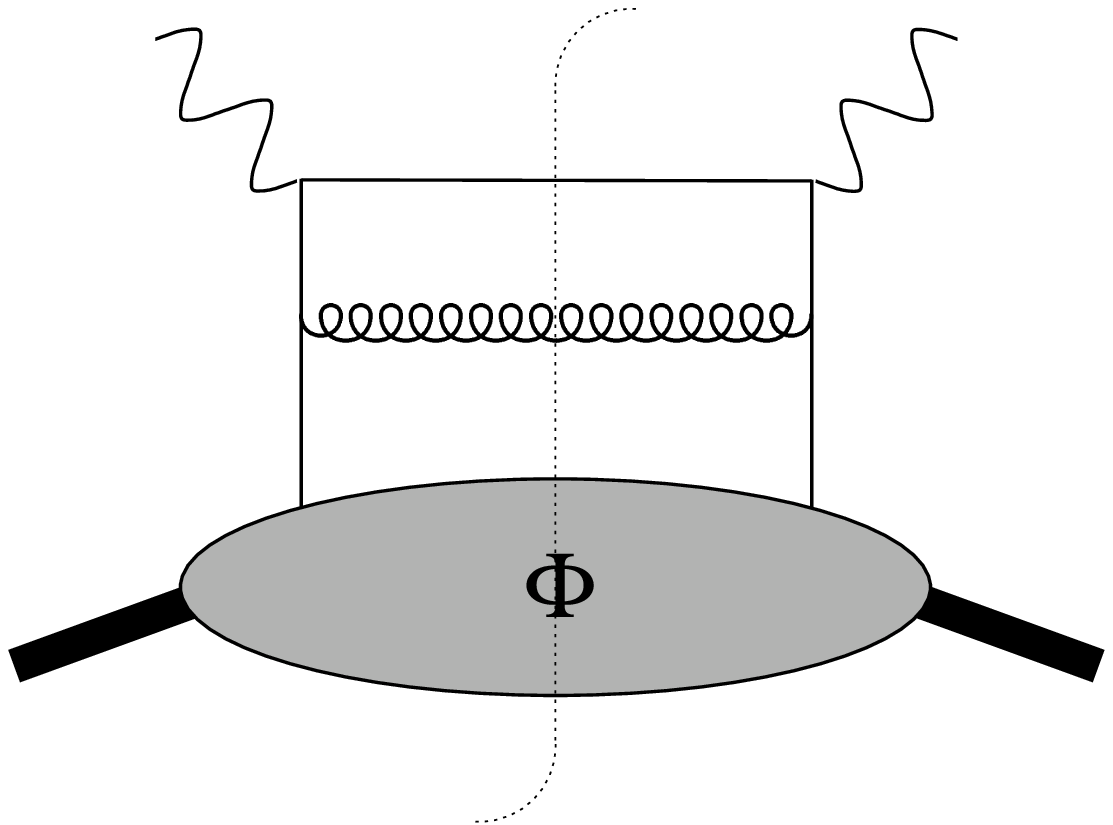}\hfil
	\includegraphics[width=5cm]{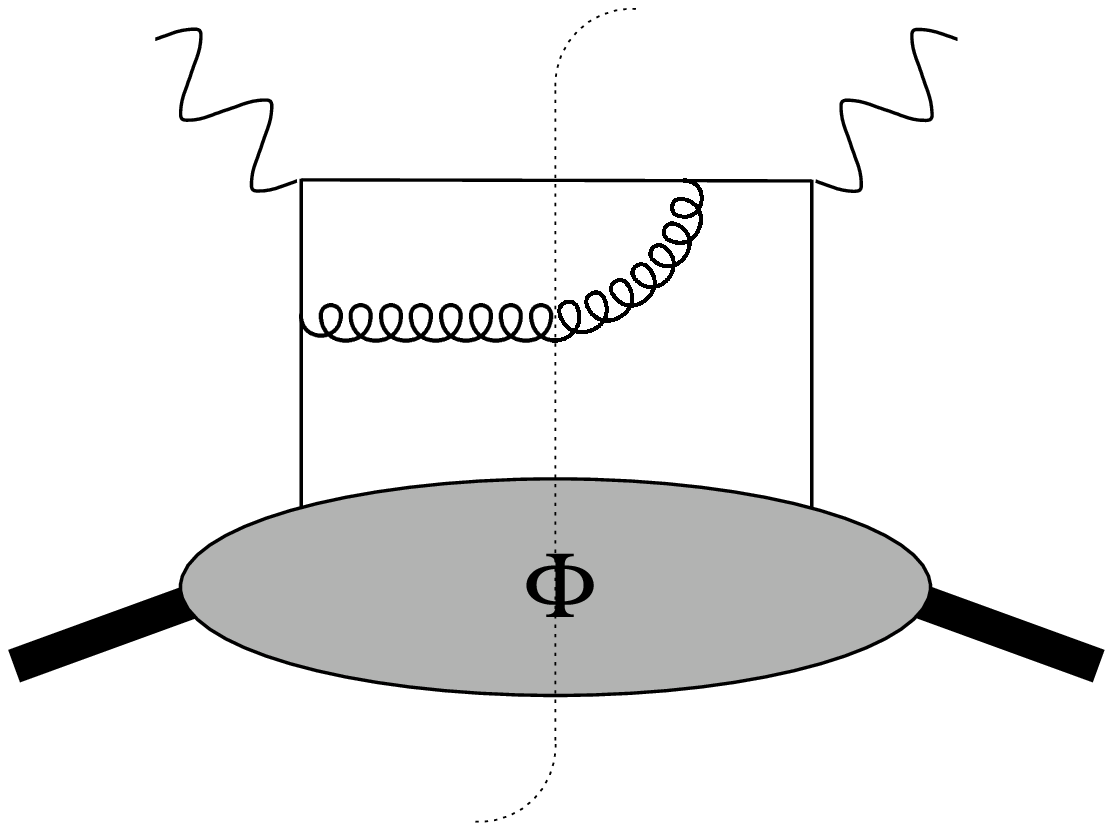}\hfil
	\includegraphics[width=5cm]{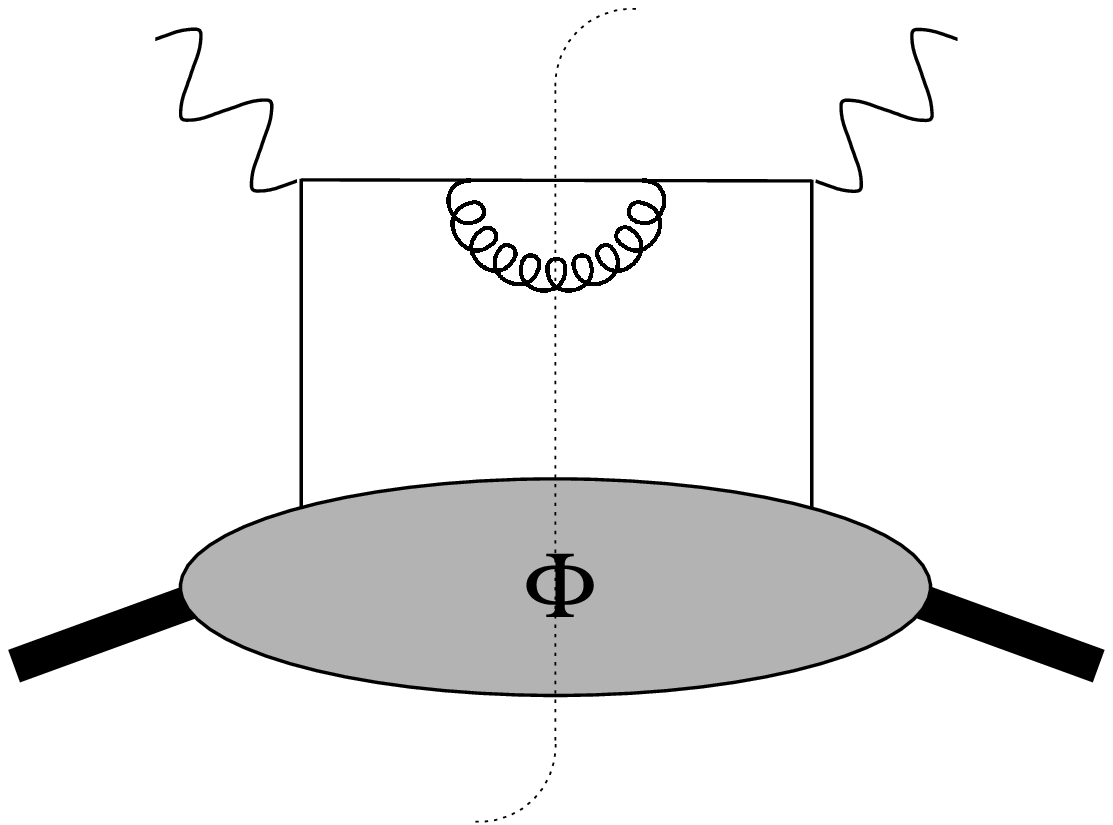}
	\caption{One gluon addition: real loop diagrams.}
	\label{f:brehm}
        \end{figure}

Now we take a brief look to perturbative corrections to the quark-quark
correlation function (see Refs.~\citen{Field} and~\citen{Peskin:1995ev}). 
They are of two kinds: virtual gluon loop
diagrams (Fig.~\ref{f:virtualloop})
and real gluon bremsstrahlung diagrams (Fig.~\ref{f:brehm}). Each
virtual diagram contains ultraviolet and infrared divergences. 
The ultraviolet divergences can be cured using standard renormalization
techniques. 
The infrared divergences cancel with analogous
divergences in the real gluon emission diagrams~\cite{Field}.

The remaining parts of the diagrams give actual $\alpha_s$ corrections to the
tree level result of the previous section. In particular, collinear
divergences give origin
to the leading-log part of the {\em evolution equations}, 
by which the parton distribution
functions (see Sec.~\vref{s:districolle}) acquire a dependence on the
scale $Q^2$~\cite{Gribov:1972ri,Altarelli:1977zs,Dokshitzer:1977sg}.

%%%%%%%%%%%%%%%%%%%%%%%%%%%%%%%%%%%%%%%%%%%%%%%%%%%%%%%%%%%%
\subsection{Leading twist part and connection with helicity formalism}
\label{s:leading}

\begin{figure}
	\centering
	\rput(2.05,3.4){\Large \boldmath ${\cal P}_+$}
	\rput(6,3.4){\Large \boldmath ${\cal P}_+$}
	\rput(3.1,4.5){\Large \boldmath ${\cal P}_-$}
	\rput(4.95,4.5){\Large \boldmath ${\cal P}_-$}
	\includegraphics[width=8cm]{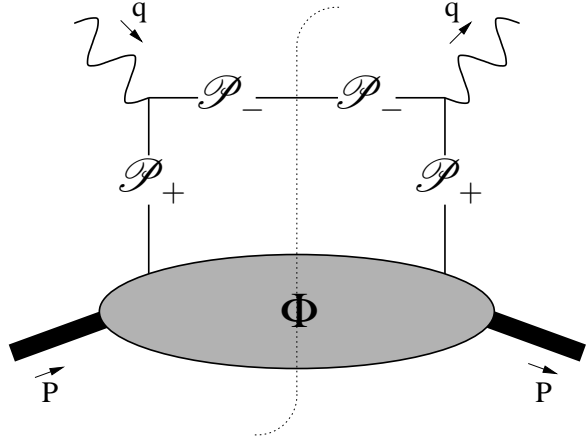}
	\caption{Graphical representation of the insertion of the good
	projectors to isolate the leading part of the hadronic tensor.}
	\label{f:project}
        \end{figure}

To identify the leading twist contributions to the cross section, 
it is convenient to define the projectors
\begin{align} 
{\cal P}_+ &= \frac{1}{2}\; \g^- \g^+, & {\cal P}_- &= \frac{1}{2} \;\g^+ \g^-.
\label{e:goodbad}
\end{align} 
Before the interaction with the virtual photon, the relevant components 
of the quark fields are
the plus components, $\psi_+ = {\cal P}_+\, \psi$. They are usually referred
to as the {\em good components}.\footnote{In the infinite momentum frame the good components are truly dominant 
and we can avoid the distinction between ``quark'' and ``good quark''.}
Vice versa, after the interaction with the virtual photon,
the relevant components of the outgoing quark fields are
the minus components, $\psi_- = {\cal P}_-\, \psi$. 
Therefore, the leading twist part of the
hadronic tensor in Eq.~\eqref{e:hadrohadro}
 can be projected out in the following way
(see Fig.~\vref{f:project})\footnote{Note that 
$\bar{(\psi_+)} = \bar{\psi}\, {\cal P_-}$ and 
$\bar{(\psi_-)} = \bar{\psi}\,{\cal P_+}$.}
\begin{equation} 
\begin{split} 
 2 M W^{\mu \nu} (q, P, S) 
 & \approx  \sum_q e_q^2\,
	\tr{\lf[{\cal P}_+\,\Phi (\xbj, S)\;{\cal P}_-\; \g^{\mu}\;
	{\cal P}_-\,\frac{P^+}{2 P \cdot q}\lf(\pslash + \qslash + m\rg) \;{\cal P}_+\; \g^{\nu}  \rg]} \\
 & =  \sum_q e_q^2\,
	\tr{\lf[{\cal P}_+\,\Phi (\xbj, S)\, \g^+ \; \frac{\g^- \g^{\mu}}{2}\,{\cal P}_-\;{\cal P}_-\,\frac{P^+}{2 P \cdot q}\lf(\pslash + \qslash + m\rg) 
	\, \g^- \; \frac{\g^+ \g^{\nu}}{2}\,{\cal P}_+  \rg]}. 
\end{split}
\end{equation}  
The differential cross section defined in Eq.~\eqref{e:crossinc} takes the form
\begin{equation} 
\begin{split} 
\lefteqn{\frac{\de^3\!\sigma}{\de \xbj \de y \de \phi_S}} \\
 &\approx \sum_q\,\frac{\alpha^2 e_q^2}{2\, s\,\xbj\,Q^2}\, L_{\mu \nu}(l, l', \lambda_e)\;
 \tr{\lf[{\cal P}_+\,\Phi (\xbj, S)\, \g^+ \; \frac{\g^- \g^{\mu}}{2}\,{\cal P}_-\;{\cal P}_-\,
		\frac{P^+}{2 P \cdot q}\lf(\pslash + \qslash + m\rg) \,
		 \g^- \; \frac{\g^+ \g^{\nu}}{2}\,{\cal P}_+  \rg]} \\
&\approx \sum_q\,\Bigl({\cal P}_+\, \Phi(\xbj, S) \g^+ \Bigr)_{ji} 
% \\
%&\quad \mbox{}\times 
\lf[\frac{\alpha^2 e_q^2}{s\,\xbj\,Q^2}\, L_{\mu \nu} (l, l',
\lambda_e) 
	\lf( \frac{\g^- \g^{\mu}}{2}\,{\cal P}_-\rg)_{il} 
	\lf( \frac{\g^+ \g^{\nu}}{2}\,{\cal P}_+\rg)_{mj}\rg] 
	\lf(\frac{{\cal P}_-}{2} \rg)_{lm},
\end{split}
\end{equation}
where we explicitly showed Dirac indices (repeated indices are summed over).
In Sec.~\vref{s:helicity}, we will see in detail how the insertion of
the projectors effectively reduces the four-dimensional Dirac space into a
two-dimensional subspace. Chiral-right and chiral-left 
 good quark spinors can
be used as a basis in this space. Therefore, it is possible to
 replace the Dirac
indices with chirality indices (of good fields).
By doing this, we put particular evidence on the connection with the
helicity/chirality formalism (see e.g.\ Refs.~\citen{Jaffe:1996wp} and~\citen{Anselmino:1996vq}). 
In fact, the cross section can be conveniently
rewritten as
\begin{equation} 
\frac{\de^3\!\sigma}{\de \xbj \de y \de \phi_S} \approx  \frac{1}{2}\sum_q\;
\Bigl({\cal P}_+\, \Phi(\xbj, S) \g^+
\Bigr)_{\chi'_{1} \chi^{\phantom'}_{1}}\;
\Bigl( \frac{\de \sigma^{eq}}{\de y} \Bigr)
	^{\chi^{\phantom'}_{1} \chi'_{1}},
\label{e:crossinca}
\end{equation} 
where the elementary electron-quark cross section is
\begin{equation} \begin{split}
\Bigl( \frac{\de \sigma^{eq}}{\de y} \Bigr)
	^{\chi^{\phantom'}_{1} \chi'_{1}} &=
	\frac{\alpha^2 e_q^2}{s \xbj Q^2}\, L_{\mu \nu} (l, l',\lambda_e) 
	\lf( \frac{\g^- \g^{\mu}}{2}\,{\cal P}_-\rg)^{\chi^{\phantom'}_{1} \chi^{\phantom'}_{2}} 
	\lf( \frac{\g^+ \g^{\nu}}{2}\,{\cal P}_+\rg)^{\chi'_{2}
\chi'_{1}} \; \delta_{\chi^{\phantom'}_{2} \chi'_{2}}
\\[3mm]
&= \frac{\alpha^2 e_q^2}{s \xbj y^2}\,
\begin{pmatrix}
A(y)+\lambda_e C(y)  & 0 \\
0& A(y)-\lambda_e C(y)
\end{pmatrix}.
\end{split}
\label{e:eqinc}
\end{equation} 
where 
\begin{align}  
A(y) &= 1 - y + \frac{y^2}{2},    & C(y) &=y \lf(1-\frac{y}{2}\rg)  .
\end{align}   

Finally, we define the matrix $F = \Bigl({\cal P}_+\, \Phi \g^+
\Bigr)^T$, i.e.\ the transpose of the
 leading-twist part of the correlation function, and we observe that  
\begin{equation} 
\begin{split}
F(x,S)_{\chi^{\phantom'}_{1} \chi'_{1}}&= \int \frac{\de \xi^-}{2 \pi \sqrt{2}} \; \e^{-\ii p \cdot \xi}  
\bra{P,S} (\psi_+)^{\dagger}_{\chi_{1}} (\xi)\; (\psi_+)_{\chi'_{1}} (0) \ket{P,S}
	\biggr\vert_{\xi^+ = \bm{\xi}_T = 0} \\
&= \frac{1}{\sqrt{2}}\sum_{X} 
\int \frac{\de^3 \! \mb{P}_{X}} {(2 \pi)^3\, 2 P_{X}^0} \,
	\bra{X}(\psi_+)_{\chi_{1}} (0) \ket{P,S}^{\ast} 
	\bra{X} (\psi_+)_{\chi'_{1}} (0) \ket{P,S} \;\delta \Bigl(\bigl(1-x\bigr)P^+-P^+_{X}\Bigr).
\end{split}
\end{equation} 
Thus, the transpose of the
correlation function describes the forward scattering of a good antiquark
off a hadron, or equivalently
the forward scattering of an antiquark off a hadron 
in the infinite momentum frame.
As any scattering matrix, 
for any antiquark-hadron state $\ket{a}$ the expectation value
$\bra{a\mspace{2mu}} M \ket{a}$ must be positive. In mathematical terms, this means that
the matrix is {\em positive semidefinite}, i.e.\  
the determinant of all the principal minors of the matrix has to be positive
or zero. 
This property will prove to be essential
in deriving bounds on the components of the correlation function, i.e.\ the
parton distribution functions.

%%%%%%%%%%%%%%%%%%%%%%%%%%%%%%%%%%%%%%%%%%%%%%%%%%%%%%%%%%%%
\section{The correlation function $\Phi$}
\markright{The correlation function $\mathit \Phi$}
\label{s:districolle}

As shown in Eq.~\eqref{e:phi1}, the quark-quark distribution correlation 
function, $\Phi$, can be expressed in terms of bilocal operators. At 
leading order in $1/Q$, it
contains all the relevant information about the nonperturbative dynamics of
the quarks
inside the hadron.
Due to its nonperturbative nature, it is not possible to calculate it from
first principles, as we don't know how the hadronic states are built up 
from the elementary quark and gluon fields. 

When considering subleading orders in a $1/Q$ expansion, quark-gluon-quark 
correlation
functions have to be considered, as we briefly mentioned in
Sec.~\vref{s:gluon}. In this case, the general
structure of the hadronic tensor becomes richer. 
In the rest of the thesis, as the analysis
will be concerned only with leading order terms, we will not deal
with quark-gluon-quark correlation functions.

To get more insight into the information contained in the correlation
function, which is a Dirac matrix, we can decompose it in a general way on
a basis of Dirac structures. Each term of the decomposition
can be a combination of the Lorentz vectors $p$ and $P$, the Lorentz
pseudovector $S$ (in case of spin-half hadrons) and the Dirac 
structures
\[ 
{\mb 1},\; \g_5,\; \g^{\mu},\; \g^{\mu}\g_5,
	\;\ii \sig^{\mu \nu}\g_5.
\]
The spin vector can only appear linearly in the 
decomposition (cf. Eq.~\eqref{e:matrixphi}). 
Moreover, each term of the full expression has to satisfy the conditions of 
Hermiticity and parity invariance
\begin{subequations} 
\begin{xalignat}{4} 
&\text{Hermiticity:} &\Phi(p,P,S)& =\g^0\, \Phi^{\dag}(p,P,S)\, \g^0,
\label{e:herm}\\             
&\text{parity:} &\Phi(p,P,S)& =\g^0\, \Phi(\tilde{p},\tilde{P},-\tilde{S})\, \g^0
\label{e:parity}
\end{xalignat} 
\end{subequations} 
where $\tilde{p}^{\nu} = \delta^{\nu \mu} p_{\mu}$ and 
so forth for the other vectors.
The most general decomposition of the correlation function $\Phi$ imposing
Hermiticity and parity invariance is~\cite{Ralston:1979ys,Mulders:1996dh}
\begin{equation} \begin{split} 
\Phi(p,P,S)  &=
                M\,A_1\,{\mb 1} + A_2\,\Pslash + A_3\,\pslash
                +\frac{A_{4}}{M}\,\sigma_{\mu \nu} P^{\mspace{2mu}\mu} p^{\nu}
                + \ii A_5\;p\cdot S\, \gamma_5
                                \\
  & \quad	+ M\,A_6 \,\Sslash\, \gamma_5
                + A_7\,\frac{p\cdot S}{M}\,\Pslash\, \gamma_5     
                + A_8\,\frac{p\cdot S}{M}\, \pslash\, \gamma_5
                + \ii A_9\,\sig_{\mu\nu}\gamma_5 S^{\mu}P^{\nu}
                              \\
  & \quad       + \ii A_{10}\,\sig_{\mu\nu}\gamma_5 S^{\mu}p^{\nu}
    +\ii A_{11}\,\frac{p\cdot S}{M^2}\,\sig_{\mu\nu}\gamma_5 P^{\mspace{2mu}\mu}p^{\nu}
	    + A_{12}\, \frac{\eps_{\mu \nu \rho \sigma}\gamma^{\mspace{2mu}\mu} P^\nu
              p^\rho S^\sigma}{M}     ,
					\label{e:decomphi1}  
\end{split} \end{equation} 
where the amplitudes $A_i$ are dimensionless real scalar functions 
$A_i=A_i(p\cdot P,p^2)$. 

The correlation function can be separated in a
{\em T-even} part 
and a {\em T-odd} part, according to the definition
\begin{subequations}
\label{e:teventodd}
\begin{align} 
\Phi^{\ast}_{\text{T-even}}(p,P,S)& =\ii \g^1 \g^3\,
\Phi_{\text{T-even}}(\tilde{p},\tilde{P},\tilde{S})\, \ii \g^1 \g^3,
\label{e:teven}\\
\Phi^{\ast}_{\text{T-odd}}(p,P,S)& = - \ii \g^1 \g^3\,
\Phi_{\text{T-odd}}(\tilde{p},\tilde{P},\tilde{S})\, \ii \g^1 \g^3.
\label{e:todd}
\end{align} 
\end{subequations}
Thus, the terms containing the amplitudes $A_4$, $A_5$ and $A_{12}$
can be classified as T-odd. 

At leading twist, we are interested in the projection ${\cal P}_+\, \Phi(\xbj,
S) \g^+$. After inserting the general decomposition of Eq.~\eqref{e:decomphi1}
into Eq.~\eqref{e:phiint}, we can project out the leading-twist components and
obtain the general expression~\cite{Bacchetta:1999kz}
\begin{equation} 
{\cal P}_+\, \Phi(x, S) \g^+ = \lf(
f_1(x) + S_L\,g_1(x)\,\gamma_5  + h_1(x)\,\gamma_5\, \Sslash_T
\rg){\cal P}_+ ,
\label{e:decomphi}
\end{equation} 
where we introduced the {\em parton distribution functions}
\begin{subequations}\label{e:distri}
\begin{align} 
f_1(x) &= \int \de^2 \! \mb{p}_T \de\! p^2 \de (2 p \cdot P)\;
\delta\lf(\mb{p}_T^2 + x^2 M^2 + p^2 -2 x p \cdot P \rg) \lf[A_2 + x A_3 \rg] , \\
g_1(x) &= \int \de^2 \! \mb{p}_T \de\! p^2 \de (2 p \cdot P)\;
\delta\lf(\mb{p}_T^2 + x^2 M^2 + p^2 -2 x p \cdot P \rg) \lf[-A_6 -
\lf(\frac{p \cdot P}{M^2} - x \rg) \lf(A_7 +x A_8 \rg)\rg],\\
h_1(x) &= \int \de^2 \! \mb{p}_T \de\! p^2 \de (2 p \cdot P)\;
\delta\lf(\mb{p}_T^2 + x^2 M^2 + p^2 -2 x p \cdot P \rg)\lf[-A_9 -x A_{10} +
\frac{\mb{p}_T^2}{2 M^2} A_{11} \rg] .
\end{align} 
\end{subequations}
The function $f_1$ is usually referred to as the unpolarized parton
distribution, and it is sometimes denoted also as simply $f$ or $q$ (where $q$ 
stands for the quark flavor). 
The function $g_1$ is the parton helicity distribution and it can be denoted
also as $\Delta f$ or $\Delta q$. Finally, the
function $h_1$ is known as the parton transversity distribution; in the
literature it is sometimes denoted as $\delta q$,  $\Delta_T q$ or $\Delta_T
f$,  although
in the original paper of Ralston and Soper~\cite{Ralston:1979ys} it was called 
$h_T$. In this thesis, we will follow the nomenclature suggested by Jaffe and
Ji~\cite{Jaffe:1992ra} and later extended in Ref.~\citen{Mulders:1996dh}, 
because it
allows a harmonious
 connection with the most general cases of transverse momentum
dependent distribution functions, as we shall see later. A thorough discussion 
on the different naming schemes is presented in Ref.~\cite{Barone:2001sp}.

The individual distribution functions can be isolated by means of the
projection
\begin{equation} 
\Phi^{[\Gamma]} \equiv \frac{1}{2} \tr\lf(\Phi \Gamma \rg),
\label{e:li}
\end{equation} 
where $\Gamma$ stands for a specific Dirac structure. In particular, we see
that
\begin{subequations} \begin{align} 
f_1(x) &=\Phi^{[\g^+]}(x), \\
g_1(x) &=\Phi^{[\g^+ \g^5]}(x), \\
h_1(x) &=\Phi^{[\ii \sig^{i +} \g^5]}(x).
\end{align} \end{subequations} 

%%%%%%%%%%%%%%%%%%%%%%%%%%%%%%%%%%%%%%%%%%%%%%%%%%%%%%%%%%%%
\subsection{Correlation function in helicity formalism}
\label{s:helicity}

We will now examine how it is possible to write the correlation function as a
matrix in the chirality space of the good quark fields $\otimes$ the spin
space of the hadron. We will work out the steps in a meticulous way, even if
we will incur the risk of introducing some redundant steps.

The correlation function is a $4\times 4$ Dirac matrix. However, due to the
presence of the projector on the good components of the quark fields,
 the leading-twist part spans only a $2\times2$ Dirac subspace.
This is evident if we express the Dirac structures of Eq.~\eqref{e:decomphi}
in the chiral or Weyl representation.
Using this representation, the correlation function reads 
\renewcommand{\arraystretch}{1}
\begin{equation}
 \Bigl({\cal P}_+\, \Phi(x, S) \g^+ \Bigr)_{ji}  = 
\begin{pmatrix}
f_1(x) + S_L\, g_1(x) &0&0& (S_x - \ii S_y)\, h_1(x)\\
0&0&0&0 \\
0&0&0&0 \\
(S_x + \ii S_y)\, h_1(x) &0&0& f_1(x) - S_L\, g_1(x)
\end{pmatrix}.
\label{e:fullphi}
\end{equation}
As shown by this explicit form, it seems that the four-dimensional
Dirac space can be reduced to a two-dimen\-sional space, 
retaining only the nonzero part of
the correlation function. The relevant part of the Dirac space is the one
corresponding to good quark fields.
To show this explicitly, 
we introduce the chiral projectors ${\cal P}_{R/L} = (1\pm \g_5)/2$ and define
the good chiral-right and
good chiral-left quark spinors, i.e.\
the normalized projections
 \renewcommand{\arraystretch}{1.5}
\begin{align}
u_{+R} &= \frac{{\cal P}_+ {\cal P}_R\, u}{|{\cal P}_+ {\cal P}_R\, u|}, &
u_{+L} &= \frac{{\cal P}_+ {\cal P}_L\, u}{|{\cal P}_+ {\cal P}_L\, u|}.
\end{align} 
Then, we can define a new matrix in the chirality space of the good quark
fields
\begin{equation}
 \Bigl({\cal P}_+\, \Phi(x, S) \g^+
\Bigr)_{\chi'_{1} \chi^{\phantom'}_{1}} \equiv
u_{+\,\chi'_1}^j \Bigl({\cal P}_+\, \Phi(x, S) \g^+ \Bigr)_{ji}\, u_{+\,\chi_1}^i
\end{equation} 
Any contraction with bad quark fields vanishes.
Explicit computation of the matrix elements yields
\begin{subequations}
\begin{align} 
 \Bigl({\cal P}_+\, \Phi(x, S) \g^+
\Bigr)_{RR} &= u_{+R}^j \Bigl({\cal P}_+\, \Phi(x, S) \g^+ \Bigr)_{ji}\,
 u_{+R}^{\dagger\,i}  = f_1(x) + S_L\, g_1(x), \\
 \Bigl({\cal P}_+\, \Phi(x, S) \g^+
\Bigr)_{LL} &= u_{+L}^j \Bigl({\cal P}_+\, \Phi(x, S) \g^+ \Bigr)_{ji}\,
 u_{+L}^{\dagger\,i} = f_1(x) - S_L\, g_1(x), \\
 \Bigl({\cal P}_+\, \Phi(x, S) \g^+
\Bigr)_{RL} &= u_{+R}^j \Bigl({\cal P}_+\, \Phi(x, S) \g^+ \Bigr)_{ji}\,
  u_{+L}^{\dagger\,i}= (S_x - \ii S_y)\, h_1(x), \\
 \Bigl({\cal P}_+\, \Phi(x, S) \g^+
\Bigr)_{LR} &= u_{+L}^j \Bigl({\cal P}_+\, \Phi(x, S) \g^+ \Bigr)_{ji}\,
  u_{+R}^{\dagger\,i}= (S_x + \ii S_y)\, h_1(x).
\end{align} 
\end{subequations}
The correlation matrix in the good quark chirality space is then
\begin{equation}
 \Bigl({\cal P}_+\, \Phi(x, S) \g^+
\Bigr)_{\chi'_{1} \chi^{\phantom'}_{1}}  = \begin{pmatrix}
f_1(x) + S_L\, g_1(x) & (S_x - \ii S_y)\, h_1(x) \\
(S_x + \ii S_y)\, h_1(x) & f_1(x) - S_L\, g_1(x)		
\end{pmatrix}.
\end{equation} 
As we could have expected, this result corresponds simply to taking the full
Dirac matrix in Weyl representation, Eq.~\eqref{e:fullphi}, and stripping off
the zeros. 
From the matrix representation in the chirality space it is clear why the
function $h_1$ is defined to be {\em chiral odd}. 

\begin{figure}
	\centering
	\rput(1.5,2.3){\large \boldmath $\chi^{\phantom'}_1$}
	\rput(5.5,2.3){\large \boldmath $\chi'_1$}
	\rput(-0.3,0){\large \boldmath $\Lambda'_1$}
	\rput(7.2,0){\large \boldmath $\Lambda^{\phantom'}_1$}
	\includegraphics[width=7cm]{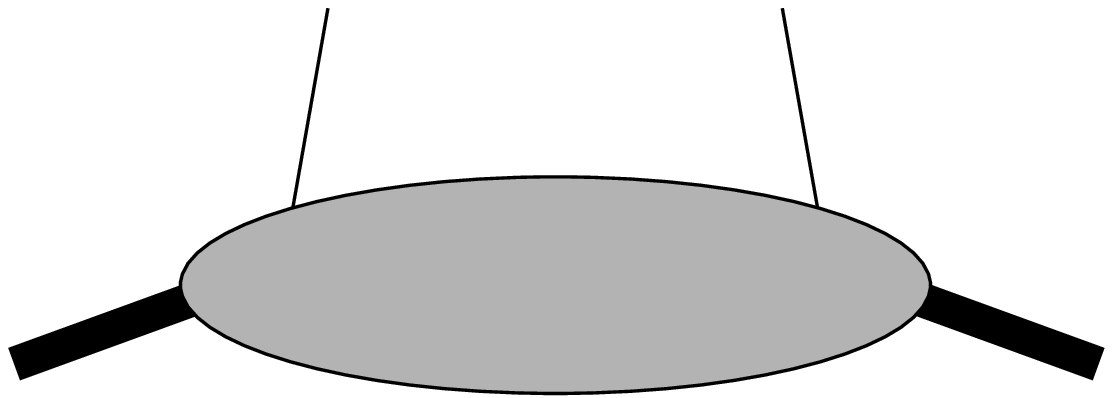}
	\caption{Illustration of the position of the indices of the
correlation matrix.}
	\label{f:phimatrix}
        \end{figure}

The correlation function is a matrix in the parton chirality space and depends
on the target spin. By introducing the helicity density matrix of the target 
\begin{equation} 
\rho(S)_{\Lambda^{\phantom'}_1 \Lambda_1'} = \frac{1}{2} \lf(1 + \mb{S}\cdot\bm{\sigma}\rg)_{\Lambda^{\phantom'}_1 \Lambda_1'} = 
	\frac{1}{2}\begin{pmatrix}
1 + S_L   & S_x - \ii S_y \\
S_x + \ii S_y & 1- S_L		
\end{pmatrix},
\label{e:rho}
\end{equation} 
we can obtain the correlation function from the trace of the helicity density
matrix and a new matrix in the 
quark chirality space $\otimes$ the target spin space:
  \begin{equation} 
 \Bigl({\cal P}_+\, \Phi(x, S) \g^+ \Bigr)_{\chi'_{1} \chi^{\phantom'}_{1}}
=  \rho(S)_{\Lambda^{\phantom'}_1 \Lambda_1'}\; \Bigl({\cal P}_+\, \Phi(x)
\g^+ \Bigr)_{\chi'_{1} \chi^{\phantom'}_{1}}^{\Lambda_1'
\Lambda^{\phantom'}_1}.
\label{e:matrixphi}
\end{equation} 
We will refer to the last term of this relation as the matrix representation
of the correlation function or, more simply, as the correlation
matrix. Fig.~\ref{f:phimatrix} shows pictorially 
the position of the spin indices.

Starting from Eq.~\eqref{e:decomphi} and using the relation
\begin{equation}
\Psi_U + S_L \Psi_L + S_x \Psi_x + S_y \Psi_y =
\rho(S)_{\Lambda^{\phantom'}_1 \Lambda_1'}\;
\begin{pmatrix} 
\Psi_U + \Psi_L   & \Psi_x - \ii \Psi_y \\
\Psi_x + \ii \Psi_y & \Psi_U- \Psi_L	
	\end{pmatrix} ^{\Lambda_1'
\Lambda^{\phantom'}_1}
\label{e:genPsi}
\end{equation} 
we can cast the correlation function in the matrix form
\begin{equation} 
\Bigl({\cal P}_+\, \Phi(x) \g^+ \Bigr)^{\Lambda_1' \Lambda^{\phantom'}_1}
= \begin{pmatrix}
\bigl(f_1(x)+g_1(x)\, \g_5 \bigr)\, {\cal P}_+ & 
		h_1(x)\, \bigl(\g_x - \ii \g_y\bigr)\, \g_5 {\cal P}_+\\ 
h_1(x)\, \bigl(\g_x + \ii \g_y\bigr)\, \g_5 {\cal P}_+ & 
	\bigl(f_1(x) - g_1(x)\, \g_5 \bigr)\, {\cal P_+}	
	\end{pmatrix}.
\end{equation} 
Finally, by expressing the Dirac structures in Weyl representation and
reducing the Dirac space as done
before, we obtain the 
matrix representation of the correlation function 
\renewcommand{\arraystretch}{1}
\begin{equation}
\addtolength{\extrarowheight}{5pt}
 \Bigl({\cal P}_+\, \Phi(x) \g^+ \Bigr)_{\chi'_{1} \chi^{\phantom'}_{1}}^{\Lambda_1' \Lambda^{\phantom'}_1} = \lf(\begin{array}{cc|cc}
f_1(x) + g_1(x) & 0 & 0 &0  \\
	0	 & f_1(x) - g_1(x) &    2 h_1(x) & 0      \\[3pt] \hline 
0     & 2 h_1(x)           &f_1(x) - g_1(x) & 0       \\
	 0     & 0          &                0     &f_1(x) + g_1(x) 
\end{array}\rg),
\label{e:phimatrix}
\end{equation} 
where the inner blocks are in the hadron
helicity space (indices $\Lambda'_{1} \Lambda^{\phantom'}_{1}$), 
while the outer matrix is in the quark chirality space (indices
$\chi'_{1} \chi^{\phantom'}_{1}$). \renewcommand{\arraystretch}{1.5}

The form of the correlation matrix can also 
be established directly from angular
momentum conservation (requiring 
$\Lambda'_{1}+\chi'_{1} =\Lambda_{1} +\chi_{1}$) and the
conditions of Hermiticity and parity invariance. 
In matrix language, the condition of parity invariance consists in~\cite{Jaffe:1996zw}
\begin{equation} 
 \Bigl({\cal P}_+\, \Phi(x) \g^+ \Bigr)_{\chi'_{1} \chi^{\phantom'}_{1}}^{\Lambda_1' \Lambda^{\phantom'}_1} = \Bigl({\cal P}_+\, \Phi(x) \g^+ \Bigr)_{-\chi'_{1}\,-\chi^{\phantom'}_{1}}^{-\Lambda_1'\, -\Lambda^{\phantom'}_1} .
\end{equation} 
The most general form of the correlation matrix complying with the previous
conditions corresponds to Eq.~\eqref{e:phimatrix}.

As mentioned at the end of Sec.~\vref{s:leading}, with transposing the quark 
chirality indices of the correlation matrix we obtain the 
scattering matrix~\cite{Bacchetta:1999kz,Bacchetta:2000zm}
\renewcommand{\arraystretch}{1}
\begin{equation}
\addtolength{\extrarowheight}{5pt}
F(x)_{\chi^{\phantom'}_{1} \chi'_{1}}^{\Lambda_1' \Lambda^{\phantom'}_1} = 
\lf(\begin{array}{cc|cc}
f_1(x) + g_1(x) & 0 & 0 & 2 h_1(x) \\
	0	 & f_1(x) - g_1(x) &    0 & 0      \\[3pt] \hline 
0     & 0           &f_1(x) - g_1(x) & 0       \\
	  2 h_1(x)     & 0          &                0     &f_1(x) + g_1(x) 
\end{array}\rg).
\label{e:Fmatrix}
\end{equation}
Note that because of the inversion of the quark indices, the lower left block
has $\chi'_{1}=R$, $\chi_{1}=L$ and vice versa for the upper right
block. 
Since this matrix must be positive semidefinite, we can readily obtain the
positivity conditions
\renewcommand{\arraystretch}{1.5}
\begin{subequations}\label{e:soffer}
\begin{align} 
f_1(x) &\ge 0 , \\
\lf\lvert g_1(x) \rg \rvert &\le f_1(x), \\
\lf\lvert h_1(x) \rg \rvert &\le \tfrac{1}{2} \bigl(f_1(x) + g_1(x) \bigr).
\end{align} 
\end{subequations}
The last relation is known as the {\em Soffer bound}~\cite{Soffer:1995ww}.

%%%%%%%%%%%%%%%%%%%%%%%%%%%%%%%%%%%%%%%%%%%%%%%%%%%%%%%%%%%%
\section{The transversity distribution function}

\begin{figure}
	\centering
	\mbox{}\\[10pt]
	\rput(0.45,1.5){ \boldmath $R$}
	\rput(3.4,1.5){ \boldmath $R$}
	\rput(-0.2,0){ \boldmath $\frac{1}{2}$}
	\rput(4.1,0){ \boldmath $\frac{1}{2}$}
	\includegraphics[width=4cm]{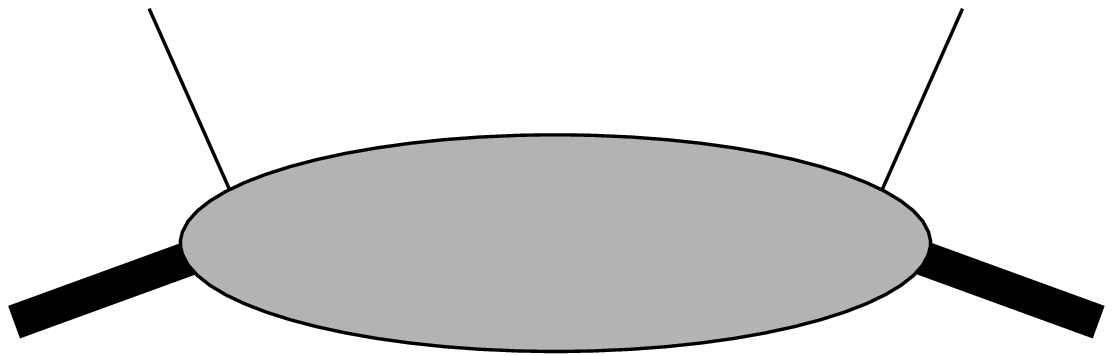}
	\rput(1.4,0.7){\Large \boldmath $+$}
	\hfil
	\rput(0.45,1.5){ \boldmath $L$}
	\rput(3.4,1.5){ \boldmath $L$}
	\rput(-0.2,0){ \boldmath $\frac{1}{2}$}
	\rput(4.1,0){ \boldmath $\frac{1}{2}$}
	\includegraphics[width=4cm]{Figures/scheleton2} \\[10pt]
	(a) \\[25pt]
	\rput(0.45,1.5){ \boldmath $R$}
	\rput(3.4,1.5){ \boldmath $R$}
	\rput(-0.2,0){ \boldmath $\frac{1}{2}$}
	\rput(4.1,0){ \boldmath $\frac{1}{2}$}
	\includegraphics[width=4cm]{Figures/scheleton2}
	\rput(1.4,0.7){\Large \boldmath $-$}
	\hfil
	\rput(0.45,1.5){ \boldmath $L$}
	\rput(3.4,1.5){ \boldmath $L$}
	\rput(-0.2,0){ \boldmath $\frac{1}{2}$}
	\rput(4.1,0){ \boldmath $\frac{1}{2}$}
	\includegraphics[width=4cm]{Figures/scheleton2} \\[10pt]
	(b) \\[5pt]
	\caption{Probabilistic interpretation of the unpolarized
distribution function $f_1$ (a), and of the helicity
distribution function $g_1$ (b).}
	\label{f:f1andg1}
        \end{figure}

\begin{figure}
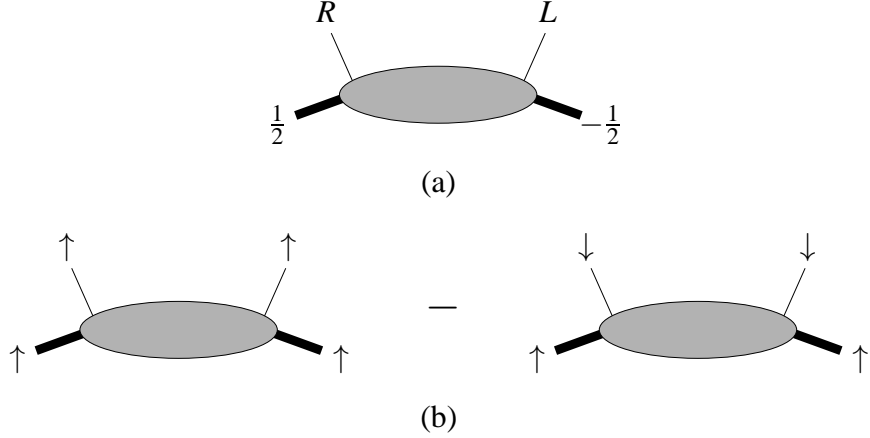

	\centering
	\mbox{}\\[10pt]
	\rput(0.45,1.5){ \boldmath $R$}
	\rput(3.4,1.5){ \boldmath $L$}
	\rput(-0.2,0){ \boldmath $\frac{1}{2}$}
	\rput(4.1,0){ \boldmath $-\frac{1}{2}$}
	\includegraphics[width=4cm]{Figures/scheleton2} \\[10pt]
	(a) \\[25pt]
	\rput(0.45,1.5){ \boldmath $\uparrow$}
	\rput(3.4,1.5){ \boldmath $\uparrow$}
	\rput(-0.2,0){ \boldmath $\uparrow$}
	\rput(4.1,0){ \boldmath $\uparrow$}
	\includegraphics[width=4cm]{Figures/scheleton2}
	\rput(1.4,0.7){\Large \boldmath $-$}
	\hfil
	\rput(0.45,1.5){ \boldmath $\downarrow$}
	\rput(3.4,1.5){ \boldmath $\downarrow$}
	\rput(-0.2,0){ \boldmath $\uparrow$}
	\rput(4.1,0){ \boldmath $\uparrow$}
	\includegraphics[width=4cm]{Figures/scheleton2} \\[10pt]
	(b) \\[5pt]
	\caption{Probabilistic interpretation of the transversity
distribution function $h_1$, in the helicity basis (a) and in the transversity 
basis (b).}
	\label{f:h1inter}
        \end{figure}

In the previous section, we wrote the forward antiquark-nucleon scattering
 matrix, $F$, in the helicity basis of the
hadron and of the good quark (to be more precise, we used the chirality basis
for the quark). 
Each entry, with indices $\chi_1 \Lambda_1'$, $\chi_1' \Lambda_1$ describes 
the product of the amplitude for the scattering of 
an antiquark with helicity (chirality)
$\chi_1'$ off a hadron with helicity $\Lambda_1$ going to anything times the
conjugate of the amplitude for antiquark with helicity
$\chi_1$ off a hadron with helicity $\Lambda_1'$ going to anything.
\begin{equation} 
F(x)_{\chi^{\phantom'}_{1} \chi'_{1}}^{\Lambda_1' \Lambda^{\phantom'}_1}
\propto \sum_{X} 
\int {\de^3 \! \mb{P}_{X}} \,
	\bra{X}(\psi_+)_{\chi_{1}} \ket{P,\Lambda_1'}^{\ast} 
	\bra{X}(\psi_+)_{\chi'_{1}} \ket{P,\Lambda_1}.
\end{equation} 
In this basis, the
probabilistic interpretation of the functions $f_1$ and $g_1$ is manifest,
since they occupy the diagonal elements of the matrix and they are therefore
connected to squares of probability amplitudes (see Fig.~\vref{f:f1andg1})
\begin{align} 
f_1(x) &= \frac{1}{2}\lf(F(x)_{R\,R}^{\smallhalf\, \smallhalf}+ F(x)_{L\,L}^{\smallhalf\, \smallhalf}\rg) &
g_1(x) &= \frac{1}{2}\lf(F(x)_{R\,R}^{\smallhalf\, \smallhalf}- F(x)_{L\,L}^{\smallhalf\, \smallhalf}\rg) 
\end{align} 
On the other hand, the transversity distribution is off-diagonal in the
helicity basis. This means that it does not describe
the square of a probability amplitude, but rather the interference between two 
different amplitudes [see Fig.~\ref{f:h1inter} (a)]
\begin{equation}
h_1(x) = \tfrac{1}{2} F(x)_{R\,L}^{\smallhalf\,{\text{\tiny $-$}\smallhalf}} 
\end{equation} 

The transversity distribution recovers a probability
interpretation if we choose the so-called {\em transversity basis}, instead of 
the helicity basis, for both quark and
hadron~\cite{Jaffe:1996zw,Jaffe:1997yz}. 
%The transversity basis is
%formed by the eigenstates of the operators 
%${\cal P}^i_{\uparrow /\downarrow}=(1 \pm \g^i \g_5)/2$.
The transversity basis is formed by the
``transverse up'' and ``transverse down'' states. They can be expressed in 
terms of chirality eigenstates 
\begin{align} 
u_{\uparrow} &= \tfrac{1}{\sqrt{2}} \lf(u_R +u_L\rg), & 
u_{\downarrow} &= \tfrac{1}{\sqrt{2}} \lf(u_R -u_L\rg).
\end{align} 
The same relation holds between the hadron transversity and helicity states.

In the new basis, the scattering matrix takes the form
\renewcommand{\arraystretch}{1}
\begin{equation}
\addtolength{\extrarowheight}{5pt}
F(x)_{\chi^{\phantom'}_{1} \chi'_{1}}^{\Lambda_1' \Lambda^{\phantom'}_1} = 
\lf(\begin{array}{cc|cc}
f_1(x) + h_1(x) & 0 & 0 & g_1(x)+ h_1(x) \\
	0	 & f_1(x) - h_1(x) &    g_1(x)- h_1(x) & 0      \\[3pt] \hline 
0     &    g_1(x)- h_1(x)        &f_1(x) - h_1(x) & 0       \\
g_1(x)+ h_1(x)     & 0          &                0     &f_1(x) + h_1(x) 
\end{array}\rg),
\label{e:Fmatrix2}
\end{equation}
and clearly the transversity distribution function can be defined as 
[Fig.~\ref{f:h1inter} (b)]
\begin{equation} 
h_1(x) = \frac{1}{2}\lf(F(x)_{\uparrow \uparrow}^{\uparrow \uparrow}-
F(x)_{\downarrow \downarrow}^{\uparrow \uparrow} \rg). 
\end{equation} 
%This probability interpretation of the transversity distribution function is
%valid even if real massless particles cannot exist in the transverse up or
%down state.

The transversity distribution has been the object of several model
calculations, using e.g.\ the bag model~\cite{Jaffe:1992ra}, 
the spectator model~\cite{Jakob:1997wg}, the
chiral soliton model~\cite{Pobylitsa:1996rs} and
others~\cite{Barone:1997un}. 
The integral of $h_1$ -- also known as the {\em tensor charge} of the nucleon
-- has been evaluated in lattice QCD~\cite{Aoki:1997pi}.
A recent review on 
the transversity distribution is presented in Ref.~\citen{Barone:2001sp}.

The transversity distribution
evolves with the energy scale in a different way as compared to the helicity 
distribution, without mixing with
gluons~\cite{Artru:1990zv,Blumlein:2001ca,Baldracchini:1981uq,Kumano:1997qp,Hirai:1998mm}.

\renewcommand{\arraystretch}{1.5}
%%%%%%%%%%%%%%%%%%%%%%%%%%%%%%%%%%%%%%%%%%%%%%%%%%%%%%%%%%%%
\subsection{Transversity and Thomas precession}

In the literature, it is common to find the statement that the difference
between the helicity distribution and the transversity distribution is
connected to relativistic effects, since boosts and rotation do not commute~\cite{Jaffe:2000kr,Jaffe:1996zw,Jaffe:1998hf,Jaffe:1997yz}.
Relativistic effects influence observable
quantities depending on the
dynamics of the system and they can therefore give important information about 
its structure. Therefore, the difference between helicity and
transversity distributions can shed light on the structure of the
nucleon and its spin.

This statement can be understood even
in the framework of purely classical relativistic 
mechanics. We will show how
relativistic effects can change an observable in 
a toy model with just a little bit of dynamical complexity. 
Of course, the model is not meant to describe a nucleon.
We will consider ``spin'' merely as a pseudovector attached to the
quark. ``Helicity'' will be the projection of the spin along the momentum of 
the
quark and ``transversity'' will be the projection transverse to the momentum of
the quark. Note that these quantities are only {\em reminescent} 
of the real helicity and
transversity.

Suppose we have a system constituted by a quark revolving in a circular orbit
with its spin aligned in the direction of the orbital axis.
To have an analogy of the helicity distribution, we boost the system 
to a velocity $\mb{v}$ along the direction of the orbital axis, and 
we measure what is the probability of finding the helicity of the quark
aligned along the axis direction. For the transversity, we
 boost the 
system to a velocity $\mb{v}$ transverse to the axis direction, 
 and measure the
probability of finding the transversity of the quark aligned along the axis
direction. 
	\begin{figure}
	\centering
	\begin{tabular}{cccc}
	\includegraphics[width=7cm]{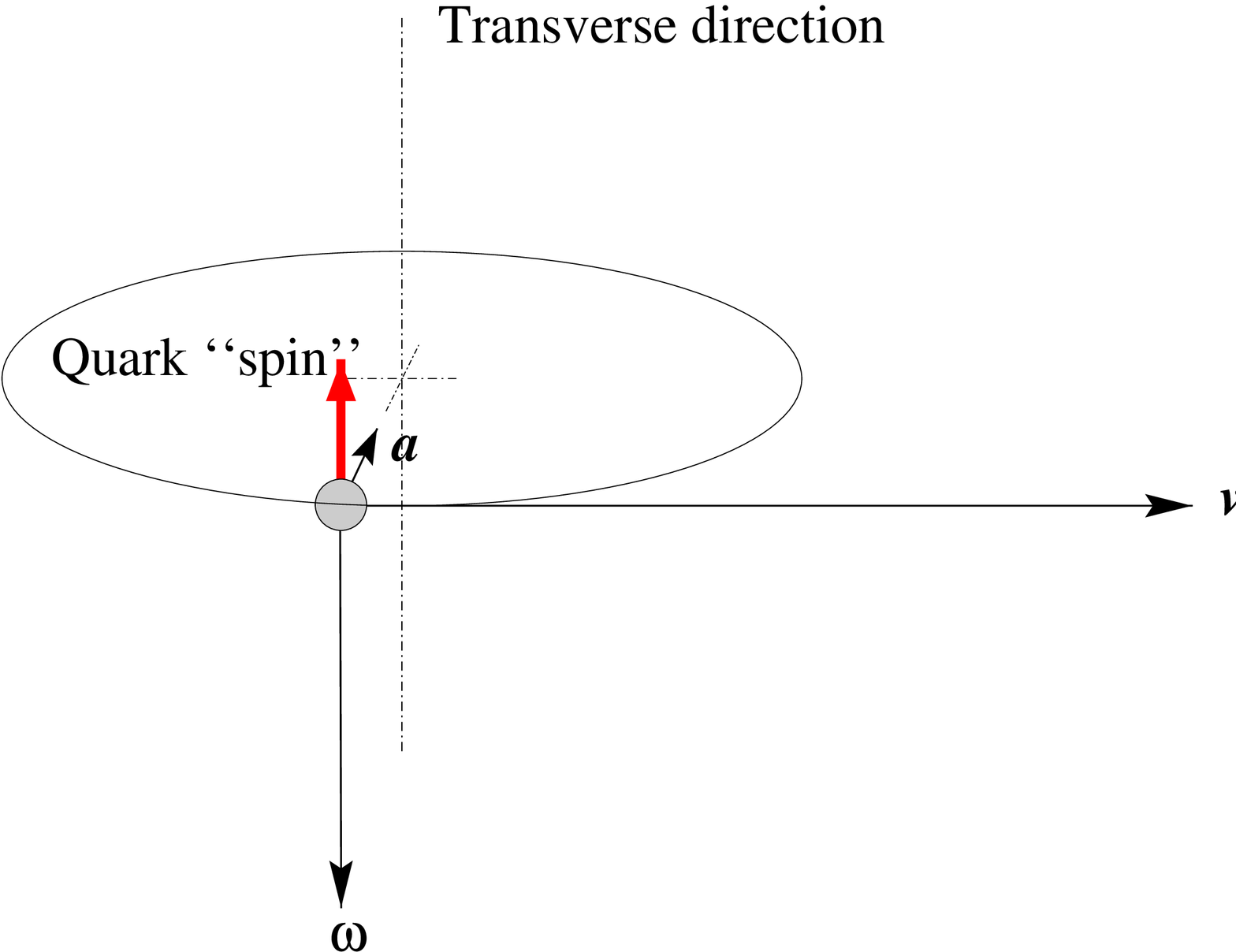}
	&&&\includegraphics[width=7cm]{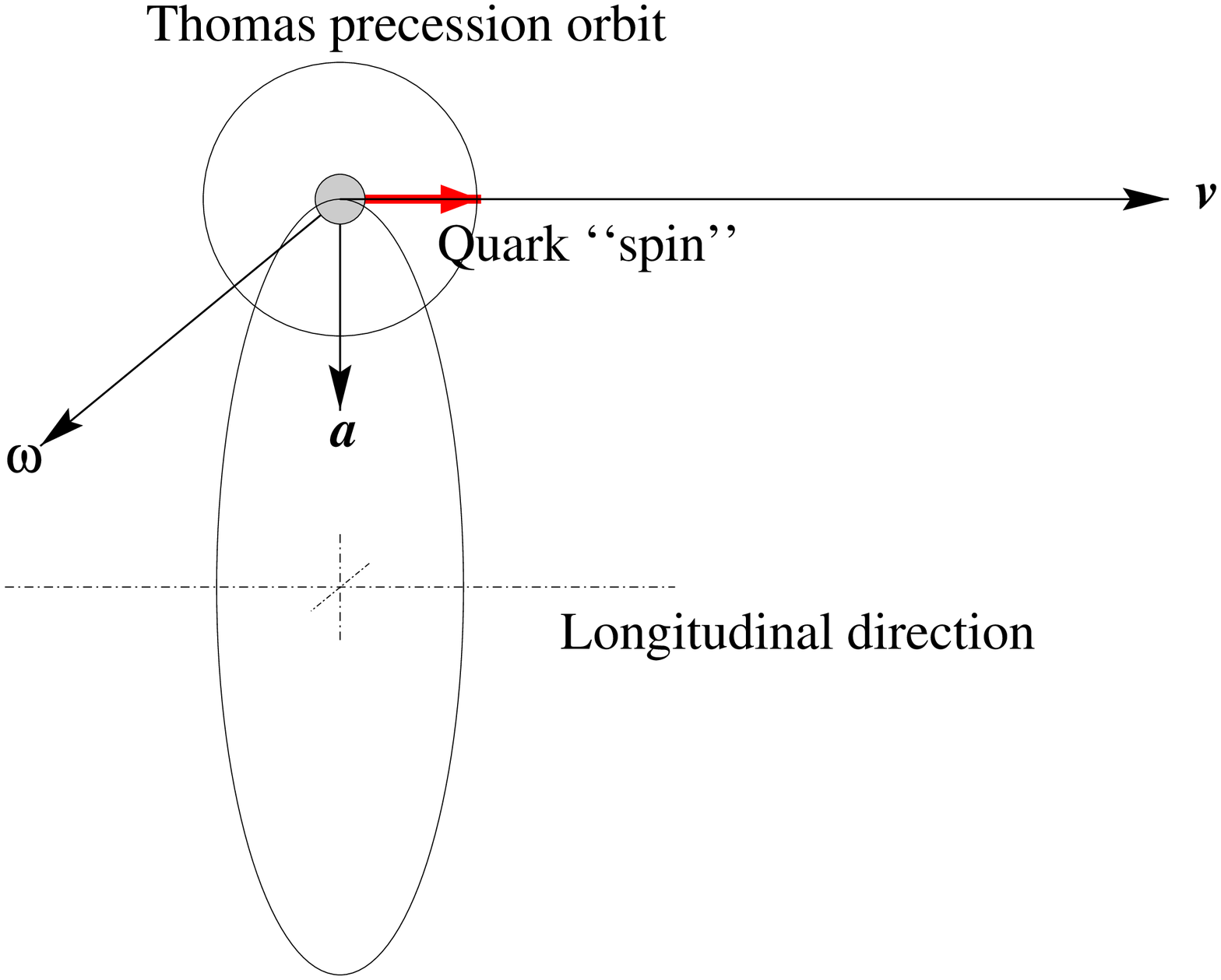} \\
	(a)&&&(b)
	\end{tabular}
	\caption{Thomas precession effect on transversity (a) and helicity
	(b). }
	\label{f:thomas}
        \end{figure}

To properly deal with this situation in a relativistic way, we have to take
into account {\em Thomas precession}, an effect which occurs whenever a
frame of reference
(in our case joined to the quark) is moving with a velocity $\mb{v}$ with 
respect to the
observer and is at the same time subject to
an acceleration $\mb{a}$~\cite{Jackson,Taylor}. 
Thomas precession causes the quark frame of
reference (and the spin of the quark with it) to {\em precess} with an angular 
velocity
\begin{equation} 
\bm{\omega}_T= \frac{\mb{v} \times \mb{a}}{2 c^2}.
\end{equation} 
In the infinite momentum frame, the speed of revolution of the quark is
negligible with respect to the overall velocity of the system, so that the
quark velocity is approximately $\mb{v}$. However we cannot
neglect its centripetal acceleration $\mb{a}$.
 
Let us first analyze the situation with transversity [Fig.~\ref{f:thomas}
(a)].
The cross product of the 
velocity of the quark and the centripetal acceleration is pointing in the
transverse direction. Thomas precession will not influence the orientation of
the spin of the quark. The transversity distribution is just 1.

The situation is dramatically different for helicity [Fig.~\ref{f:thomas}
(b)]. The cross product of 
the quark velocity and its acceleration is still pointing in a transverse
direction, which is now orthogonal to the spin of the quark. This causes
the spin of the quark to precess around a transverse axis. The net
helicity of the quark will be zero and so will be the helicity distribution.

In a simpler system, for instance if the quark would not have any orbital
angular momentum, helicity and transversity distributions would be the
same. 
In conclusion, from this example we see that, due to relativistic effects, 
the difference between two apparently similar observables can reveal something 
important about the structure of a system.

%%%%%%%%%%%%%%%%%%%%%%%%%%%%%%%%%%%%%%%%%%%%%%%%%%%%%%%%%%%%
\section{Inclusion of transverse momentum}
\label{s:inclusion}

So far we have been concerned with the {\em integrated} correlation function,
defined in Eq.~\eqref{e:phiint}, the only relevant one in totally inclusive
deep inelastic scattering. In the next chapters, we will analyze also
semi-inclusive scatterings, where we will need to consider the
transverse-momentum dependent version of the correlation function, i.e.\
\begin{equation} \begin{split} 
\Phi_{ji}(x, \mb{p}_T, S) &= \int \de p^-  \;\Phi_{ji}(p,P,S)
\biggr\vert_{p^+ = x P^+} \\
&=\int \frac{\de \xi^- \de^2 \! \bm{\xi}_T}{(2 \pi)^3} \; \e^{-\ii p \cdot \xi}  
\bra{P,S} \bbar{\psi}_i (\xi)	\; \psi_j (0) \ket{P,S}
	\biggr\vert_{\xi^+ = 0}.
\end{split}
\label{e:phinoint} \end{equation}  

Starting from the general decomposition presented in Eq.~\eqref{e:decomphi1},
the leading order part of the 
transverse-momentum dependent correlation function becomes
\begin{equation} \begin{split}  
{\cal P}_+\, \Phi(x, \mb{p}_\st, S) \g^+ &=
\Biggl\{
f_1(x,\bm p^2_\st)
+ \ii\,h_1^\perp(x,\bm p^2_\st)\,\frac{\pslash_\st}{M}
%\\
%&\quad 
+ S_L\,g_{1L}(x,\bm p^2_\st)\,\gamma_5
+ S_L\,h_{1L}^\perp(x,\bm p^2_\st)\,
\gamma_5\,\frac{\pslash_\st}{M}
\\
&\quad +
f_{1\st}^\perp(x,\bm p^2_\st)\,\frac{\epsilon_{\st \rho \sigma}
S_\st^\rho p_\st^\sigma }{M}
+ g_{1\st}(x,\bm p^2_\st)\,\frac{\bm p_\st\cdot\bm S_\st}{M}
\,\gamma_5
\\
&\quad +
h_{1\st}(x,\bm p^2_\st)\,\gamma_5\,\Sslash_\st
+ h_{1\st}^\perp(x,\bm p^2_\st)\,\frac{\bm p_\st\cdot\bm S_\st}{M}
\,\gamma_5\,\frac{\pslash_\st}{M}
\Biggr\}\, {\cal P}_+\,.
\label{e:phifull}
\end{split} \end{equation}  
The definition of the parton distribution functions in terms of the amplitudes
$A_i$, introduced in Eq.~\eqref{e:decomphi},
 can be found elsewhere~\cite{Levelt,Tangerman,Jakob:1997wg}.

For any transverse-momentum dependent distribution function, 
it will turn out to be convenient to define the notation
\begin{subequations} \begin{align} 
f^{\lf(1/2\rg)}(x,\mb{p}_T^2) &\equiv \frac{\lf\lvert\mb{p}_T\rg\rvert}{2M}\;f(x,\mb{p}^2_T), \\
f^{(n)}(x,\mb{p}^2_T) &\equiv \lf(\frac{{\mb p}_T^2}{2M^2}\rg)^n 
f(x,\mb{p}^2_T), 
\end{align} \end{subequations} 
for $n$ integer.
We also need to introduce the function
\begin{equation} 
h_1(x,\mb{p}^2_T) \equiv h_{1T} (x,\mb{p}^2_T)
        + h_{1T}^{\perp (1)}(x,\mb{p}^2_T).
\label{e:h1trans}
\end{equation} 

The connection with the integrated distribution functions defined in
Eq.~\eqref{e:distri} is
\begin{subequations}
\begin{align} 
f_1(x) &= \int \de^2 \!{\mb p}_T \;f_{1}(x,\mb{p}^2_T),  \\
g_1(x) &= \int \de^2 \!{\mb p}_T \;g_{1L}(x,\mb{p}^2_T),  \\
h_1(x) &= \int \de^2 \!{\mb p}_T \;h_1(x,\mb{p}^2_T). 
\end{align} 
\end{subequations}

Note that the distribution functions $h_1^{\perp}$ and $f_{1 \st}^{\perp}$ are
T-odd. At
first, this class of functions was supposed to vanish due to time-reversal
invariance~\cite{Ralston:1979ys}. Sivers~\cite{Sivers:1990cc,Sivers:1991fh} 
was the first one to consider an observable arising from the T-odd
distribution function $f_{1 \st}^{\perp}$, since then called the {\em
Sivers function}. The complete analysis of leading-twist T-odd distribution
functions and the introduction of $h_1^{\perp}$ were carried out by Boer and
Mulders~\cite{Boer:1998nt}. 
A proof of the nonexistence of T-odd distribution
functions was suggested by Collins in Ref.~\citen{Collins:1993kk}, but recently it
has been repudiated by the same author~\cite{Collins:2002kn} after Brodsky,
Hwang and Schmidt~\cite{Brodsky:2002cx} explicitly obtained a nonzero Sivers
function in the context of a simple model. The question at the moment awaits
clarification, but it is likely that T-odd distribution functions will stir a
lot of interest in the near future, together with
T-odd fragmentation functions, of which we shall abundantly speak in the next
chapters.

As done in the previous section, we can express the transverse momentum
dependent correlation function as a matrix in the parton chirality space
$\otimes$ target helicity space. 
To simplify the formulae, it is useful to identify the T-odd functions
as imaginary parts of some of the T-even functions, which become then complex
scalar functions. The following redefinitions are required:\footnote{From a
rigorous point of view, it would be better to introduce new functions, e.g.\
$\tilde{g}_{1T}$ and $\tilde{h}_{1L}^{\perp}$, but this would overload the
notation.}
\begin{align} 
g_{1T} + \ii f_{1T}^{\perp} &\rightarrow g_{1T}, &  
h_{1L}^{\perp} + \ii h_{1}^{\perp} &\rightarrow h_{1L}^{\perp}.  
\label{e:g1T&h1lperp}
\end{align} 
The resulting correlation matrix
is~\cite{Bacchetta:1999kz,Bacchetta:2000zm}
\begin{equation} 
\renewcommand{\arraystretch}{2}
%\addtolength{\extrarowheight}{10pt}
F(x, \mb{p}_T)
_{\chi^{\phantom'}_{1} \chi'_{1}}^{\Lambda_1' \Lambda^{\phantom'}_1} = 
\lf(\begin{array}{cc|cc}
f_1 + g_{1L} &
\dfrac{\vert \mb{p}_\st\vert}{M}\,\e^{-\ii\phi_p}\,g_{1T}&
\dfrac{\vert \mb{p}_\st\vert}{M}\,\e^{\ii\phi_p}\,h_{1L}^{\perp\ast}&
2\, h_1\\
\dfrac{\vert \mb{p}_\st\vert}{M}\,\e^{\ii\phi_p}\,g_{1T}^\ast&
f_1 - g_{1L} & 
\dfrac{\vert \mb{p}_\st\vert^2}{M^2}\e^{2\ii\phi_p}\,h_{1T}^\perp &
-\dfrac{\vert \mb{p}_\st\vert}{M}\,\e^{\ii\phi_p}\,h_{1L}^{\perp}\\[3pt] \hline
\dfrac{\vert \mb{p}_\st\vert}{M}\,\e^{-\ii\phi_p}\,h_{1L}^{\perp}&
\dfrac{\vert \mb{p}_\st\vert^2}{M^2}\e^{-2\ii\phi_p}\,h_{1T}^\perp &
f_1 - g_{1L} &
-\dfrac{\vert \mb{p}_\st\vert}{M}\,\e^{-\ii\phi_p}\,g_{1T}^\ast  \\
 2\,h_1&
-\dfrac{\vert \mb{p}_\st\vert}{M}\,\e^{-\ii\phi_p}\,h_{1L}^{\perp\ast}& 
-\dfrac{\vert \mb{p}_\st\vert}{M}\,\e^{\ii\phi_p}\,g_{1T} &
f_1 + g_{1L}
\end{array}\rg),
\label{e:Fmatrixfull}
\end{equation} 
where
for sake of brevity we did not explicitly indicate the 
$x$ and $\bm p_\st^2$ dependence of the distribution functions
 and where $\phi_p$ is the azimuthal angle of the transverse momentum vector.
\renewcommand{\arraystretch}{1.5}

The distribution matrix is clearly Hermitean.
Notice that by introducing the transverse momentum of the quark, the angular
momentum conservation requirement becomes less constraining and we can have non
zero contributions in all the entries of the scattering matrix. To be more
specific, when an exponential $\e^{\ii l' \phi_p}$ appears in the matrix, we
have to take into account $l'$ units of angular momentum in the final
state. The condition of angular momentum conservation becomes then
$\Lambda'_{1}+\chi'_{1}+l' =\Lambda_{1} +\chi_{1}$. Also the condition of
parity conservation is influenced by the presence of orbital angular momentum
and becomes
\begin{equation} 
F(x,\mb{p}_T)
_{\chi^{\phantom'}_{1} \chi'_{1}}^{\Lambda_1' \Lambda^{\phantom'}_1} = 
(-1)^{l'}\,F(x,\mb{p}_T)
_{-\chi^{\phantom'}_{1}\, -\chi'_{1}}^{-\Lambda_1'\, -\Lambda^{\phantom'}_1}  
\biggr\rvert_{l'\rightarrow - l'}.
\end{equation}

The positivity of the matrix is not hampered by the introduction of the
transverse momentum dependence, since
\begin{equation} 
\begin{split}
F(x,\mb{p}_T,S)_{\chi^{\phantom'}_{1} \chi'_{1}}^{\Lambda_1' \Lambda^{\phantom'}_1}&= \int \frac{\de \xi^- \de^2\! \bm{\xi}_T}{(2 \pi)^3 \sqrt{2}} \; \e^{-\ii p \cdot \xi}  
\bra{P,\Lambda'_1} (\psi_+)^{\dagger}_{\chi_{1}} (\xi)\; (\psi_+)_{\chi'_{1}} (0) \ket{P,\Lambda_1}
	\biggr\vert_{\xi^+  = 0} \\
&= \frac{1}{\sqrt{2}}\sum_{X} 
\int \frac{\de^3 \! \mb{P}_{X}} {(2 \pi)^3\, 2 P_{X}^0} 
	\bra{X}(\psi_+)_{\chi_{1}} (0) \ket{P,\Lambda_1'}^{\ast} 
	\bra{X}(\psi_+)_{\chi'_{1}} (0) \ket{P,\Lambda_1} \\
&\quad \times \delta \Bigl(\bigl(1-x\bigr)P^+-P^+_{X}\Bigr)\;\delta^{(2)}\Bigl(\mb{p}_T - \mb{P}_{X\,T}\Bigr).
\end{split}
\end{equation} 
Bounds to insure positivity of any matrix element can be
obtained by looking at the one-dimensional and two-dimensional 
subspaces and at 
the eigenvalues of the full matrix.\footnote{Cf. Ref.~\citen{Lu:1997ae} for an 
earlier discussion on positivity bounds for
transverse momentum dependent {\em structure functions}.}
The one-dimensional subspaces give the trivial bounds 
\begin{align} 
f_1(x,\mb p_\st^2) &\ge 0 \, , &  
\lf\lvert g_{1L}(x,\mb p_\st^2)\rg\rvert &\le f_1(x,\mb p_\st^2) \,.
\end{align} 
From the two-dimensional subspaces we get
\begin{subequations}
\begin{align} 
\lf\lvert h_1 \rg\rvert &\le
\frac{1}{2}\left( f_1 + g_{1L}\right)
\le f_1,
\label{e:Soffer}\\
\lf\lvert h_{1T}^{\perp(1)}\rg\rvert &\le
\frac{1}{2}\left( f_1 - g_{1L}\right)
\le f_1,
\\
 \lf\lvert g_{1T}^{(1)}\rg\rvert^2
&\le \frac{\mb p_\st^2}{4M^2}
\left( f_1 + g_{1L}\right)
\left( f_1 - g_{1L}\right)
\le  \lf(f_1^{(1/2)}\rg)^2,
\\
 \lf\lvert h_{1L}^{\perp (1)}\rg\rvert^2
&\le \frac{\mb p_\st^2}{4M^2}
\left( f_1 + g_{1L}\right)
\left( f_1 - g_{1L}\right)
\le \lf(f_1^{(1/2)}\rg)^2,
\end{align} 
\end{subequations}
where, once again, we did not explicitly indicate the $x$ and $\mb p_\st^2$ 
dependence to avoid too heavy a notation.
Besides the Soffer bound of Eq.~\eqref{e:Soffer}, now extended to include the
transverse momentum dependence, new bounds
for the distribution functions are found. 
%In particular, one sees that 
%functions like
%$g_{1T}^{(1)}(x,\mb p_\st^2)$ and $h_{1L}^{\perp (1)}(x,\mb p_\st^2)$ 
%are proportional to $\lf\lvert \mb p_\st\rg\rvert$ for small
%$p_\st$. 

The positivity bounds can be sharpened even further by imposing the positivity
of the eigenvalues of the correlation matrix. The complete analysis has been
accomplished in Ref.~\citen{Bacchetta:1999kz} (see also
Ref.~\citen{Bacchetta:2000zm}).

%%%%%%%%%%%%%%%%%%%%%%%%%%%%%%%%%%%%%%%%%%%%%%%%%%%%%%%%%%%%
\section{Cross section and spin asymmetries}

The cross section for inclusive deep inelastic scattering at leading twist is
expressed by Eq.~\eqref{e:crossinca}. Extracting the target helicity density
matrix as done in Eq.~\eqref{e:matrixphi} and using the matrix $F$, 
the equation becomes
\begin{equation} 
 \frac{\de^3\!\sigma}{\de \xbj \de y \de \phi_S} \approx  \sum_q\;
\rho(S)_{\Lambda^{\phantom'}_1 \Lambda_1'}\; F(x)_{\chi^{\phantom'}_{1} \chi'_{1}}^{\Lambda_1'
\Lambda^{\phantom'}_1}
\Bigl( \frac{\de \sigma^{eq}}{\de y} \Bigr)
	^{\chi^{\phantom'}_{1} \chi'_{1}} \frac{1}{2}.
\end{equation} 
Inserting the expressions of Eq.~\eqref{e:eqinc}, \eqref{e:rho} and
\eqref{e:Fmatrix} into the previous expression leads to the result
\begin{equation} 
\frac{\de^3\!\sigma}{\de \xbj \de y \de \phi_S} \approx 
\frac{2 \alpha^2 }{s \xbj y^2}\,\sum_q e_q^2\,\lf[ \lf(1-y+\frac{y^2}{2}\rg) f_1^q(\xbj)
+ \lambda_e\, S_L\,\lf(y-\frac{y^2}{2}\rg)\, g_1^q(\xbj)\rg],
\label{e:crossdistr}
\end{equation} 
where the index $q$ denotes the quark flavor.

The transversity distribution {\em does not appear} in the cross section for
totally inclusive deep inelastic scattering at leading twist. 
The reason is that it is a chiral 
odd object and in any observable it must be connected to another
chiral odd ``probe''. In inclusive deep inelastic scattering, what probes the
structure of the correlation function is the elementary photon-quark
scattering, which conserves chirality.
 We will see in the next chapters how semi-inclusive deep
inelastic scattering provides the necessary chiral odd partners for the
transversity distribution (i.e.\ chiral odd fragmentation functions).

The r.h.s.\ of Eq.~\eqref{e:crossdistr} is independent of the azimuthal angle
$\phi_S$, so that we can integrate the cross section over this angle. The 
result is 
\begin{equation} 
\frac{\de^2\!\sigma}{\de \xbj \de y} \approx
\frac{4 \pi \alpha^2}{s \xbj y^2}\,\sum_q e_q^2\,\lf[ \lf(1-y+\frac{y^2}{2}\rg) f_1^q(\xbj)
+ \lambda_e S_L\,\lf(y-\frac{y^2}{2}\rg)\, g_1^q(\xbj)\rg].
\label{e:crossint}
\end{equation} 

If the spin of the target is oriented along the
direction of the electron beam, it will have its longitudinal component
oriented along the $-z$ direction, that is $S_L$ will
be negative. On the contrary, orienting the spin of the target in the opposite 
direction will produce a positive $S_L$. 
Summing the cross sections obtained with opposite polarization, we isolate the
unpolarized part of the cross section
\begin{equation}
\de^2\!\sigma_{UU}\equiv\frac{1}{2}\lf(
\de^2\!\sigma_{\rightarrow \leftarrow} + \de^2\!\sigma_{\rightarrow \rightarrow}\rg)
\approx  
\frac{4 \pi \alpha^2}{s \xbj y^2}\,\lf(1-y+\frac{y^2}{2}\rg)\,\sum_q e_q^2\, f_1^q(\xbj).
\label{e:uuf1}
\end{equation} 
The first subscript indicates the polarization of the beam, while the second
stands for the polarization of the target. The letter $U$
stands for unpolarized. The right arrow means polarization
along the beam direction, the left arrow means the opposite. 
Subtracting the cross section we obtain the polarized part 
\begin{equation} 
\de^2\!\sigma_{LL} \equiv \frac{1}{2}\lf(
\de^2\!\sigma_{\rightarrow \leftarrow} - \de^2\!\sigma_{\rightarrow
\rightarrow}\rg)\approx 
\frac{4 \pi \alpha^2}{s \xbj y^2}\,\lvert \lambda_e \rvert\,
\lvert S_L \rvert\,\lf(y-\frac{y^2}{2}\rg)\,
\sum_q e_q^2\, g_1^q(\xbj).
\label{e:llg1}
\end{equation} 
where now the subscript $L$ specifies that longitudinal polarizations of beam
and target are required. 
We can define the longitudinal double spin asymmetry
\begin{equation} 
A_{LL} (\xbj,y) \equiv \frac{\de^2\!\sigma_{LL}}{\de^2\!\sigma_{UU}}
 \approx \lvert \lambda_e \rvert\, \lvert S_L \rvert
 \,\frac{(1/{\xbj y^2})\,\bigl(y-y^2/2\bigr)\,\sum_q e_q^2\, g_1^q(\xbj)}
{(1/{\xbj y^2})\,\bigl(1-y+ y^2/2\bigr)\,\sum_q e_q^2\, f_1^q(\xbj)}.
\end{equation} 

Note that, since the beam direction 
does not exactly correspond to the virtual photon direction, the degree of
{\em longitudinal} polarization of the target
will be somewhat smaller than the degree of
polarization {\em along the beam direction},
while a small transverse polarization
will arise~\cite{Oganessyan:1998ma,Kotzinian:1999dy}. These effects are anyway
$1/Q$ suppressed.

%%%%%%%%%%%%%%%%%%%%%%%%%%%%%%%%%%%%%%%%%%%%%%%%%%%%%%%%%%%%
\section{Summary}

In this chapter, we introduced the hadronic tensor, containing the 
information on the
structure of hadronic targets
 in deep inelastic scattering [cf.\ Eq.~\eqref{e:crossinc}].
We studied the hadronic tensor in the
framework of the parton model at tree level and we came 
to the introduction of a quark-quark
correlation function [cf. Eqs.~\eqref{e:phiint} and~\eqref{e:hadrophi}].
The correlation function can be parametrized in terms of parton distribution
functions.
In particular, at leading order in $1/Q$ (leading twist) we introduced 
the unpolarized distribution function,
 $f_1$, the helicity distribution function, $g_1$,
and the transversity distribution function, $h_1$ [Eq.~\eqref{e:decomphi}].

We demonstrated how the leading twist part of the correlation function can be
cast in the form of a forward scattering matrix [cf.\ Eq.~\eqref{e:Fmatrix}]. 
Exploiting the positivity
of this matrix, we derived relations
among the distribution functions [Eq.~\eqref{e:soffer}].
We also discussed the probabilistic interpretation of the distribution
function, with a particular emphasis on the transversity distribution.

We repeated the analysis of the correlation function 
introducing partonic transverse momentum. In this case the
decomposition of the correlation function contains eight distribution
functions [Eq.~\eqref{e:phifull}]. We have seen that 
each entry of the corresponding scattering matrix is nonzero 
[Eq.~\eqref{e:Fmatrixfull}], indicating that the full quark spin structure in
a polarized nucleon is accessible if transverse momentum is included.
The connection with the helicity formalism and consequently the extraction
of positivity
bounds on transverse momentum dependent distribution functions are among the
original results of our work. 

Finally, we expressed the cross section for inclusive deep inelastic
scattering at leading order in $1/Q$ 
in terms of distribution functions [Eq.~\eqref{e:crossdistr}]. 
We showed that the helicity distribution can be accessed by measuring 
the longitudinal double spin asymmetry, but we concluded that the transversity 
distribution is {\em not accessible} in inclusive deep inelastic
scattering. 

In the next chapter, we are going to see how the transversity distribution can 
be measured in one-particle inclusive deep inelastic scattering, in
combination with the Collins fragmentation function.

\renewcommand{\quot}{%
\parbox{7cm}{Gather up the fragments that remain, that nothing be lost.} \\
  John 6:12
}

%%%%%%%%%%%%%%%%%%%%%%%%%%%%%%%%%%%%%%%%%%%%%%%%%%%%%%%%%%%%%%%%%%%%%%%%%%%%
\chapter[Fragmentation functions and the Collins function]{Fragmentation functions \\ and the Collins function}
\label{c:collins}

In the previous chapter we analyzed totally inclusive deep inelastic
scattering as a means to probe the quark structure of the nucleons. We
concluded that some aspects of this structure -- 
notably the transversity distribution
and the transverse momentum distribution -- are not accessible in
this kind of measurement.
It is therefore desirable to turn the attention to a more complex case,
that of one-particle inclusive deep inelastic scattering, where one of the
fragments produced in the collision is detected. In this case, we
will need to introduce some new nonperturbative objects, the {\em
fragmentation functions}.

%%%%%%%%%%%%%%%%%%%%%%%%%%%%%%%%%%%%%%%%%%%%%%%%%%%%%%%%%%%%
\section{One-particle inclusive deep inelastic scattering}

\begin{figure}
	\centering
	\rput(2.55,5.61){$'$}
	\rput(7,0.44){\scriptsize ${\cal X'}$}
	\includegraphics[width=7cm]{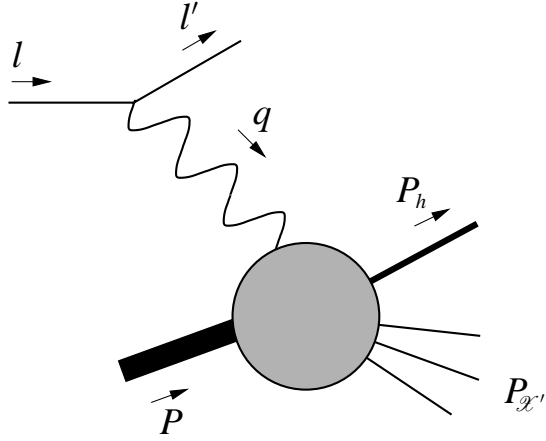}
	\caption{One-particle inclusive deep inelastic scattering.}
	\label{f:1p}
        \end{figure}

In one-particle inclusive deep inelastic scattering a high energy electron
collides on a target nucleon via the exchange of a photon with a high
virtuality. The target breaks up and several hadrons are produced. One of the
produced hadrons is detected in coincidence with the scattered electron (see
Fig.\ \vref{f:1p}).  
As a result of the hardness of the collision, the final state should consist
of two well separated ``clusters'' of particles, one is represented by the
debris of the target, broken by the collision, and the other is represented by
the hadrons formed and ejected by the hard interaction with the virtual
photon~\cite{Trentadue:1994ka}. The former are called {\em target fragments}
and the latter {\em current fragments}. We will take into consideration only
events where the tagged final state hadron belongs to the current fragments.

%%%%%%%%%%%%%%%%%%%%%%%%%%%%%%%%%%%%%%%%%%%%%%%%%%%%%%%%%%%%
\subsection{Kinematics}

As for inclusive scattering, we denote with $l$ and $l'$ the momenta of the
electron before and after the scattering, with $P$, $M$ and $S$ the momentum,
mass and spin of the target nucleon, with $q$ the momentum of the exchanged
photon. Moreover, we introduce the momentum $P_h$ and the mass $M_h$ of the
outgoing hadron.

We define the variables
\begin{align}
\xbj&=\frac{Q^2}{2 P \cdot q}, & y&= \frac{P \cdot q}{P \cdot l}, 
& z_h&=\frac{P \cdot P_h}{P \cdot q}.
\label{e:xyandz}
\end{align} 
We will assume that the following conditions 
hold: $Q^2, P\cdot q,
P \cdot P_h \gg M^2, M_h^2$, and $\xbj, z_h$ fixed. 

For a distinction between
the current and the target fragments, the 
rapidity separation should be taken into account. This separation is quite
clear for high values of $z_h$.
For low values of $z_h$, separating the two clusters becomes
arduous. In events with higher $W^2$, the limit of $z_h$ 
can be pushed lower~\cite{Mulders:2000jt}.

In the frame of reference defined in Sec.~\vref{s:kin1}, the target momentum
and the virtual photon momentum have no transverse components. The outgoing
hadron's momentum can be parametrized as
\begin{equation} 
P_h^{\mspace{2mu}\mu} = \lf[\frac{z_h Q}{A \sqrt{2}},\; \frac{A \lf(M_h^2 + \lvert
\mb{P}_{h \perp}\rvert^2\rg)}{z_h Q \!\sqrt{2}} ,\;
\mb{P}_{h \perp}\rg].
\end{equation}
To avoid the introduction of further hard scales, it is required that 
$P_{h \perp}^2 \ll Q^2$.

The frame of reference we adopted is the more natural one from the
experimental point of view, as the longitudinal direction corresponds (up to
order $1/Q$) to the beam direction and the hadron's transverse momentum
corresponds to what it is actually measured in the lab.  
However, as it will become clear later, in order to preserve a symmetry
between the distribution and fragmentation functions, it is convenient to use
a different frame of reference where the target and outgoing hadron momenta
are collinear, while the photon acquires a transverse component.
To distinguish between the two frames of reference when needed, from now on
 we will use the subscript
$T$ when denoting a transverse component in the new frame, while we will use
the subscript $\perp$ to denote transverse
components in the former frame. 

In the $T$ frame, the external momenta are 
\begin{subequations}\label{e:momenta1p}
\begin{align} 
P^{\mspace{2mu}\mu}& = \lf[\frac{\xbj M^2}{A Q \sqrt{2}},\; \frac{A Q}{\xbj \!\sqrt{2}} ,\;
\mb{0}\rg]  =\lf[\frac{M^2}{2 P^+},\; P^+ ,\; \mb{0}\rg], \\
P_h^{\mspace{2mu}\mu} &= \lf[\frac{z_h Q}{A \sqrt{2}},\; \frac{A M_h^2}{z_h Q \!\sqrt{2}} ,\;
\mb{0}\rg] = \lf[ P_h^-,\; \frac{M_h^2}{2 P_h^-},\; {\bf 0} \rg], 
\label{e:pht}\\
q^{\mspace{2mu}\mu}& = \lf[\frac{Q}{A \sqrt{2}},\; -\frac{A\, \lf(Q^2-\lvert
\mb{q}_{T}\rvert^2\rg)}{Q\sqrt{2}} ,\; \mb{q}_T \rg]
\approx \lf[\frac{P_h^-}{z_h},\; - \xbj P^+
,\; \mb{q}_T\rg], \label{e:qt}
\end{align} 
\end{subequations}
with $P_h^- \approx Q^2 z_h/(2P^+\xbj)$. 
The connection with the transverse momentum components of the photon in the
$T$ frame and of the outgoing hadron in the $\perp$ frame
is~\cite{Mulders:1996dh} 
\begin{equation} 
\mb{q}_T = - \frac{\mb{P}_{h \perp}}{z_h}.
\end{equation}

%%%%%%%%%%%%%%%%%%%%%%%%%%%%%%%%%%%%%%%%%%%%%%%%%%%%%%%%%%%%
\subsection{The hadronic tensor}
\label{s:hadro1p}

The cross section for one-particle inclusive electron-nucleon scattering can
be written as
\begin{equation} 
\frac{2 E_h \de^6\! \sigma}{\de^3\! \mb{P}_h \de \xbj \de y  \de \phi_S}
= \frac{\alpha^2}{2  s \xbj  Q^2}\, L_{\mu \nu} (l,l',\lambda_e) \,2 M W^{\mu \nu}(q,P,S,P_h),
\end{equation} 
or equivalently as
\begin{equation} 
\frac{\de^6\!\sigma}{\de \xbj \de y \de z_h \de \phi_S \de^2\! \mb{P}_{h \perp}}
= \frac{\alpha^2}{4  z_h  s \xbj Q^2} \, 
	L_{\mu \nu}(l, l', \lambda_e)\; 2M W^{\mu \nu}(q, P, S,P_h).
\label{e:cross1pinc}
\end{equation} 
To obtain the previous formula, we made use of the relation
$\de^3\! P_h/2 E_h \approx \de z_h \de^2\! \mb{P}_{h \perp}/2 z_h$.

The hadronic tensor for one-particle inclusive scattering is defined as 
\begin{align} 
2 M W^{\mu \nu}(q,P, S,P_h) 
&=  \frac{1}{(2 \pi)^4} \sum_{\cal X'}  
\int \frac{\de^3 \! \mb{P}_{\cal X'}} {2 P_{\cal X'}^0}
  \;2 \pi\,  \delta^{(4)}\Bigl(q+P-P_{\cal X'}-P_h\Bigr)\, H^{\mu \nu}(P,
S,P_{\cal X'},P_h), \\
H^{\mu \nu}(P,S,P_{\cal X'},P_h) &=
\bra{P,S} J^{\mu} (0) \ket{P_h, {\cal X'}} \bra{P_h, {\cal X'}} J^{\nu} (0) \ket{P,S}.
\end{align} 
By integrating over the momentum of the final-state hadron and 
summing over all
possible hadrons, we recover the hadronic tensor for totally 
inclusive scattering
\begin{equation}
\begin{split} 
\sum_{h} \int \frac{\de^3 \! P_{h}}{2 P_{h}^0}\;  2 M W^{\mu \nu}(q,P, S,P_h) 
&= \frac{1}{2 \pi}
\sum_{\cal X} \int \frac{\de^3 \! P_{\cal X}} {2 P_{\cal X}^0} 
\;2 \pi\,  \delta^{(4)}\Bigl(q+P-P_{\cal X}\Bigr) \,
H^{\mu \nu} (P, S,P_{\cal X}) \\
&\equiv  2 M W^{\mu \nu} (q,P, S),
\end{split}
\end{equation} 
where now the state $\cal X$ indicates the sum of the states $\cal X'$ and
$h$,  
and $P_{\cal X} = P_{\cal X'}+P_h$.
Note that we did not include any dependence of the hadronic tensor on the
polarization of the final state hadron. The reason is that this polarization
is usually 
measured through the decay of the final state hadron into other hadrons.
In this sense, this case 
falls within 
the context of two-particle (or even three-particle)
inclusive scattering, as we will show in the next chapter.

\begin{figure}
	\centering
	\includegraphics[width=8cm]{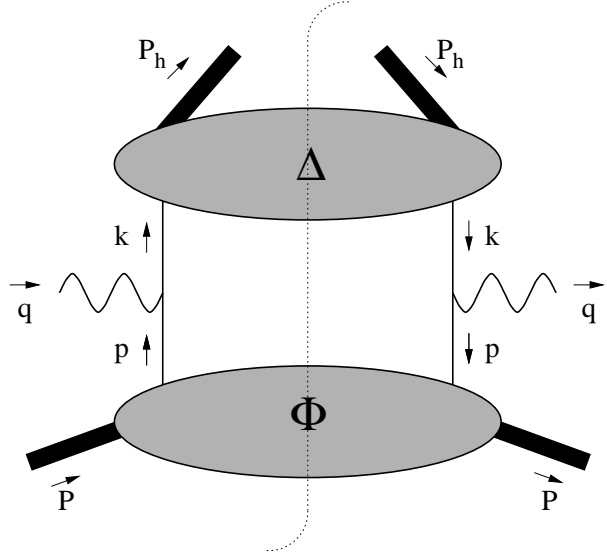}
	\caption{The bull diagram, describing the hadronic tensor at tree
	level.} 
	\label{f:semiinc}
        \end{figure}

In the spirit of the parton model,
the virtual photon strikes a  quark inside the
nucleon. In the case of current fragmentation, the tagged final state hadron
comes from the fragmentation of the struck quark. The scattering process can
then be factorized in two soft hadronic parts connected by a hard scattering
part, as shown in
Fig.~\vref{f:semiinc}.

We need to introduce a parametrization for the vectors
\begin{subequations}\label{e:pk1p}
\begin{align} 
p^{\mspace{2mu}\mu} &= \lf[\frac{p^2 + |\mb{p}_T|^2}{2 x P^+},\; x P^+ ,\;\mb{p}_T\rg], \\
k^{\mspace{2mu}\mu} &= \lf[\frac{P_h^-}{z},\;\frac{z \lf(k^2 + |\mb{k}_T|^2\rg)}{2 P_h^-},\;\mb{k}_T\rg].
\end{align} 
\end{subequations}
Without expliciting including the antiquark contributions, 
the hadronic tensor can be written at
tree level as
\begin{equation}
2 M W^{\mu \nu}(q,P,S, P_h) 
= \sum_q e_q^2  
\int \de^4 \! p \de^4 \! k \;\delta^{(4)}\lf(p+q-k\rg)\, 
\tr \lf(\Phi(p,P,S)\, \g^{\mspace{2mu}\mu} \,\Delta(k,P_h)\, \g^{\nu} \rg),
\end{equation} 
where $\Phi$ is the correlation function defined in Eq.~\eqref{e:phi1} and
$\Delta$,
\begin{equation} 
\begin{split}
\Delta_{kl}(k, P_h)
 &= \frac{1}{(2\pi)^4} \int \de^4 \! \xi\;\e^{\ii k \cdot \xi }\,
\bra{0} \psi_k (\xi) \ket{P_h}
	\bra{P_h} \bbar{\psi}_l (0) \ket{0} \\
&=\sum_{Y} 
\int \frac{\de^3 \! \mb{P}_{Y}} {(2 \pi)^3\, 2 P_{Y}^0}\;\bra{0} \psi_k (0)
\ket{P_h,Y}  
	\bra{P_h,Y} \bbar{\psi}_l (0) \ket{0}
\;\delta^{(4)}\Bigl(k-P_h-P_{Y}\Bigr),
\end{split}
\label{e:delta}
\end{equation} 
is a new correlation function we need to introduce to describe the
fragmentation process~\cite{Collins:1982uw}.

Neglecting terms which are $1/Q$ suppressed, we obtain the compact expression
\begin{equation} 
 2 M W^{\mu \nu} (q, P,S,P_h) 
= 	4 z_h \,{\cal I} \Bigl[ \tr{\lf(\Phi (\xbj,\mb{p}_T, S)\, \g^{\mspace{2mu}\mu}\,
		\Delta (z_h, \mb{k}_T) \,\g^{\nu}  \rg)} \Bigr],
\label{e:1phadrontrans}
\end{equation} 
where, as we shall do very often, we used the shorthand notation
\begin{equation}
{\cal I}\Bigl[\dotsb \Bigr] \equiv
  \int \de^2 \! \mb{p}_T \de^2 \mb{k}_T \,\delta^{(2)}\Bigl(\mb{p}_T
	+\mb{q}_T-\mb{k}_T\Bigr) \;\Bigl[\dotsb \Bigr]
= \int \de^2 \! \mb{p}_T \de^2 \mb{k}_T \,\delta^{(2)}\Bigl(\mb{p}_T
	-\frac{\mb{P}_{h \perp}}{z}-\mb{k}_T\Bigr) \;\Bigl[\dotsb \Bigr],
\label{e:convolution}
\end{equation} 
and where we introduced the integrated correlation functions
\begin{subequations} 
\begin{align}  
\Phi (x, \mb{p}_T, S) &\equiv \int \de p^- \Phi (p,P, S)\Bigr \vert_{p^+ = x P^+}, \\
\Delta (z, \mb{k}_T) &\equiv \frac{1}{4z} \int \de k^+ \Delta (k,P_h)
		\Bigr \vert_{k^- = P_h^- / z}.
\label{e:deltaint}
\end{align}  
\end{subequations} 

Often in experimental situations the transverse momentum of the outgoing
hadron is not measured. This corresponds to integrating 
over the outgoing hadron's transverse momentum, $\mb{P}_{h \perp}$. 
As we already pointed out, this
transverse momentum squared 
has to be small compared to $Q^2$. Experimentally, this
can be insured by imposing a cut-off on the data. Due to the $\delta$
distribution in Eq.~\eqref{e:convolution}, this condition implies also
$(\mb{p}_T+\mb{k}_T)^2 \ll Q^2$.
Considerations as the one discussed in Sec.~\vref{s:hadro1}, 
can be applied to the perturbative and intrinsic components of the transverse
momenta. 

The integration of the cross section yields
\begin{equation} 
\frac{\de^4\! \sigma}{\de \xbj \de y \de z_h \de \phi_S} 
 =  \frac{\alpha^2}{4 z_h s \xbj\,Q^2} \, 
	L_{\mu \nu}(l, l', \lambda_e)\; 2M W^{\mu \nu}(q, P, S),
\end{equation} 
where
\begin{subequations} \begin{align} 
 2 M W^{\mu \nu} (q, P, S) 
 & =  4 z_h \, \tr{\lf(\Phi (\xbj, S)\, \g^{\mspace{2mu}\mu}\,
		\Delta (z_h) \,\g^{\nu}  \rg)}, \\
\Phi (x, S) &\equiv \int \de p^- \de^2\! \mb{p}_T \,
	\Phi (p,P, S)\Bigr \vert_{p^+ = x P^+} ,  \\
\Delta (z) &\equiv \frac{z}{4} \int \de k^+ \de^2\! \mb{k}_T\,
	\Delta (k,P_h)
		\Bigr \vert_{k^- = P_h^- / z}. 
\label{e:deltaint2}
\end{align} \end{subequations}

%%%%%%%%%%%%%%%%%%%%%%%%%%%%%%%%%%%%%%%%%%%%%%%%%%%%%%%%%%%%
\subsection{Leading twist part and connection with helicity formalism}
\label{s:1pheltwist}

Using the projectors $\cal P_+$ and $\cal P_-$ defined in
Eq.~\eqref{e:goodbad}, we can isolate the leading order part of the hadronic
tensor, analogously to what was done in Sec.~\vref{s:leading}
\begin{equation} 
\begin{split} 
 2 M W^{\mu \nu} (q, P, S,P_h) 
 & \approx  4 z_h \, 
	{\cal I} \Biggl[\tr{\lf({\cal P}_+\,\Phi (\xbj,\mb{p}_T, S)\;{\cal P}_-\; \g^{\mspace{2mu}\mu}\;
	{\cal P}_-\,\Delta (z_h,\mb{k}_T) \;{\cal P}_+\; \g^{\nu}  \rg)}\Biggr] \\
 & =  4 z_h \, {\cal I} \lf[
	\tr{\lf({\cal P}_+\,\Phi (\xbj,\mb{p}_T, S)\, \g^+ \; \frac{\g^- \g^{\mspace{2mu}\mu}}{2}\,{\cal P}_-\;{\cal P}_-\,
	\Delta (z_h,\mb{k}_T)\, \g^- \; \frac{\g^+ \g^{\nu}}{2}\,{\cal P}_+  \rg)}\rg]. 
\end{split}
\end{equation}
The differential cross section becomes
\begin{equation} 
\begin{split} 
\lefteqn{\frac{\de^6\!\sigma}{\de \xbj \de y \de z_h \de \phi_S \de^2\!\mb{P}_{h\perp}}} \\
 &\approx \sum_q\,\frac{\alpha^2 e_q^2}{s \xbj Q^2}\, L_{\mu \nu}(l, l', \lambda_e)\;{\cal I} \lf[
 \tr{\lf({\cal P}_+\,\Phi (\xbj, \mb{p}_T, S)\, \g^+ \; \frac{\g^- \g^{\mspace{2mu}\mu}}{2}\,{\cal P}_-\;{\cal P}_-\,\Delta (z_h, \mb{k}_T)\,
		 \g^- \; \frac{\g^+ \g^{\nu}}{2}\,{\cal P}_+  \rg)}\rg] \\
&\approx \sum_q\,{\cal I} \lf[\Bigl({\cal P}_+\, \Phi(\xbj, \mb{p}_T, S) \g^+ \Bigr)_{ij} 
% \\
%&\quad \mbox{}\times 
\;\frac{\alpha^2 e_q^2}{s \xbj Q^2}\, L_{\mu \nu} (l, l',
\lambda_e) 
	\lf( \frac{\g^- \g^{\mspace{2mu}\mu}}{2}\,{\cal P}_-\rg)_{jl} 
	\lf( \frac{\g^+ \g^{\nu}}{2}\,{\cal P}_+\rg)_{mi} 
	\Bigl( {\cal P}_-\,\Delta (z_h, \mb{k}_T)\,
		 \g^- \Bigr)_{lm}\rg].
\end{split}
\end{equation}
As we already discussed in Sec.~\vref{s:helicity}, the restriction to the
leading order allows us to reduce the four dimensional Dirac space to the
two-dimensional subspace of good fields. Writing all the components of the
cross section in the chirality space of the good fields, we obtain
\begin{equation} \begin{split} 
\frac{\de^6\!\sigma}{\de \xbj \de y \de z_h \de \phi_S \de^2\! \mb{P}_{h \perp}} &= 
  \rho(S)_{\Lambda^{\phantom'}_1 \Lambda_1'}\;{\cal I} \lf[ \Bigl({\cal
P}_+\, \Phi(\xbj, \mb{p}_T) \g^+ \Bigr)_{\chi'_{1} \chi^{\phantom'}_{1}}^{\Lambda_1'
\Lambda^{\phantom'}_1}
 \Bigl( \frac{\de \sigma^{eq}}{\de y} \Bigr)
	^{\chi^{\phantom'}_{1} \chi'_{1} ; \,\chi^{\phantom'}_{2} \chi'_{2}}
	\Bigl({\cal P}_-\,\Delta(z_h)\, \g^- \Bigr)
	_{\chi'_{2} \chi^{\phantom'}_{2}}\rg] \\
&\equiv  \rho(S)_{\Lambda^{\phantom'}_1 \Lambda_1'}\;{\cal I} \lf[
F(\xbj, \mb{p}_T)_{\chi^{\phantom'}_{1} \chi'_{1}}^{\Lambda_1' \Lambda^{\phantom'}_1}
\Bigl( \frac{\de \sigma^{eq}}{\de y} \Bigr)
	^{\chi^{\phantom'}_{1} \chi'_{1} ; \,\chi^{\phantom'}_{2} \chi'_{2}}
	D(z_h, \mb{k}_T)_{\chi'_{2} \chi^{\phantom'}_{2}}\rg]. 
\label{e:1pcross}
\end{split} \end{equation}  
The elementary electron-quark scattering matrix is 
\renewcommand{\arraystretch}{1}
\begin{equation} \begin{split} 
\Bigl( \frac{\de \sigma^{eq}}{\de y} \Bigr)
	^{\chi^{\phantom'}_{1} \chi'_{1} ; \,\chi^{\phantom'}_{2} \chi'_{2}} &=
	\frac{\alpha^2 e_q^2}{s \xbj Q^2}\, L_{\mu \nu} (l, l',\lambda_e) 
	\lf( \frac{\g^- \g^{\mspace{2mu}\mu}}{2}\,{\cal P}_-\rg)^{\chi^{\phantom'}_{1} \chi^{\phantom'}_{2}} 
	\lf( \frac{\g^+ \g^{\nu}}{2}\,{\cal P}_+\rg)^{\chi'_{2}
\chi'_{1} } \\[3mm]
&=\frac{2 \alpha^2 e_q^2}{s \xbj y^2}\,
\lf(\begin{array}
%{c@{\hspace{1cm}}c@{\hspace{5mm}}|@{\hspace{5mm}}c@{\hspace{1cm}}c}
{cc|cc}
A(y)+\lambda_e C(y)  & 0 & 0 &-B(y)  \\
	0	 & 0&    0 & 0      \\ \hline 
0     & 0          &0 & 0       \\
-B(y)	   & 0          &                0     &A(y)-\lambda_e C(y)
\end{array}\rg),
\end{split}
\label{e:eq} \end{equation}  \renewcommand{\arraystretch}{1.5}
where
\begin{align} 
A(y) &= 1 - y + \frac{y^2}{2},    & B(y) &= (1-y), & C(y) &=y \lf(1-\frac{y}{2}\rg)  .
\end{align}  
The internal blocks have indices
$\chi'_{1}\; \chi^{\phantom'}_{1}$ and the outer matrix has indices
$\chi'_{2}\; \chi^{\phantom'}_{2}$. 
Notice the difference between Eq.~\eqref{e:eq} and Eq.~\eqref{e:eqinc}: in the 
latter, the presence of the Kronecker delta was identifying two of the 
chirality
indices, thus reproducing a true scattering matrix. Here, we don't have just a 
quark line connecting the two scattering amplitudes, but rather the
correlation function $\Delta$. For this reason, we cannot identify the
chirality indices of the outgoing quark. Strictly speaking, this is not a
scattering matrix anymore, but a scattering amplitude times the conjugate of a 
different scattering amplitude~\cite{Anselmino:1996vq}. 
However, for conciseness we follow the 
notation of, e.g., Ref.~\citen{Jaffe:1998hf}.

The leading order part of the correlation function corresponds to the 
transition probability density
for the process of a quark decaying into hadrons
\begin{equation} 
\begin{split}
\Bigl({\cal P}_-\,\Delta(z, \mb{k}_T)\, \g^- \Bigr)
	_{\chi'_{2} \chi^{\phantom'}_{2}}\hspace{-1.5mm}\equiv D(z, \mb{k}_T)_{\chi'_{2}
\chi^{\phantom'}_{2}}\hspace{-1.5mm} &=
\int \frac{\de \xi^+ \de^2\! \bm{\xi}_T}{(2 \pi)^3 \sqrt{2}} \; \e^{\ii k \cdot \xi}  
\bra{0} (\psi_-)_{\chi'_{2}} (\xi)\ket{P_h}
	\bra{P_h} (\psi_-)^{\dagger}_{\chi_{2}} (0) \ket{0}
	\biggr\vert_{\xi^- = 0} \\
&= \frac{1}{\sqrt{2}}\sum_{Y} 
\int \frac{\de^3 \! \mb{P}_{Y}} {(2 \pi)^3\, 2 P_{Y}^0}\;
\bra{0} (\psi_-)_{\chi'_{2}} (0) \ket{P_h,Y}  
\bra{0} (\psi_-)_{\chi_{2}} (0) \ket{P_h,Y}^{\ast}  \\
&\quad \times \delta
\Bigl(\bigl(1/z-1\bigr)P_h^--P^-_{Y}\Bigr)\;\delta^{(2)}\lf(\mb{k}_T -
\mb{P}_{Y T}\rg).
\end{split}
\end{equation} 
Analogously to the distribution case, 
the fragmentation correlation matrix is positive semidefinite, allowing us to
set bounds on the fragmentation functions.

%%%%%%%%%%%%%%%%%%%%%%%%%%%%%%%%%%%%%%%%%%%%%%%%%%%%%%%%%%%%
\section{The correlation function $\Delta$}
\markright{The correlation function $\mathit \Delta$}

While the distribution correlation function describe 
the {\em confinement} of partons
inside hadrons, the fragmentation correlation function describes the
way a virtual parton  ``decays'' into a hadron plus something else, i.e.\ $q^*
\rightarrow h  Y$. 
This process is referred to as {\em hadronization}.
It is a clear
manifestation of color confinement: the asymptotic physical states detected in 
experiment must be color neutral, so that quarks have to evolve into 
hadrons.\footnote{Note that on the way to the final state hadrons, 
the color carried by the
initial quark can be neutralized without breaking factorization, 
for instance via soft gluon
contributions.}  Nowadays, we can count on reliable phenomenological descriptions of
hadronization, such as the Lund model.  
On the other side, understanding
it from first principles, as well as including  
spin degrees of freedom, is very difficult. 
 
As on the distribution side the quark-quark correlation function is sufficient
to describe the dynamics at leading order in $1/Q$, also for
fragmentation it is sufficient to consider quark-quark correlation functions.

The procedure for generating a complete decomposition of the correlation
functions closely follows what has been done on the distribution side in
Sec.~\vref{s:districolle}. It is necessary to combine the Lorentz vectors $k$
and $P_h$ with a basis of structures in Dirac space, and impose the condition
of Hermiticity and parity invariance. The outcome is
\begin{equation} 
\Delta(k,P_h)  =
                M_h\,B_1\,{\mb 1} + B_2\,\Pslash_h + B_3\,\kslash
                +\frac{B_{4}}{M_h}\,\sigma_{\mu \nu} P_h^{\mspace{2mu}\mu} k^{\nu}.
					\label{e:decomdelta1}  
\end{equation}
The amplitudes  $B_i$ are dimensionless real scalar functions 
$B_i=B_i(k\cdot P_h,k^2)$. The T-even and T-odd part of the correlation
function $\Delta$ can be defined in analogy to Eqs.~\eqref{e:teventodd}. 
According to 
those definitions, the last term can be classified as T-odd.

At leading twist, we are interested in the projection 
${\cal P}_-\,\Delta(z, \mb{k}_T)\, \g^- $. The insertion of the decomposition
given in Eq.~\eqref{e:decomdelta1} into Eq.~\eqref{e:deltaint} and the
subsequent projection of the leading-twist component leads to
\begin{equation}  
{\cal P}_-\,\Delta(z, \mb{k}_T)\, \g^- = \frac{1}{2}\lf(
D_1(z,z^2 \mb{k}^2_\st)
+ i\,H_1^\perp(z,z^2 \bm k^2_\st)\,\frac{\kslash_\st}{M_h} \rg) {\cal P}_-\,,
\label{e:decompdelta}
\end{equation} 
where we introduced the {\em parton fragmentation functions}
\begin{subequations}
\begin{align} 
D_1 (z, z^2 \mb{k}_T^2) &=\frac{1}{2z}\int \de k^2 \de (2 k \cdot P_h)\;\delta\lf(\mb{k}_T^2 + \frac{M_h^2}{z^2} + k^2 -\frac{2 k \cdot P_h}{z} \rg)\; \lf[B_2 + \frac{1}{z} B_3 \rg] , \\
H_1^{\perp} (z, z^2 \mb{k}_T^2) &=\frac{1}{2z} \int \de k^2 \de (2 k \cdot P_h)\;\delta\lf(\mb{k}_T^2 + \frac{M_h^2}{z^2} + k^2 -\frac{2 k \cdot P_h}{z} \rg)\; \lf[-B_4\rg].
\end{align} 
\end{subequations}
The fragmentation function $H_1^{\perp}$ is known with the name of {\em
Collins function}~\cite{Collins:1993kk}. 

The individual fragmentation functions can be isolated by means of the
projection\footnote{The absence of the factor 1/2 in Eq.~\eqref{e:qui} as 
compared to Eq.~\eqref{e:li} is due to the absence of an averaging over 
initial states.}
\begin{equation} 
\Delta^{[\Gamma]} \equiv \tr\lf(\Delta \Gamma \rg),
\label{e:qui}
\end{equation} 
where $\Gamma$ stands for a specific Dirac structure. In particular, we see
that
\begin{subequations} \label{e:projfrag}
\begin{align} 
D_1 (z, z^2 \mb{k}_T^2) &=\Delta^{[\g^-]}(z, \mb{k}_T) ,\\
\frac{\eps_T^{ij} k_{T\,j}}{M_h}H_1^{\perp} (z, z^2 \mb{k}_T^2) &=\Delta^{[\ii 
\sigma^{i -} \g_5]}(z, \mb{k}_T) .
\end{align} \end{subequations} 
As we have done with the distribution functions, it will be helpful to introduce the
notation
\begin{subequations}
\begin{align} 
D^{\lf(1/2\rg)}(z,z^2 \mb{k}^2_T) &\equiv
\frac{\lf\lvert\mb{k}_T\rg\rvert}{2M_h}\;D(z,z^2 \mb{k}^2_T), \\
D^{(n)}(z,z^2 \mb{k}^2_T) &\equiv \lf(\frac{{\mb
k}_T^2}{2M_h^2}\rg)^n D(z,z^2 \mb{k}^2_T), 
\end{align} 
\end{subequations}
for $n$ integer.

%%%%%%%%%%%%%%%%%%%%%%%%%%%%%%%%%%%%%%%%%%%%%%%%%%%%%%%%%%%%
\subsection{Correlation function in helicity formalism}
\label{s:helicityfrag}

\begin{figure}
	\centering
	\rput(-0.25,0.1){\large \boldmath $\chi'_2$}
	\rput(5.3,0.1){\large \boldmath $\chi^{\phantom'}_2$}
	\includegraphics[width=5cm]{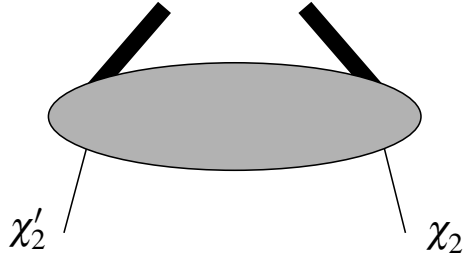}\\
	\caption{Illustration of the position of the indices of the
correlation matrix.}
	\label{f:deltamatrix}
        \end{figure}

Expressing the Dirac matrices of Eq.~\eqref{e:decompdelta}
in the chiral or Weyl representation, as done in Sec.~\vref{s:helicity}, we get
for the leading twist part of the correlation function the expression
\renewcommand{\arraystretch}{1}
\begin{equation} 
\Bigl({\cal P}_-\,\Delta(z, \mb{k}_T)\, \g^- \Bigr)
	_{kl}=\frac{1}{2}
\begin{pmatrix}
0&0&0&0 \\
0 & D_1(z,z^2 \mb{k}_T^2)&\ii \e^{\ii \phi_k} \,\dfrac{\lf\lvert \mb{k}_T \rg \rvert}{M_h}\,H_1^{\perp}(z,z^2 \mb{k}_T^2) &0  \\
0& -\ii \e^{-\ii \phi_k}\, \dfrac{\lf\lvert \mb{k}_T \rg
\rvert}{M_h}\,H_1^{\perp}(z,z^2 \mb{k}_T^2) & D_1(z,z^2 \mb{k}_T^2) & 0\\
0&0&0&0
\end{pmatrix}.
\end{equation}
Restricting ourselves to the subspace of good quark fields and using the
chirality basis, we can rewrite the correlation function as
\renewcommand{\arraystretch}{1.5}
\begin{equation} 
D(z, \mb{k}_T)_{\chi'_{2} \chi^{\phantom'}_{2}}=
\frac{1}{2}\begin{pmatrix}
D_1(z,z^2 \mb{k}_T^2)&\ii \e^{\ii \phi_k} \,\dfrac{\lf\lvert \mb{k}_T \rg \rvert}{M_h}\,H_1^{\perp}(z,z^2 \mb{k}_T^2)  \\
 -\ii \e^{-\ii \phi_k} \,\dfrac{\lf\lvert \mb{k}_T \rg
\rvert}{M_h}\,H_1^{\perp}(z,z^2 \mb{k}_T^2) & D_1 (z,z^2 \mb{k}_T^2)
\end{pmatrix}.
\label{e:Dmatrix}
\end{equation}
Fig.~\vref{f:deltamatrix} shows diagrammatically the position of the indices 
of the correlation
function.
Besides being Hermitean, 
the matrix fulfills the properties of angular momentum
conservation and parity invariance. Because of the presence of factors
$\e^{\ii l \phi_k}$, we have to take into account $l$ units of angular
momentum in the initial state, therefore the condition of angular momentum
conservation is $\chi'_{2}=\chi_{2}+l$ and the condition of parity invariance
is
\begin{equation} 
D(z, \mb{k}_T)_{\chi'_{2} \chi^{\phantom'}_{2}}= 
(-1)^{l}\,D(z, \mb{k}_T)_{-\chi'_{2}\; -\chi^{\phantom'}_{2}}
\Bigr\rvert_{l\rightarrow - l}.
\end{equation} 

Positivity of the correlation matrix implies the bounds
\begin{subequations}
\begin{align} 
D_1(z,z^2 \mb{k}_T^2) &\ge 0, \\
\lf \lvert H_1^{\perp (1)}(z,z^2 \mb{k}_T^2)\rg \rvert &
\le D_1^{(1/2)} (z,z^2 \mb{k}_T^2) . \label{e:boundcollins}
\end{align} 
\end{subequations}

%%%%%%%%%%%%%%%%%%%%%%%%%%%%%%%%%%%%%%%%%%%%%%%%%%%%%%%%%%%%
\section{Cross section and asymmetries}

In order to obtain the cross section for one-particle inclusive deep inelastic
scattering,
the expressions for the spin density matrix of the target, Eq.~\eqref{e:rho},
for the distribution correlation matrix,
Eq.~\eqref{e:Fmatrix}, and for the fragmentation correlation matrix
Eq.~\eqref{e:Dmatrix} have to be inserted in the formula for the cross
section
\begin{equation} 
\frac{\de^6\!\sigma}{\de \xbj \de y \de z_h \de \phi_S \de^2\! \mb{P}_{h \perp}} = 
\sum_q\, \rho(S)_{\Lambda^{\phantom'}_1 \Lambda_1'}\;{\cal I} \lf[
F(\xbj, \mb{p}_T)_{\chi^{\phantom'}_{1} \chi'_{1}}^{\Lambda_1' \Lambda^{\phantom'}_1}
\Bigl( \frac{\de \sigma^{eq}}{\de y} \Bigr)
	^{\chi^{\phantom'}_{1} \chi'_{1} ; \,\chi^{\phantom'}_{2} \chi'_{2}}
	D(z_h, \mb{k}_T)_{\chi'_{2} \chi^{\phantom'}_{2}}\rg].
\label{e:1pcrossdouble}
\end{equation} 

\begin{figure}
	\centering
	\includegraphics[width=13cm]{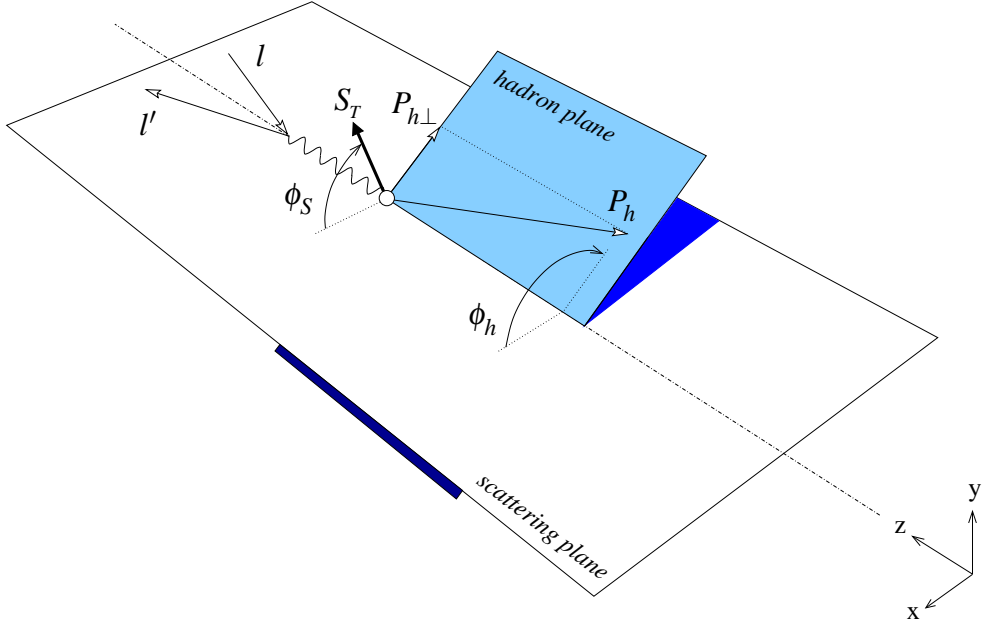}
	\rput(-9.7,7.5){$l$}
	\rput(-11.2,6.6){$l'$}
	\rput(-6.75,4.0){$\phi_h$}
	\rput(-9.2,5.6){$\phi_S$}
	\rput(-4.9,5.5){$P_h$}
	\rput(-7.65,6.8){$P_{h\perp}$}
	\caption{Description of the vectors and angles involved in one-particle inclusive deep inelastic scattering.}
	\label{f:planes2}
        \end{figure}

Instead of presenting the full cross section, we turn
directly to sum and differences of polarized cross sections. As in the
previous chapter, we will use the symbols $\rightarrow$ to indicate 
 polarization along the beam direction and
$\leftarrow$ opposite to it. We will also use $\uparrow$ to indicate
transverse polarization in the direction specified by the angle $\phi_S$, and
$\downarrow$ opposite to it. The subscript $U$ will denote unpolarized
particles, while $L$ and $T$  will denote 
longitudinally and transversely polarized particles. The first subscript
describes always the beam polarization and the second subscript the target
polarization. 
Fig.~\vref{f:planes2} gives a pictorial description of the vectors and angles
involved in one-particle inclusive deep inelastic scattering.
The unpolarized cross section reads~\cite{Boer:1999uu}
\begin{multline}
\frac{\de^6\!\sigma_{UU}}{\de \xbj \de y \de z_h \de \phi_S \de^2\! \mb{P}_{h
\perp}} = 
\frac{2 \alpha^2 }{s \xbj y^2}\,\Biggl\{
A(y)\sum_q e_q^2\,
	{\cal I}\lf[f_1^q(\xbj, \mb{p}_T^2) D_1^q(z_h, z_h^2 \mb{k}_T^2) \rg] \\
- B(y) \cos{2\phi_h}\,\sum_q e_q^2\,
	{\cal I}\lf[\frac{2(\mb{p}_T \cdot \h)(\mb{k}_T
 \cdot \h) - \mb{p}_T\cdot \mb{k}_T}{M M_h}\,h_1^{\perp
q}(\xbj, \mb{p}_T^2) H_1^{\perp q}(z_h, z_h^2 \mb{k}_T^2) \rg]\Biggr\}.
\label{e:dUU}
\end{multline}  
In the following equations, we will omit to explicitly indicate the variables
in which the cross section is differential and the variables 
the distribution and fragmentation functions depend on.
We define the following polarized cross sections differences
\begin{subequations}
\begin{align}
\de^6\!\sigma_{LL}&\equiv \frac{1}{2} \lf(\de^6\!\sigma_{\rightarrow \leftarrow} -
\de^6\!\sigma_{\rightarrow \rightarrow}\rg), & 
\de^6\!\sigma_{UL}&\equiv \frac{1}{2} \lf(\de^6\!\sigma_{U\leftarrow} -
\de^6\!\sigma_{U\rightarrow}\rg),\\
\de^6\!\sigma_{LT}&\equiv \frac{1}{2} \lf(\de^6\!\sigma_{\rightarrow \uparrow} -
\de^6\!\sigma_{\rightarrow \downarrow}\rg), &
\de^6\!\sigma_{UT}&\equiv \frac{1}{2} \lf(\de^6\!\sigma_{U\uparrow} -
\de^6\!\sigma_{U\downarrow}\rg).
\end{align} 
\end{subequations}
for which we obtain the following expressions in terms of distribution and
fragmentation functions~\cite{Boer:1999uu}
\begin{subequations}
\begin{align} 
\de^6\!\sigma_{LL}  &= \sum_{q} \frac{2 \alpha^2 e_q^2}{s \xbj
y^2}\,\lvert \lambda_e \rvert\,\lvert S_L \rvert\,C(y)\,
{\cal I}\lf[g_1^q D_1^q \rg], \\
\de^6\!\sigma_{UL}  &= \sum_{q} \frac{2 \alpha^2 e_q^2}{s \xbj y^2}
\,\lvert S_L \rvert\, B(y) \sin{2\phi_h}\,
{\cal I}\lf[\frac{2(\mb{p}_T \cdot
\h)(\mb{k}_T
 \cdot \h) - \mb{p}_T\cdot \mb{k}_T}{M M_h}\,h_{1L}^{\perp
q} H_1^{\perp q}\rg], 
\\
\de^6\!\sigma_{LT}&= \sum_{q} \frac{2 \alpha^2 e_q^2}{s \xbj
y^2}\,\lvert \lambda_e \rvert\,\sst\,C(y)\,\cos(\phi_h -\phi_S){\cal I}\lf[\frac{\mb{p}_T \cdot
\h}{M}\,g_{1T}^q D_1^q \rg], \\
\begin{split}
\de^6\!\sigma_{UT}&= \sum_{q} \frac{2 \alpha^2 e_q^2}{s \xbj y^2}\,\sst\Biggl\{
B(y) \sin \lf(\phi_h +\phi_S\rg){\cal I}\lf[\frac{\mb{k}_T\cdot\h}{M_h}\, h_{1}^{q} H_1^{\perp q} \rg] 
\\&\quad  
+A(y) \sin\lf(\phi_h -\phi_S\rg) {\cal I}\lf[\frac{\mb{p}_T\cdot\h}{M}\, f_{1T}^{\perp q} D_1^q \rg]+B(y) \sin\lf(3\phi_h -\phi_S\rg) \\
&\quad \times{\cal
I}\lf[\frac{4(\mb{p}_T\cdot\h)^2(\mb{k}_T\cdot\h)-2(\mb{p}_T\cdot\h)(\mb{p}_T\cdot\mb{k}_T)-\mb{p}_T^2(\mb{k}_T\cdot\h)}
{2 M^2 M_h}\, h_{1T}^{\perp q} H_1^{\perp q} \rg]\Biggr\}.
\label{e:dUT}
\end{split} 
\end{align} 
\end{subequations}

%%%%%%%%%%%%%%%%%%%%%%%%%%%%%%%%%%%%%%%%%%%%%%%%%%%%%%%%%%%%
\subsection{Transverse momentum measurements}

Due to the presence of an observable transverse momentum, i.e.\ that of the
final hadron, one-particle inclusive deep inelastic scattering gives the
possibility of extracting some information on the transverse momentum of
partons. It is convenient to introduce the transverse momentum of the hadron
with respect to the quark, $\mb{K}_T = -z \mb{k}_T$. In experiments where jets 
can be identified, it corresponds to the transverse
momentum of the detected hadron 
with respect to the jet axis.
The simplest quantity to be measured is
\begin{equation} \begin{split} 
\bigl\langle \mb{P}_{h \perp}^2 (\xbj, z_h)\bigr\rangle  &\equiv\frac{\int  \de \phi_S\, \de^2\! \mb{P}_{h\perp}\,
	\mb{P}_{h \perp}^2\; \de^6\sigma_{UU}}
	{\int  \de \phi_S\, \de^2\! \mb{P}_{h\perp}\;
	\de^6\sigma_{UU}}  \\
%&= \frac{\sum_q e_q^2\, \int \de^2\! \mb{P}_{h\perp}\,
%	\mb{P}_{h \perp}^2\, {\cal I}\lf[f_1^q(\xbj,
%\mb{p}_T^2) D_1^q(z_h, z_h^2 \mb{k}_T^2) \rg] }{\sum_q e_q^2\,f_1^q(\xbj)
%D_1^q(z_h) } \\
&= \frac{\sum_q e_q^2\, \lf[z_h^2 \,\bigl\langle \mb{p}_T^2 (\xbj)\bigr\rangle + \bigl\langle
\mb{K}_T^2 (z_h) \bigr\rangle \rg] f_1^q(\xbj)
D_1^q(z_h)}{\sum_q e_q^2\,f_1^q(\xbj)
D_1^q(z_h) }, 
\end{split} \end{equation}  
where
\begin{align} 
\bigl\langle \mb{p}_T^2 (\xbj) \bigr\rangle &\equiv \frac{\int \de^2\! \mb{p}_T
\,\mb{p}_T^2\, f_1^q(\xbj,
\,\mb{p}_T^2)}{f_1^q(\xbj)}, & \bigl\langle \mb{K}_T^2  (z_h)\bigr\rangle &\equiv
\frac{\int \de^2\! \mb{K}_T
\,\mb{K}_T^2\, D_1^q(z_h,
\mb{K}_T^2)}{D_1^q(z_h)}.
\end{align} 
Assuming that all quark flavors have the same transverse momentum distribution
we reach the result
\begin{equation} 
\bigl\langle \mb{P}_{h \perp}^2  (\xbj, z_h)\bigr\rangle = z_h^2 \,\bigl\langle \mb{p}_T^2 (\xbj)
\bigr\rangle + \bigl\langle \mb{K}_T^2 (z_h) \bigr\rangle.
\label{e:phperpavg}
\end{equation} 

%%%%%%%%%%%%%%%%%%%%%%%%%%%%%%%%%%%%%%%%%%%%%%%%%%%%%%%%%%%%
\subsection{Transversity measurements}

At the end of the previous chapter, we concluded that in inclusive deep
inelastic scattering it is not possible to measure the transversity
distribution function, $h_1$. 
In one-particle inclusive deep inelastic scattering, we see from
Eq.~\eqref{e:dUT} 
that it is possible for the transversity distribution to appear in an
observable in connection with a chiral-odd fragmentation function, i.e.\
 the Collins
function. 

It is convenient to introduce the angle $\phi\equiv\phi_h + \phi_S$. As
specified before, we
define the azimuthal angles with reference to the electron scattering
plane. On the other side, it is possible to choose the transverse component of
the target spin as the reference axis for the measurement of  azimuthal
angles. When expressing angles with respect to the target spin, we will use
a superscript $S$. The relations between the angles in the two different
systems is  $\phi_S = -\phi_l^S$, $\phi_h =\phi_h^S -\phi_l^S$, and, in
particular, $\phi\equiv \phi_h^S -2 \phi_l^S$.

The relevant quantity to be measured is the azimuthal 
single transverse spin asymmetry
\begin{equation} \begin{split} 
\left \langle \sin{\phi} \right \rangle_{UT} (\xbj,y,z_h)
% \nn \\
%&\qquad=& 	
& \equiv \frac{\int  \de \phi_S\, \de^2\! \mb{P}_{h\perp}\,
	\sin{\phi}\; \de^6\! \sigma_{UT}}
	{\int  \de \phi_S\, \de^2\! \mb{P}_{h\perp}\;
	\de^6\!\sigma_{UU}} \\
&= |\mb{S}_{\st}|\,\frac{(1/\xbj y^2)\, B(y)\,\sum_q e_q^2\,\int\de^2\! \mb{P}_{h\perp}\, {\cal I}\lf[({\mb{k}_T\cdot\h/2M_h})\, h_{1}^{q} H_1^{\perp q} \rg]} 
{(1/\xbj y^2)\,A(y)\sum_q e_q^2\, f_1^q
(\xbj)\,D_1^q(z_h)}\,.
\end{split} \end{equation}  
In this expression,  the 
transverse momenta of $h_1$ and $H_1^{\perp}$ are
entangled in a convolution integral~\cite{Ralston:1979ys}. To simplify the
situation, we have to make some assumptions on the transverse
momentum dependence of the distribution and fragmentation function.

The simplest example is to suppose there is no intrinsic transverse momentum
of the partons inside the target~\cite{Anselmino:2001js,Anselmino:1999pw}, i.e.
\begin{equation}	
h_1 (x,\mb{p}_{\st}^2) \approx h_1 (x) \, \frac{\delta \bigl(\mb{p}_{\st}^2\bigr)}{\pi}\, .
\end{equation}
Under this assumption, the pion transverse momentum with respect to
the virtual photon is entirely due
to the fragmentation process, i.e.\, $\mb{P}_{h\perp}=\mb{K}_T$,
and the convolution can be disentangled 
\begin{equation} 
\left\langle \sin{\phi} \right\rangle_{UT} (\xbj,y,z_h)
%\nn \\
%&\qquad \approx& 
\approx |\mb{S}_{\st}|\frac{(1/\xbj y^2)\, 
B(y)\,\sum_q e_q^2\, h_1^q (\xbj)\,H_1^{\perp (1/2)q}(z_h)}
{(1/\xbj y^2)\,A(y)\sum_q e_q^2\, f_1^q
(\xbj)\,D_1^q(z_h)}\, ,
\label{e:asym1}
\end{equation} 
where the approximation sign reminds us that the equality is 
assumption dependent. 

Another possibility is to assume that the transverse momentum distribution
is Gaussian-like in both the distribution and fragmentation side, i.e.\
\begin{align} 
h_1 (x,\mb{p}_{\st}^2) &\approx h_1 (x) \, \frac{\e^{-\mb{p}_{\st}^2\big/\lf\langle\mb{p}_{\st}^2(x)\rg\rangle}}{\pi\,
\bigl\langle\mb{p}_{\st}^2(x)\bigr\rangle},
&
H_1^{\perp} (z,\mb{K}_{\st}^2) &\approx H_1^{\perp} (z) \, \frac{\e^{-\mb{K}_{\st}^2\big/\lf\langle\mb{K}_{\st}^2(z)\rg\rangle}}{\pi\,
\bigl\langle\mb{K}_{\st}^2(z)\bigr\rangle}.
\end{align} 
The convolution becomes~\cite{Mulders:1996dh}
\begin{equation} 
{\cal I}\lf[\frac{\mb{k}_T\cdot\h}{2 M_h}\, h_{1} H_1^{\perp} \rg]
\approx 
h_1 (\xbj)\, H_1^{\perp} (z_h)\; \frac{
\bigl\langle\mb{K}_{\st}^2(z_h)\bigr\rangle}{\bigl\langle\mb{P}_{h
\perp}^2(\xbj,z_h)\bigr\rangle} \,\frac{\pperp}{z_h M_h}\,\frac{\e^{-\mb{P}_{h
\perp}^2\big/\lf\langle\mb{P}_{h \perp}^2(\xbj,z_h)\rg\rangle}}{\pi\,
\bigl\langle\mb{P}_{h \perp}^2(\xbj,z_h)\bigr\rangle} ,
\end{equation} 
where $\bigl\langle\mb{P}_{h \perp}^2\bigr\rangle$ is given by
Eq.~\eqref{e:phperpavg}. 
Using the relation
$\langle|\mb{a}_T|\rangle^2= \langle\mb{a}_T^2\rangle \pi/4$, valid for
Gaussian distributions, we can carry out the integration
over $\mb{P}_{h \perp}$ and obtain
\begin{equation} \begin{split} 
\int \de^2\! \mb{P}_{h\perp}\, {\cal I}\lf[\frac{\mb{k}_T\cdot\h}{2 M_h}\,
h_{1}^{q} H_1^{\perp q} \rg] & = \frac{
\bigl\langle\mb{K}_{\st}^2\bigr\rangle}{\bigl\langle\mb{P}_{h
\perp}^2\bigr\rangle} \,\dfrac{\bigl\langle\pperp\bigr\rangle}{2 z_h 
M_h}\,h_1^q (\xbj)\,H_1^{\perp q}(z_h) 
\\ & 
= \frac{\bigl\langle\mb{K}_{\st}^2\bigr\rangle\sqrt{\pi}/2}{2 z_h  M_h
\sqrt{\bigl\langle\mb{P}_{h \perp}^2\bigr\rangle}}
\,h_1^q (\xbj)\,H_1^{\perp q}(z_h) \\
& = \frac{\bigl\langle|\mb{K}_{\st}|\bigr\rangle}{2 z_h M_h
\sqrt{1 + z_h^2 {\bigl\langle\mb{p}_{\st}^2\bigr\rangle} \big/ {\bigl\langle\mb{K}_{\st}^2\bigr\rangle}}}
\,h_1^q (\xbj)\,H_1^{\perp q}(z_h),
\end{split} \end{equation}  
where the $\xbj$ and $z_h$ dependence of the average transverse momenta
squared is understood. 
The asymmetry becomes
\begin{equation}
\left\langle \sin{\phi} \right\rangle_{UT} (\xbj,y,z_h)\approx
 |\mb{S}_{\st}|\,\dfrac{(1/\xbj y^2)\, 
B(y)\,\sum_q e_q^2\,h_1^q (\xbj)\,H_1^{\perp (1/2) q}(z_h)\,\Big/ 
\sqrt{1 + z_h^2 {\bigl\langle\mb{p}_{\st}^2\bigr\rangle} \big/ {\bigl\langle\mb{K}_{\st}^2\bigr\rangle}}}
{(1/\xbj y^2)A(y)\sum_q e_q^2\, f_1^q
(\xbj)\,D_1^q(z_h)}
\, ,
\label{e:asym1gaus}
\end{equation}  
where the approximation sign reminds us that the equality is 
assumption dependent. 

If we want to disentangle the convolution integral of Eq.~\eqref{e:dUT} 
without making any assumption on the intrinsic transverse 
momentum distribution, we need to weight the integral with the
magnitude of the pion transverse momentum~\cite{Kotzinian:1997wt}. 
This procedure results in 
the azimuthal transverse spin asymmetry
\begin{equation} \begin{split} 
\bigg\langle \frac{|\mb{P}_{h\perp}|}{M_h} \sin{\phi}
\bigg\rangle_{UT} (\xbj,y,z_h) 
% \nn \\
%&\qquad=& 
 &\equiv \frac{\int  \de \phi_S \de^2\! \mb{P}_{h\perp}
\,(|\mb{P}_{h\perp}|/ M_h)\,\sin{\phi}\;
(\de^6\sigma_{U\uparrow} - \de^6\sigma_{U\downarrow})}
{\int  \de \phi_S \de^2\! \mb{P}_{h\perp}
(\de^6\sigma_{U\uparrow} + \de^6\sigma_{U\downarrow})}  \\
&= |\mb{S}_{\st}|\,\frac{(1/\xbj y^2)\,
	 B(y)\,z_h\,\sum_q e_q^2\, h_1^q (\xbj)\,H_1^{\perp (1)q}(z_h)}
{(1/\xbj y^2)\,A(y)\,\sum_q e_q^2\, f_1^q
(\xbj)\,D_1^q(z_h)}\;. 
\label{e:asym2}
\end{split} \end{equation}  
We achieved an assumption-free factorization of the $\xbj$ dependent transversity
distribution and the $z_h$ dependent Collins function. The measurement of this
asymmetry 
requires binning the cross section according to the magnitude of the 
pion transverse momentum.
On the other side, this asymmetry represents potentially
the cleanest method to measure the transversity distribution together with
the Collins function. 
Moreover, it turns out that it is possible to study the evolution of the
moment $H_1^{\perp (1)}$ with the energy scale~\cite{Henneman:2001ev}, without 
incurring 
complications due to Sudakov factors~\cite{Boer:2001he}.
However, the inclusion of transverse momentum raises delicate issues related
to color gauge invariance, factorization and
evolution~\cite{Kundu:2001pk,Ji:2002aa}. 

The Collins function can be measured also in  $e^+ e^-$ annihilation into two
hadrons belonging to two different jets~\cite{Boer:1998qn,Boer:1997mf}. 
We will only briefly mention this
issue at the end of Chap.~\ref{c:model}. The relevance of this measurement is
clear, since an
independent measurement of the Collins function  would be extremely useful to
pin down the transversity distribution, despite the problems with relating
measurements at different energy scales~\cite{Boer:2001he}.

%%%%%%%%%%%%%%%%%%%%%%%%%%%%%%%%%%%%%%%%%%%%%%%%%%%%%%%%%%%%
%\section{The Collins function in $e^+ e^-$ annihilation}

%%%%%%%%%%%%%%%%%%%%%%%%%%%%%%%%%%%%%%%%%%%%%%%%%%%%%%%%%%%%
\section{Summary}

In this chapter we studied one-particle inclusive deep inelastic scattering at 
leading twist, including transverse momenta and neglecting the 
polarization of the final state hadron.
To describe the process $q^* \rightarrow h  Y$ we introduced the fragmentation
correlation function  of Eq.~\eqref{e:delta}. At leading order in $1/Q$, the
correlation function can be decomposed in two terms, containing the
unpolarized fragmentation function $D_1$, and the chiral-odd, T-odd
Collins fragmentation function, $H_1^{\perp}$ [Eq.~\eqref{e:decomdelta1}].

Similarly to what was done in Chap.~\ref{c:trans}, we cast the correlation 
function in the form of a forward decay matrix in the quark chirality space
[Eq.~\eqref{e:Dmatrix}]. Since the matrix is positive semidefinite, we were
able to suggest for the first time 
a bound on the Collins function [Eq.~\eqref{e:boundcollins}].

The Collins function is extremely important for possible transversity
measurements. Being chiral odd, it appears in the single transverse spin
asymmetry  of Eq.~\eqref{e:dUT}
 in a convolution with the transversity distribution.
The convolution can be disentangled by measuring the weighted azimuthal
asymmetry of Eq.~\eqref{e:asym2}. This is therefore a first way to observe the 
transversity distribution.

This method to access transversity poses some problems. First of all, at
present we 
have no convincing information on the magnitude of the Collins function. This 
is a problem common to {\em all} polarized fragmentation functions, and in
particular to T-odd functions. We postpone this discussion to
Chap.~\ref{c:model}, where we will mention the scarce information we have on 
the Collins function, we will attempt to estimate it in a consistent model, and
we will address the question if T-odd fragmentation functions could
offer a good chance to tackle the transversity.

The second problem presented by the Collins function is the need of including 
transverse momentum in the analysis. The study of transverse momentum
dependent distribution and fragmentation functions is {\em per se} a very
intriguing subject, but one has to deal with theoretical subtleties (e.g.\ 
Sudakov
factors, transverse gauge links, evolution equations),
phenomenological difficulties (e.g.\ describing the transverse momentum
dependence of the functions) and experimental challenges (e.g.\ measuring a
weighted azimuthal asymmetry).

In the next chapter, we will examine two-particle inclusive DIS and we will
show that the transversity distribution can appear in connection with two
different fragmentation functions, even if the cross section is integrated
over the transverse momentum of the outgoing hadron.

\renewcommand{\quot}{%
\parbox{7cm}{We often think that when we have completed our study of
one, we know all about two, because ``two'' is ``one and one''. We forget that
we have still to make a study of ``and''.}\\   Sir A. Eddington
}

%%%%%%%%%%%%%%%%%%%%%%%%%%%%%%%%%%%%%%%%%%%%%%%%%%%%%%%%%%%%%%%%%%%%%%%%%%%%
\chapter[Two-hadron fragmentation functions]
{Two-hadron\\ fragmentation functions}
\label{c:twohadron}

To observe transversity, an alternative process is represented by two-particle
inclusive deep inelastic scattering, where we can introduce
two-hadron fragmentation functions. 
The transverse spin of the target can be 
correlated via a transversely polarized
quark to the relative transverse
momentum of the hadronic pair instead of
the transverse momentum of
the outgoing hadron, as in the case of the Collins function. 
This provides a way in which the 
transversity can be probed without
including partonic transverse momenta, thus avoiding several
complications and subtleties.

Two-hadron fragmentation functions and their relevance for transversity
measurements have been partially analyzed
in some
articles~\cite{Collins:1994kq,Jaffe:1998hf,Jaffe:1998pv}. 
The most complete treatment of these functions
 has been carried out by Bianconi 
et al.\ in Ref.~\citen{Bianconi:1999cd}. Model calculations have been performed
in Refs.~\citen{Collins:1994ax,Bianconi:1999uc,Radici:2001na}. 
Two-pion fragmentation functions have been studied also in the context of 
$e^+ e^-$ annihilation with a somewhat different formalism~\cite{Artru:1996zu}.

In an apparently independent context, semi-inclusive production of spin-one
hadrons (e.g. $\rho$, $\omega$, $\phi$) has been also studied and proposed as 
an alternative method to measure the transversity
distribution~\cite{Efremov:1982sh,Ji:1994vw,Anselmino:1996vq,Bacchetta:2000jk}. 
However, to measure the
polarization of the outgoing vector meson (e.g. $\rho^0$) it is necessary 
to measure the four-momenta of the decay products (e.g. $\pi^+ \pi^-$). Thus, 
the reaction
$ep \rightarrow e \rho^0 {\cal X} (\rho^0 \rightarrow \pi^+ \pi^-)$ is
 just a part of the more general reaction $ep
\rightarrow e \pi^+ \pi^- {\cal X}$, 
namely the part where the total invariant
mass of the pion pair is equal to the $\rho$ mass.

In this chapter, we will go along the same route presented in Ref.~\citen{Bianconi:1999cd}
and we will complement that treatment with the study of the partial
wave expansion of two-hadron fragmentation functions. This step will prove to
be essential to unravel the connection with spin-one fragmentation functions
and interference fragmentation functions.

%%%%%%%%%%%%%%%%%%%%%%%%%%%%%%%%%%%%%%%%%%%%%%%%%%%%%%%%%%%%
\section{Two-particle inclusive deep inelastic scattering}

\begin{figure}
	\centering
	\rput(2.55,5.61){$'$}
	\rput(7.1,0.44){\scriptsize ${\cal X''}$}
	\includegraphics[width=7cm]{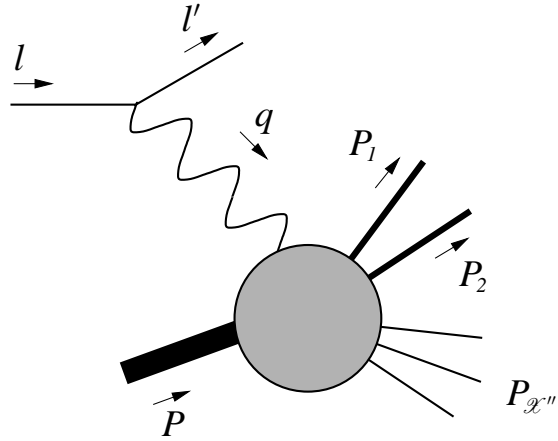}
	\caption{Two-particle inclusive deep-inelastic scattering.}
	\label{f:2p}
        \end{figure}

In two-particle inclusive deep inelastic scattering, {\em two} of the hadrons
belonging to the current fragmentation region are detected in coincidence with 
the scattered electron, as shown schematically in Fig.~\vref{f:2p}.

Some experimental results on two pion production are already
available~\cite{Cohen:1982zg,Arneodo:1986tc,Aubert:1983un}. 
More recent data are
available only on exclusive production of two
pions~\cite{Breitweg:1998ed,Breitweg:1998nh,Breitweg:1999fm,Ackerstaff:2000bz,Adloff:1999kg}
and on two-hadron production in $e^+ e^-$
annihilation~\cite{Abreu:1995rg,Abreu:1997wd,Ackerstaff:1997kd,Ackerstaff:1997kj,Abbiendi:1999bz}.

\begin{figure}
	\centering
	\includegraphics[width=9cm]{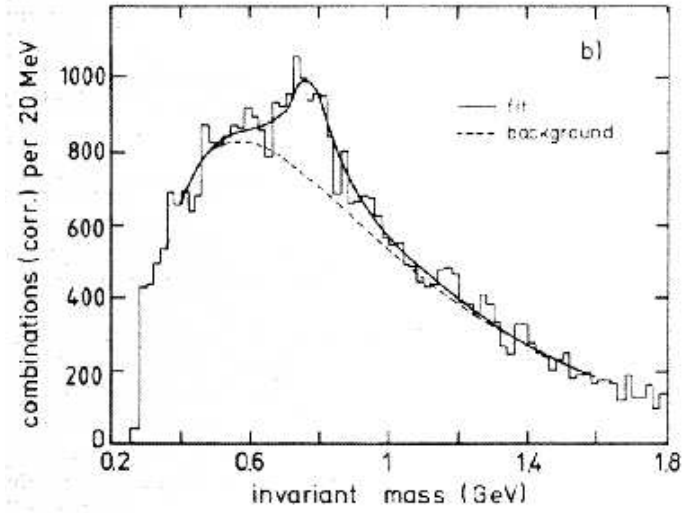}
	\caption{Typical invariant mass spectrum of pion pairs produced in
	semi-inclusive DIS~\protect{\cite{Aubert:1983un}}. The $\rho$ peak is 
	visible.}
	\label{f:spectrum}
        \end{figure}

If we consider only low invariant masses, hadron pairs are produced mainly in
the $s$-wave channel, with a typically smooth distribution over the invariant
mass, or in the $p$-wave channel, via a spin-one resonance, with its typical
Breit-Wigner invariant mass distribution.
This is the case of the production of two pions (which can proceed through a
$\rho(770)$ resonance), two kaons ($\phi(1020)$ resonance), a pion
and a kaon ($K^{*}(892)$ resonance). 
Fig.~\ref{f:spectrum} shows the typical 
invariant-mass spectrum of pion pairs in two-particle inclusive DIS.

The situation is different for proton-pion production, where on top of 
the smooth $s$-wave continuum the sharp spin-half $\Lambda(1115)$ resonance
appears. Since the $\Lambda$ undergoes a parity-violating weak decay 
the formalism we will develop in this chapter will prove to be
inadequate.
Moreover, the resonance is so sharp that it is more appropriate and easier
to
study it in the framework of polarized single hadron production. 
For these reasons, 
although polarized $\Lambda$ production 
represents a very nice way to
observe the transversity distribution, we will not
touch this subject, for which it is possible to consult 
a vast literature~\cite{Cortes:1992ja,Artru:1990zv,Anselmino:1996id,Anselmino:1996vq,Anselmino:2000ga,Anselmino:2000mb,Anselmino:2000vs,Anselmino:2001ps,Anselmino:2001xc,Anselmino:2002rr,deFlorian:1998am}.

%%%%%%%%%%%%%%%%%%%%%%%%%%%%%%%%%%%%%%%%%%%%%%%%%%%%%%%%%%%%%%%%%%%%%%%%%%%%
\subsection{Kinematics}
\label{s:kinematics}

We assume to detect two hadrons 
with masses $M_1$ and $M_2$ and momenta $P_1$ and $P_2$. We denote the
center-of-mass momentum as $P_h$ and the total invariant 
mass of the system as $M_h$. 
We maintain the parametrization of the vectors $P$, $P_h$ and
$q$ as given in Eqs.~\eqref{e:momenta1p} and Eqs.~\eqref{e:pk1p}, but we
introduce also the semi-difference of the two hadrons' momenta, $R$. We
parametrize the new momenta according to\footnote{The connection with the 
variable $\xi$ used in Ref.~\citen{Bianconi:1999cd} is $\xi=\frac{1}{2}(1+\zeta)$.}
\begin{subequations}\label{e:momenta}
\begin{align}
%P_h^{\mspace{2mu}\mu} &= \biggl[ P_h^-,\; \frac{M_h^2}{2 P_h^-},\; {\bf 0} \biggr], \\
R^{\mu} &= \biggl[ \frac{\zeta}{2}\, P_h^-,\; 
	\frac{\lf(M_1^2-M_2^2\rg) - \frac{\zeta}{2}\,M_h^2}{2 P_h^-},\; 
			\mathbold{R}_{T} \biggr], \\
P_1^{\mspace{2mu}\mu} = \frac{P_h^\mu}{2}+R^\mu &= \biggl[ \frac{(1+\zeta)}{2}\, P_h^-,\; 
	\frac{\lf(M_1^2-M_2^2\rg) + \frac{(1-\zeta)}{2}\,M_h^2}{2 P_h^-},\; 
			\mathbold{R}_{T} \biggr], \\
P_2^{\mspace{2mu}\mu} = \frac{P_h^\mu}{2}-R^\mu &= \biggl[ \frac{(1-\zeta)}{2}\, P_h^-,\; 
	\frac{\lf(M_1^2-M_2^2\rg) + \frac{(1+\zeta)}{2}\,M_h^2}{2 P_h^-},\; 
			-\mathbold{R}_{T} \biggr]. 
%\\
%k^{\mspace{2mu}\mu} &= \biggl[ \frac{1}{z}\, P_h^-,\; 
%	z \frac{k^2 - k_{T}^2}{2 P_h^-},\; 
%			\mathbold{k}_{T} \biggr].
\end{align}
\end{subequations}
The total invariant mass squared, $M_h^2$, has to be small compared to $Q^2$.
For later convenience, we compute the scalar quantities
\begin{subequations}
\begin{align} 
R^2 &= \frac{\lf(M_1^2 + M_2^2\rg)}{2} -\frac{M_h^2}{4}, \\
\mb{R}_{T}^2 &= \frac{1}{2}\,\biggl[\frac{(1-\zeta)(1+\zeta)}{2} M_h^2 - (1- \zeta)
M_1^2 -(1+\zeta)M_2^2\biggr], \\
P_h \cdot R &= \frac{\lf(M_1^2 - M_2^2\rg)}{2}, \\
P_h \cdot k &= \frac{M_h^2}{2z} +
z\, \frac{k^2 + \lvert\mb{k}_T\rvert^2}{2},  \\
k \cdot R &= \frac{\lf(M_1^2-M_2^2\rg) - \frac{\zeta}{2}\,M_h^2}{2 z} + z\, \zeta\;
\frac{k^2 + \lvert\mb{k}_T\rvert^2}{4} - \mb{k}_T \cdot \mb{R}_T.
\end{align} 
\end{subequations} 

%%%%%%%%%%%%%%%%%%%%%%%%%%%%%%%%%%%%%%%%%%%%%%%%%%%%%%%%%%%%
\subsection{The hadronic tensor}

For a process with two outgoing hadrons we can define the cross section
\begin{equation} 
\frac{2 E_1\, 2 E_2 \de^9\! \sigma}{\de^3\! \mb{P}_1 \de^3\! \mb{P}_2 
\de \xbj \de y  \de \phi_S}
= \frac{\alpha^2}{2 s \xbj Q^2}\, L_{\mu \nu} (l,l',\lambda_e) 
\,2 M W^{\mu \nu}(q,P,S,P_1,P_2),
\end{equation} 
which can be rewritten in terms of the different variables as
\begin{equation} 
\frac{\de^9\! \sigma}{\de \zeta \de M_h^2 \de \phi_R \de z_h\de^2\! \mb{P}_{h \perp} \de \xbj
\de y  \de \phi_S}= \frac{\alpha^2}{32\, z_h s \xbj Q^2}\,  L_{\mu \nu} (l,l',\lambda_e) \, 
2 M W^{\mu \nu}(q,P,S,P_1,P_2).
\label{e:crosstrans}
\end{equation} 
The angle $\phi_R$ is 
the azimuthal angle of the vector $\mb{R}_T$ with respect to 
the lepton plane, measured in a plane perpendicular to the direction of the
outgoing hadron. Neglecting $1/Q$ corrections, it can be identified with the
azimuthal angle measured in a plane perpendicular to the virtual photon
direction or to the lepton beam. 
Notice the extra factor 8 in the denominator compared to the one-particle
inclusive case.
To obtain the previous formula, we made use of the steps
\begin{equation} 
\begin{split} 
\frac{\de^3\! \mb{P}_1 \de^3\! \mb{P}_2 }{2 E_1 \, 2 E_2} &= \frac{\de^3\! \mb{P}_h \de^3\! \mb{R}}
{E_h^2 \lf(1-4 E_R^2 / E_h^2 \rg)} 
\approx \frac{\de z_h\de^2\! \mb{P}_{h \perp} \de \zeta \de^2\! \mb{R}_T }{z^2 (1- \zeta^2)} \\
& = \frac{\de z_h\de^2\! \mb{P}_{h \perp} \de \zeta \de^2\! \mb{R}_T}{2 z_h(1- \zeta^2)} 
 = \frac{\de z_h\de^2\! \mb{P}_h \de \zeta \de M_h^2 \de \phi_R}{16 z_h}\,.
\end{split} 
\end{equation} 

The hadronic tensor for two-particle inclusive scattering is defined as
\begin{align} 
\begin{split}
2 M W^{\mu \nu}(q,P, S,P_1,P_2)  
%\\ &\quad 
&= 
\frac{1}{(2 \pi)^7} \sum_{\cal X''}\int \frac{\de^3 \! \mb{P}_{\cal X''}}{(2 \pi)^3\, 2
P_{X''}^0} \; \\
&\quad \times (2 \pi)^4 \,  \delta^{(4)}\bigl(q+P-P_{\cal X''}-P_1 - P_2\bigr)
\, H^{\mu \nu}(P, S,P_{\cal X''},P_1,P_2),
\end{split}
\\
H^{\mu \nu}(P,S,P_{\cal X''},P_1,P_2) &=
\bra{P,S} J^{\mu} (0) \ket{P_1, P_2, {\cal X''}} \bra{P_1, P_2, {\cal X''}} J^{\nu} (0) \ket{P,S}.
\end{align}
With this definition,
it is possible to recover the hadronic tensor for one-particle
inclusive scattering by integration over the second hadron
\begin{equation} 
\begin{split}
\lefteqn{\sum_{h_2}
\int \frac{\de^3 \! \mb{P}_{2}}{2 P_{2}^0}\; 
	2 M W^{\mu \nu}(q,P, S,P_1,P_2)} \\
&=\sum_{h_2} \int \frac{\de^3 \! \mb{P}_{2}} {(2 \pi)^3\, 2 P_{2}^0} 
 \sum_{\cal X''}\int \frac{\de^3 \! \mb{P}_{\cal X''}} {(2 \pi)^3\, 2 P_{X''}^0}
  \;  \delta^{(4)}\bigl(q+P-P_{\cal X''}-P_2-P_1\bigr)\, 
H^{\mu \nu}(P, S,P_{\cal X''},P_1,P_2)  
\\
&=\sum_{\cal X'}
 \int \frac{\de^3 \! \mb{P}_{\cal X'}} {(2 \pi)^3\, 2 P_{X'}^0}
  \;  \delta^{(4)}\bigl(q+P-P_{\cal X'}-P_1\bigr)\, 
\int \frac{\de^3 \! \mb{P}_{2}} {(2 \pi)^3\, 2 P_{2}^0}\,
H^{\mu \nu}(P, S,P_{\cal X'}-P_2,P_1,P_2)  
\\
&= \sum_{\cal X'}
\int \frac{\de^3 \! \mb{P}_{\cal X'}} {(2 \pi)^3\, 2 P_{\cal X'}^0}
  \;  \delta^{(4)}\bigl(q+P-P_{\cal X'}-P_1\bigr)\, H^{\mu \nu}(P, S,P_{\cal X'},P_1) 
	\equiv 2 M W^{\mu \nu} (q,P, S,P_1)
\end{split}
\end{equation} 
where the state $\cal X'$ represents the sum of the states $\cal X''$ plus all 
possible states of the 
second hadron, and $P_{\cal X'} = P_{\cal X''}+P_2$.

In analogy to what we presented in Sec.~\vref{s:hadro1p}, at leading order in
$1/Q$ the hadronic tensor can be expressed in terms of correlation functions 
as
\begin{equation} 
 2 M W^{\mu \nu} (q, P,S,P_1,P_2) =  
	32 z_h\, \,{\cal I}  \Bigl[  \tr{\lf(\Phi (\xbj,\mb{p}_T, S)\; \g^{\mspace{2mu}\mu}\;
		\Delta (z_h, \mb{k}_T, \zeta, M_h^2, \phi_R) \;\g^{\nu}  \rg)} \Bigr],
\label{e:hadrontrans}
\end{equation} 
where the fragmentation correlation function has been generalized to 
include also the dependence on the vector $R$
\begin{equation}
\Delta (z, \zeta, M_h^2, \phi_R, \mb{k}_T) \equiv \frac{1}{32 z} \int \de k^+
\Delta (k, P_h, R)
		\Bigr \vert_{k^- = P_h^- / z}.
\label{e:anotherdelta}
\end{equation} 

The fully differential cross section might be a little bit too complex for
experimental measurements. To simplify the situation, we can perform the
integration over the transverse part of the center-of-mass momentum, 
$\mb{P}_{h \perp}$. The integrated cross section is
\begin{equation} 
\frac{\de^7\! \sigma}{\de \zeta \de M_h^2 \de \phi_R \de z_h \de \xbj
\de y  \de \phi_S} = \frac{\alpha^2}{32 z_h s \xbj Q^2}\, L_{\mu \nu} (l,l',\lambda_e) \, 
2 M W^{\mu \nu}(q,P,S,P_1,P_2),
\end{equation} 
where
\begin{subequations} \begin{align} 
 2 M W^{\mu \nu} (q,P,S,P_1,P_2) 
 & =  32 z_h\, \tr{\lf[\Phi (\xbj, S)\; \g^{\mspace{2mu}\mu}\;
		\Delta (z_h, \zeta, M_h^2, \phi_R) \;\g^{\nu}  \rg]}. 
\\
%\Phi (x, S) &\equiv \int \de p^- \de^2\! \mb{p}_T \,
%	\Phi (p,P, S)\Bigr \vert_{p^+ = x P^+} ,  \\
\Delta (z, \zeta, M_h^2, \phi_R) &\equiv \frac{z}{32} \int \de k^+ \de^2\! \mb{k}_T\,
	\Delta (k,P_h,R)
		\Bigr \vert_{k^- = P_h^- / z}. 
\end{align} \end{subequations} 

To identify the leading twist part of the hadronic tensor and write down the
cross section, we can trivially repeat the steps described in
Sec.~\vref{s:1pheltwist}.

%%%%%%%%%%%%%%%%%%%%%%%%%%%%%%%%%%%%%%%%%%%%%%%%%%%%%%%%%%%%
\section{The correlation function $\Delta$}
\markright{The correlation function $\mathit \Delta$}
\label{s:delta2p}

The most general expansion of the quark-quark correlation matrix $\Delta$,
respecting Hermiticity and parity invariance, is~\cite{Bianconi:1999cd}\footnote{Note that we
always use $M_h$ to render the amplitudes dimensionless, in contrast to what
is done in Ref.~\citen{Bianconi:1999cd}. }
\begin{equation} \begin{split} 
\Delta(k,P_h,R)  &=
                M_h\,C_1\,{\mb 1} 
	+ C_2\,\Pslash_h + C_3\,\Rslash + C_4\,\kslash  \\
&\quad 	+ \frac{C_5}{M_h}\,\sigma_{\mu \nu} P_h^{\mspace{2mu}\mu} k^{\nu}
	+ \frac{C_6}{M_h}\,\sigma_{\mu \nu} R^{\mu} k^{\nu}
	+ \frac{C_7}{M_h}\,\sigma_{\mu \nu} P_h^{\mspace{2mu}\mu} R^{\nu}
        + \frac{C_8}{M_h^2}\,\g_5 \eps^{\mspace{2mu}\mu \nu \rho \sigma}
		\gamma_\mu P_{h \nu} R_\rho k_\sigma.
					\label{e:decomdelta2p}  
\end{split} \end{equation}  
The amplitudes $C_i$ are dimensionless real scalar functions 
$C_i=C_i(k \cdot P_h, k^2,  R^2, k \cdot R)$. 
The last four terms are T-odd.

We are going to consider first the case when no hadron transverse momentum,
$\mb{P}_{h \perp}$, is
detected and postpone the complete case to the second part of the
chapter. 
We insert the general decomposition of Eq.~\eqref{e:decomdelta2p} into 
Eq.~\eqref{e:deltaint2} and extract the leading-twist projection 
\begin{equation}  
{\cal P}_-\,\Delta(z, \zeta, M_h^2, \phi_R)\, \g^- = \frac{1}{8 \pi}\lf(
D_1(z, \zeta, M_h^2 )
+ i\,H_1^\newangle(z,\zeta, M_h^2 )\,\frac{\Rslash_T}{M_h} \rg) {\cal P}_-\,,
\label{e:decomdelta2}
\end{equation} 
The prefactor has been chosen to have a better connection with one-hadron
results, i.e.\ integrated over $\zeta$, $M_h^2$ and $\phi_R$.
We introduced the {\em two-hadron} parton fragmentation functions
\begin{subequations}
\begin{align} 
D_1 (z, \zeta, M_h^2) &=\frac{z}{2}\int \de^2\!\mb{k}_T \de k^2 \de (2 k \cdot
P_h)\;\delta\lf(\mb{k}_T^2 + \frac{M_h^2}{z^2} + k^2 -\frac{2 k \cdot P_h}{z}
\rg)\; \lf[C_2 + \frac{\zeta}{2} C_3 + \frac{1}{z} C_4 \rg] , \\
H_1^{\newangle} (z, \zeta, M_h^2) &=\frac{z}{2} \int \de^2\!\mb{k}_T \de k^2
\de (2 k \cdot P_h)\;\delta\lf(\mb{k}_T^2 + \frac{M_h^2}{z^2} + k^2 -\frac{2 k
\cdot P_h}{z} \rg)\; \lf[-C_7 + \frac{1}{z} C_6 \rg].
\end{align} 
\end{subequations}
The fragmentation function $H_1^{\newangle}$ 
is chiral-odd and T-odd. Like the
Collins function, it can be used as a partner for the transversity
distribution, as we will see in the next section. Notice that this function
does not require the presence of partonic transverse momentum. Because of
this, its evolution equations could be simpler than the ones of the Collins
function. Since this function has the same operator structure as the
transversity, it has been suggested that it could have the same evolution
equations~\cite{Boer:2001zr,Stratmann:2001pt,Boer:2001zw}.  However, 
the situation could be complicated by the presence of the dependence on the
variables $\zeta$ and $M_h^2$, which are not present in single particle
functions~\cite{Sukhatme:1980vs}.

%%%%%%%%%%%%%%%%%%%%%%%%%%%%%%%%%%%%%%%%%%%%%%%%%%%%%%%%%%%%
\subsection{Correlation function in helicity formalism}

Once again, expressing the Dirac matrices of Eq.~\eqref{e:decomdelta2}
in the chiral or Weyl representation, we
obtain 
\renewcommand{\arraystretch}{1}
\begin{equation} 
\Bigl({\cal P}_-\,\Delta(z, \zeta, M_h^2, \phi_R)\, \g^- \Bigr)
	_{kl}=\frac{1}{8 \pi}
\begin{pmatrix}
0&0&0&0 \\
0 & D_1(z,\zeta, M_h^2)&\ii \e^{\ii \phi_R} \,\dfrac{\lf\lvert \mb{R}_T \rg
\rvert}{M_h}\,H_1^{\newangle}(z, \zeta, M_h^2) &0  \\
0& -\ii \e^{-\ii \phi_R} \,\dfrac{\lf\lvert \mb{R}_T \rg
\rvert}{M_h}\,H_1^{\newangle}(z,\zeta, M_h^2) & D_1(z, \zeta, M_h^2) & 0\\
0&0&0&0
\end{pmatrix}.
\end{equation}
\renewcommand{\arraystretch}{1.5}
As already observed before, we can restrict ourselves to the subspace of 
good quark fields and adopt the
chirality basis, and 
rewrite the correlation function as
\begin{equation} 
D(z, \zeta, M_h^2, \phi_R)_{\chi'_{2} \chi^{\phantom'}_{2}}=
\frac{1}{8 \pi}\begin{pmatrix}
D_1(z,\zeta, M_h^2)&\ii \e^{\ii \phi_R} \dfrac{\lf\lvert \mb{R}_T \rg \rvert}{M_h}H_1^{\newangle}(z,\zeta, M_h^2)  \\
 -\ii \e^{-\ii \phi_R} \dfrac{\lf\lvert \mb{R}_T \rg
\rvert}{M_h}H_1^{\newangle}(z,\zeta, M_h^2) & D_1 (z,\zeta, M_h^2)
\end{pmatrix}.
\label{e:Dmatrix2}
\end{equation}
From the positivity of the previous matrix, we can derive bounds for the
two-hadron fragmentation functions defined above:
\begin{subequations} \label{e:boundstwohadron1}
\begin{align} 
D_1(z,\zeta, M_h^2) &\ge 0, \\
\frac{\lf\lvert \mb{R}_T \rg
\rvert}{M_h} \lf \lvert H_1^{\newangle}(z,\zeta, M_h^2)\rg \rvert &
\le D_1 (z,\zeta, M_h^2).
\end{align} 
\end{subequations}

%%%%%%%%%%%%%%%%%%%%%%%%%%%%%%%%%%%%%%%%%%%%%%%%%%%%%%%%%%%%
\section{Cross section and asymmetries}
\begin{figure}
	\centering
	\includegraphics[width=13cm]{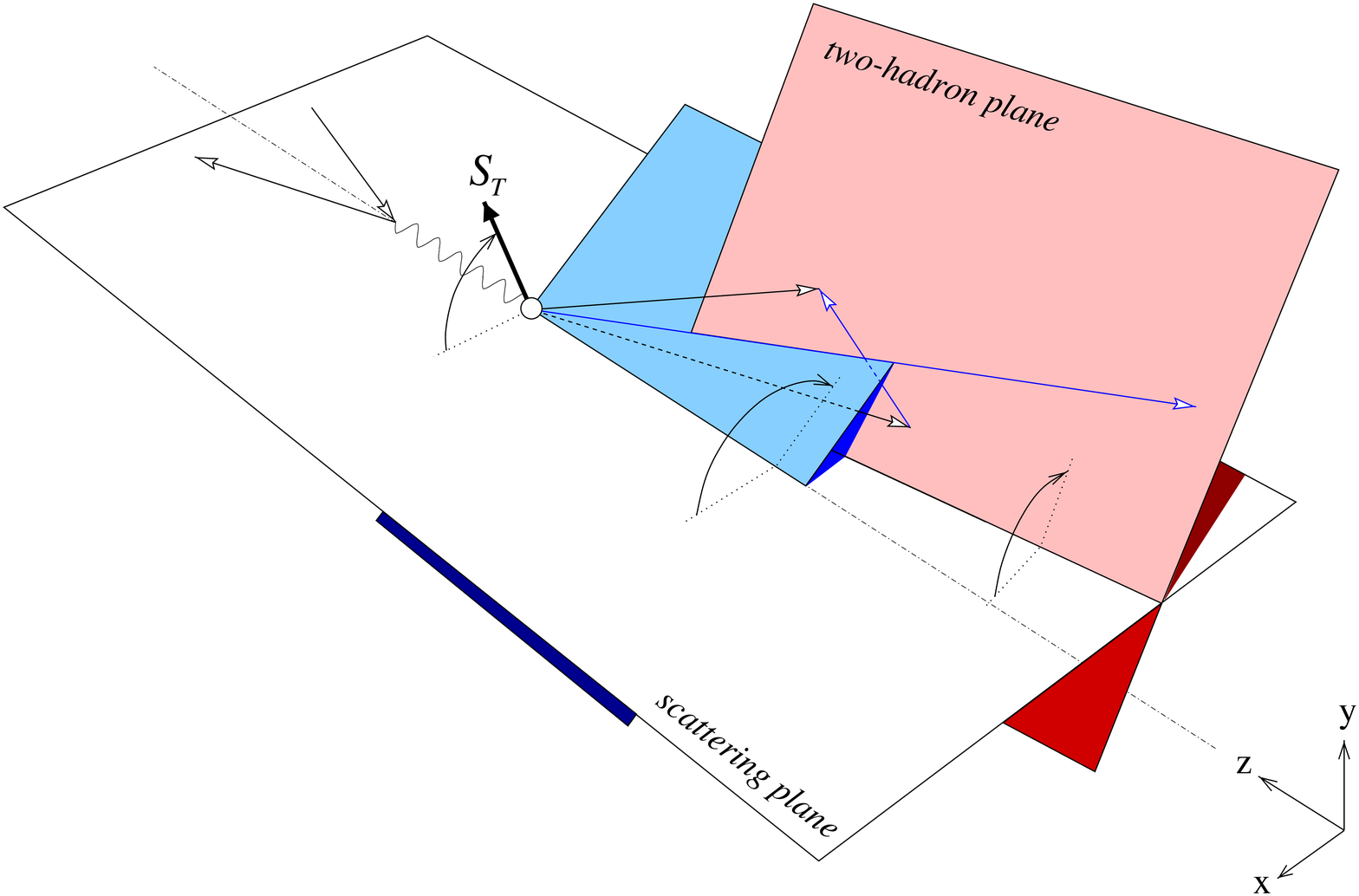}
	\rput(-9.7,7.5){$l$}
	\rput(-11.2,6.6){$l'$}
	\rput(-6.75,4.0){$\phi_h$}
	\rput(-9.2,5.6){$\phi_S$}
	\rput(-3.85,3.3){$\phi_R$}
	\rput(-4.6,5.4){$2R$}
	\rput(-1.6,5){$P_h$}
	\rput(-5.4,6.1){$P_1$}
	\rput(-4.2,4.2){$P_2$}
	\caption{Description of the vectors and angles involved in two-particle inclusive deep inelastic scattering.}
	\label{f:planes3}
        \end{figure}

In Fig.~\vref{f:planes3}, we give a pictorial description of the vectors and 
angles involved in one-particle inclusive deep inelastic scattering.
The differential cross section integrated over the
center-of-mass transverse momentum reads
\begin{equation} 
\frac{\de^7\! \sigma}{\de \zeta \de M_h^2 \de \phi_R \de z_h \de \xbj
\de y  \de \phi_S}
= \sum_q\, \rho(S)_{\Lambda^{\phantom'}_1 \Lambda_1'}\;
F(\xbj)_{\chi^{\phantom'}_{1} \chi'_{1}}^{\Lambda_1' \Lambda^{\phantom'}_1}
\Bigl( \frac{\de \sigma^{eq}}{\de y} \Bigr)
	^{\chi^{\phantom'}_{1} \chi'_{1} ; \,\chi^{\phantom'}_{2} \chi'_{2}}
	\,D(z_h, \zeta, M_h^2, \phi_R)_{\chi'_{2} \chi^{\phantom'}_{2}}. 
\label{e:2pcross}
\end{equation}   
Inserting into the previous equation
the formulae obtained for the distribution correlation matrix, Eq.~\eqref{e:Fmatrix}, the
elementary cross section, Eq.~\eqref{e:eq}, and the two-hadron
fragmentation matrix, Eq.~\eqref{e:Dmatrix2}, we obtain the
following result\footnote{The different prefactor appearing
in Ref.~\citen{Radici:2001na} is due to the
 different definition of the hadronic tensor and of the fragmentation
functions.}  
\begin{multline}
\frac{\de^7\! \sigma}{\de \zeta \de M_h^2 \de \phi_R \de z_h \de \xbj
\de y  \de \phi_S} =
\frac{2 \alpha^2 }{4 \pi s \xbj y^2}\,\sum_q e_q^2\,\Biggl[
A(y)\,
f_1^q(\xbj) D_1^q (z, \zeta, M_h^2) \\
%&\quad 
+ \lambda_e \,S_L \,C(y)\, g_1^q(\xbj) D_1^q (z, \zeta, M_h^2)
%\\	&\quad 	
+B(y)\,|\mb{S}_{\perp}|\,\frac{|\mb{R}_T|}{M_h}\, \sin(\phi_R + \phi_S)\, h_1^q(\xbj) 
H_1^{\newangle q}(z, \zeta, M_h^2)\Biggr]. 
\label{e:crosstwohadron}
\end{multline}

%%%%%%%%%%%%%%%%%%%%%%%%%%%%%%%%%%%%%%%%%%%%%%%%%%%%%%%%%%%%
\subsection{Transversity measurements}

To 
access transversity in two-particle inclusive deep inelastic scattering it
is required to measure the azimuthal single transverse spin
asymmetry~\cite{Radici:2001na} 
\begin{equation} \begin{split} 
\left \langle \sin(\phi_R+\phi_S) \right \rangle_{UT} (\xbj,y,z_h)
&\equiv \frac{\int  \de \phi_S\, \de \phi_R\, \de \zeta\, \de M_h^2
	\, \sin(\phi_R+\phi_S)\, \de^7\! \sigma_{UT}}
	{\int  \de \phi_S\, \de \phi_R\, \de \zeta\, \de M_h^2\,
	\de^7\!\sigma_{UU}} \\
&= |\mb{S}_{\st}|\frac{(1/\xbj y^2)\,
	 B(y)\,\sum_q e_q^2\, h_1^q (\xbj)\,
	\int \de \zeta\, \de M_h^2\,\dfrac{|\mb{R}_T|}{2M_h} \, H_1^{\newangle q}(z_h, \zeta, M_h^2)}
{(1/\xbj y^2)\,A(y)\,\sum_q e_q^2\, f_1^q
(\xbj)\,\int \de \zeta\, \de M_h^2\, D_1^q(z_h, \zeta, M_h^2)}\;. 
\end{split} \end{equation}  
The most valuable characteristic of this asymmetry is that it does not require 
the measurement of the center of mass transverse momentum, so that the
complications connected to the inclusion of partonic transverse momenta can be 
avoided. Apart from the usual variables $\xbj$, $y$, $z_h$, the only
other variable to be measured is the angle $\phi_R + \phi_S$. 
In case the transverse spin
direction is used as a reference axis, instead of the lepton scattering plane, 
then the angle to be measured is $\phi_R^S - 2 \phi_l^S$.

%%%%%%%%%%%%%%%%%%%%%%%%%%%%%%%%%%%%%%%%%%%%%%%%%%%%%%%%%%%%
\section{Partial wave expansion}

Up to now, we did not make any study of the inner structure of the
two-hadron fragmentation functions. It is useful to expand them in partial
waves, because if we restrict ourselves to systems with low invariant masses, 
the dominant contributions come only from the lowest
harmonics, i.e. $s$ and $p$ waves.

%%%%%%%%%%%%%%%%%%%%%%%%%%%%%%%%%%%%%%%%%%%%%%%%%%%%%%%%%%%%
\subsection{Center-of-mass parameters}

\begin{figure}
	\centering
	\includegraphics[width=6cm]{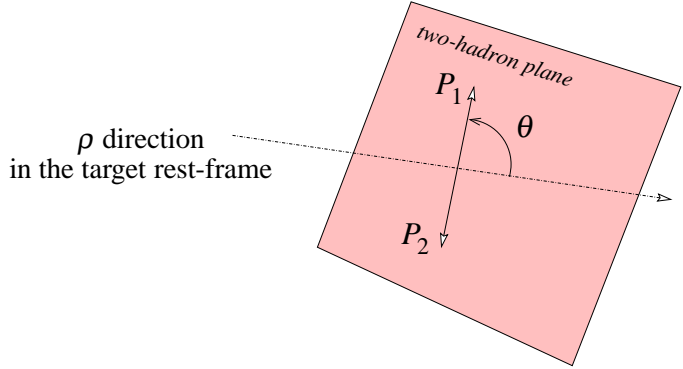}
	\rput(-7.3,3){\small $\rho$ direction}
	\rput(-7.3,2.6){\small in the target rest-frame}
	\rput(-2.2,3.2){$\theta$}
	\rput(-3.2,3.7){$P_1$}
	\rput(-3.65,1.7){$P_2$}
	\caption{Description of the polar angle $\theta$, 
	in the center-of-mass frame of the hadron couple.}
	\label{f:theta}
        \end{figure}

The partial-wave expansion can be performed only in the frame of reference of
the center of mass of the hadron pair. 
As a first step, we need to express all vectors in this
frame, i.e.\
%Maintaining the orientation of the
%axes typical of a $T$ frame of reference\footnote{Keep in mind that in these
%frames the $z$ direction is opposite to the direction of the outgoing hadron.},\;
%the momenta can be expressed in the center-of-mass frame as
\begin{subequations}\label{e:common}
\begin{align}
P_h^{\mspace{2mu}\mu} &\overset{\rm cm}{=} \Biggl[ \frac{M_h}{\sqrt{2}},\; \frac{M_h}{\sqrt{2}},\; 0,\; 0\Biggr], \\
\begin{split}
R^{\mu} &\overset{\rm cm}{=} \Biggl[ \frac{\sqrt{M_1^2+ \rr^2}-\sqrt{M_2^2+ \rr^2}-2 \rr
\cos{\theta}}{2\sqrt{2}},\;  \\
& \qquad \frac{\sqrt{M_1^2+ \rr^2}-\sqrt{M_2^2+ \rr^2}+2 \rr
\cos{\theta}}{2\sqrt{2}},\;  
%\nn \\ & \qquad 
\rr \sin{\theta} \cos{\phi_R} ,\; \rr \sin{\theta} \sin{\phi_R} \Biggr],
\end{split}
\\
P_1^{\mspace{2mu}\mu} &\overset{\rm cm}{=} \Biggl[ \frac{\sqrt{M_1^2+ \rr^2}- \rr
\cos{\theta}}{\sqrt{2}},\;
%\nn \\ & \qquad 
\frac{\sqrt{M_1^2+ \rr^2}+ \rr
\cos{\theta}}{\sqrt{2}},\;
%\nn \\ & \qquad 
\rr \sin{\theta} \cos{\phi_R} ,\;
 \rr \sin{\theta} \sin{\phi_R} \Biggr], \\
P_2^{\mspace{2mu}\mu} &\overset{\rm cm}{=}  \Biggl[ \frac{\sqrt{M_2^2+ \rr^2}- \rr
\cos{\theta}}{\sqrt{2}},\;
%\nn \\ & \qquad 
\frac{\sqrt{M_2^2+ \rr^2}+ \rr
\cos{\theta}}{\sqrt{2}},\;
%\nn \\ & \qquad
-\rr \sin{\theta} \cos{\phi_R} ,\; -\rr \sin{\theta} \sin{\phi_R} \Biggr],
\end{align}
\end{subequations}
where
\begin{equation} 
\rr= \frac{1}{2} \sqrt{M_h^2 -2 \lf(M_1^2 + M_2^2\rg)+ \lf(M_1^2 -
M_2^2\rg)^2 \bigg/ M_h^2}.
\label{e:rr1} 
\end{equation} 
The polar angle $\theta$ is illustrated in Fig.~\vref{f:theta}.
The variable $\zeta$ is connected to the center-of-mass variable
$\cos{\theta}$ in the following way:
\begin{equation}
\zeta \equiv \frac{2 R^-}{P_h^-} \overset{\rm cm}{=} \frac{1}{M_h}
\lf(\sqrt{M_1^2+
\rr^2}-\sqrt{M_2^2+ \rr^2}-2 \rr \cos{\theta}\rg).
\end{equation} 
If $M_1=M_2$, notice that $R$ is purely spacelike and
\begin{align} 
\rr & =-|R|=\frac{1}{2}\sqrt{M_h^2 -4 M_1^2}, & \zeta &= - \frac{1}{M_h}
\sqrt{M_h^2 - 4 M_1^2}\, \cos{\theta}.
\label{e:rr2} 
\end{align}

%%%%%%%%%%%%%%%%%%%%%%%%%%%%%%%%%%%%%%%%%%%%%%%%%%%%%%%%%%%%
\subsection{The correlation function $\Delta$ with partial wave-expansion}
\markright{The correlation function $\mathit \Delta$ with partial
wave-expansion} 
\label{s:deltapw}

To perform the partial-wave expansion of the fragmentation functions, let us
first express the correlation function $\Delta$ in terms of center-of-mass
parameters 
\begin{equation} 
{\cal P}_-\,\Delta(z, \zeta(\cos \theta), M_h^2, \phi_R)\, \g^- =  
 \frac{1}{8 \pi}\lf(D_1\bigl(z, \zeta(\cos{\theta}), M_h^2\bigr) 
+ \ii H_1^{\newangle}\bigl(z, \zeta(\cos{\theta}), M_h^2\bigr)
\,\sin{\theta}\, \frac{\rr}{M_h}\,\nslash_{\phi_R} 
\rg){\cal P}_- \,,
\end{equation} 
where
$n_{\phi_R}^{\mu} =\bigl[ 0,\; 0, \; \cos{\phi_R} ,\; \sin{\phi_R} \bigr]$. 
In general, 
we can expand the fragmentation functions on the basis of the Legendre
polynomials in the following way
\begin{equation}
D_1 \bigl(z, \zeta(\cos{\theta}), M_h^2\bigr) = \sum_n D_{1n} \bigl(z, M_h^2\bigr)\,
P_n (\cos{\theta})
\end{equation} 
with
\begin{equation} 
D_{1n} \bigl(z, M_h^2\bigr) = \int_{-1}^1 \de \cos{\theta} \;P_n (\cos{\theta})\,
D_1 \bigl(z, \zeta(\cos{\theta}), M_h^2\bigr). 
\end{equation}

Considering only two-hadron systems with a low invariant mass, we assume we
can truncate the expansion in order to include only $s$-wave and $p$-wave
contributions  to the correlation function. The connection between the
correlation function $\Delta$ in terms of the variable $\zeta$ and in terms of 
the variable $\cos \theta$ is
\begin{equation} 
\Delta(z, \cos \theta, M_h^2, \phi_R) = \frac{2 \rr}{M_h}\,
	\Delta(z, \zeta, M_h^2, \phi_R),
\end{equation} 
to take into account the fact that $\de \zeta = 2 \rr / M_h \de \cos \theta$. 
Then we can write the partial-wave expanded correlation function as
\begin{equation} \begin{split} 
{\cal P}_-\,\Delta(z, \cos \theta, M_h^2, \phi_R)\, \g^- 
& \approx 
 \frac{1}{8 \pi}\Biggl[D_{1, UU}\bigl(z, M_h^2\bigr)	
	+D_{1, UL}^{sp}\bigl(z, M_h^2\bigr)\, \cos{\theta}
+D_{1, LL}^{pp}\bigl(z, M_h^2\bigr)\,\frac{1}{4}\lf(3 \cos^2{\theta} -1\rg) 
\\ & \quad
+ \ii \lf(H_{1, UT}^{\newangle \, sp}\bigl(z, M_h^2\bigr) 
	+ H_{1, LT}^{\newangle \, pp}\bigl(z, M_h^2\bigr) \cos{\theta}
\rg)\,\sin{\theta}\,  \frac{\rr}{M_h}\,\nslash_{\phi_R} 
\Biggr]\,{\cal P}_-\,,  
\end{split} 
\label{e:deltapwexpanded}
\end{equation}  
The correlation function can be consequently written in matrix form as
\begin{multline} 
D(z, \cos \theta, M_h^2, \phi_R)_{\chi'_{2} \chi^{\phantom'}_{2}} \approx \\
\frac{1}{8 \pi}
%\lf(
%\rule{0mm}{8.09mm}
%\begin{smallmatrix}
\begin{pmatrix}
D_{1, UU} + D_{1, UL}^{sp} \cos{\theta} + D_{1, LL}^{pp}\,\frac{1}{4}\lf(3 \cos^2{\theta} -1\rg) &
	+ \ii \lf(H_{1, UT}^{\newangle \, sp}
	+ H_{1, LT}^{\newangle \, pp}
		\cos{\theta}\rg)\sin{\theta}\,\dfrac{\rr}{M_h}\,
		  \e^{\ii \phi_R}  \\
- \ii \lf(H_{1, UT}^{\newangle \, sp}
+ H_{1, LT}^{\newangle \, pp}
\cos{\theta}\rg)\sin{\theta}\,\dfrac{\rr}{M_h}\,
   \e^{-\ii  \phi_R}  &
	D_{1, UU} + D_{1, UL}^{sp} \cos{\theta} + D_{1, LL}^{pp}\,\frac{1}{4}\lf(3 \cos^2{\theta} -1\rg) 
%\end{smallmatrix}\rg)
\end{pmatrix}.
\label{e:Dpw}
\end{multline}  
Here, it is not yet clear what are the motivations to assign such names to the
functions. It will become clear after we distinguish the $s$- and $p$-wave
contributions in the rest of the section and also after we study spin-one
fragmentation functions in Chap.~\ref{c:spinone}.

The information encoded in the correlation function can be expressed in a
different way. Depending on the angular
momentum of the system, the angular distribution of the two hadrons is
characterized by a specific bilinear combination of spherical harmonics. If
the angular momentum can be only 0 or 1, as in the case of our truncated
partial wave expansion, the angular distribution of the two hadrons can be
fully described by the decay matrix 
\begin{multline}
%\addtolength{\extrarowheight}{5pt}
{\cal D}(\theta, \phi_R)_{jm,j'm'}  = Y_{j}^{m}\, Y_{j'}^{m'\, \ast} = \\
 \frac{1}{4 \pi} \lf(\begin{array}{c|ccc}
	1 & -\sqrt{{\frac{3}{2}}} \sin{\theta} \,\e^{\ii \phi_R} & \sqrt{3} \cos{\theta}
			& \sqrt{{\frac{3}{2}}} \sin{\theta} \,\e^{- \ii \phi_R} 
\\[3pt] \hline
-\sqrt{{\frac{3}{2}}} \sin{\theta} \,\e^{- \ii \phi_R} &
	\frac{3}{2} \sin^2{\theta} &
 		-\frac{3}{\sqrt{2}} \cos{\theta} \sin{\theta} \,\e^{- \ii \phi_R} &
			- \frac{3}{2}\sin^2{\theta} \,\e^{- 2 \ii \phi_R} 
\\
\sqrt{3} \cos{\theta} &
   -\frac{3}{\sqrt{2}} \cos{\theta} \sin{\theta} \,\e^{ \ii \phi_R} &
	3 \cos^2{\theta} &
		\frac{3}{\sqrt{2}} \cos{\theta} \sin{\theta} \,\e^{- \ii \phi_R}
\\
\sqrt{{\frac{3}{2}}} \sin{\theta} \,\e^{\ii \phi_R} &
	- \frac{3}{2}\sin^2{\theta} \,\e^{ 2 \ii \phi_R} &
		\frac{3}{\sqrt{2}} \cos{\theta} \sin{\theta} \,\e^{\ii \phi_R} &
			\frac{3}{2} \sin^2{\theta} 	
\end{array}\rg).	
\label{e:decaymatrix}
\end{multline}
The angular momentum indices, $j$ and $j'$, run only from 0 to 1. 
For convenience, we split the matrix into blocks: 
the upper-left block ($j=j'=0$) refers to the pure $s$-wave component, 
the lower-right block ($j=j'=1$)
describes the $p$-wave component, and it is subdivided according to the value
of $m, m' = +1, 0, -1$,
Finally, the off-diagonal blocks describe the $sp$ interference.

The original correlation function can now be expressed as the trace of
the decay matrix and a fragmentation matrix in the quark chirality space
$\otimes$ the hadronic system angular momentum space 
\begin{equation} 
D(z, \cos \theta, M_h^2, \phi_R)_{\chi'_{2} \chi^{\phantom'}_{2}} =
D(z, M_h^2)_{\chi'_{2} \chi^{\phantom'}_{2}}^{j'm',jm}
{\cal D}(\theta,\phi_R)_{jm,j'm'}.
\end{equation} 
The solution for the fragmentation matrix is the $8 \times 8$ matrix
\begin{equation} 
D(z, M_h^2)_{\chi'_{2} \chi^{\phantom'}_{2}}^{j'm',jm} = 
\frac{1}{8}\,\begin{pmatrix}
A_{j'm',jm} & B_{j'm',jm} \\
B_{j'm',jm}^{\dagger} & A_{j'm',jm}
\end{pmatrix},
\label{e:Dzmh}
\end{equation} 
where the inner blocks, spanning the hadronic angular momentum space, read
explicitly
\begin{subequations}
\begin{align} 
\addtolength{\extrarowheight}{5pt}
A_{j'm',jm} &= \lf(\begin{array}{c|ccc}
D_{1, UU}^{ss} & 0 & \frac{2}{\sqrt{3}}D_{1, UL}^{sp} & 0 \\[3pt] \hline
0 &  D_{1, UU}^{pp} -\frac{1}{3}D_{1, LL}^{pp} & 0 & 0 \\
\frac{2}{\sqrt{3}} D_{1, UL}^{sp} & 0 
	& D_{1, UU}^{pp} +\frac{2}{3} D_{1, LL}^{pp} & 0\\
 0 & 0 & 0 & D_{1, UU}^{pp} -\frac{1}{3}D_{1, LL}^{pp} 
\end{array}\rg), \\
B_{j'm',jm} &=\lf(\begin{array}{c|ccc}
0 & 0 & 0 &  \ii \frac{2 \sqrt{2}}{\sqrt{3}} \frac{\rr}{M_h} H_{1,UT}^{\newangle \, sp}
\\[3pt] \hline
- \ii \frac{2 \sqrt{2}}{\sqrt{3}}\frac{\rr}{M_h} H_{1,UT}^{\newangle \, sp} & 0 &
- \ii \frac{2 \sqrt{2}}{3}\frac{\rr}{M_h} H_{1, LT}^{\newangle \, pp} & 0 \\
0 & 0 & 0 & \ii \frac{2 \sqrt{2}}{3}\frac{\rr}{M_h} H_{1,LT}^{\newangle \, pp} 
\\
0 & 0 & 0 & 0
\end{array}\rg).
\end{align} 
\label{e:blockB}
\end{subequations}

The choice of the indices in the names of the fragmentation
functions is connected to their position in the matrix. The $ss$ functions are 
typical of unpolarized two-hadron fragmentation. The $sp$ sector describes the 
interference between $s$- and $p$-wave fragmentation; this is the 
sector studied by Jaffe et al. in Refs.~\citen{Jaffe:1998hf} and \citen{Tang:1998wp}. 
The $pp$ functions
correspond to the spin-one fragmentation functions studied 
in Refs.~\citen{Ji:1994vw,Anselmino:1996vq,Bacchetta:2000jk}. We will take a
look at them from a different point of view in the next chapter and
also the choice of the subscripts will become more transparent.

Note that the matrix
fulfills the properties
 of Hermiticity, conservation of angular momentum 
(requiring 
$m+\chi'_{2} =m' +\chi_{2}$) and parity invariance~\cite{Jaffe:1998hf}
\begin{equation} 
D_{\chi'_{2}\chi^{\phantom'}_{2}}^{j'm',jm}
=D_{{-\chi'_{2}}\, {-\chi^{\phantom'}_{2}}}^{j'\,{-m'},j\, {-m}}.
\end{equation} 
 The imaginary entries of the matrix correspond to T-odd
functions. 

In principle the fragmentation matrix could contain more
functions, without violating any symmetry, but after tracing it with the decay
matrix, they would vanish. In this sense, only part of the full information
contained in the fragmentation
matrix can be analyzed through a parity-conserving process~\cite{Bourrely:1980mr}. 
We shall come back to
this issue after we studied spin-one fragmentation functions in the next
chapter.  

Finally, if we trace the fragmentation matrix with the decay matrix, we obtain
exactly the correlation function expanded in partial waves,
Eq.~\eqref{e:Dpw} , except for the
fact that the unpolarized fragmentation function turns out to be expressed in
terms of pure $s$- and $p$-wave contributions, i.e
\begin{equation}
D_{1, UU}\bigl(z, M_h^2\bigr) = \frac{1}{4} D_{1, UU}^{ss}\bigl(z, M_h^2\bigr)
+ \frac{3}{4} D_{1, UU}^{pp}\bigl(z, M_h^2\bigr).
\end{equation} 
In any cross section, these two contributions are merged together and they are
kinematically indistinguishable, unless a specific behavior of the invariant
mass dependence can be assumed. For instance, the $p$-wave contribution could
be due to the existence of a resonance, emerging over a background of
continuum $s$-wave states~\cite{Collins:1994kq}.

%%%%%%%%%%%%%%%%%%%%%%%%%%%%%%%%%%%%%%%%%%%%%%%%%%%%%%%%%%%%
\subsection{Positivity bounds on partial-wave fragmentation functions}

The fragmentation matrix, Eq.~\eqref{e:Dzmh}, 
has to be positive definite, thus allowing us to
set positivity bounds on the two-hadron fragmentation functions.
From the positivity of the diagonal matrix elements, we obtain the bounds
\begin{subequations}\label{e:pwbound1}
\begin{align} 
D_{1, UU}^{ss}(z, M_h^2) &\ge 0, \\
D_{1, UU}^{pp}(z, M_h^2) &\ge 0, \label{e:D1boundpw}\\
-\frac{3}{2}\,D_{1, UU}^{pp}(z, M_h^2) &\leq D_{1, LL}^{pp}(z, M_h^2) \leq 
3\,D_{1, UU}^{pp}(z, M_h^2), \label{e:B1boundpw}
\end{align} 
\end{subequations}
while from the positivity of the two-dimensional minors we get
\begin{subequations}\label{e:pwbound2}
\begin{align} 
\lf \lvert D_{1, UL}^{sp}\rg \rvert &\le \sqrt{\frac{3}{4}\, D_{1, UU}^{ss} \lf(
D_{1, UU}^{pp} +\frac{2}{3}\, D_{1, LL}^{pp} \rg)} \leq \frac{3}{2}\, D_{1, UU}, \\
\frac{\rr}{M_h} \lf \lvert H_{1,UT}^{\newangle \, sp}\rg \rvert &\le \sqrt{\frac{3}{8}\, D_{1, UU}^{ss} \lf(
D_{1, UU}^{pp} -\frac{1}{3}\, D_{1, LL}^{pp} \rg)} \leq \frac{3}{2}\, D_{1,
UU},  \\
\frac{\rr}{M_h} \lf \lvert H_{1,LT}^{\newangle \, pp}\rg \rvert &\le
\frac{3}{2 \sqrt{2}}\sqrt{\lf(
D_{1, UU}^{pp} +\frac{2}{3}\, D_{1, LL}^{pp} \rg)\lf(
D_{1, UU}^{pp} -\frac{1}{3}\, D_{1, LL}^{pp} \rg)} \leq \frac{9}{8}\, D_{1, UU}.
\label{e:H1LTbound} 
\end{align} 
\end{subequations}

%%%%%%%%%%%%%%%%%%%%%%%%%%%%%%%%%%%%%%%%%%%%%%%%%%%%%%%%%%%%
\subsection{Cross section and asymmetries with partial-wave expansion}

\begin{figure}
	\centering
	\begin{tabular}{ccc}
	\raisebox{1cm}{\includegraphics[width=7cm]{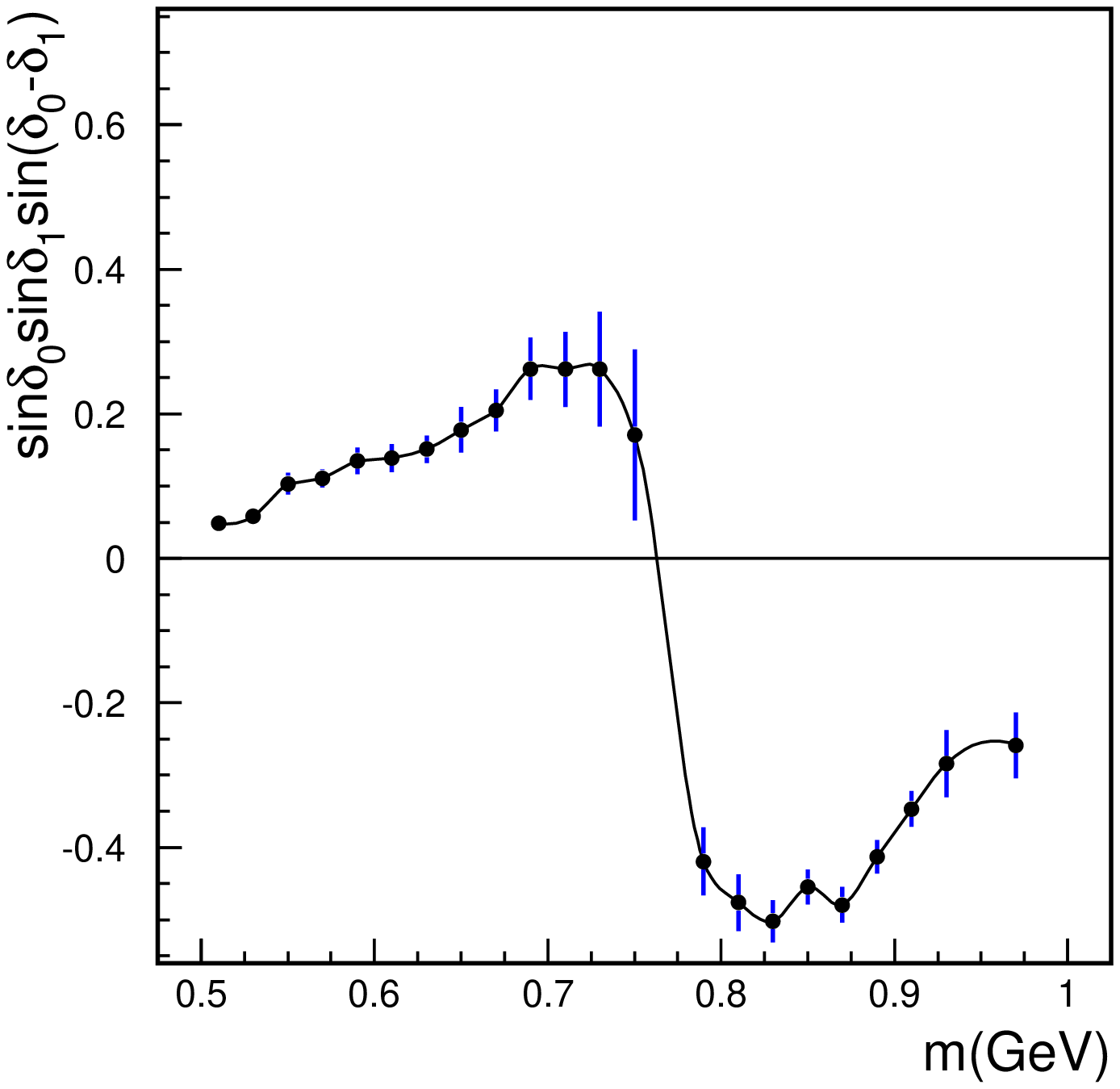}}
	&\hspace{1.5cm}&
	\includegraphics[width=6.5cm]{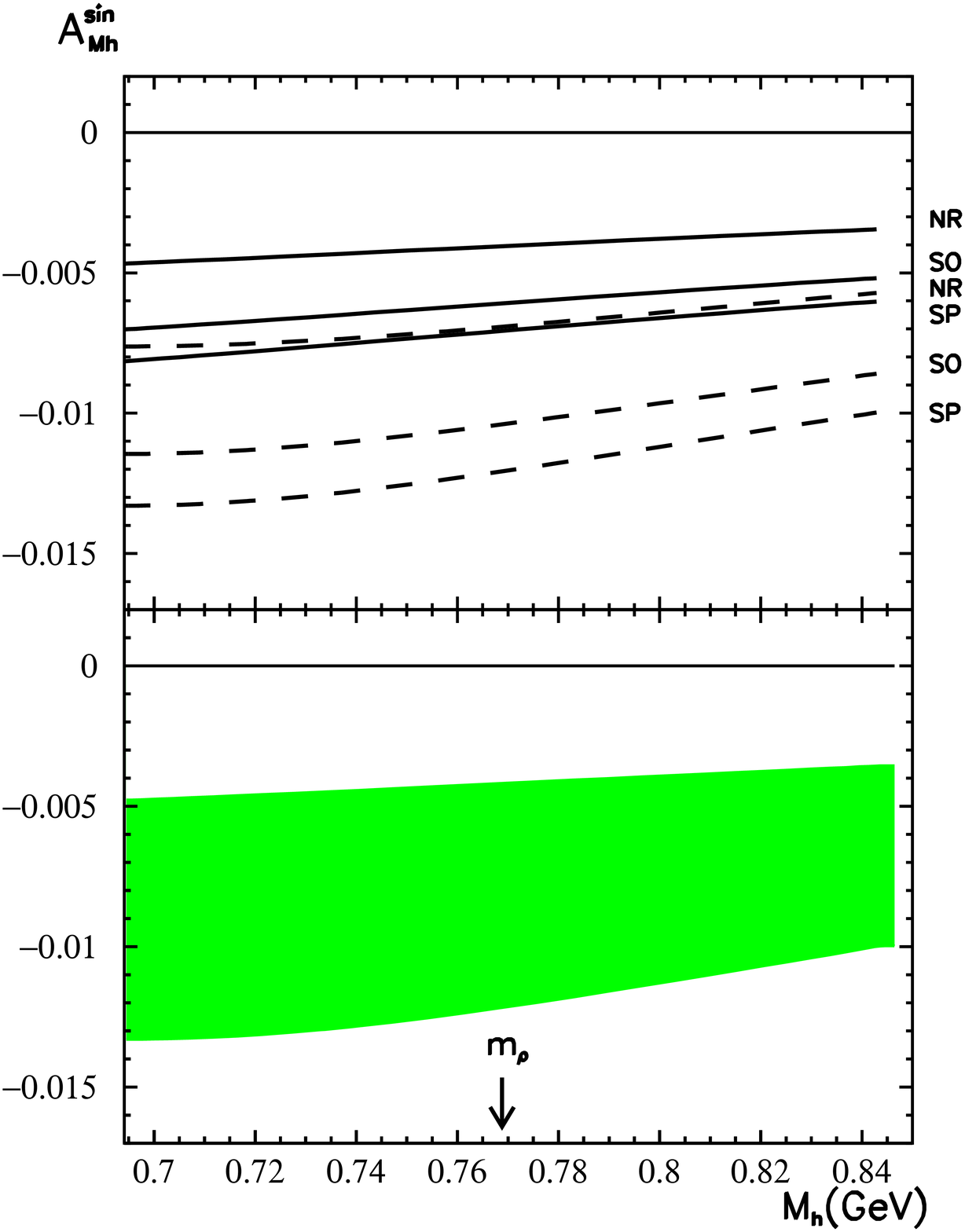}
%        \psframe[linewidth=1pt](-5.76,0.745)(-0.54,7.8)
%	\qline(-5.75,4.27)(-0.555,4.27)
%	\psline[linewidth=0.5pt](-5.75,3.9)(-0.55,3.9)
%	\psline[linewidth=0.5pt](-5.75,7.41)(-0.55,7.41)
	\\
	(a) & &(b)
	\end{tabular}
	\caption{Two different models of the invariant mass behavior of the
$sp$ interference fragmentation function: (a) Jaffe et
al.~\cite{Jaffe:1998hf}, (b)  Radici et al.~\cite{Radici:2001na}.}
	\label{f:jaffemass}
        \end{figure}

Thanks to the partial-wave expansion, 
the cross section at leading twist can be expressed in a factorized way~\cite{Jaffe:1996wp,Jaffe:1998hf}
\begin{multline}
\frac{\de^7\! \sigma}{\de \cos\theta \de M_h^2 \de \phi_R \de z_h \de \xbj
\de y  \de \phi_S} =  \\
\sum_q\, \rho(S)_{\Lambda^{\phantom'}_1 \Lambda_1'}\;
F(\xbj)_{\chi^{\phantom'}_{1} \chi'_{1}}^{\Lambda_1' \Lambda^{\phantom'}_1}
\Bigl( \frac{\de \sigma^{eq}}{\de y} \Bigr)
	^{\chi^{\phantom'}_{1} \chi'_{1} ; \,\chi^{\phantom'}_{2} \chi'_{2}}
	D(z_h,M_h^2)_{\chi'_{2} \chi^{\phantom'}_{2}}^{j'm',jm}
	\,{\cal D}(\theta,\phi_R)_{jm,j'm'}. 
\label{e:2pcrosspw}
\end{multline}
The unpolarized and polarized parts of the cross section are~\footnote{The
distribution and fragmentation functions are understood to have a flavor
index $q$.}
\begin{subequations}\label{e:cross2ppw} \begin{align} 
\begin{split}
\de^7\! \sigma_{UU} &= 
\sum_q \frac{\alpha^2 e_q^2}{2\pi s \xbj y^2}\,
	A(y)\,f_1(\xbj) \\
	&\quad\times\biggl(D_{1, UU}\bigl(z_h, M_h^2\bigr)	
		+\cos{\theta}\, D_{1, UL}^{sp}\bigl(z_h, M_h^2\bigr) 
		+\frac{1}{4}\lf(3 \cos^2{\theta} -1\rg) 
			D_{1, LL}^{pp}\bigl(z_h, M_h^2\bigr)\biggr)
\end{split}
\\
\begin{split}
\de^7\! \sigma_{LL} &=
\sum_q \frac{\alpha^2 e_q^2}{2\pi s \xbj y^2}\,
\lambda_e\, S_L\, C(y)\, g_1(\xbj) \\
	&\quad\times\biggl(D_{1, UU}\bigl(z_h, M_h^2\bigr)	
		+\cos{\theta}\, D_{1, UL}^{sp}\bigl(z_h, M_h^2\bigr) 
		+\frac{1}{4}\lf(3 \cos^2{\theta} -1\rg) 
			D_{1, LL}^{pp}\bigl(z_h, M_h^2\bigr)\biggr)
\end{split}
\\
\begin{split}
\de^7\! \sigma_{UT} &=
\sum_q \frac{\alpha^2 e_q^2}{2\pi s \xbj y^2}\,
B(y)\,|\mb{S}_{\perp}|\,\frac{\rr}{M_h}\, \sin(\phi_R + \phi_S)
 \,\sin{\theta}\, h_1(\xbj)	\\
	&\quad \times \biggl(H_{1, UT}^{\newangle \, sp}\bigl(z_h, M_h^2\bigr) 
	+\cos{\theta}\, H_{1, LT}^{\newangle \, pp}\bigl(z_h, M_h^2\bigr)
\biggr). 
\label{e:asymmtwohadron}
\end{split}
\end{align} \end{subequations} 
From the partial wave analysis we see in particular that the transversity
distribution can be matched with two different chiral-odd, T-odd fragmentation 
functions, one pertaining to the interference between the $s$- and $p$-wave
channels of two-hadron production, the second being a purely $p$-wave
effect. This considerations agree with the past literature on interference
fragmentation functions~\cite{Jaffe:1998hf,Tang:1998wp} and spin-one 
fragmentation functions~\cite{Ji:1994vw,Anselmino:1996vq,Bacchetta:2000jk}.
A priori, we don't know what is the magnitude of these functions, nor their
behavior with respect to the invariant mass of the system. 
Jaffe et al.~\cite{Jaffe:1998hf} studied one of
the possible mechanisms that could generate an $sp$ interference 
fragmentation function in $\pi \pi$ production. They separated the
production of the pion pair (which they did not evaluate) from a $\pi
\pi$ rescattering process, which determines the T-odd character of the
fragmentation function and implies a peculiar behavior with respect to the
invariant mass, shown in Fig.~\ref{f:jaffemass} (a). 
A different model was applied by Radici et
al.~\cite{Radici:2001na}: the expected invariant mass behavior turns out to be 
very different, as shown in Fig.~\ref{f:jaffemass} (b), and the magnitude of the effect is estimated to be of the
order of 1\%.
For what concerns $pp$ fragmentation functions, 
at the moment there are no estimates of their magnitude and behavior. 
However, in general $p$-wave production of two
hadrons becomes significant only when they come from the decay of a
spin-one resonance. Because of this, we can expect that the invariant mass
shape of these functions corresponds to a Breit-Wigner curve peaked at the
resonance mass.

If we integrate over $\cos \theta$, with the polar angle ranging
from 0 to $\pi$, the only
surviving contributions to the cross sections are
\begin{subequations} \begin{align}  
\de^6\! \sigma_{UU} &= 
\sum_q \frac{\alpha^2 e_q^2}{\pi s \xbj y^2}\,
	A(y)\,f_1(\xbj)\, D_{1, UU}\bigl(z_h, M_h^2\bigr)	
\\
\de^6\! \sigma_{LL} &=
\sum_q \frac{\alpha^2 e_q^2}{\pi s \xbj y^2}\,
\lambda_e\, S_L\, C(y)\, g_1(\xbj)\, D_{1, UU}\bigl(z_h, M_h^2\bigr)	
\\
\de^6\! \sigma_{UT} &=
\sum_q \frac{\alpha^2 e_q^2}{4  s \xbj y^2}\,
B(y)\,|\mb{S}_{\perp}|\,\frac{\rr}{M_h}\, \sin(\phi_R + \phi_S)
 \,h_1(\xbj)\,H_{1, UT}^{\newangle \, sp}\bigl(z_h, M_h^2\bigr). 
\label{e:asym3}
\end{align} \end{subequations} 
It is possible to integrate over $\cos \theta$ with the polar angle going
 from $-\pi/2$ to 
$\pi/2$. While the first two cross sections vanish, the 
transverse cross section would turn out to be
\begin{equation} 
\de^6\! \sigma_{UT} =
\sum_q \frac{\alpha^2 e_q^2}{4 s \xbj y^2}\,
B(y)\,|\mb{S}_{\perp}\,|\frac{\rr}{M_h}\, \sin(\phi_R + \phi_S)
 \,h_1(\xbj)\,\lf(H_{1, UT}^{\newangle \, sp}\bigl(z_h,
M_h^2\bigr)+ \frac{4}{3\pi} H_{1, LT}^{\newangle \, pp}\bigl(z_h, M_h^2\bigr) \rg). 
\label{e:asym4}
\end{equation} 
While the asymmetry in Eq.~\eqref{e:asym3} is almost identical to the one
discussed in Ref.~\citen{Jaffe:1998hf}, the last asymmetry contains a 
contribution of pure 
$p$ waves that, as already mentioned,  will become particularly relevant
 when the two hadrons are produced 
through a
spin-one resonance.

%%%%%%%%%%%%%%%%%%%%%%%%%%%%%%%%%%%%%%%%%%%%%%%%%%%%%%%%%%%%
\section{The correlation function $\Delta$ with transverse momentum}

We turn now to 
the correlation functions with the dependence on the transverse
momenta, Eqs.~\eqref{e:phinoint} and \eqref{e:anotherdelta}. 
Consequently, their general decomposition is richer than before. We
already know what is the form of the correlation function $\Phi$
[Eq.~\eqref{e:phifull}].  For what
concerns the correlation function $\Delta$, the inclusion of transverse
momentum results in 
\begin{multline}
{\cal P}_-\,\Delta(z, \zeta, M_h^2, \phi_R, \mb{k}_T)\, \g^- =
\frac{1}{8 \pi}\Biggl( D_1\bigl(z, \zeta, M_h^2, \mb{k}_T^2, \mb{k}_T \cdot \mb{R}_T\bigr) 
+ \ii H_1^{\newangle \prime}\bigl(z, \zeta,M_h^2, \mb{k}_T^2, \mb{k}_T \cdot
\mb{R}_T\bigr) \,\frac{\Rslash_T}{M_h} \\ 
+ \ii H_1^{\perp}\bigl(z, \zeta,M_h^2, \mb{k}_T^2, \mb{k}_T \cdot \mb{R}_T\bigr) \,\frac{\kslash_T}{M_h} 
+ G_1^{\perp}\bigl(z, \zeta,M_h^2, \mb{k}_T^2, \mb{k}_T \cdot \mb{R}_T\bigr) \,\frac{\eps_T^{\mspace{2mu}\mu \nu} R_{T
\mu} k_{T \nu}}{M_h^2}\, \g_5 
\Biggr)\,{\cal P}_- .
\end{multline}
Once again, we can express the correlation function as a matrix in the
chirality space of the quark
\begin{equation} 
D(z, \zeta, M_h^2, \phi_R, \mb{k}_T)_{\chi'_{2} \chi^{\phantom'}_{2}} =  
\frac{1}{8 \pi}
\begin{pmatrix}
D_1 + \dfrac{\lf\lvert\mb{k}_T \times \mb{R}_T\rg\rvert}{M_h^2}\,G_1^{\perp} &  
	\ii \lf(\e^{\ii \phi_R} \dfrac{\lf\lvert \mb{R}_T \rg\rvert}{M_h}\,H_1^{\newangle \prime}
	+ \e^{\ii \phi_k} \dfrac{\lf\lvert \mb{k}_T \rg\rvert}{M_h}\,H_1^{\perp}\rg)\\
-\ii \lf(\e^{-\ii \phi_R} \dfrac{\lf\lvert \mb{R}_T \rg\rvert}{M_h}\,H_1^{\newangle \prime} 
	 + \e^{-\ii \phi_k}\dfrac{\lf\lvert \mb{k}_T \rg\rvert}{M_h}\,H_1^{\perp}\rg)& 
		D_1 - \dfrac{\lf\lvert\mb{k}_T \times \mb{R}_T\rg\rvert}{M_h^2}\,G_1^{\perp}		
\end{pmatrix}.
\label{e:Dmatrix2b}
\end{equation} 
We can easily obtain positivity bounds on the
transverse-momentum dependent functions
\begin{align}	
\frac{\lf\lvert\mb{k}_T \times \mb{R}_T\rg\rvert}{M_h^2} 
\lf\lvert G_1^{\perp} \rg \rvert & \le 
D_1, \\
\frac{\lf\lvert\mb{R}_T\rg\rvert^2}{M_h^2}\, \lf(H_1^{\newangle\prime}\rg)^2 
	+ \frac{\lf\lvert\mb{k}_T\rg\rvert^2}{M_h^2}\, \lf(H_1^{\perp}\rg)^2 
		+ \frac{2 \,\mb{k}_T \cdot \mb{R}_T}{M_h^2}\,H_1^{\newangle \prime}H_1^{\perp}
& \le
D_1^2 - \frac{\lf\lvert\mb{k}_T \times \mb{R}_T\rg\rvert^2}{M_h^4}\, \lf(G_1^{\perp}\rg)^2,
\end{align} 
where it is understood that all the functions are dependent on the variables
$z,\, \zeta,\, M_h^2,\, \mb{k}_T^2,\,  \mb{k}_T \cdot \mb{R}_T$.

Note that if we integrate over $\mb{k}_T$, the fragmentation function
$G_1^{\perp}$ will disappear, while parts of 
the functions $H_1^{\newangle \prime}$ and
$H_1^{\perp}$ will merge into a single function, $H_1^{\newangle}$, and we
will recover the results of Sec.~\vref{s:delta2p}.

%%%%%%%%%%%%%%%%%%%%%%%%%%%%%%%%%%%%%%%%%%%%%%%%%%%%%%%%%%%%
\section{Cross section and asymmetries with transverse momentum}

Prior to integration over the 
center-of-mass transverse momentum
we have the nine-fold cross section
\begin{multline}
\frac{\de^9\! \sigma}{\de \zeta \de M_h^2 \de \phi_R \de z_h\de^2\! \mb{P}_{h \perp} \de \xbj
\de y  \de \phi_S}
= \\
\sum_q\, \rho(S)_{\Lambda^{\phantom'}_1 \Lambda_1'}\;{\cal I} \lf[
F(\xbj, \mb{p}_T)_{\chi^{\phantom'}_{1} \chi'_{1}}^{\Lambda_1' \Lambda^{\phantom'}_1}
\Bigl( \frac{\de \sigma^{eq}}{\de y} \Bigr)
	^{\chi^{\phantom'}_{1} \chi'_{1} ; \,\chi^{\phantom'}_{2} \chi'_{2}}
	\,D(z_h, \mb{k}_T, \zeta, M_h^2, \phi_R)_{\chi'_{2} \chi^{\phantom'}_{2}}\rg], 
\label{e:2pcrosswithpt}
\end{multline} 
To simplify the notation, we introduce the projection $\mb{a}_T \wedge
\mb{b}_T = a_i \eps_T^{ij} b_j$. 
Inserting 
the formulae obtained for the distribution correlation matrix, 
Eq.~\eqref{e:Fmatrix}, the
elementary cross section, Eq.~\eqref{e:eq}, and the two-hadron
fragmentation matrix, Eq.~\eqref{e:Dmatrix2b}, we obtain the
following result\footnote{The $UU$ and $UT$ cross sections correspond
 to the one calculated 
in Ref.~\citen{Radici:2001na}
except for a difference in the overall coefficient -- 
an extra factor 2 in the denominator, due to the use of the
variable $\zeta$ instead of $\xi$, the lack of a factor $(2 \pi)^3$ in the
denominator, due to the different definition of the hadronic tensor, an extra
factor $2/\pi$ due to the different definitions of the fragmentation functions--
the use of $M_h$ in the denominators
instead of $M_1$, $M_2$ or $M_1+M_2$, and a factor 2 difference in the
definition of the coefficient $C(y)$.}
%\begin{subequations}
\begin{align}	
\begin{split}
\label{e:xsectOO}
\de^9\!\sigma_{UU}  
&=
\sum_{q} \frac{\alpha^2 e_q^2}{2\pi s \xbj y^2}\,\Biggl\{ 
A(y)\,{\cal I}\left[f_1 \, D_1\right]
-B(y)\,\frac{|{\mb{R}}_\st|}{M_h}\,\cos(\phi_h+\phi_R)\,
   {\cal I}\left[\frac{\mb{p}_T \cdot \h}{M}\,
     h_1^{\perp} \, H_1^{\newangle \prime}\right]
\\ & \quad 
  +B(y)\,\frac{|{\mb{R}}_\st|}{M_h}\,\sin(\phi_h+\phi_R)\,
   {\cal I}\left[\frac{\h \wedge \mb{p}_T}{M}\,
     h_1^{\perp} \, H_1^{\newangle \prime}\right]
\\&\quad
-B(y)\,\cos(2\phi_h)\,
   {\cal I}\left[\frac{2(\mb{p}_T \cdot \h)(\mb{k}_T
 \cdot \h) - \mb{p}_T\cdot \mb{k}_T}{M M_h}\,h_1^{\perp} \, H_1^{\perp}\right]
\\ & \quad
+B(y)\,\sin(2\phi_h)\,
   {\cal I}\left[\frac{(\mb{p}_T \cdot \h)(\h \wedge \mb{k}_T) +(\mb{k}_T \cdot \h)(\h \wedge \mb{p}_T)}{M M_h}\,h_1^{\perp} \, H_1^{\perp}\right]\Biggr\},
\end{split} \\
\begin{split} 
\label{e:xsectLO}
\de^9\!\sigma_{LU}  
&= 
- \sum_{q} \frac{\alpha^2 e_q^2}{2\pi s \xbj y^2}\,\lf\lvert \lambda_e \rg\rvert\,
C(y)\,\frac{|{\mb{R}}_\st|}{M_h}\,\Biggl\{\sin(\phi_h-\phi_R)\,
   {\cal I}\left[\frac{\mb{k}_T\cdot\h}{M_h}\,
     f_1 \, G_1^{\perp}\right]
\\ & \quad 
+ \cos(\phi_h-\phi_R)\,
   {\cal I}\left[\frac{\h \wedge \mb{k}_T}{M_h}\,
     f_1 \, G_1^{\perp}\right]\Biggr\},
\end{split} \\
\begin{split} 
\label{e:xsectOL}
\de^9\!\sigma_{UL}  
&=
\sum_{q} \frac{\alpha^2 e_q^2}{2\pi s \xbj y^2}\,\lf\lvert S_L \rg\rvert\Biggl\{
-A(y)\,\frac{|{\mb{R}}_\st|}{M_h}\,\sin(\phi_h-\phi_R)\,
   {\cal I}\left[\frac{\mb{k}_T\cdot\h}{M_h}\,
     g_{1L} \, G_1^{\perp}\right] 
\\ &\quad
-A(y)\,\frac{|{\mb{R}}_\st|}{M_h}\,\cos(\phi_h-\phi_R)\,
   {\cal I}\left[\frac{\h \wedge \mb{k}_T}{M_h}\,
     g_{1L} \, G_1^{\perp}\right] 
\\ &\quad
+B(y)\,\frac{|{\mb{R}}_\st|}{M_h}\,\sin(\phi_h+\phi_R)\,
   {\cal I}\left[\frac{\mb{p}_T\cdot\h}{M}\,
     h_{1L}^{\perp} \, H_1^{\newangle \prime}\right]\\
&\quad
+B(y)\,\frac{|{\mb{R}}_\st|}{M_h}\,\cos(\phi_h+\phi_R)\,
   {\cal I}\left[\frac{\h \wedge \mb{p}_T}{M}\,
     h_{1L}^{\perp} \, H_1^{\newangle \prime}\right]\\
&\quad
+B(y)\,\sin(2\phi_h)\,
   {\cal I}\left[\frac{2(\mb{p}_T \cdot \h)(\mb{k}_T \cdot \h) 
	- \mb{p}_T\cdot \mb{k}_T}{M M_h}\,
     h_{1L}^{\perp} \, H_1^{\perp}\right]
\\
&\quad
+B(y)\,\cos(2\phi_h)\,
   {\cal I}\left[\frac{(\mb{p}_T \cdot \h)(\h \wedge \mb{k}_T) +(\mb{k}_T \cdot \h)(\h \wedge \mb{p}_T)}{M M_h}\,
     h_{1L}^{\perp} \, H_1^{\perp}\right]\Biggr\},
\end{split} \\
\begin{split} 
\label{e:xsectLL}
\de^9\!\sigma_{LL}  
&=
\sum_{q} \frac{\alpha^2 e_q^2}{2\pi s \xbj y^2}\,\lf\lvert \lambda_e \rg\rvert\,
\lf\lvert S_L \rg\rvert\, C(y)\,
   {\cal I}\left[g_{1L} \, D_1\right],
\end{split} \\
\begin{split} 
\label{e:xsectOT}
\de^9\!\sigma_{UT}  
&=
\sum_{q} \frac{\alpha^2 e_q^2}{2\pi s \xbj y^2}\,\sst\,
A(y)\,\Biggl\{\frac{|{\mb{R}}_\st|}{M_h}\,
   \sin(\phi_R-\phi_{S})\,
   {\cal I}\left[\frac{\mb{p}_T\cdot \mb{k}_T}{2 M M_h}
     g_{1T} \, G_1^{\perp}\right]
\\ &\quad
-\frac{|{\mb{R}}_\st|}{M_h}\,
   \cos(\phi_R-\phi_{S})\,
   {\cal I}\left[\frac{(\mb{p}_T \cdot \h)(\h \wedge \mb{k}_T) -(\mb{k}_T \cdot \h)(\h \wedge \mb{p}_T)}{2 M M_h}
     g_{1T} \, G_1^{\perp}\right]
\\ &\quad
-\frac{|{\mb{R}}_\st|}{M_h}\,
   \sin(2\phi_h-\phi_R-\phi_{S})\,
   {\cal I}\left[\frac{2(\mb{p}_T \cdot\h)(\mb{k}_T \cdot \h) 
	- \mb{p}_T\cdot \mb{k}_T}{2M M_h}\,
     g_{1T} \, G_1^{\perp}\right]
\\ &\quad
-\frac{|{\mb{R}}_\st|}{M_h}\,
   \cos(2\phi_h-\phi_R-\phi_{S})\,
   {\cal I}\left[\frac{(\mb{p}_T \cdot \h)(\h \wedge \mb{k}_T) +(\mb{k}_T \cdot \h)(\h \wedge \mb{p}_T)}{2M M_h}\,
     g_{1T} \, G_1^{\perp}\right]\\
&\quad
+ \sin(\phi_h-\phi_{S})\,
   {\cal I}\left[\frac{\mb{p}_T\cdot\h}{M}\,
     f_{1T}^{\perp} \, D_1\right]
%\\
%&\quad
+ \cos(\phi_h-\phi_{S})\,
   {\cal I}\left[\frac{\h \wedge \mb{p}_T}{M}\,
     f_{1T}^{\perp} \, D_1\right]\Biggr\}
\\ &\quad
+\sum_{q} \frac{\alpha^2 e_q^2}{2\pi s \xbj y^2}\,\sst\,
B(y)\,\Biggl\{\sin(\phi_h+\phi_{S})\,
   {\cal I}\left[\frac{\mb{k}_T\cdot\h}{M_h}\,
     h_1 \, H_1^{\perp}\right]
\\ &\quad
+\cos(\phi_h+\phi_{S})\,
   {\cal I}\left[\frac{\h \wedge \mb{k}_T}{M_h}\,
     h_1 \, H_1^{\perp}\right]
%\\ &\quad
+\frac{|{\mb{R}}_\st|}{M_h}\,\sin(\phi_R+\phi_{S})\,
   {\cal I}\left[h_1 \, H_1^{\newangle \prime}\right]
+\sin(3\phi_h-\phi_{S}) \\
&\quad \times
   {\cal I}\left[\frac{4(\mb{p}_T\cdot\h)^2(\mb{k}_T\cdot\h)
	-2(\mb{p}_T\cdot\h)(\mb{p}_T\cdot\mb{k}_T)
	-\mb{p}_T^2(\mb{k}_T\cdot\h)}{2 M^2 M_h}\,
     h_{1T}^{\perp} \, H_1^{\perp}\right]\\
&\quad
+\cos(3\phi_h-\phi_{S})\, 
%\\
%&\quad \times
   {\cal I}\Biggl[\Biggl(\frac{
%(\mb{k}_T\cdot\h)\lf((\mb{p}_T\cdot\h)^2 + 2
%(\mb{p}_T\cdot\h)(\h \wedge \mb{p}_T) \rg) -(\h \wedge \mb{k}_T)(\h \wedge \mb{p}_T)^2
2(\mb{p}_T\cdot\h)^2(\h \wedge \mb{k}_T)
	+2(\mb{k}_T\cdot\h)(\mb{p}_T\cdot\h)(\h \wedge \mb{p}_T)}{2 M^2 M_h}
\\ & \quad -\frac{
	\mb{p}_T^2(\h \wedge \mb{k}_T)
}{2 M^2 M_h}\Biggr)\,
     h_{1T}^{\perp} \, H_1^{\perp}\Biggr]
%\\&\quad
+\frac{|{\mb{R}}_\st|}{M_h}\,
    \sin(2\phi_h+\phi_R-\phi_{S})\,
   {\cal I}\left[\frac{2(\mb{p}_T \cdot \h)^2
		- \mb{p}_T^2}{2 M^2}\,
     h_{1T}^{\perp} \, H_1^{\newangle \prime}\right]
\\
&\quad
+\frac{|{\mb{R}}_\st|}{M_h}\,
    \cos(2\phi_h+\phi_R-\phi_{S})\,
   {\cal I}\left[\frac{(\mb{p}_T \cdot \h)(\h \wedge \mb{p}_T)}{2 M^2}\,
     h_{1T}^{\perp} \, H_1^{\newangle \prime}\right]
\Biggr\},
\end{split} \\
\begin{split} 
\label{e:xsectLT}
\de^9\!\sigma_{LT}  
&=
\sum_{q} \frac{\alpha^2 e_q^2}{2\pi s \xbj y^2}\,\sst\,C(y)\,\Biggl\{ 
\cos(\phi_h-\phi_{S})\,
   {\cal I}\left[\frac{\mb{p}_T\cdot\h}{M}\,
     g_{1T} \, D_1 \right]\\
&\quad
-\sin(\phi_h-\phi_{S})\,
   {\cal I}\left[\frac{\h \wedge \mb{p}_T}{M}\,
     g_{1T} \, D_1 \right]\\
&\quad
-\frac{|{\mb{R}}_\st|}{M_h}\,
    \cos(\phi_R-\phi_{S})\,
   {\cal I}\left[\frac{\mb{p}_T\cdot \mb{k}_T}{2 M M_h}
     f_{1T}^{\perp} \, G_1^{\perp}\right]\\
&\quad
+\frac{|{\mb{R}}_\st|}{M_h}\,
    \cos(2\phi_h-\phi_R-\phi_{S})\,
   {\cal I}\left[\frac{2(\mb{p}_T \cdot \h)(\mb{k}_T \cdot \h) 
		- \mb{p}_T\cdot \mb{k}_T}{2 M M_h}\,
     f_{1T}^{\perp} \, G_1^{\perp}\right]
\\
&\quad
-\frac{|{\mb{R}}_\st|}{M_h}\,
    \sin(\phi_R-\phi_{S})\,
   {\cal I}\left[\frac{(\mb{p}_T \cdot \h)(\h \wedge \mb{k}_T) -(\mb{k}_T \cdot \h)(\h \wedge \mb{p}_T)}{2 M M_h}
     f_{1T}^{\perp} \, G_1^{\perp}\right]\\
&\quad
+\frac{|{\mb{R}}_\st|}{M_h}\,
    \sin(2\phi_h-\phi_R-\phi_{S})\,
   {\cal I}\left[\frac{(\mb{p}_T \cdot \h)(\h \wedge \mb{k}_T) +(\mb{k}_T \cdot \h)(\h \wedge \mb{p}_T)}{2 M M_h}\,
     f_{1T}^{\perp} \, G_1^{\perp}\right]
\Biggr\}.
\end{split}
\end{align}
%\end{subequations}
 
%%%%%%%%%%%%%%%%%%%%%%%%%%%%%%%%%%%%%%%%%%%%%%%%%%%%%%%%%%%%
\section{Partial-wave expansion with transverse momentum}

To proceed with the partial wave expansion of transverse-momentum dependent
functions, it turns out to be convenient
 to rewrite the correlation function in a somewhat
different way, i.e.
\begin{equation} 
D(z, \zeta, M_h^2, \phi_R, \mb{k}_T)_{\chi'_{2} \chi^{\phantom'}_{2}} =  
\frac{1}{8 \pi}
\begin{pmatrix}
D_1 + \dfrac{\lf\lvert\mb{k}_T \times \mb{R}_T\rg\rvert}{M_h^2}\,G_1^{\perp} &  
	\ii \e^{\ii \phi_k} \,\dfrac{\lf\lvert \mb{k}_T \rg\rvert}{M_h}
		\,H_1^{\perp \newangle}\\
-\ii \e^{-\ii \phi_k}\,\dfrac{\lf\lvert \mb{k}_T \rg\rvert}{M_h}
		\,\lf(H_1^{\perp \newangle}\rg)^{\ast}& 
		D_1 - \dfrac{\lf\lvert\mb{k}_T \times \mb{R}_T\rg\rvert}{M_h^2}\,G_1^{\perp}		
\end{pmatrix}.
\end{equation}
The functions $H_1^{\newangle}$ and $H_1^{\perp}$ have been merged into the
complex function 
\begin{equation} \begin{split}
H_1^{\perp \newangle}\lf(z, \zeta, M_h^2, \mb{k}_T^2,\e^{\ii(\phi_k -
\phi_R)} \rg) &= H_1^{\perp}\lf(z, \zeta, M_h^2, \mb{k}_T^2, \cos(\phi_k -
\phi_R)\rg) 
\\ &\quad
+\e^{\ii(\phi_R-\phi_k)}\,H_1^{\newangle}
\lf(z, \zeta, M_h^2, \mb{k}_T^2, \cos(\phi_k -
\phi_R)\rg).
\label{e:h1perpang}
\end{split} \end{equation}   
Because of the presence of the azimuthal angles, this function can be 
expanded in spherical harmonics, provided we retain only those terms which are 
consistent with the definition in Eq.~\eqref{e:h1perpang}.
If we wish to include 
only the $s$- and $p$-wave contributions, the partial wave 
expansion of the function $H_1^{\perp \newangle}$ takes the form
\begin{equation} \begin{split}   
H_1^{\perp \newangle} &=  H_{1, UU}^{\perp} + H_{1, UL}^{\perp\,sp}\cos{\theta}
 + H_{1, LL}^{\perp\,pp}\,\frac{1}{4}\lf(3\cos^2{\theta} -1\rg)
%\\ &\quad 
+ \e^{\ii(\phi_k - \phi_R)} \sin \theta 
	\lf(H_{1,UT}^{\perp\,sp} +  H_{1, LT}^{\perp\,pp}\cos{\theta} \rg)
\\ &\quad 
+ \e^{-\ii(\phi_k - \phi_R)} \sin \theta 
	\lf(H_{1,UT}^{\perp\,sp} +  H_{1, LT}^{\perp\,pp}\cos{\theta} 
+ \frac{\rr}{\lf\lvert \mb{k}_T \rg\rvert}\,H_{1, UT}^{\newangle \, sp \prime} + \frac{\rr}{\lf\lvert \mb{k}_T \rg\rvert}\,H_{1,LT}^{\newangle \, pp \prime}
		\cos{\theta}	\rg)
\\ &\quad 
+ \e^{2\ii(\phi_k - \phi_R)} \sin^2 \theta \,H_{1, TT}^{\perp\,pp}
+ \e^{-2\ii(\phi_k - \phi_R)} \sin^2 \theta \lf(H_{1, TT}^{\perp\,pp}+
\frac{\rr}{\lf\lvert \mb{k}_T \rg\rvert}\,H_{1, TT}^{\newangle \, pp}\rg)
\end{split} \end{equation}  
Note that the functions on
the right-hand side depend only on the variables $z$,
 $M_h^2$, $\mb{k}_T^2$.

The expansion of the other two functions, $D_1$ and $G_1^{\perp}$, is
considerably simpler, because they can depend only on  $\cos(\phi_h - \phi_R)$
\begin{subequations} \begin{align} 
%\begin{split}
D_1 &= D_{1, UU} + D_{1, UL}^{sp}\cos{\theta} + D_{1, LL}^{pp}\,\frac{1}{4}\lf(3
\cos^2{\theta} -1\rg)   
+ \cos(\phi_k - \phi_R)\,\sin{\theta} 
\lf(D_{1,UT}^{\perp\,sp} +  D_{1, LT}^{\perp\,pp}\cos{\theta}\rg)
\nn \\ &\quad 
+\cos(2\phi_k - 2\phi_R)\,\sin^2\theta\,D_{1, TT}^{\perp\,pp} , 
%\end{split} 
\\
G_1^{\perp}& = G_{1, UT}^{\perp\,sp} +  G_{1, LT}^{\perp\,pp}\cos{\theta}
 + \cos(\phi_k - \phi_R)\,\sin{\theta} \, G_{1, TT}^{\perp\,pp},
\end{align} \end{subequations} 
where the functions on
the right-hand side depend only on the variables $z$,
 $M_h^2$, $\mb{k}_T^2$.

As a conclusive remark, note that it would have been possible to expand the 
functions $H_1^{\newangle}$ and $H_1^{\perp}$ in a way similar to what we have 
done for $D_1$ and $G_1^{\perp}$, but 
some care is required to properly treat the component 
$H_{1,TT}^{\newangle \,pp}$,  
which is shared by the two functions. This is
why we preferred to take an alternative way. In any case, for completeness we
give also the expansions of 
$H_1^{\newangle}$ and $H_1^{\perp}$ separately:
\begin{subequations} \begin{align} 
%\begin{split}
H_1^{\perp} &= H_{1, UU}^{\perp} + H_{1, UL}^{\perp\,sp}\cos{\theta}
 + H_{1, LL}^{\perp\,pp}\,\frac{1}{4}\lf(3\cos^2{\theta} -1\rg)
+ 2 \cos(\phi_k - \phi_R)\,\sin{\theta} 
\lf(H_{1,UT}^{\perp\,sp} +  H_{1, LT}^{\perp\,pp}\cos{\theta} \rg) 
\nn \\ &\quad 
+2 \cos (2\phi_k - 2\phi_R)\,\sin^2\theta\,H_{1, TT}^{\perp\,pp} 
-\sin^2\theta\, \frac{\rr}{\ktt}\,H_{1, TT}^{\newangle\,pp}, 
%\end{split} 
\\
H_1^{\newangle}& = H_{1, UT}^{\newangle\,sp\prime} +  H_{1, LT}^{\newangle\,pp 
\prime}\cos{\theta}
 + 2 \cos(\phi_k - \phi_R)\,\sin{\theta} \, H_{1, TT}^{\newangle\,pp}.
\end{align} \end{subequations} 
As before, the functions on
the right-hand side depend only on the variables $z$, $M_h^2$, $\mb{k}_T^2$.

As we have done in Sec.~\vref{s:deltapw}, 
it is possible to rewrite the correlation function as
a trace between the decay matrix, Eq.~\eqref{e:decaymatrix}, and a
fragmentation matrix in the quark chirality space
$\otimes$ the hadronic system angular momentum space
\begin{equation} 
D(z, \cos \theta, M_h^2, \phi_R, \mb{k}_T)_{\chi'_{2} \chi^{\phantom'}_{2}} =
D(z, M_h^2, \mb{k}_T^2)_{\chi'_{2} \chi^{\phantom'}_{2}}^{j'm',jm}
{\cal D}_{jm,j'm'}(\theta,\phi_R).
\end{equation}
Once again, the result is a $8 \times 8$ matrix
\begin{equation} 
D(z, M_h^2, \mb{k}_T^2)_{\chi'_{2} \chi^{\phantom'}_{2}}^{j'm',jm} = 
\frac{1}{8}
\begin{pmatrix}
A_{j'm',jm} & B_{j'm',jm} \\
B_{j'm',jm}^{\dagger} & C_{j'm',jm}
\end{pmatrix}.
\label{e:Dzmhkt}
\end{equation} 
The inner blocks span the space of the orbital angular momentum of the
hadronic system and read
%\begin{flalign} 
%\addtolength{\extrarowheight}{5pt}
\begin{multline*}
A_{j'm',jm} = 
\\
\lf(\begin{array}{>{\scriptstyle}c|>{\scriptstyle}c>{\scriptstyle}c>{\scriptstyle}c}
D_{1, UU}^{ss} 
& 
-\sqrt{\frac{2}{3}}\e^{\ii\phi}
	\lf(D_{1,UT}^{\perp\,sp}
		+\ii \frac{\ktt \rr}{M_h^2} G_{1, UT}^{\perp\,sp}\rg) 
& 
\frac{2}{\sqrt{3}}D_{1, UL}^{sp} 
& 
\sqrt{\frac{2}{3}}\e^{-\ii\phi}
	\lf(D_{1,UT}^{\perp\,sp}
		-\ii \frac{\ktt \rr}{M_h^2} G_{1, UT}^{\perp\,sp}\rg) 
\\[3pt] \hline
-\sqrt{\frac{2}{3}}\e^{-\ii\phi}
	\lf(D_{1,UT}^{\perp\,sp}
		-\ii \frac{\ktt \rr}{M_h^2}G_{1, UT}^{\perp\,sp}\rg)
&
  D_{1, UU}^{pp} -\frac{1}{3}D_{1, LL}^{pp} 
& 
-\frac{\sqrt{2}}{3} \e^{-\ii\phi}\lf(D_{1, LT}^{\perp\,pp} 
	- \ii\frac{\ktt \rr}{M_h^2} G_{1, LT}^{\perp\,pp} \rg)
& 
-\frac{2}{3} \e^{-2\ii\phi}\lf(2 D_{1, TT}^{\perp\,pp}
	-\ii\frac{\ktt \rr}{M_h^2} G_{1, TT}^{\perp\,pp}  \rg) 
\\
\frac{2}{\sqrt{3}} D_{1, UL}^{sp} 
& 
-\frac{\sqrt{2}}{3} \e^{\ii\phi}\lf(D_{1, LT}^{\perp\,pp} 
	+\ii\frac{\ktt \rr}{M_h^2} G_{1, LT}^{\perp\,pp}\rg)  	
& 
 D_{1, UU}^{pp} +\frac{2}{3} D_{1, LL}^{pp} 
& 
\frac{\sqrt{2}}{3} \e^{-\ii\phi}\lf(D_{1, LT}^{\perp\,pp} 
	- \ii\frac{\ktt \rr}{M_h^2} G_{1, LT}^{\perp\,pp} \rg)
\\
\sqrt{\frac{2}{3}}\e^{\ii\phi}
	\lf(D_{1,UT}^{\perp\,sp}
		+\ii \frac{\ktt \rr}{M_h^2} G_{1, UT}^{\perp\,sp}\rg) 
& 
-\frac{2}{3} \e^{2\ii\phi}\lf(2 D_{1, TT}^{\perp\,pp}
	+\ii\frac{\ktt \rr}{M_h^2} G_{1, TT}^{\perp\,pp}  \rg)
& 
\frac{\sqrt{2}}{3} \e^{\ii\phi}\lf(D_{1, LT}^{\perp\,pp} 
	+\ii\frac{\ktt \rr}{M_h^2} G_{1, LT}^{\perp\,pp}\rg) 
& 
 D_{1, UU}^{pp} -\frac{1}{3}D_{1, LL}^{pp} 
\end{array}\rg), 
\end{multline*}
%\begin{multline*}
\begin{equation*} 
B_{j'm',jm} =
%\\
\ii \frac{\ktt}{M_h}
\lf(\begin{array}
{>{\scriptstyle}c|>{\scriptstyle}c>{\scriptstyle}c>{\scriptstyle}c}
\e^{\ii\phi} H_{1, UU}^{\perp\,ss}
& 
 -\frac{4}{\sqrt{6}} \e^{2\ii\phi} H_{1,UT}^{\perp\,sp}
& 
 \frac{2}{\sqrt{3}} \e^{\ii\phi} H_{1,UL}^{\perp\,sp}
& 
 \frac{4}{\sqrt{6}}\lf( \frac{\rr}{\ktt} H_{1,UT}^{\newangle \, sp \prime}
	+ H_{1,UT}^{\perp \, sp} \rg)
\\[3pt] \hline
-\frac{4}{\sqrt{6}}\lf( \frac{\rr}{\ktt} H_{1,UT}^{\newangle \, sp \prime}
	+ H_{1,UT}^{\perp \, sp} \rg)
& 
 \e^{\ii\phi}\lf( H_{1, UU}^{\perp\,pp}  -\frac{1}{3}H_{1, LL}^{\perp\,pp}\rg)
& 
- \frac{2 \sqrt{2}}{3}\lf(\frac{\rr}{\ktt} H_{1, LT}^{\newangle \, pp \prime} 
		+ H_{1, LT}^{\perp \, pp}\rg)
& 
 -\frac{8}{3}\e^{-\ii\phi}\lf(\frac{\rr}{\ktt} H_{1, TT}^{\newangle \, pp} 
		+ H_{1, TT}^{\perp \, pp}\rg)
\\
\frac{2}{\sqrt{3}} \e^{\ii\phi} H_{1,UL}^{\perp\,sp}
& 
- \frac{2\sqrt{2}}{3}\e^{2\ii\phi}  H_{1, LT}^{\perp \, pp}
& 
 \e^{\ii\phi}\lf( H_{1, UU}^{\perp\,pp}  +\frac{2}{3}H_{1, LL}^{\perp\,pp}\rg)
& 
 \frac{2 \sqrt{2}}{3}\lf(\frac{\rr}{\ktt}H_{1, LT}^{\newangle \, pp \prime} 
		+ H_{1, LT}^{\perp \, pp}\rg)
\\
  \frac{4}{\sqrt{6}} \e^{2\ii\phi} H_{1,UT}^{\perp\,sp}
& 
  -\frac{8}{3}\e^{3\ii\phi} H_{1, TT}^{\perp \, pp} 
& 
  \frac{2\sqrt{2}}{3}\e^{2\ii\phi}  H_{1, LT}^{\perp \, pp}
& 
  \e^{\ii\phi}\lf( H_{1, UU}^{\perp\,pp}  -\frac{1}{3}H_{1, LL}^{\perp\,pp}\rg)
\end{array}\rg).
%\end{multline*} 
\end{equation*}
The block $C$ of the fragmentation matrix can be obtained from the block $A$
by imposing parity invariance relations.

The $ss$ sector of the matrix 
contains functions analogous to the ones 
we discussed in the previous chapter in 
the context of single-particle 
unpolarized fragmentation. The $pp$ sector corresponds to the
spin-one fragmentation functions studied in Ref.~\citen{Bacchetta:2000jk}. We will
discuss them again in the next chapter. The $sp$ interference sector has never 
been studied with the inclusion of transverse momentum.
 
Similar considerations to the ones discussed after Eqs.~\eqref{e:blockB} hold
for the transverse momentum dependent fragmentation matrix. In particular, 
the matrix
fulfills the properties of Hermiticity, conservation of angular momentum
($m+\chi'_{2}=m' +\chi_{2}+l$) and parity invariance
\begin{equation} 
D_{\chi'_{2}\chi^{\phantom'}_{2}}^{j'm',jm}
=(-1)^{l}D_{{-\chi'_{2}}\, {-\chi^{\phantom'}_{2}}}^{j'\,{-m'},j\;{-m}}.
\end{equation}
The imaginary parts of the matrix represent T-odd functions. Note that the
off-diagonal blocks, which are chiral-odd, can only contain T-odd functions.
The matrix could in principle contain other functions, but they are lost when
tracing with the decay matrix and thus cannot be analyzed by a parity
conserving decay process.

From the fact that the matrix $D(z, M_h^2, \mb{k}_T^2)$ is positive
semi-definite, it is possible to obtain positivity bounds on the fragmentation
functions, as we have done already before. 

%%%%%%%%%%%%%%%%%%%%%%%%%%%%%%%%%%%%%%%%%%%%%%%%%%%%%%%%%%%%
\section{Summary}
 
In this chapter we analyzed two-particle inclusive deep inelastic scattering,
at leading order in $1/Q$, without and with partonic transverse momentum.
For the description of the fragmentation side, we introduced 
a correlation function $\Delta$ that depends on
the center-of-mass momentum of the two hadrons and on their relative momentum.

In the first part of the chapter we neglected partonic transverse momentum.
We distinguished one chiral-even T-even and one chiral-odd T-odd 
fragmentation
function [Eq.~\eqref{e:decomdelta2}]. They are complicated objects depending on three variables, whereas
single-particle fragmentation functions depend on one variable only. We
derived positivity bounds on these functions [Eqs.~\eqref{e:boundstwohadron1}] and we computed the cross section
of two-particle inclusive scattering [Eq.~\eqref{e:crosstwohadron}].

Since two-particle systems with a low invariant mass are usually produced in
the $s$ and $p$ waves, we performed a partial wave expansion of the
fragmentation functions taking into consideration only these two lowest
modes.
We split the chiral-even fragmentation function into three contributions,
pertaining to the pure $s$ waves, pure $p$ waves and $sp$ interference.  
We split the chiral-odd function into two contributions, typical of 
pure $p$ waves and $sp$ interference. These five new fragmentation functions
depend only on two variables, namely $z$ and the invariant mass squared of the 
system [Eq.~\eqref{e:deltapwexpanded}].
Employing the usual helicity formalism, we discussed positivity bounds on the
new fragmentation functions [Eqs.~\eqref{e:pwbound1} and~\eqref{e:pwbound2}].
We computed all non vanishing spin asymmetries of two-particle inclusive DIS
at leading twist [Eqs.~\eqref{e:cross2ppw}]. In particular, we showed that
in the single
transverse spin asymmetry of Eq.~\eqref{e:asymmtwohadron} 
the transversity distribution appears in
connection with both chiral-odd fragmentation functions. Integrating the
asymmetry over different ranges of the polar angle $\theta$, we presented two
distinct ways to access the transversity distribution. The first one 
[Eq.~\eqref{e:asym3}] involves
only the $sp$ interference term and corresponds to the one discussed in
Ref.~\citen{Jaffe:1998hf}, while the second one [Eq.~\eqref{e:asym4}] involves also the pure $p$ term, 
which could have an
entirely different physical origin and should be particularly relevant in the presence of a spin-one resonance. These two asymmetries are in all respects
two distinct ways to access the transversity distribution function. The second 
one has never been clearly indicated in the literature.

In the rest of the chapter, we repeated the analysis of the correlation
function including partonic transverse momentum and also in this case we
applied a partial wave expansion.

\renewcommand{\quot}{%
\parbox{6cm}{Looks like we've made it,\\
	Look how far we've come, my baby.\\ 
	You are still the one that I love, \\
	The only one I dream of.}\\[2mm] S. Twain  
}

%%%%%%%%%%%%%%%%%%%%%%%%%%%%%%%%%%%%%%%%%%%%%%%%%%%%%%%%%%%%%%%%%%%%%%%%%%%%
\chapter{Spin one}
\label{c:spinone}

In the previous chapter we examined the production of two hadrons in the 
$s$ and $p$
waves. As we already mentioned, the $p$-wave sector of the analysis should
overlap with the description of the fragmentation functions for spin-one
mesons. The reason is that the polarization of the vector meson can be
analyzed by measuring its decay in two other particles, as
in the case of a $\rho$ meson decaying into two pions.

In this chapter, we will take a small detour from the mainstream of the
thesis, and we shall deal with spin-one targets and spin-one fragments. In 
the first part, we will examine what is necessary for the description of a
spin-one object and what are the differences with the spin-half case. We will
introduce distribution functions for quarks inside spin-one objects, with and
without transverse momentum. 

Then we will turn our attention to spin-one hadrons in the final state and we
will examine the fragmentation functions for a quark producing a spin-one
hadron. In the end, we will check whether the connection with two-hadron
fragmentation can be established and if there are any differences between the
two situations.

%%%%%%%%%%%%%%%%%%%%%%%%%%%%%%%%%%%%%%%%%%%%%%%%%%%%%%%%%%%%
\section{Spin-one targets}

In recent years some attention has been devoted to distribution functions
characterizing  spin-one targets, starting from the work of
Hoodbhoy, Jaffe and Manohar~\cite{Hoodbhoy:1989am}. 
Unfortunately, the only
available spin-one target is the deuteron, which is essentially a weakly bound
system of two spin-half hadrons. In fact, deuteron targets are often used in
deep inelastic scattering with the major purpose of extracting the neutron
distribution functions. However, the spin-one structure of the deuteron in
itself can be very interesting~\cite{Nikolaev:1997jy,Bora:1998pi,Edelmann:1997qe,Umnikov:1997qv}.
A large amount of work is already present on deuteron structure functions,
especially in Drell-Yan processes~\cite{Hino:1998ww,Hino:1999fh}.

%%%%%%%%%%%%%%%%%%%%%%%%%%%%%%%%%%%%%%%%%%%%%%%%%%%%%%%%%%%%%%%%%%%%%%%%%%%%
\subsection{Spin density matrix and spin tensor}
\label{s:spinone}

The description of particles with spin can be attained by using a spin density
matrix ${\bm \rho}$ in the rest frame of the particle.
The parametrization of the density matrix for a spin-$J$ particle is
conveniently  performed with the introduction of irreducible spin tensors up
to rank $2J$.
For example, we have already seen [cf. Eq.~\eqref{e:rho}] that
 the density matrix of a spin-half particle can be 
decomposed on a Cartesian basis of $2 \times 2$ matrices, 
formed by the identity matrix and the three Pauli matrices,
\begin{equation}
{\bm \rho}=\frac{1}{2}\left(\mbox{\bf 1} + S^i {\bm \sigma}^i \right),
\end{equation}
where we introduced the (rank-one) spin vector $S^i$.

To parametrize the density matrix of a spin-one particle we can choose a 
Cartesian basis of
$3\times 3$ matrices, formed by the identity matrix,
three spin matrices ${{\bm{\Sigma}}}^i$
(generalization of the Pauli matrices 
to the three-dimensional case) and five extra matrices ${{\bm{\Sigma}}}^{ij}$. 
These last ones
can be built using bilinear combinations of the spin matrices. In three 
dimensions these combinations are no longer dependent on the spin matrices 
themselves, as would be for the Pauli matrices.
We choose them to be (see Refs.~\citen{Bourrely:1980mr,Elliot} and 
\citen{Madison:1971} for a comparison)
\begin{equation}
{{\bm{\Sigma}}}^{ij}=\frac{1}{2}\left({{\bm{\Sigma}}}^i {{\bm{\Sigma}}}^j 
	+{{\bm{\Sigma}}}^j {{\bm{\Sigma}}}^i\right)
		-\frac{2}{3}\,{\bf 1} \; \delta^{ij}.
\end{equation}

With these preliminaries, we can write the spin density matrix as
\begin{equation}
{\bm \rho}=\frac{1}{3}\left({\bf 1} + \frac{3}{2}\, S^i {{\bm{\Sigma}}}^i 
		+ 3\, T^{ij} {{\bm{\Sigma}}}^{ij}\right),
\label{e:density}
\end{equation}
where we introduced the symmetric traceless rank-two spin tensor $T^{ij}$.
We choose the following  way of parametrizing the spin vector and tensor
in the rest frame of the hadron,
\begin{subequations} \label{e:vectorandtensor} \begin{align} 
{\mb S}&= \left(S_{T}^x, S_{T}^y, S_{L}\right),    \label{e:vector}\\
{\mb T}&= \frac{1}{2}\left(\begin{array}{ccc}
	{S_{LL}}+{S_{TT}^{xx}} 
		& {S_{TT}^{xy}}	& {S_{LT}^{x}} \\
	{S_{TT}^{xy}}& 	{S_{LL}}-{S_{TT}^{xx}}
					& {S_{LT}^{y}}  \\
	 {S_{LT}^{x}}	& {S_{LT}^{y}}	& -2{S_{LL}} \\
	\end{array}\right).
\label{e:tensor}
\end{align} \end{subequations} 
The parameter $S_{LL}$ is often called {\em
alignment}~\cite{Bourrely:1980mr,Elliot,Efremov:1982vs,Efremov:1994xf}. 

Although we introduced the spin vector and tensor in the particle rest frame,
it is possible to introduce covariant generalizations of them,  
satisfying
the conditions $P_{\mu}S^{\mu}=0$ and $P_{\mu}T^{\mu \nu}=0$,
where $P_{\mu}$ is the momentum of the hadron. In a
collinear frame, the
parametrizations of the covariant spin vector and tensor will be
\begin{subequations}\label{e:covariantST}
\begin{align}
S^{\mu} &= \biggl[-S_L\,\frac{M}{2 P^+},\; S_L\,\frac{P^+}{M},
	\; \mb{S}_T \biggr], \\
T^{\mu \nu} &=\frac{1}{2}
	\begin{bmatrix}
	-S_{LL}\frac{M^2}{2 (P^+)^2} & S_{LL} & 
		- S_{LT}^x \frac{M}{2P^+} &- S_{LT}^y \frac{M}{2P^+} \\
	S_{LL} & -2 S_{LL}\frac{(P^+)^2}{M^2} &
		 S_{LT}^x \frac{P^+}{M} & S_{LT}^y \frac{P^+}{M} \\
	- S_{LT}^x \frac{M}{2P^+} & S_{LT}^x \frac{P^+}{M} &
		S_{TT}^{xx} +S_{LL} & S_{TT}^{xy} \\
	- S_{LT}^y \frac{M}{2P^+} & S_{LT}^y \frac{P^+}{M} &
		 S_{TT}^{xy} & -S_{TT}^{xx} +S_{LL} \\
	\end{bmatrix}. \label{e:covarianttensor}
\end{align} \end{subequations}
In the particle rest frame (where $P^+ = M/\sqrt{2}$) they correspond to
Eqs.~\eqref{e:vectorandtensor} 
(note that here they are written in light-cone coordinates). 

Inserting in Eq.~\eqref{e:density} the spin vector, Eq.~\eqref{e:vector}, 
and the spin tensor,
Eq.~\eqref{e:tensor},
the explicit form of the spin density matrix $\rho$
turns out to be
\begin{equation}
{\bm \rho}=\left(\begin{array}{ccc}
\frac{1}{3}-\frac{S_{LL}}{2}+\frac{S_L}{2} &
	 \frac{S_{LT}^{x} - \ii S_{LT}^{y}}{2\sqrt{2}}
			+\frac{S_{T}^{x} - \ii S_{T}^{y}}{2\sqrt{2}} &
				\frac{S_{TT}^{xx} - \ii S_{TT}^{xy}}{2}\\
\frac{S_{LT}^{x} + \ii S_{LT}^{y}}{2\sqrt{2}}
	+\frac{S_{T}^{x} + \ii S_{T}^{y}}{2\sqrt{2}} &
		\frac{1}{3}+ S_{LL} &
			-\frac{S_{LT}^{x} - \ii S_{LT}^{y}}{2\sqrt{2}}
				+\frac{S_{T}^{x} - \ii S_{T}^{y}}{2\sqrt{2}}\\
\frac{S_{TT}^{xx} + \ii S_{TT}^{xy}}{2} &
	-\frac{S_{LT}^{x} +\ii S_{LT}^{y}}{2\sqrt{2}}
			+\frac{S_{T}^{x} + \ii S_{T}^{y}}{2\sqrt{2}}&
				\frac{1}{3}-\frac{S_{LL}}{2}-\frac{S_L}{2}
\end{array}\right). 
\label{e:rhospin1}
\end{equation}

The treatment of spin-one particles can be done equivalently by introducing
the complex polarization vector $\bm{\eps}$, and its covariant generalization
$\eps^{\mspace{2mu}\mu}$. The two formalisms can be related by means of the
formulae~\cite{Elliot}
\begin{align}	
\mb{S} &= \im \lf(\bm{\eps}^{\ast} \times \bm{\eps} \rg), &
T_{ij} &= \frac{1}{3} \delta_{ij} - \re \lf(\eps^{\ast}_i \eps_j\rg),
\end{align}	
or for the covariant generalizations
\begin{align}	
S_{\mu} &=-\eps_{\mu \alpha \beta \gamma} P^{\alpha} \im \lf(\eps^{\ast \beta}
\eps^{\gamma} \rg), &
T_{\mu \nu} &=- \re \lf(\eps^{\ast}_{\mu} \eps_{\nu}\rg) -\frac{1}{3}\lf(g_{\mu 
\nu} - \frac{P_{\mu} P_{\nu}}{P^2}\rg) \delta_{ij} .
\end{align}

%%%%%%%%%%%%%%%%%%%%%%%%%%%%%%%%%%%%%%%%%%%%%%%%%%%%%%%%%%%%%%%%%%%%%%%%
\subsection{Interpretation of the components of the spin tensor}
\label{a:tensor}

A particular component of the spin tensor measures a combination of
probabilities of finding the system in a certain spin state (defined in the
particle rest frame). 
As ``analyzing'' spin states we can choose the eigenstates of the spin vector
operator in a particular direction. 
We can write the spin vector operator in terms of polar and azimuthal angles,
\begin{equation}
{{\bm{\Sigma}}}^i \hat n_i = {{\bm{\Sigma}}}_x \cos{\vartheta}\cos{\varphi}
			+{{\bm{\Sigma}}}_y \cos{\vartheta}\sin{\varphi} 
			+{{\bm{\Sigma}}}_z \sin{\vartheta},
\end{equation}
and we can denote its eigenstates as $\ket{m,(\vartheta,\varphi)}$, $m$
being their magnetic quantum number.
The probability of finding one of these states can be calculated as
\begin{equation}
{\cal P}\left[m,(\vartheta,\varphi)\right]= 
	\tr \left\{{\bm \rho} \, 
		\ket{m,(\vartheta,\varphi)}
		\bra{m,(\vartheta,\varphi)}\right\}.
\end{equation}
From the explicit formula of the density matrix, Eq.~\eqref{e:rhospin1}, one
can compute
\begin{subequations}\label{e:probabilities}
\begin{gather}
S_{LL} =- \tfrac{1}{3}\,{\cal P}\bigl[1,{(0,0)}\bigr]
	 -\tfrac{1}{3}\,{\cal P}\bigl[-1,{(0,0)}\bigr] +\tfrac{2}{3}\,{\cal P}\bigl[0,{(0,0)}\bigr], 
\label{e:alignmentprob} \\
%end{equation} 
\begin{align}
S_{LT}^{x} &= {\cal P}\left[0,{\bigl(-\tfrac{\pi}{4},0\bigr)}\right] 
		- {\cal P}\left[0,{\bigl(\tfrac{\pi}{4},0\bigr)}\right], &
S_{LT}^{y} &= {\cal P}\left[0,{\bigl(-\tfrac{\pi}{4},\tfrac{\pi}{2}\bigr)}\right]
		- {\cal P}\left[0,{\bigl(\tfrac{\pi}{4},\tfrac{\pi}{2}\bigr)}\right], 
\\
S_{TT}^{xx} &= {\cal P}\left[0,{\bigl(\tfrac{\pi}{2},-\tfrac{\pi}{4}\bigr)}\right] 
		- {\cal P}\left[0,{\bigl(\tfrac{\pi}{2},\tfrac{\pi}{4}\bigr)}\right], &
S_{TT}^{xy} &= {\cal P}\left[0,{\bigl(\tfrac{\pi}{2},\tfrac{\pi}{2}\bigr)}\right] 
		- {\cal P}\left[0,{\bigl(\tfrac{\pi}{2},0\bigr)}\right]. 
\end{align}
\end{gather}
\end{subequations}
Below, we suggest a diagrammatic interpretation of these probability
combinations.  
Arrows represent spin states
$m=+1$ and $m=-1$ in the direction of the arrow itself, while dashed lines
denote spin state $m=0$ in the direction of the line itself.
\begin{eqnarray*}
	& \makebox[1cm][c]{\makebox[4cm][r]{\parbox[c]{3cm}
		{\includegraphics[width=3cm]{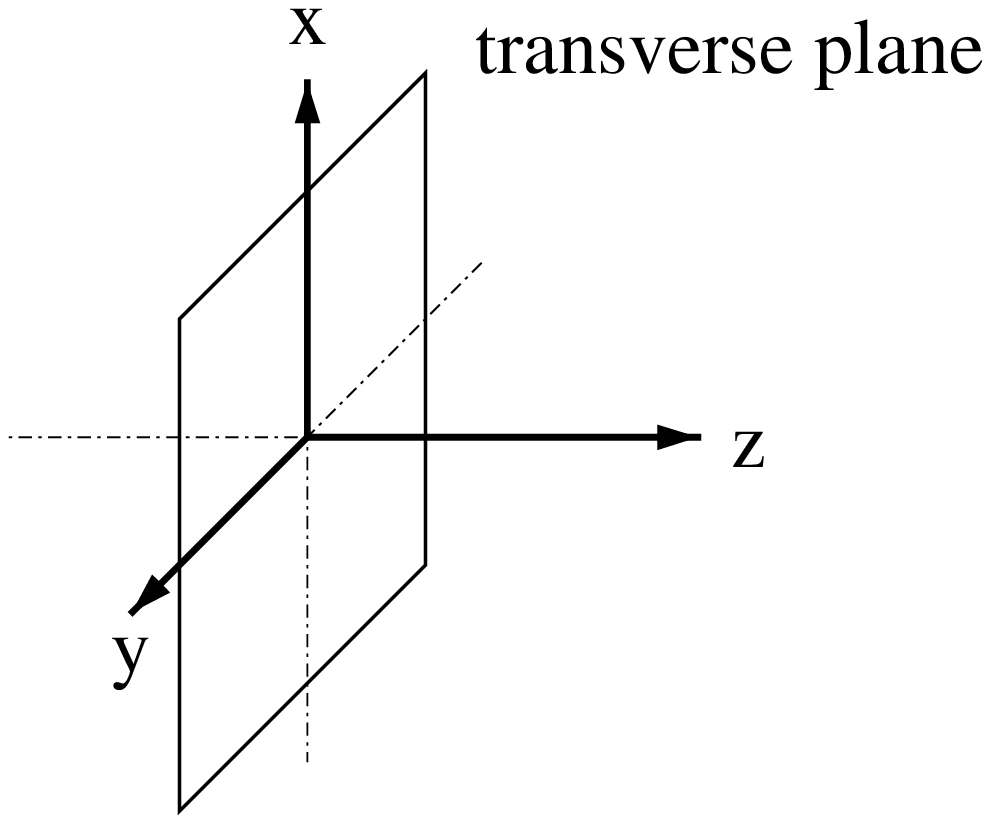}}}} &\nn \\
& \makebox[1cm][c]{$S_{LL}=$
    \raisebox{0.24cm}{\parbox[c]{6cm}{\includegraphics[width=6cm]{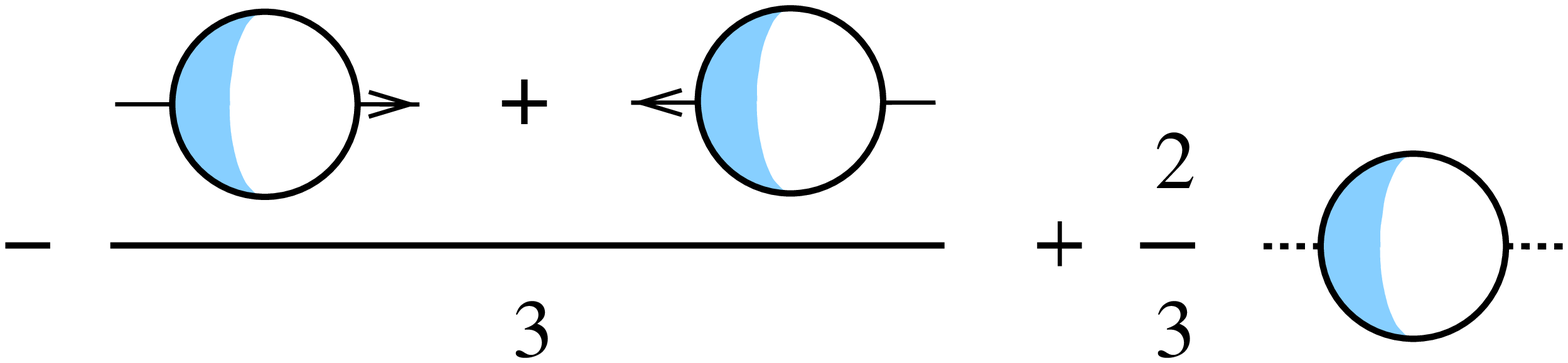}}}}& \\
S_{LT}^{x}=
   \makebox[4cm][c]{\parbox[c]{3.5cm}{\includegraphics[width=3.5cm]{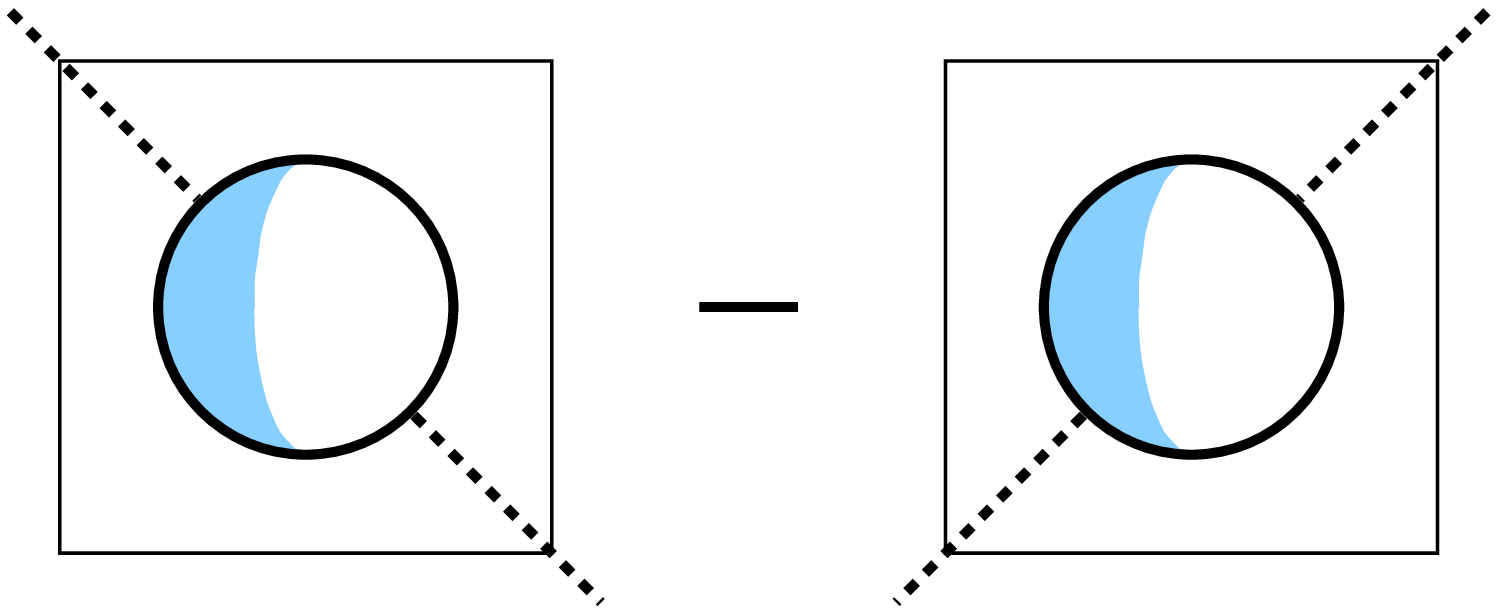}}} &&
S_{LT}^{y}=
	\parbox[c]{4cm}{\includegraphics[width=4cm]{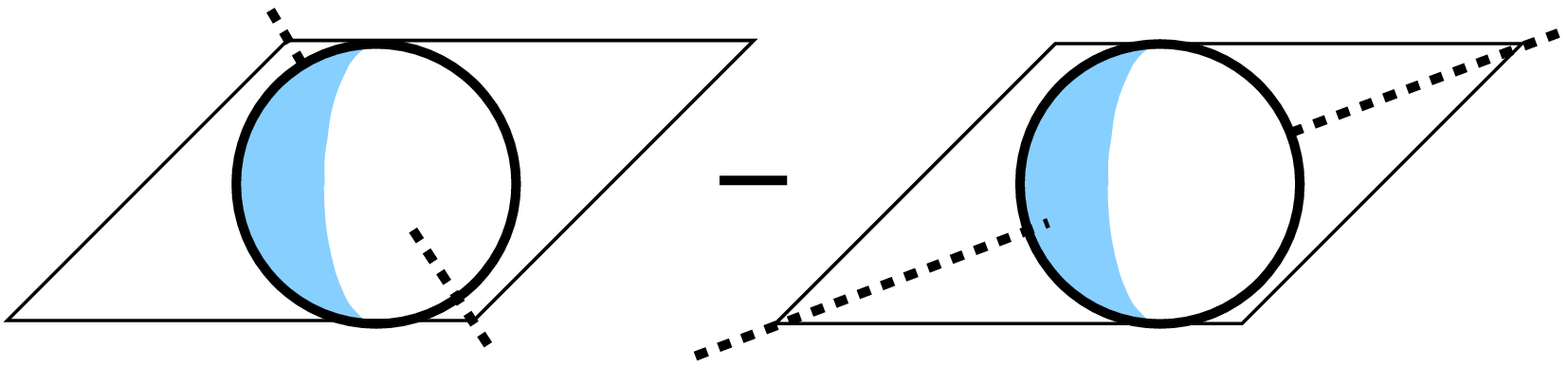}}	\\
S_{TT}^{xy}=
       \makebox[4cm][c]{\parbox[c]{3cm}{\includegraphics[width=3cm]{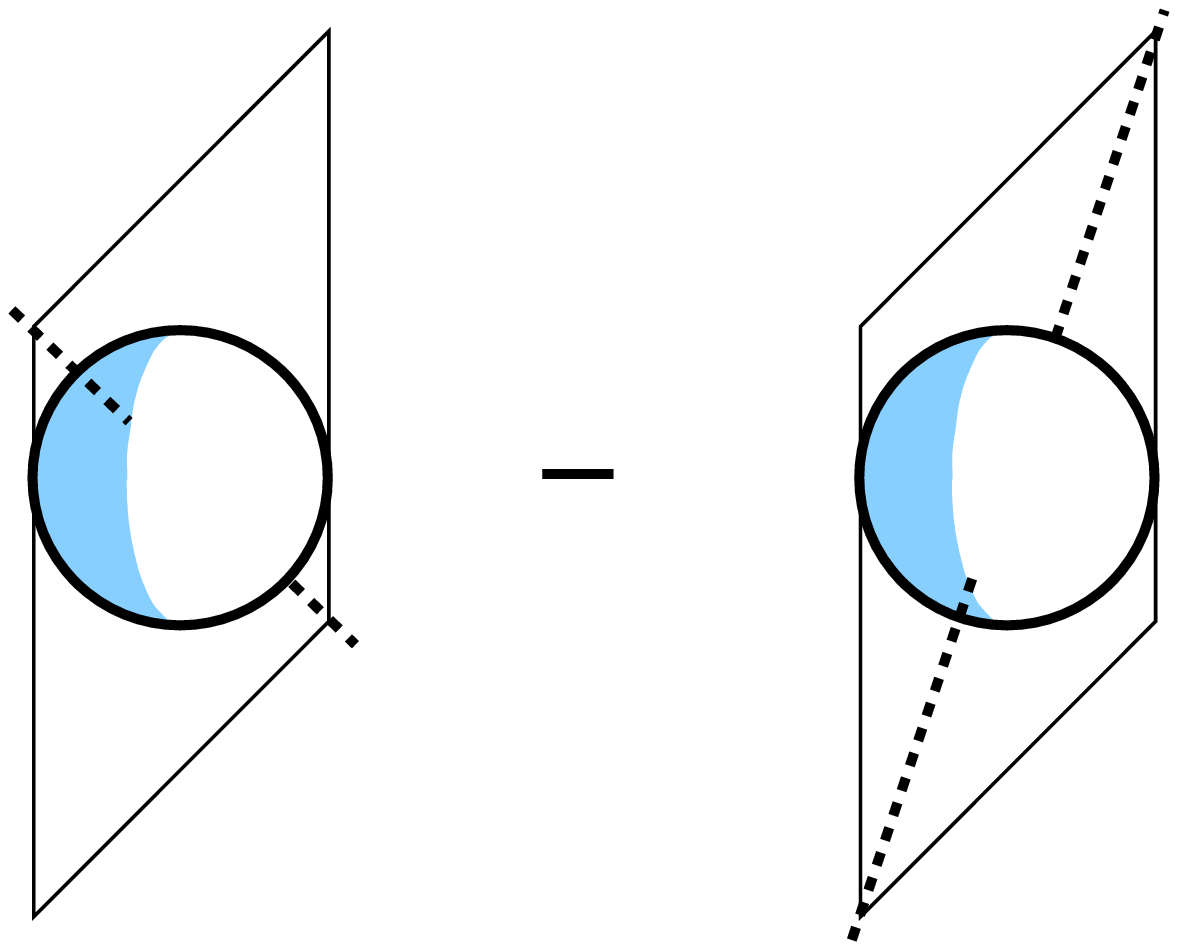}}}&& 
S_{TT}^{xx}=
	\makebox[4cm][c]{\parbox[c]{3cm}{\includegraphics[width=3cm]{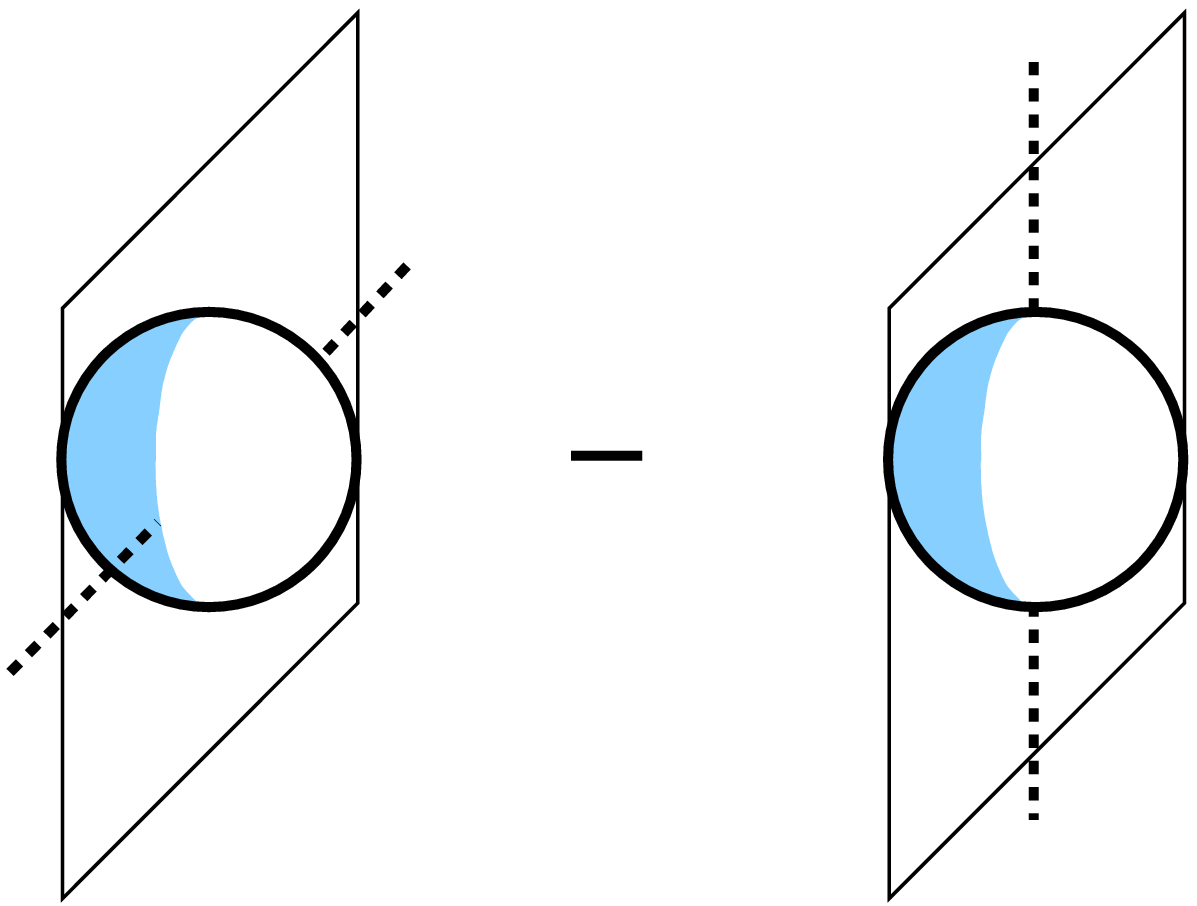}}}.
\end{eqnarray*} 

The probabilistic interpretations suggest straightforward  bounds 
on the values the spin tensor parameters can achieve, namely
\begin{align}	
-\frac{1}{3} &\leq S_{LL} \leq \frac{2}{3}, & -1 &\leq S_{LT}^{i} \leq 1, 
& -1 \leq S_{TT}^{ij} \leq 1,
\end{align} 
where $i,j=x,y$.
Finally, it is possible to define a total degree of polarization
\begin{equation} \begin{split} 
d &= \sqrt{\frac{3}{4}\, S^i S_i  
		+ \frac{3}{2}\, T^{ij} T_{ij}}  \\
  &= \left\{\frac{3}{4}\left[S_L^2 + (S_T^x)^2 + (S_T^y)^2 \right]
%\right.
%\nn \\ && \qquad\left.\mbox{}
		+ \frac{3}{4}\left[3\,S_{LL}^2 
			+ (S_{LT}^x)^2 +(S_{LT}^y)^2
			+(S_{TT}^{xx})^2+(S_{TT}^{xy})^2
			\right]\right\}^{1/2},
\end{split} \end{equation}  
whose value ranges between 0 and 1.

%%%%%%%%%%%%%%%%%%%%%%%%%%%%%%%%%%%%%%%%%%%%%%%%%%%%%%%%%%%%
\section{The correlation function $\Phi$}
\markright{The correlation function $\mathit \Phi$}
\label{s:phispinone}

The correlation function $\Phi$ has to fulfill the condition of 
Hermiticity and parity invariance. For spin-one hadrons, they are slightly
different from the spin-half case, because of the presence of the spin tensor
\begin{subequations} 
\begin{xalignat}{4} 
&\text{Hermiticity:} &\Phi(p,P,S)& =\g^0 \Phi^{\dag}(p,P,S,T) \g^0,
\label{e:hermspin1}\\             
&\text{parity:} &\Phi(p,P,S)& =\g^0
\Phi(\tilde{p},\tilde{P},-\tilde{S},\tilde{T}) \g^0 
\label{e:parityspin1}
\end{xalignat}
\end{subequations} 
where $\tilde{p}^{\nu} = \delta^{\nu \mu} p_{\mu}$ and 
so forth for the other vectors, and $\tilde{T}^{\mu \nu} = \delta^{\mu \sigma}
\delta^{\nu \rho} T_{\sigma \rho}$.

The most general decomposition of the correlation function $\Phi$ for spin-one 
hadrons is\footnote{In Ref.~\citen{Hino:1999qi} a similar decomposition of the
correlation function was attempted, but in an incorrect way.}
\begin{equation} \begin{split} 
\Phi(p,P,S,T)  &= 
                M\,A_1\,{\bm 1} + A_2\,\Pslash + A_3\,\pslash
                +\left(\frac{A_{4}}{M}\,\sigma_{\mu \nu} P^{\mu} p^{\nu}\right)
                +\left( \ii A_5\;p\cdot S\, \gamma_5\right)
                                 \\
  & 		+ M\,A_6 \,\Sslash\, \gamma_5
                + A_7\,\frac{p\cdot S}{M}\,\Pslash\, \gamma_5     
                + A_8\,\frac{p\cdot S}{M}\, \pslash\, \gamma_5
                + \ii A_9\,\sig_{\mu\nu}\gamma_5 S^{\mu}P^{\nu}
                               \\
  &            + \ii A_{10}\,\sig_{\mu\nu}\gamma_5\, S^{\mu}p^{\nu}
    +\ii A_{11}\,\frac{p\cdot S}{M^2}\,\sig_{\mu\nu}\gamma_5 P^{\mu}p^{\nu}
	    +\left( A_{12}\, \frac{\eps_{\mu \nu \rho \sigma}\gamma^\mu P^\nu
              p^\rho S^\sigma}{M}\right)                      \\
  & 		+\frac{A_{13}}{M}\,p_{\mu}p_{\nu}T^{\mu \nu}\,{\bm 1}
		+\frac{A_{14}}{M^2}\;p_{\mu}p_{\nu}T^{\mu \nu}\Pslash
		+\frac{A_{15}}{M^2}\;p_{\mu}p_{\nu}T^{\mu \nu}\pslash  \\
  & 		+\left(\frac{A_{16}}{M^3}\;p_{\mu}p_{\nu}T^{\mu \nu}
			\sig_{\rho \sigma} P^{\rho}p^{\sig}\right) 
    + A_{17}\;p_{\mu}T^{\mu \nu}\g_{\nu}
    +\left(\frac{A_{18}}{M}\,\sig_{\nu \rho}P^{\rho}\,p_{\mu}T^{\mu\nu}\right)
		 \\
  & 		
    +\left(\frac{A_{19}}{M}\,\sig_{\nu \rho}p^{\rho}p_{\mu}T^{\mu\nu}\right)
    +\left(\frac{A_{20}}{M^2}\,\eps_{\mu \nu \rho \sigma} 
		\g^{\mu} \g_5 P^{\nu} p^{\rho}\, p_{\tau}T^{\tau \sig} \right).
					\label{e:decomphispinone}  
\end{split} \end{equation}  
The amplitudes $A_i$ are real functions $A_i=A_i(p\cdot P,p^2)$. 
The terms with the
amplitudes $A_4$, $A_5$, $A_{12}$, $A_{16}$, $A_{18}$, $A_{19}$ and
$A_{20}$ (included between parentheses) constitute
 the T-odd
part of the correlation function, according to the definition
\begin{equation} 
\Phi^{\ast}_{\text{T-odd}}(p,P,S) = - \ii \g^1 \g^3\,
\Phi_{\text{T-odd}}(\tilde{p},\tilde{P},\tilde{S},\tilde{T}\,) \ii \g^1 \g^3.
\label{e:toddspinone}
\end{equation}

The leading-twist part of the correlation function,  
parametrized in terms of five distribution functions, 
is~\cite{Bacchetta:2000jk,Bacchetta:2001rb}
\begin{equation} 
{\cal P}_+\, \Phi(x, S,T) \g^+ =\Bigl(
		f_1(x) +
		g_{1}(x)\,S_{L}\,\g_5 +
		h_{1}(x)\,\g_5 \Sslash_T +
		b_{1}(x)\,S_{LL} +
		\ii\,\underline{h_{1LT}}(x)\,\Sslash_{LT}\Bigr)\,
		{\cal P}_+.		 
\label{e:phispin1} 
\end{equation} 
The underlined function, $h_{1LT}$, is T-odd.
Apart from the usual distribution functions defined in
Eq.~\eqref{e:distri},  we introduced the spin-one parton distribution functions
\begin{subequations}
\begin{align}
\begin{split} 
b_1(x) &= \int \de^2 \! \mb{p}_T \de p^2 \de (2 p \cdot P)\;
\delta\lf(\mb{p}_T^2 + x^2 M^2 + p^2 -2 x p \cdot P \rg)
\\ & \quad \times
\lf\{
\frac{1}{4 M^2}
	\lf[\lf(\frac{p^2+\mb{p}_T^2}{M x} - M x \rg)^2 - 2 \mb{p}_T^2\rg]
\lf(A_{14} + x\, A_{15} \rg) -\frac{p^2 + \mb{p}_T^2 - M^2 x^2}{2 M^2 x} \,A_{17} 
\rg\}, 
\end{split}
\\
\begin{split}
h_{1LT}(x) &= \int \de^2 \! \mb{p}_T \de p^2 \de (2 p \cdot P)\;
\delta\lf(\mb{p}_T^2 + x^2 M^2 + p^2 -2 x p \cdot P \rg)
\\ & \quad \times
\lf\{\frac{1}{4 M}
	\lf(\frac{p^2+\mb{p}_T^2}{M x} - M x \rg)\lf(-A_{18} -x\,
A_{19}\rg)\rg\}.
\end{split}
\end{align} 
\end{subequations}
Note that the distribution function $b_1$, introduced
in Ref.~\citen{Hoodbhoy:1989am}, was called in a different way 
in Refs.~\citen{Bacchetta:2000jk} and \citen{Bacchetta:2001rb} 
to follow a more systematic naming of the
functions, especially in view of the inclusion of transverse momentum, as we
will do in Sec.~\ref{s:transspinone}. Finally, note that the function
$h_{1LT}$ is T-odd, but does not require the presence of intrinsic transverse
momentum, a feature that is absent in T-odd distribution functions for
spin-half targets at leading order in $1/Q$. It could be interesting to study
this object in order to 
clarify the role of intrinsic transverse momentum in generating
T-odd effects in distribution functions~\cite{Brodsky:2002cx,Ji:2002aa}.

%%%%%%%%%%%%%%%%%%%%%%%%%%%%%%%%%%%%%%%%%%%%%%%%%%%%%%%%%%%%
\subsection{Correlation function in the helicity formalism}
\label{s:helicityspinone}

The only difference from the analysis of Sec.~\vref{s:helicity} is
that we have to deal with a more complex spin density matrix and with a
$3\times3$ target spin space. The connection with the correlation function and 
its matrix representation is
\begin{equation}
\Psi(S,T) = \rho(S,T)_{\Lambda^{\phantom'}_1 \Lambda_1'} 
\Psi^{\Lambda_1'\Lambda^{\phantom'}_1},
\end{equation} 
where
\renewcommand{\arraystretch}{1}
\begin{subequations} \begin{align} 
\begin{split} 
\Psi(S,T) &= \Psi_{U} + S_L\,\Psi_{L} - 3\,S_{LL}\,\Psi_{LL}  
		+ \tfrac{1}{\sqrt 2}\,(S_T^x + \ii \, S_T^y)\,\Psi_{T}
		+ \tfrac{1}{\sqrt 2}\,(S_T^x - \ii \, S_T^y)\,\Psi_{T}^{\ast}
 \\
&		\quad+ \tfrac{1}{\sqrt 2}\,(S_{LT}^x+\ii\,S_{LT}^y)\,\Psi_{LT}
		+ \tfrac{1}{\sqrt 2}\,(S_{LT}^x-\ii\,S_{LT}^y)\,\Psi_{LT}^{\ast}
 \\
&		\quad+ \tfrac{1}{2}\,(S_{TT}^{xx}+\ii\,S_{TT}^{xy})\,\Psi_{TT}
		+ \tfrac{1}{2}\,(S_{TT}^{xx}-\ii\,S_{TT}^{xy})\,\Psi_{TT}^{\ast}
,
\end{split} \\ \nn \\
\Psi^{\Lambda_1'\Lambda^{\phantom'}_1} &= 
\begin{pmatrix}
\Psi_{U} + \Psi_{L} + \Psi_{LL} & \Psi_{T} + \Psi_{LT} & \Psi_{TT} \\ \\
\Psi_{T}^{\ast} + \Psi_{LT}^{\ast} & \Psi_{U} - 2\,\Psi_{LL} &
	\Psi_{T} - \Psi_{LT} \\ \\
\Psi_{TT}^{\ast} & \Psi_{T}^{\ast} - \Psi_{LT}^{\ast} &
	\Psi_{U} - \Psi_{L} + \Psi_{LL}
\end{pmatrix}.
\end{align} \end{subequations}
Eventually, the $6 \times 6$ leading-twist scattering matrix turns out to be
\footnote{Note the difference of sign in the imaginary components of the
matrix with respect to Ref.~\citen{Bacchetta:2001rb}.}
\renewcommand{\arraystretch}{1.5}
\addtolength\arraycolsep{-8pt} 
\begin{multline}
F(x)_{\chi^{\phantom'}_{1} \chi'_{1}}^{\Lambda_1' \Lambda^{\phantom'}_1} =  \\
\lf(\begin{array}{@{\hspace{3pt}}ccc@{\hspace{3pt}}|@{\hspace{3pt}}ccc@{\hspace{3pt}}}
{f_1} + {g_1}-\frac{1}{3}\,b_1& 0 & 0 & 0 & 
	\sqrt{2}\,({h_1}-\ii \,{h_{1LT}})& 0 \\
0 & {f_1} +\frac{2}{3}\,b_1 & 0 & 0 & 0 &\sqrt{2}\,({h_1}+\ii\, {h_{1LT}}) \\
0 & 0 &{f_1} - {g_1}-\frac{1}{3}\,b_1 & 0 & 0 & 0 \\ 
\hline
0 & 0 & 0 & {f_1} - {g_1}-\frac{1}{3}\,b_1& 0 & 0 \\
\sqrt{2}\,({h_1}+\ii\, {h_{1LT}}) & 0 & 0 & 0 &{f_1} +\frac{2}{3}\,b_1 & 0 \\
0 &\sqrt{2}\,({h_1}-\ii\, {h_{1LT}}) & 0 & 0 & 0 & 
		{f_1} + {g_1}-\frac{1}{3}\,b_1
\end{array}\rg),
\label{e:Fspinone}
\end{multline}
where the functions on the right-hand side depend on the variable $x$ only.
The matrix $F(x)$ 
is Hermitean, parity invariant and conserves angular momentum 
($\Lambda'_{1}+\chi'_{1} =\Lambda_{1} +\chi_{1}$).

 From the positivity of the diagonal elements we obtain the bounds:
\addtolength\arraycolsep{8pt} 
\begin{subequations} \label{e:boundspinone1} \begin{align} 
f_1 (x) &\geq  0 \\
-\frac{3}{2}\,f_1(x) &\leq b_1(x) \;\leq\; 3\,f_1 (x) \\
|g_1 (x)|&\leq f_1 (x)-\frac{1}{3}\,b_1(x)\;\leq\;\frac{3}{2}\,f_1(x),
\end{align} \end{subequations}  
while positivity of two-dimensional minors gives the
bound  
\begin{equation} \begin{split} 
\bigl(h_1(x)\bigr)^2 + \bigl(h_{1LT}(x)\bigr)^2
	&\leq \frac{1}{2}\lf(f_1(x)+\frac{2}{3}\,b_1(x)\rg) 
		\lf(f_1(x)+g_1(x)-\frac{1}{3}\,{b_1(x)}\rg) \\
	&\leq \frac{1}{4}
	\lf(\frac{3}{2}\,f_1(x) + g_1(x) \rg)^2  
		\leq \frac{9}{8}\, \bigl(f_1(x)\bigr)^2
,
\label{e:soffer2}
\end{split} \end{equation}  
This bound is a generalization of the Soffer bound~\cite{Soffer:1995ww} 
and must be fulfilled by
any spin-one target. If we assume that T-odd distribution functions
vanish due to time-reversal invariance, 
then the bound can be reduced to
\begin{equation} \begin{split} 
|h_1(x)|&\leq \sqrt{\frac{1}{2}\lf(f_1(x)+\frac{2}{3}\,b_1(x)\rg) 
		\lf(f_1(x)+g_1(x)-\frac{1}{3}{b_1(x)}\rg)} \\
	&\leq \frac{1}{2}\lf(\frac{3}{2}\,f_1(x) + g_1(x)\rg)
		\leq \frac{3}{2\sqrt{2}}\,f_1(x)
.
\end{split} \end{equation}  

%%%%%%%%%%%%%%%%%%%%%%%%%%%%%%%%%%%%%%%%%%%%%%%%%%%%%%%%%%%%
\subsection{Inclusion of transverse momentum}
\label{s:transspinone}

We now consider the correlation function unintegrated over the parton
transverse momentum, as defined in Eq.~\eqref{e:phinoint}.
For convenience, the correlation function can be decomposed in
several terms in relation to the polarization state of the target, i.e.\
$\Phi = \Phi_U + \Phi_L +\Phi_T + \Phi_{LL} +\Phi_{LT} + \Phi_{TT}$. 
To leading order in $1/Q$,  these terms can be decomposed 
as (T-odd terms are underlined)
\begin{subequations} \label{e:phispin1withtrans} \begin{align} 
{\cal P}_+\, \Phi_U (x, \mb{p}_\st, S, T) \g^+ & = \left\{
	f_1{} + \ii
  \underline{h_1^{\perp}}{}\,
		\frac{\pslash_T}{M} \right\}{\cal P}_+\,,\\
%%%%%%%%%%%%%%%%
{\cal P}_+\, \Phi_L (x, \mb{p}_\st, S, T) \g^+ & = S_L\, \left\{
	g_{1L}{}\,\g_5  +
	h_{1L}^{\perp}{}\,
		\g_5 \frac{\pslash_T}{M}\right\}{\cal P}_+\,,\\
%%%%%%%%%%%%%%%%
\begin{split}
{\cal P}_+\, \Phi_T (x, \mb{p}_\st, S, T) \g^+ & = \left\{
	g_{1T}{}\,\frac{\mb{S}_T\cdot\mb{p}_T}{M}\,\g_5 +
	h_{1T}{}\,\g_5 \Sslash_T \right.  \\
&\left.\quad +
	h_{1T}^{\perp}{}\,\frac{\mb{S}_T \cdot \mb{p}_T}{M}\, 
	     \g_5 \frac{\pslash_T}{M} 
%\right.  \\
%&&\left.\mbox{}+
  +\underline{f^{\perp}_{1T}}{}\,\frac{\epsilon_{\st\,\rho \sigma}
 S_\st^\rho p_\st^\sigma}{M}\right\}{\cal P}_+\,,
\end{split} \\
%%%%%%%%%%%%%%%%
{\cal P}_+\, \Phi_{LL} (x, \mb{p}_\st, S, T) \g^+ & = S_{LL} \,\left\{
	b_{1}{}+ \ii
  \underline{h_{1LL}^{\perp}}{}\,\frac{\pslash_T}{M}
		\right\}{\cal P}_+\,, \\
%%%%%%%%%%%%%%%%
\begin{split}
{\cal P}_+\, \Phi_{LT} (x, \mb{p}_\st, S, T) \g^+ & = \left\{
	f_{1LT}{}\,\frac{\mb{S}_{LT} \cdot \mb{p}_T}{M}+
  \underline{g_{1LT}}{}\,
	\eps_T^{\rho \sigma}S_{LT\,\rho}\frac{p_{T\,\sigma}}{M} \,\g_5
			\right.  \\ 
&\left.\quad+ \ii
  \underline{h'_{1LT}}{}
    \, \Sslash_{LT}
%			\right.  \\ 
%&\left.\quad
  +\ii \underline{h_{1LT}^{\perp}}{}
	\, \frac{\mb{S}_{LT} \cdot \mb{p}_T}{M}\,
			\frac{\pslash_T}{M} 
		 \right\}{\cal P}_+\,,
\end{split}\\
%%%%%%%%%%%%%%%%
\begin{split}
{\cal P}_+\, \Phi_{TT} (x, \mb{p}_\st, S, T) \g^+ & = \left\{
   f_{1TT}{}\,
	\frac{\mb{p}_T \cdot \mb{S}_{TT} \cdot \mb{p}_T}{M^2}
%		 \right.  \\ 
%&&\left.\mbox{}
-\underline{g_{1TT}}{}\, \eps_T^{\rho \sig} S_{TT\,\sig \lambda}
	\frac{p_T^{\lambda} p_{T\,\rho}}{M^2}\,\g_5 \right.  \\
&\left.\quad- \ii
  \underline{h'_{1TT}}{}
		\, \g^{\sig}\, S_{TT\,\sig \lambda}
	\frac{p_T^{\lambda}}{M}
%			\right.  \\ 
%&&\left.\mbox{}
 +\ii \underline{h_{1TT}^{\perp}}{}\,
		\frac{\mb{p}_T \cdot \mb{S}_{TT} \cdot \mb{p}_T}{M^2}\,
		\frac{\pslash_T}{M} \right\}{\cal P}_+\,.
\end{split}
\end{align} \end{subequations}
All the distribution functions on the right-hand side are understood to depend 
on $x$ and $\mb{p}_T^2$.

Similarly to the function $h_1$, defined in Eq.~\eqref{e:h1trans}, 
it is convenient to define the functions
\begin{subequations} \begin{align} 
h_{1LT}(x, {p}^2_T)
       &= h'_{1LT} (x,{p}^2_T)
	+ h_{1LT}^{\perp (1)}(x,{p}^2_T), \\
h_{1TT}(x,{p}^2_T) 
	&= h'_{1TT} (x,{p}^2_T)
	+ h_{1TT}^{\perp (1)}(x,{p}^2_T).
\end{align} \end{subequations} 
In the rest of this Section, unless
otherwise specified, all the functions are understood to depend on the 
variables $x$ and $\mb{p}_T^2$. 

As we have already done in Eq.~\eqref{e:g1T&h1lperp}, to simplify the
notation it is preferable to introduce new complex functions with a real part,
corresponding to a T-even function, and an imaginary part, corresponding to a
T-odd function. To avoid the introduction of new names, we will simply call
the new functions with the name of their real part, in the following way
\begin{align} 
g_{1T}+\ii \, f_{1T}^{\perp} &\rightarrow g_{1T}, & 
f_{1LT}+\ii \, g_{1LT} &\rightarrow f_{1LT},  \nn \\
h_{1}+\ii \, h_{1LT} &\rightarrow h_{1}, &
f_{1TT}+\ii \, g_{1TT}^{\perp} &\rightarrow f_{1TT}, \\
h_{1T}^{\perp}+\ii  \,h_{1LT}^{\perp} &\rightarrow h_{1T}^{\perp}.\nn
\end{align}

Following steps analogous to the previous section, we can reconstruct the
complete $6 \times 6$ scattering matrix. 
\begin{equation} 
F(x, \mb{p}_T)
_{\chi^{\phantom'}_{1} \chi'_{1}}^{\Lambda_1' \Lambda^{\phantom'}_1}=
\begin{pmatrix}
A_{\Lambda_1' \Lambda^{\phantom'}_1} & B_{\Lambda_1' \Lambda^{\phantom'}_1} \\
B_{\Lambda_1' \Lambda^{\phantom'}_1}^{\dagger} & C_{\Lambda_1' \Lambda^{\phantom'}_1}
\end{pmatrix}.
\end{equation} 
The inner blocks span the target spin space and they are
\begin{align} 
A_{\Lambda_1' \Lambda^{\phantom'}_1} & = 
\begin{pmatrix}
{f_1} + {g_1}-\frac{1}{3}\,b_1& 
	\frac{\sqrt{2}}{2}\,\frac{\vert \mb{p}_\st\vert}{M}\,\e^{-\ii\phi_p}
	\,\lf(g_{1T}+f_{1LT}\rg) & 
\frac{1}{2}\,\frac{\vert \mb{p}_\st\vert^2}{M^2}\,\e^{-2\ii\phi_p}
	\,f_{1TT} \\
\frac{\sqrt{2}}{2}\,\frac{\vert \mb{p}_\st\vert}{M}\,\e^{\ii\phi_p}
	\,\lf(g_{1T}^*+f_{1LT}^*\rg) & {f_1} +\frac{2}{3}\,b_1 & 
\frac{\sqrt{2}}{2}\,\frac{\vert \mb{p}_\st\vert}{M}\,\e^{-\ii\phi_p}
	\,\lf(g_{1T}-f_{1LT}\rg) \\
\frac{1}{2}\,\frac{\vert \mb{p}_\st\vert^2}{M^2}\,\e^{2\ii\phi_p}
	\,f_{1TT}^* & 
\frac{\sqrt{2}}{2}\,\frac{\vert \mb{p}_\st\vert}{M}\,\e^{\ii\phi_p}
	\,\lf(g_{1T}^*-f_{1LT}^*\rg)
	&{f_1} - {g_1}-\frac{1}{3}\,b_1 
\end{pmatrix},  \\
B_{\Lambda_1' \Lambda^{\phantom'}_1} & = 
\begin{pmatrix}
\frac{\vert \mb{p}_\st\vert}{M}\,\e^{\ii\phi_p}\lf[h_{1L}^{\perp} - \ii
\lf(h_{1}^{\perp} - \frac{1}{3} h_{1LL}^{\perp}\rg)\rg] & 
	\sqrt{2}\,{h_1^*}&
	-2 \ii \frac{\vert \mb{p}_\st\vert}{M}\,\e^{-\ii\phi_p} h_{1TT}  \\
\frac{\sqrt{2}}{2}\,\frac{\vert \mb{p}_\st\vert^2}{M^2}\,\e^{2\ii\phi_p}
\,h_{1T}^{\perp *} & 
-\ii \frac{\vert \mb{p}_\st\vert}{M}\,\e^{\ii\phi_p}\lf(h_{1}^{\perp} +
	\frac{2}{3}  h_{1LL}^{\perp}\rg)
 &\sqrt{2}\,{h_1} \\
-\ii \frac{\vert \mb{p}_\st\vert^3}{M^3}\,\e^{3\ii\phi_p}
\,h_{1TT}^{\perp}  & 
\frac{\sqrt{2}}{2}\,\frac{\vert \mb{p}_\st\vert^2}{M^2}\,\e^{2\ii\phi_p}
\,h_{1T}^{\perp} & 
\frac{\vert \mb{p}_\st\vert}{M}\,\e^{\ii\phi_p}\lf[-h_{1L}^{\perp} - \ii
\lf(h_{1}^{\perp} - \frac{1}{3} h_{1LL}^{\perp}\rg)\rg] 
\end{pmatrix}.
\end{align} 
This matrix is Hermitean. Conservation of angular momentum is guaranteed
($\Lambda'_{1}+\chi'_{1}+l' =\Lambda_{1} +\chi_{1}$). The matrix 
is also parity invariant, i.e.
\begin{equation} 
F(x,\mb{p}_T)
_{\chi^{\phantom'}_{1} \chi'_{1}}^{\Lambda_1' \Lambda^{\phantom'}_1} = 
(-1)^{l'}\,F(x,\mb{p}_T)
_{-\chi^{\phantom'}_{1}\, -\chi'_{1}}^{-\Lambda_1'\, -\Lambda^{\phantom'}_1}  
\biggr\rvert_{l'\rightarrow - l'}.
\end{equation}
In fact, 
 block $C$ of the fragmentation matrix can be obtained from block $A$
by imposing parity invariance relations.

Since the matrix $F(x, \mb{p}_T)$ is positive semidefinite, we can
extract bounds on the distribution functions.
If we fully exploit the positivity of the scattering matrix, 
we can write several relations involving an
increasing number of different functions. We feel this to be an excessive task
if compared to the exiguity of information we have on the functions involved.
Therefore, here 
we choose to focus only on the relations stemming from positivity of the 
two-dimensional minors of the matrix.

Because of the symmetry properties of the matrix, 
only nine
independent inequality relations between the different functions are 
produced:\footnote{Note the difference in Eq.~\eqref{e:mistake} compared to
Eq.~(38) of 
Ref.~\citen{Bacchetta:2001rb}, due to an error in the latter.}
\begin{subequations}\label{e:spinoneboundswithtrans} \begin{align}	
\vert h_{1} \vert^2 & \leq 
	\frac{1}{2}\,\left(f_1+\frac{2\,b_{1}}{3}\right) 
		\left(f_1+g_1-\frac{b_{1}}{3}\right) ,  \\
\frac{\vert p_{T}\vert^2}{2 M^2} \; \vert g_{1T} + f_{1LT} \vert^2 & \leq
	\left(f_1+\frac{2\,b_{1}}{3}\right) 
		\left(f_1+g_1-\frac{b_{1}}{3}\right) ,  \\
\frac{\vert p_{T}\vert^2}{2 M^2} \; \vert g_{1T} - f_{1LT} \vert^2 & \leq
	\left(f_1+\frac{2\,b_{1}}{3}\right) 
		\left(f_1-g_1-\frac{b_{1}}{3}\right) ,  \\
\frac{\vert p_{T}\vert^4}{2 M^4} \; \vert h_{1T}^{\perp}\vert^2 & \leq
	\left(f_1+\frac{2\,b_{1}}{3}\right) 
		\left(f_1-g_1-\frac{b_{1}}{3}\right) ,  \\
\frac{\vert p_{T}\vert^6}{M^6} \;{h_{1TT}^{\perp}}^2 & \leq 
	\left(f_1-g_1-\frac{b_{1}}{3}\right)^2 ,  \\
\frac{\vert p_{T}\vert^2}{4 M^2} 
	\;\left(h_{1}^{\perp} +\frac{2\, h_{1LL}^{\perp}}{3}\right)^2 & \leq 
	\left(f_1+\frac{2\,b_{1}}{3}\right)^2 ,  \\
\frac{\vert p_{T}\vert^2}{M^2} \;h_{1TT}^2 & \leq 
	\frac{1}{4}\,\left(f_1+g_1-\frac{b_{1}}{3}\right)^2 , \label{e:mistake}  \\
\frac{\vert p_{T}\vert^2}{M^2} \; \left[{h_{1L}^{\perp}}^2 +
	\left(h_{1}^{\perp} -\frac{h_{1LL}^{\perp}}{3}\right)^2\right] & \leq
	\left(f_1+g_1-\frac{b_{1}}{3}\right)
	\left(f_1-g_1-\frac{b_{1}}{3}\right), \\	
\frac{\vert p_{T}\vert^4}{4 M^4} \; \vert f_{1TT} \vert^2 & \leq
	\left(f_1+g_1-\frac{b_{1}}{3}\right)
	\left(f_1-g_1-\frac{b_{1}}{3}\right).
\end{align} \end{subequations}

%%%%%%%%%%%%%%%%%%%%%%%%%%%%%%%%%%%%%%%%%%%%%%%%%%%%%%%%%%%%
\section{Cross sections and asymmetries for spin-one targets}

We will not dwell very much on the analysis of spin-one targets. The
only available one is the deuteron, which is a system of two spin-half hadrons 
connected by nuclear interaction. In the approximation of independent
scattering off the nucleons, the distribution function $b_1$ 
is expected to be
small. However, calculations of $b_1$ are available in the literature 
(see e.g.\ 
Ref.~\citen{Umnikov:1997qv}). Some calculations show that 
this function could be sizeable in the low-$x$
region~\cite{Nikolaev:1997jy,Bora:1998pi,Edelmann:1997qe}, due to nuclear
shadowing effects.
New precision measurements of the deuteron 
structure functions (e.g.\ from HERMES) may provide an experimental test of
this expectation, as already
observed in several
papers~\cite{Anselmino:1995gn,Kumano:1993zm,Efremov:1994xf}. 

We will focus only on totally inclusive deep inelastic scattering off a
spin-one target at leading twist. 
As we have seen in Chap.~\ref{c:trans}, in this type of
process we can access only chiral-even,
transverse momentum independent functions.
Spin-one targets are characterized by three such functions, $f_1$, $g_1$ and
$b_1$. Therefore, we should be able to define three independent
asymmetries. In addition to the indices $\rightleftarrows$ to denote the 
$\pm 1$
polarization along the direction of the beam,
with the index 0 we will denote when the target is polarized
along 
the direction of the beam {\em but} with magnetic number equal to 0.  

The total cross section integrated over the azimuthal angle $\phi_S$ is
analogous to Eq.~\eqref{e:crossint}
\begin{equation} 
\frac{\de^2\!\sigma}{\de \xbj \de y} =
\frac{4 \pi \alpha^2}{s \xbj y^2}\,\sum_q e_q^2\,\lf[A(y)\lf(f_1^q(\xbj)+S_{LL}\,b_1^q(\xbj)\rg)
+ \lambda_e S_L\,B(y)\, g_1^q(\xbj)\rg].
\label{e:crossint2}
\end{equation} 
A few words on the value of the polarization coefficients are needed.
When the target is polarized {\em purely} in the $\pm 1$ state along the beam
direction, then $S_L = \mp 1$ and $S_{LL} = -1/3$ [cf. Eq.~\eqref{e:alignmentprob}]. In case the polarization is 
not complete, 
then a possible notation could be $S_L = \mp \lvert S_L \rvert$ and 
$S_{LL} = -\lvert S^1_{LL} \rvert /3$, with 
$\lvert S_L \rvert = \lvert S^1_{LL}
\rvert < 1$. When the target is polarized purely in the 0 state along the
beam, then $S_L = 0$ and $S_{LL} = 2/3$. In case the polarization is not
complete,  
we can use the notation $S_{LL} = 2 \lvert S^0_{LL} \rvert /3$, where the
coefficient $\lvert S^0_{LL}\rvert$ is smaller than 1, and can be
different from $\lvert S^1_{LL}\rvert$.
If we just sum the cross section with opposite vector polarization $S_L$, in
analogy to Eq.~\eqref{e:uuf1},  we
will obtain 
 \begin{equation}
\frac{1}{2}\lf(
\de^2\!\sigma_{\rightarrow \leftarrow} + \de^2\!\sigma_{\rightarrow \rightarrow}\rg)
=  
\frac{4 \pi \alpha^2}{s \xbj y^2}\,A(y)\,\sum_q e_q^2\,
\lf(f_1^q(\xbj)-\frac{\lvert S^1_{LL}\rvert}{3} b_1^q (\xbj) \rg),
\label{e:spin1cross1}
\end{equation} 
where a contamination of tensor polarization is still present. 
Thus, the unpolarized part of the cross section has to be properly defined as
\begin{equation}
\de^2\!\sigma_{UU}\equiv\frac{1}{3}\lf(
\de^2\!\sigma_{\rightarrow \leftarrow} + \de^2\!\sigma_{\rightarrow\, 0} +\de^2\!\sigma_{\rightarrow \rightarrow}\rg)
=  
\frac{4 \pi \alpha^2}{s \xbj y^2}\,A(y)\,\sum_q e_q^2\, f_1^q(\xbj).
\label{e:spin1cross2}\end{equation} 
The vector polarized part of the cross section corresponds exactly to
Eq.~\eqref{e:llg1}. We rewrite it here for convenience
\begin{equation} 
\de^2\!\sigma_{LL} \equiv \frac{1}{2}\lf(
\de^2\!\sigma_{\rightarrow \leftarrow} - \de^2\!\sigma_{\rightarrow
\rightarrow}\rg)= 
\frac{4 \pi \alpha^2}{s \xbj y^2}\,\lvert \lambda_e \rvert\,
\lvert S_L \rvert\,B(y)\,
\sum_q e_q^2\, g_1^q(\xbj).
\label{e:spin1cross3}\end{equation} 
Finally, we can define a tensor polarized part of the cross section
\begin{equation}
\de^2\!\sigma_{U^{\,L}_{\,L}}\equiv\frac{1}{6}\lf(2 \de^2\!\sigma_{\rightarrow \,0} 
-\de^2\!\sigma_{\rightarrow \leftarrow}  - \de^2\!\sigma_{\rightarrow \rightarrow}\rg)
=  
\frac{4 \pi \alpha^2}{s \xbj y^2}\,\lvert S^0_{LL}\rvert\,A(y)\,\sum_q e_q^2\, b_1^q(\xbj).
\label{e:spin1cross4}\end{equation} 
Therefore, the appropriate asymmetry to measure the spin-one distribution
function $b_1$ is
\begin{equation} 
A_{U^{\,L}_{\,L}} (\xbj,y) \equiv \frac{\de^2\!\sigma_{U^{\,L}_{\,L}}}{\de^2\!\sigma_{UU}}
 = \lvert S^0_{LL} \rvert
 \,\frac{(1/{\xbj y^2})\,A(y)\,\sum_q e_q^2\, b_1^q(\xbj)}
{(1/{\xbj y^2})\,A(y)\,\sum_q e_q^2\, f_1^q(\xbj)}.
\end{equation} 

%%%%%%%%%%%%%%%%%%%%%%%%%%%%%%%%%%%%%%%%%%%%%%%%%%%%%%%%%%%%
\section{Spin-one hadrons in the final state}

Instead of pointing the attention to spin-one targets,
 it is possible to
analyze spin-one final state hadrons in
semi-inclusive deep inelastic scattering or in $e^+\,e^-$ annihilation. 
This idea was first considered by
Efremov and Teryaev~\cite{Efremov:1982vs} (and more recently rediscussed in Ref.~\cite{Schafer:1999am}). A systematic study was
accomplished by Ji~\cite{Ji:1994vw}, 
who singled out two new fragmentation functions,
the function $\hat{b}_1$, analogous to the distribution 
function $b_1$, and  the T-odd function
$\hat{h}_{\overline{1}}$. These new
fragmentation functions can be observed in the production 
of vector mesons,
e.g.\ $\rho$, $K^{\ast}$, $\phi$.
However, these functions require polarimetry on the final-state meson, which
can be done by studying the angular 
distribution of its
decay products (e.g.\ $\pi^+ \, \pi^-$ in the case of $\rho^0$ meson). In this
sense, vector meson production represents just a specific 
contribution to the more general 
case of two-particle production 
near the vector 
meson mass.

%%%%%%%%%%%%%%%%%%%%%%%%%%%%%%%%%%%%%%%%%%%%%%%%%%%%%%%%%%%%
\subsection{The decay of a spin-one hadron}
\label{s:decayspinone}

We want to describe the decay of a spin-one hadron with momentum $P_h$ and
mass $M_h$ into two unpolarized particles with momenta $P_1$, $P_2$, and
masses $M_1$, $M_2$. We assume the same parametrization of the momenta as
presented in Chap.~\ref{c:twohadron}, in particular Eqs.~\eqref{e:momenta}.
In general, the angular distribution of the decay products of a spin-one
hadron into two unpolarized hadrons is
\begin{equation}
W(\cos\theta,\phi_R)= {\rho}^{\Lambda_2' \Lambda^{\phantom'}_2}\,{\cal D}(\theta,\phi_R)_{1\Lambda^{\phantom'}_2,1\Lambda_2'},
\label{e:w}
\end{equation}
where $\theta$ and $\phi_R$ are the polar and azimuthal angles of the vector
$R=(P_1 - P_2)/2$ in the decaying particle rest frame, as 
defined in Chap.~\ref{c:twohadron}, Eqs.~\eqref{e:common}, and they correspond to 
the angles usually measured in
experiments~\cite{Breitweg:1999fm,Adloff:1999kg,Ackerstaff:2000bz}. 
The decay matrix, ${\cal D}$, has been defined in
Eq.~\eqref{e:decaymatrix}. In this chapter, we are only interested in the
$j=1$ sector, therefore for convenience we define a
spin-one decay matrix
\begin{equation}
{R}(\theta,\phi_R)_{\Lambda^{\phantom'}_2 \Lambda_2'} \equiv
  {\cal D}(\theta,\phi_R)_{1\Lambda^{\phantom'}_2,1\Lambda_2'}.
\label{e:Rdecaymatrix}
\end{equation} 
As can be checked by explicit comparison, this matrix 
can be rewritten as
\begin{equation}
{\bm R}(\theta,\phi_R) = 
	\frac{1}{4 \pi} \left[\bm 1 + 3\, {{\bm{\Sigma}}}_{ij}
	\left(\frac{1}{3}\, \delta^{ij}
	-\hat R_{\rm cm}^i \hat R_{\rm cm}^j
	\right)\right],
\label{e:R2}
\end{equation}
where $\hat{\mb{R}}_{\rm cm}=\mb{R}_{\rm cm}/\rr$. 
Notice that $\rr$ stands for the modulus of the vector $\mb{R}$ in the
center-of-mass frame of reference. We know from Eq.~\eqref{e:rr1} or
Eq.~\eqref{e:rr2} how to express it in terms of $M_1$, $M_2$ and $M_h$.

In general, the decay matrix can be expressed in terms of {\em analyzing
powers}: 
\begin{equation}
{\bm R}(\theta,\phi_R) = \frac{1}{4 \pi}
	\left({\bm 1} + \frac{3}{2}\,{{\bm{\Sigma}}}_i\, A^i(\theta,\phi_R)+
		 3\, {{\bm{\Sigma}}}_{ij}\, A^{ij}(\theta,\phi_R) \right),
\label{e:R3}
\end{equation}
and the decay distribution can be obtained  accordingly as
\begin{equation}
W(\cos\theta,\phi_R)= \frac{1}{4 \pi}\left(1 + \frac{3}{2}\,S_i\, A^i 
			+ 3\, T_{ij}\, A^{ij} \right).
\label{e:w2}
\end{equation}
By comparing Eq.~\eqref{e:R2} with Eq.~\eqref{e:R3} we can identify
\begin{align} 
A^i&=0,  &
A^{ij} &= \frac{1}{3}\, \delta^{ij}-\hat{R}_{\rm cm}^i \hat{R}_{\rm cm}^j,
\end{align} 
from which we observe that the vector analyzing powers are zero~\cite{Bourrely:1980mr}.\footnote{The
vector analyzing powers can be different from zero in a decay that does not
conserve parity.}

The tensor analyzing power can be written in a covariant form:  
\begin{equation}
A^{\mu \nu} = \frac{1}{\rr^2}\,R^{\mu}R^{\nu}
   -\frac{1}{3}\left(g^{\mu \nu} -\frac{P_h^{\mu}P_h^{\nu}}{M_{h}^2}\right).
\end{equation}
We can use a parametrization of the tensor analyzing power analogous to that
of the spin tensor,
Eq.~\eqref{e:covarianttensor}, provided we exchange the plus and minus
components and we replace $P$ and $M$ with $P_h$ and $M_h$. Then, we can write 
the parameters of the tensor in terms of the angles $\theta$ and $\phi_R$:
\begin{subequations}
\label{e:analyzingpowers}
\begin{gather}
A_{LL}=\frac{1}{3} \left(3 \cos^2{\theta}-1\right), \\
\begin{align}
A_{LT}^x&=-\sin{2\theta} \;\cos{\phi_R}, &
			 A_{LT}^y &=-\sin{2\theta} \;\sin{\phi_R},\\
A_{TT}^{xx}&=-\sin^2{\theta}\; \cos{2\phi_R}, & 
			A_{TT}^{xy}&=-\sin^2{\theta}\; \sin{2\phi_R}.
\end{align} 
\end{gather}
\end{subequations}

Substituting the explicit form of the tensor analyzing power 
in Eq.~\eqref{e:w2}, 
we obtain the decay distribution (cf. Ref.~\citen{Schilling:1970um})
\begin{equation} \begin{split} 
W(\cos \theta, \phi_R)
	&= \frac{3}{8\pi}\Biggl(\frac{2}{3}
	+S_{LL}(3 \cos^2{\theta}-1)
	-S_{LT}^{x} \sin{2\theta} \;\cos{\phi_R} 
		-S_{LT}^{y} \sin{2\theta}\; \sin{\phi_R}  \\ 
&\quad -S_{TT}^{xx}\sin^2{\theta}\; \cos{2\phi_R}
	-S_{TT}^{xy}\sin^2{\theta}\; \sin{2\phi_R}\Biggr).
\end{split} \end{equation}  

%
%In the case where the polar axis is chosen in the 
%direction of the virtual photon, 
%in order to determine the relevant invariant quantity for $S_L$,
%$S_{LL}$, $S_{LT}^\mu$ and $S_{TT}^{\mu\nu}$, we construct the
%covariant comparison as in Eq.~\eqref{covtransf}, using the relation
%between $g_\perp^{\mu\nu}$ and $g_T^{\mu\nu}$. It is then easy to
%find, for any hadron (neglecting order  $1/Q^2$ corrections),
%\begin{eqnarray}
%S_{L} &=& \frac{M\,(S\cdot q)}{P\cdot q},\\
%S_{T}^\mu &=& S_{\perp}^\mu - S_{L}\,\frac{P_{\perp}^\mu}{M},\\
%  \\
%S_{LL} 
%&=& -\frac{M^2\,(q^\rho T_{\rho\sigma}q^\sigma)}{(P\cdot q)^2} ,\\ 
%\frac{1}{2}\,S_{LT}^\mu 
%&=& \frac{M\,(g_\perp^{\mu\rho}T_{\rho\sigma}q^\sigma)}{P\cdot q} -
%+S_{LL}\,\frac{P_{\perp}^\mu}{M} ,\\ 
%\frac{1}{2}\,S_{TT}^{\mu\nu} 
%&=& g_\perp^{\mu\rho}T_{\rho\sigma}g_\perp^{\sigma\nu}
%- \frac{1}{2}\,\frac{P_{\perp}^{\{\mu}S_{LT}^{\nu\}}}{M}
%+\,S_{LL}\,\frac{P_{\perp}^\mu P_{\perp}^\nu}{M^2} 
% \\
%\mbox{}\hspace{1cm}
%&=& g_\perp^{\mu\rho}T_{\rho\sigma}g_\perp^{\sigma\nu}
%-\frac{P_{\perp}^{\{\mu}g_\perp^{\nu\}\rho}
%T_{\rho\sigma}q^\sigma}{P\cdot q}
%- \,S_{LL}\,\frac{P_{\perp}^\mu P_{\perp}^\nu}{M^2} .
%\end{eqnarray}

%%%%%%%%%%%%%%%%%%%%%%%%%%%%%%%%%%%%%%%%%%%%%%%%%%%%%%%%%%%%
\section{The correlation function $\Delta$}
\markright{The correlation function $\mathit \Delta$}
\label{s:deltaspinone}

The decomposition of the correlation function $\Delta$ is analogous to that of 
$\Phi$, just replacing $p$, $P$, $A_i(p\cdot P,p^2)$ with $k$, $P_h$, $B_i(k\cdot
P_h,k^2)$ and the spin vector and tensor, $S^{\mu}$ and $ T^{\mu \nu}$, with the
vector and tensor analyzing powers, $A^{\mu}$ and $ A^{\mu \nu}$.
Therefore, the leading-twist part of the correlation functions, 
${\cal P}_-\,\Delta(z, A)\, \g^-$,
can be 
parametrized in terms of five fragmentation functions
\begin{equation} \begin{split} 
{\cal P}_-\,\Delta(z, \cos{\theta},\phi_R)\, \g^-&=\frac{1}{8 \pi}\,\Bigl(
		D_1(z) +
		G_{1}(z)\,A_{L}\,\g_5 +
		H_{1}(z)\,\g_5 \Aslash_T +
		B_{1}(z)\,A_{LL} +
		\ii\,H_{1LT}(z)\,\Aslash_{LT}\Bigr)\,
		{\cal P}_-,		 
\label{e:deltaspin1}
\end{split} \end{equation}  
The prefactor has been chosen to have a better connection to the unpolarized
results, i.e.\ integrated over $\cos\theta$ and $\phi_R$. 
We retained the terms containing the vector analyzing powers, although we
know that they vanish in the case of a parity-conserving decay.
The function $H_{1LT}$ is T-odd and corresponds to the function
$\hat{h}_{\overline{1}}$ of Ref.~\citen{Ji:1994vw}. This function evolves with 
the energy scale in the same way as the transversity fragmentation 
function~\cite{Stratmann:2001pt}.

The matrix representation of the leading part of $\Delta(z)$ can be obtained
almost directly from the result of the matrix $F(x)$, shown in 
Eq.~\eqref{e:Fspinone}, provided we apply the correct replacements. The result 
is
\addtolength\arraycolsep{-11pt} 
\begin{multline}
D(z)_{\chi'_{2} \chi^{\phantom'}_{2} }^{\Lambda_2' \Lambda^{\phantom'}_2} = \\
\frac{1}{6}\lf(\begin{array}{@{\hspace{3pt}}ccc@{\hspace{3pt}}|@{\hspace{3pt}}ccc@{\hspace{3pt}}}
{D_1} + {G_1}-\frac{1}{3}\,B_1& 0 & 0 & 0 & 
	\sqrt{2}\,({H_1}+\ii \,{H_{1LT}})& 0 \\
0 & {D_1} +\frac{2}{3}\,B_1 & 0 & 0 & 0 &\sqrt{2}\,({H_1}-\ii\, {H_{1LT}}) \\
0 & 0 &{D_1} - {G_1}-\frac{1}{3}\,B_1 & 0 & 0 & 0 \\ 
\hline
0 & 0 & 0 & {D_1} - {G_1}-\frac{1}{3}\,B_1& 0 & 0 \\
\sqrt{2}\,({H_1}-\ii\, {H_{1LT}}) & 0 & 0 & 0 &{D_1} +\frac{2}{3}\,B_1 & 0 \\
0 &\sqrt{2}\,({H_1}+\ii\, {H_{1LT}}) & 0 & 0 & 0 & 
		{D_1} + {G_1}-\frac{1}{3}\,B_1
\end{array}\rg). \\
\label{e:Dspinone}
\end{multline}
Conservation of angular momentum in the fragmentation process
requires $\Lambda_{2}+\chi'_{2} =\Lambda'_{2} +\chi_{2}$. 
\addtolength\arraycolsep{11pt}

At this point, we can compare the matrix we obtained with the pure $p$-wave
sector of the matrix $D(z, M_h^2)$ in Eq.~\eqref{e:Dzmh}. We notice that
the two matrix have an almost identical form, after we identify
\begin{subequations} \label{e:twohadandspinone} \begin{align} 
D_1 & = \frac{3}{4}\, D_{1, UU}^{pp}, \\
B_1 & = \frac{3}{4}\, D_{1, LL}^{pp}, \\
H_{1LT} & = \frac{3}{4}\,\lf(-\frac{2}{3}\frac{\rr}{M_h} H_{1,LT}^{\newangle \,
pp}\rg). 
\end{align} \end{subequations} 
There are, however, a couple of differences.
The first
difference is that two-hadron fragmentation functions depend also on the
total invariant mass squared, while spin-one fragmentation functions do not,
since they are supposed to 
be nonzero only at the mass of the hadron we are observing. 
More appropriately,
since we are usually observing a resonance
we should assume that spin-one fragmentation functions have to be  multiplied
by an invariant mass distribution, e.g.\ a Breit-Wigner curve peaked at 
the resonance mass.

The second difference is that in the analysis of two-hadron fragmentation we
missed the components $G_1$ and $H_1$, typical of vector
polarization. The reason is that a parity
violating decay is required to analyze these components. 
In fact, if we trace the matrix of
Eq.~\eqref{e:Dspinone} with a  decay matrix such as the one defined in
Eq.~\eqref{e:Rdecaymatrix},
the vector polarization functions, $G_1$ and $H_1$, 
would not appear in the decay distribution. When studying two-hadron
fragmentation, we imposed parity invariance from the very beginning -- from
Eq.~\eqref{e:decomdelta2p} -- thus missing any contribution of vector 
polarization.   

We are not going to analyze the features of the correlation function $\Delta$
when partonic transverse momentum is included, as they essentially reproduce
those of the correlation function $\Phi$, studied in
Sec.~\vref{s:transspinone}, modulo the replacements we discussed at the
beginning of this section.

%%%%%%%%%%%%%%%%%%%%%%%%%%%%%%%%%%%%%%%%%%%%%%%%%%%%%%%%%%%%
\subsection{Positivity bounds on spin-one fragmentation functions}

The bounds stemming from the positivity of the
fragmentation matrix are very similar to the ones we discussed for spin-one
distribution functions in Sec.~\vref{s:helicityspinone}. From the diagonal
elements we get
\addtolength\arraycolsep{8pt} 
\begin{subequations} \begin{align} 
D_1 (z) &\geq  0, \label{e:D1boundspinone}\\
-\frac{3}{2}\,D_1(z) &\leq B_1(z) \;\leq\; 3\,D_1 (z), \label{e:B1boundspinone}\\
|G_1 (z)|&\leq D_1 (z)-\frac{1}{3}\,B_1(z)\;\leq\;\frac{3}{2}\,D_1(z).
\label{e:boundG1}
\end{align} \end{subequations}
Eqs.~\eqref{e:D1boundspinone} and \eqref{e:B1boundspinone} correspond to the
Eqs.~\eqref{e:D1boundpw} and \eqref{e:B1boundpw}.
From positivity of the two-dimensional minors we get
\begin{equation}
\bigl(H_{1LT}(z)\bigr)^2 + \bigl(H_1(z)\bigr)^2 \leq
	\frac{1}{2}\left(D_1(z)+\frac{2}{3}\,B_1(z)\right) 
	\left(D_1(z)+G_1(z)-\frac{1}{3}B_1(z)\right) .
\label{e:boundH1LT.}
\end{equation} 
Because of the lack of information on vector polarized fragmentation
functions, the bound in Eq.~\eqref{e:boundG1} does not have a direct relevance 
for experiments, but it can be very useful to test the consistency of model
calculations. The bound in Eq.~\eqref{e:boundH1LT.}
can be used in a less restrictive version~\cite{Bacchetta:2001rb}
\begin{equation} 
\left\lvert H_{1LT}(z)\right\rvert
	\leq	\sqrt{\left(D_1(z)+\frac{2}{3}\,B_1(z)\right) 
		\left(D_1(z)-\frac{1}{3}\, B_1(z)\right)} \leq
	\frac{3}{2 \sqrt{2}}\, D_1(z) ,
\end{equation} 
corresponding to Eq.~\eqref{e:H1LTbound}.

%%%%%%%%%%%%%%%%%%%%%%%%%%%%%%%%%%%%%%%%%%%%%%%%%%%%%%%%%%%%
\section{Cross section and asymmetries for spin-one production}
\label{s:spin1cross}

\begin{figure}
	\centering
	\includegraphics[width=8cm]{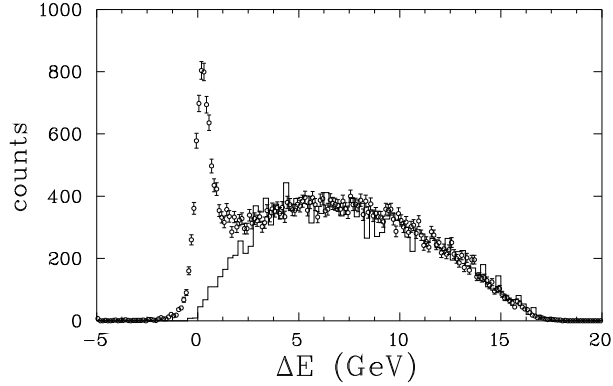}
	\caption{Distribution of the energy absorbed by the target in 
        $\rho$ production events~\cite{Ackerstaff:2000bz}.}
	\label{f:deltae}
        \end{figure}

We consider one-particle inclusive DIS events where the
target consists of a spin-half hadron and the fragment is a spin-one
hadron.
We assume the polarization of the final state hadron is analyzed by means of a 
two-particle, parity-conserving decay. It is worthwhile to mention that the
experimental 
analysis of spin-one final state polarization has been already performed
for {\em exclusive} leptoproduction of vector
mesons~\cite{Breitweg:1998ed,Breitweg:1998nh,Breitweg:1999fm,Ackerstaff:2000bz,Adloff:1999kg} 
and in $e^+ e^-$
annihilation~\cite{Abreu:1995rg,Abreu:1997wd,Ackerstaff:1997kd,Ackerstaff:1997kj,Abbiendi:1999bz}.
Fig.~\vref{f:deltae} shows a typical distribution 
of the energy absorbed by the target in 
$\rho$ production events. The peak represents the exclusive events, 
while the rest of the distribution contains deep inelastic events, which have 
always been excluded from data analyses.

Let us first analyze the cross section integrated over the transverse momentum 
of the outgoing hadron. It will be differential in six variables
\begin{equation} 
\frac{\de^6\! \sigma}{\de \cos\theta \de \phi_R \de z_h \de \xbj
\de y  \de \phi_S} =  
\rho(S)_{\Lambda^{\phantom'}_1 \Lambda_1'}\;
F(\xbj)_{\chi^{\phantom'}_{1} \chi'_{1}}^{\Lambda_1' \Lambda^{\phantom'}_1}
\Bigl( \frac{\de \sigma^{eq}}{\de y} \Bigr)
	^{\chi^{\phantom'}_{1} \chi'_{1} ; \,\chi^{\phantom'}_{2} \chi'_{2}}
D(z_h)_{\chi'_{2} \chi^{\phantom'}_{2}}^{\Lambda_2' \Lambda^{\phantom'}_2}\,
	{R}(\theta,\phi_R)_{\Lambda^{\phantom'}_2 \Lambda_2'}, 
\label{e:spinonecross}
\end{equation} 
where $\theta$ and $\phi_R$ are the decay angles discussed in
Sec.~\vref{s:decayspinone}.  

The unpolarized and polarized parts of the cross section are
\begin{subequations} \begin{align} 
%\begin{split}
\de^6\! \sigma_{UU} &= 
\sum_q \frac{\alpha^2 e_q^2}{2 \pi s \xbj y^2}\,
	A(y)\,f_1^q(\xbj)\, 
%\\	&\quad\times
\biggl(D_1^q(z_h)	
		+\frac{1}{3}\lf(3 \cos^2{\theta} -1\rg) 
			B_1^q(z_h)\biggr),
%\end{split}
\\
%\begin{split}
\de^6\! \sigma_{LL} &=
\sum_q \frac{\alpha^2 e_q^2}{2 \pi s \xbj y^2}\,
\lambda_e\, S_L\, C(y)\, g_1^q(\xbj)\,
% \\&\quad\times
\biggl(D_1^q(z_h)	
		+\frac{1}{3}\lf(3 \cos^2{\theta} -1\rg) 
			B_1^q(z_h)\biggr), \label{e:llspinone}
%\end{split}
\\
%\begin{split}
\de^6\! \sigma_{UT} &=-
\sum_q \frac{\alpha^2 e_q^2}{2 \pi s \xbj y^2}\,
B(y)\,|\mb{S}_{\perp}| \, \sin 2 \theta\, \sin(\phi_R + \phi_S)
 \, h_1^q(\xbj)	
%\\	&\quad \times 
\, H_{1LT}^q(z_h). \label{e:utspinone}
%\end{split}
\end{align} \end{subequations} 
The unpolarized cross section corresponds to Eq.~(58)
of Ref.~\citen{Bacchetta:2000jk} (there the unpolarized fragmentation function has
been neglected), while Eq.~\eqref{e:llspinone} 
corresponds to Eq.~(61) of Ref.~\citen{Bacchetta:2000jk} integrated over
the hadron transverse momentum and finally Eq.~\eqref{e:utspinone} 
corresponds to
Eq.~(63) of Ref.~\citen{Bacchetta:2000jk}.\footnote{An overall minus sign is
missing in Eq.~(63) of Ref.~\citen{Bacchetta:2000jk}}
At the same time, note the correspondence with the $pp$
sector of Eqs.~\eqref{e:cross2ppw}, after we apply the identifications 
of Eqs.~\eqref{e:twohadandspinone}.

The asymmetry of Eq.~\eqref{e:utspinone} contains the transversity
distribution multiplied with the chiral-odd T-odd function $H_{1LT}$. This
observable is just a part of the one we presented in Eq.~\eqref{e:asymmtwohadron}, but this is 
a good point to stress once more a couple of details. The function
$H_{1LT}$ is {\em different} from  the $sp$ interference fragmentation
function $H_{1, UT}^{\newangle \, sp}$. 
Although a proper analysis of this function has to be done in the
framework of two-particle fragmentation, it is important to realize that it
enjoys the characteristics 
of a single-particle function. For instance, there is
little doubt about the fact that it will be strongly peaked at the mass of
some spin-one resonance. On the contrary, 
we don't know what is the invariant mass
behavior of $sp$ interference fragmentation functions. Moreover, its physical
origin could be more similar to that of the Collins function than to that of
$sp$ interference fragmentation
functions. Since we have some indications that the Collins function is 
possibly sizable, there is hope that also $H_{1LT}$ will turn out to be large, 
maybe larger than $sp$ interference fragmentation functions. However, it must
be remembered that $H_{1LT}$ does not depend on partonic transverse momentum.

If we don't integrate over the transverse momentum of the outgoing hadron, the 
the cross section will be differential
in eight variables 
\begin{multline}
\frac{\de^8\! \sigma}{\de \cos\theta \de \phi_R \de z_h \de^2\! \mb{P}_{h \perp} \de \xbj
\de y  \de \phi_S} =  \\
\rho(S)_{\Lambda^{\phantom'}_1 \Lambda_1'}\;{\cal I} \lf[
F(\xbj, \mb{p}_T)_{\chi^{\phantom'}_{1} \chi'_{1}}^{\Lambda_1' \Lambda^{\phantom'}_1}
\Bigl( \frac{\de \sigma^{eq}}{\de y} \Bigr)
	^{\chi^{\phantom'}_{1} \chi'_{1} ; \,\chi^{\phantom'}_{2} \chi'_{2}}
	D(z_h, \mb{k}_T)_{\chi'_{2} \chi^{\phantom'}_{2}}^{\Lambda_2' \Lambda^{\phantom'}_2}\rg]\,
	{R}(\theta,\phi_R)_{\Lambda^{\phantom'}_2 \Lambda_2'} . 
\label{e:spinonecrosswithtrans}
\end{multline}
Obviously, the cross sections are much more complex than before. The following 
formulae correspond to the ones listed in Ref.~\citen{Bacchetta:2000jk}, 
if we 
replace the tensor polarization components with the analyzing powers of
Eqs.~\eqref{e:analyzingpowers}. However, here we show also
the unpolarized contributions and
we examine also the terms with T-odd distribution
functions.\footnote{Moreover, in
Ref.~\citen{Bacchetta:2000jk} there are a few typos and
there is a systematic error sign whenever the factors $|S_{hLT}|$ and
$|S_{hTT}|$  are used.}

%%%%%%%%%%%%%%%%%%%%%%%%%%%%%%%%%%%%%%%%%%%%%%%%%%%%%%%%%%%%
\subsubsection{Unpolarized lepton beam and unpolarized target} 

\begin{equation} 
\begin{split}
\label{e:crosssecOO}
\de^8\!\sigma_{UU}  
&=
%%%%%%%%%%%%%%%%%%%%%%%%%%%%%%%%%%%%%%%%%%%%%%%%%%%%%%%%%%%%
\sum_{q} \frac{\alpha^2 e_q^2}{2 \pi s \xbj y^2}\,
A(y)\,\Biggl\{{\cal I}\left[f_1 \, D_1\right]
+ \frac{1}{3}\lf(3 \cos^2{\theta} -1\rg) {\cal I}\left[f_1 \, B_1\right] 
\\ & \quad
- \sin 2\theta\, \cos(\phi_h - \phi_R) 
	\,{\cal I}\left[\frac{\mb{k}_T\cdot\h}{M_h}\,f_1 \, D_{1LT}\right]
%\\ & \quad
- \sin^2\theta\, \cos(2\phi_h - 2\phi_R) 
\\ & \quad \times
	{\cal I}\left[\frac{2(\mb{k}_T \cdot \h)^2
		- \mb{k}_T^2}{M_h^2} \,f_1 \, D_{1TT}\right]\Biggr\}
%%%%%%%%%%%%%%%%%%%%%%%%%%%%%%%%%%%%%%%%%%%%%%%%%%%%%%%%%%%%
%\\ & \quad
+\sum_{q} \frac{\alpha^2 e_q^2}{2 \pi s \xbj y^2}\,
B(y)\,
%\\ & \quad \times
 \Biggl\{-\frac{1}{3}\lf(3 \cos^2{\theta} -1\rg) \cos 2\phi_h
\\ & \quad \times
	{\cal I} \left[\frac{2(\mb{p}_T \cdot \h)(\mb{k}_T
 \cdot \h) - \mb{p}_T\cdot \mb{k}_T}{M M_h}\,  h_1^{\perp} \, H_{1LL}^{\perp}\right] 
%\\ & \quad
+\sin 2\theta\, \cos (\phi_h + \phi_R) 
\\ & \quad \times
	{\cal I}\left[\frac{\mb{p}_T\cdot\h}{M}\, h_1^{\perp}\, H_{1LT}\right]
+\sin^2\theta\, \cos 2\phi_R \,
	{\cal I}\left[\frac{\mb{p}_T\cdot\mb{k}_T}{M M_h}\, h_1^{\perp}\,
H_{1TT}\right]
%\\ & \quad
+\sin 2\theta\, \cos (3 \phi_h - \phi_R) 
\\ & \quad \times 
	{\cal I}\left[\frac{4(\mb{k}_T\cdot\h)^2(\mb{p}_T\cdot\h)
	-2(\mb{k}_T\cdot\h)(\mb{p}_T\cdot\mb{k}_T)
	-\mb{k}_T^2(\mb{p}_T\cdot\h)}{2 M M_h^2}\,
     h_1^{\perp} \, H_{1LT}^{\perp}\right]
\\ & \quad
+\sin^2 \theta\, \cos (4 \phi_h - 2 \phi_R) 
%\\ & \quad  \times 
	\,{\cal I}\Biggl[\Biggl(\frac{\lf[\mb{k}_T^2 -
4(\mb{k}_T\cdot\h)^2\rg]\lf[\mb{p}_T\cdot\mb{k}_T-4(\mb{k}_T\cdot\h)(\mb{p}_T\cdot\h)\rg]}{2 M M_h^3}
\\ & \quad
	-\frac{8(\mb{k}_T\cdot\h)^3(\mb{p}_T\cdot\h)}{2 M M_h^3}\Biggr)\,
     h_1^{\perp} \, H_{1TT}^{\perp}\Biggr]
\Biggr\},
\end{split}
\end{equation} 

%%%%%%%%%%%%%%%%%%%%%%%%%%%%%%%%%%%%%%%%%%%%%%%%%%%%%%%%%%%%
\subsubsection{Polarized lepton beam and unpolarized target} 

\begin{equation} 
\begin{split} 
\label{e:crosssecLO}
\de^8\!\sigma_{LU}  
&= 
- \sum_{q} \frac{\alpha^2 e_q^2}{2 \pi s \xbj y^2}\,\lambda_e\,
C(y)\,\Biggl\{
\sin 2\theta\, \sin(\phi_h - \phi_R)
	\,{\cal I}\left[\frac{\mb{k}_T\cdot\h}{M_h}\,f_1 \, G_{1LT}\right]
\\ & \quad
+\sin^2 \theta\, \sin(2\phi_h - 2\phi_R)
   \,{\cal I}\left[\frac{2(\mb{k}_T \cdot \h)^2
		- \mb{k}_T^2}{M_h^2}\,
     f_1 \, G_{1TT}\right]\Biggr\},
\end{split}
\end{equation} 

%%%%%%%%%%%%%%%%%%%%%%%%%%%%%%%%%%%%%%%%%%%%%%%%%%%%%%%%%%%%
\subsubsection{Unpolarized lepton beam and longitudinally polarized target} 

\begin{equation} \begin{split}
\label{e:crosssecOL}
\de^8\!\sigma_{UL}  
&=
-\sum_{q} \frac{\alpha^2 e_q^2}{2 \pi s \xbj y^2}\,\lf\lvert S_L \rg\rvert
\,A(y)\,
\Biggl\{\sin 2\theta \,\sin(\phi_h-\phi_R)\,
   {\cal I}\left[\frac{\mb{k}_T\cdot\h}{M_h}\,
     g_{1L} \, G_{1LT}\right] 
\\ & \quad 
+ \sin^2 \theta\, \sin(2\phi_h - 2\phi_R)
	\,{\cal I}\left[\frac{2(\mb{k}_T \cdot \h)^2
		- \mb{k}_T^2}{M_h^2}\,
     g_{1L} \, G_{1TT}\right] 
\Biggr\}
%%%%%%%%%%%%%%%%%%%%%%%%%%%%%%%%%%%%%%%%%%%%%%%%%%%%%%%%%%%%
\\ & \quad
-\sum_{q} \frac{\alpha^2 e_q^2}{2 \pi s \xbj y^2}\,\lf\lvert S_L \rg\rvert
\,B(y)\,
\Biggl\{-\frac{1}{3}\lf(3 \cos^2{\theta} -1\rg)\, \sin 2\phi_h
\\ & \quad
	\times {\cal I} \left[\frac{2(\mb{p}_T \cdot \h)(\mb{k}_T
 \cdot \h) - \mb{p}_T\cdot \mb{k}_T}{M M_h}\,  h_{1L}^{\perp} \, H_{1LL}^{\perp}\right] 
%\\ & \quad
+\sin 2\theta \,\sin(\phi_h+\phi_R)
\\ & \quad \times
   {\cal I}\left[\frac{\mb{p}_T\cdot\h}{M}\,
     h_{1L}^{\perp} \, H_{1LT}\right] 
%\\ & \quad 
+ \sin^2 \theta\, \sin 2\phi_R 
   \,{\cal I}\left[\frac{\mb{p}_T\cdot\mb{k}_T}{M M_h}\,
     h_{1L}^{\perp} \, H_{1TT}\right] 
%\\ & \quad 
+\sin 2\theta \,\sin(3 \phi_h - \phi_R)
\\ & \quad
\times   {\cal I}\left[\frac{4(\mb{k}_T\cdot\h)^2(\mb{p}_T\cdot\h)
	-2(\mb{k}_T\cdot\h)(\mb{p}_T\cdot\mb{k}_T)
	-\mb{k}_T^2(\mb{p}_T\cdot\h)}{2 M M_h^2}\,
     h_{1L}^{\perp} \, H_{1LT}^{\perp}\right]
\\ & \quad
+\sin^2 \theta\, \sin (4 \phi_h - 2 \phi_R) 
%\\ & \quad \times
	{\cal I}\Biggl[\Biggl(\frac{\lf[\mb{k}_T^2 -
4(\mb{k}_T\cdot\h)^2\rg]\lf[\mb{p}_T\cdot\mb{k}_T-4(\mb{k}_T\cdot\h)(\mb{p}_T\cdot\h)\rg]}{2 M M_h^3}
\\ & \quad 
	-\frac{8(\mb{k}_T\cdot\h)^3(\mb{p}_T\cdot\h)}{2 M M_h^3}\Biggr)\,
     h_{1L}^{\perp} \, H_{1TT}^{\perp}\Biggr]
\Biggr\},
\end{split}
\end{equation} 

%%%%%%%%%%%%%%%%%%%%%%%%%%%%%%%%%%%%%%%%%%%%%%%%%%%%%%%%%%%%
\subsubsection{Polarized lepton beam and longitudinally polarized target} 

\begin{equation} 
\begin{split}
\label{e:crosssecLL}
\de^8\!\sigma_{LL}  
&=
\sum_{q} \frac{\alpha^2 e_q^2}{2 \pi s \xbj y^2}\,\lambda_e\,
\lf\lvert S_L \rg\rvert\, C(y)\,
\Biggl\{{\cal I}\left[g_{1L} \, D_1\right]
+ \frac{1}{3}\lf(3 \cos^2{\theta} -1\rg) {\cal I}\left[g_{1L} \, B_1\right] 
\\ & \quad
- \sin 2\theta\, \cos(\phi_h - \phi_R) 
	\,{\cal I}\left[\frac{\mb{k}_T\cdot\h}{M_h}\,g_{1L} \, D_{1LT}\right]
\\ & \quad
- \sin^2\theta\, \cos(2\phi_h - 2\phi_R) 
	\,{\cal I}\left[\frac{2(\mb{k}_T \cdot \h)^2
		- \mb{k}_T^2}{M_h^2} \,g_{1L} \, D_{1TT}\right]\Biggr\},
\end{split} \end{equation} 

%%%%%%%%%%%%%%%%%%%%%%%%%%%%%%%%%%%%%%%%%%%%%%%%%%%%%%%%%%%%
\subsubsection{Unpolarized lepton beam and  transversely polarized target} 

%\begin{equation} \begin{split}
\begin{align} 
\label{e:crosssecOT}
\de^8\!\sigma_{UT}  
&=
\sum_{q} \frac{\alpha^2 e_q^2}{2 \pi s \xbj y^2}\,\sst\,A(y)\,
\Biggl\{
\sin 2\theta \sin(\phi_R - \phi_S)
	\,{\cal I}\left[\frac{(\mb{p}_T \cdot \mb{k}_T)}{2 M M_h} 
		\,g_{1T} \, G_{1LT}\right]
\nn \\ & \quad
-\sin 2\theta \sin(2 \phi_h - \phi_R - \phi_S)
	\,{\cal I}\left[\frac{2(\mb{p}_T \cdot \h)(\mb{k}_T
 \cdot \h) - \mb{p}_T\cdot \mb{k}_T}{2 M M_h}
		\,g_{1T} \, G_{1LT}\right]
\nn \\ & \quad
-\sin^2 \theta \sin(\phi_h - 2 \phi_R + \phi_S)
	\,{\cal I}\left[\frac{2 (\mb{k}_T\cdot\h) (\mb{p}_T\cdot\mb{k}_T)
	- \mb{k}_T^2(\mb{p}_T\cdot\h)}{2 M M_h^2} 
		\,g_{1T} \, G_{1TT}\right]
\nn \\ & \quad
-\sin^2 \theta \sin(3 \phi_h - 2 \phi_R - \phi_S)
%\nn \\ & \quad \times 
	\,{\cal I}\Biggl[\Biggl(\frac{4(\mb{k}_T\cdot\h)^2(\mb{p}_T\cdot\h)
	-2(\mb{k}_T\cdot\h)(\mb{p}_T\cdot\mb{k}_T)}{2 M M_h^2}
\nn \\ & \quad 
	-\frac{\mb{k}_T^2(\mb{p}_T\cdot\h)}{2 M M_h^2}\Biggr)
		\,g_{1T} \, G_{1TT}\Biggr]
%\nn \\ & \quad
+\sin(\phi_h - \phi_S)
	\,{\cal I}\left[\frac{\mb{p}_T \cdot \h}{M}
		\,f_{1T}^{\perp} \, D_1\right]
%\nn \\ & \quad
+\frac{1}{3}\lf(3 \cos^2{\theta} -1\rg) 
\nn \\ & \quad \times
\sin(\phi_h - \phi_S)
%\nn \\ & \quad \times
	\,{\cal I}\left[\frac{\mb{p}_T \cdot \h}{M}
		\,f_{1T}^{\perp} \, B_1\right]
%\nn \\ & \quad
-\sin 2\theta \sin(\phi_R - \phi_S)
	\,{\cal I}\left[\frac{(\mb{p}_T \cdot \mb{k}_T)}{2 M M_h} 
		\,f_{1T}^{\perp} \, D_{1LT}\right]
\nn \\ & \quad
-\sin 2\theta \sin(2 \phi_h - \phi_R - \phi_S)
	\,{\cal I}\left[\frac{2(\mb{p}_T \cdot \h)(\mb{k}_T
 \cdot \h) - \mb{p}_T\cdot \mb{k}_T}{2 M M_h}
		\,f_{1T}^{\perp} \, D_{1LT}\right]
\nn \\ & \quad
+\sin^2 \theta \sin(\phi_h - 2 \phi_R + \phi_S)
	\,{\cal I}\left[\frac{2 (\mb{k}_T\cdot\h) (\mb{p}_T\cdot\mb{k}_T)
	- \mb{k}_T^2(\mb{p}_T\cdot\h)}{2 M M_h^2} 
		\,f_{1T}^{\perp} \, D_{1TT}\right]
\nn \\ & \quad
-\sin^2 \theta \sin(3 \phi_h - 2\phi_R - \phi_S)
%\nn \\ & \quad
	\,{\cal I}\Biggl[\Biggl(\frac{4(\mb{k}_T\cdot\h)^2(\mb{p}_T\cdot\h)
	-2(\mb{k}_T\cdot\h)(\mb{p}_T\cdot\mb{k}_T)}{2 M M_h^2}
\nn \\ & \quad 
	-\frac{\mb{k}_T^2(\mb{p}_T\cdot\h)}{2 M M_h^2}\Biggr)
		\,f_{1T}^{\perp} \, D_{1TT}\Biggr]\Biggr\}
%%%%%%%%%%%%%%%%%%%%%%%%%%%%%%%%%%%%%%%%%%%%%%%%%%%%%%%%%%%%
%\nn \\ & \quad
+\sum_{q}\frac{\alpha^2 e_q^2}{2 \pi s \xbj y^2}\,
B(y)\,
%\nn \\ & \quad \times
\Biggl\{\frac{1}{3}\lf(3 \cos^2{\theta} -1\rg) \sin(\phi_h + \phi_S)
\nn \\ & \quad \times
	{\cal I} \left[\frac{\mb{k}_T \cdot \h}{M_h}\,  h_1 \, H_{1LL}^{\perp}\right] 
%\nn \\ & \quad
-\sin 2\theta\, \sin (\phi_R + \phi_S) 
	\,{\cal I}\left[h_1\, H_{1LT}\right]
+\sin^2\theta\, 
\\ & \quad \times
\sin (\phi_h -2\phi_R-\phi_S) 
%\nn \\ & \quad \times
	\,{\cal I}\left[\frac{\mb{k}_T \cdot \h}{M_h}\, h_1\,
H_{1TT}\right]
%\nn \\ & \quad
-\sin 2\theta\, 
%\nn \\ & \quad \times
\sin (2 \phi_h - \phi_R+ \phi_S) 
\nn \\ & \quad \times
	{\cal I}\left[\frac{2(\mb{k}_T \cdot \h)^2
		- \mb{k}_T^2}{2 M_h^2} \,
     h_1 \, H_{1LT}^{\perp}\right]
%\nn \\ & \quad
-\sin^2 \theta\, \sin (3 \phi_h - 2 \phi_R+\phi_S) 
\nn \\ & \quad \times
	{\cal I}\left[\frac{4(\mb{k}_T \cdot \h)^3
		- 3 \mb{k}_T^2 (\mb{k}_T \cdot \h)}{2 M_h^3} \,
     h_1 \, H_{1TT}^{\perp}\right]
%\nn \\ & \quad
%%%%%%%%%%%%%%%%%%%%%%%%%%%%%%%%%%%%%%%%%%%%%%%%%%%%%%%%%%%%
+\frac{1}{3}\lf(3 \cos^2{\theta} -1\rg) \sin(3 \phi_h - \phi_S)
\nn \\ & \quad \times 
{\cal I} \left[\frac{4(\mb{p}_T\cdot\h)^2(\mb{k}_T\cdot\h)
	-2(\mb{p}_T\cdot\h)(\mb{p}_T\cdot\mb{k}_T)
	-\mb{p}_T^2(\mb{k}_T\cdot\h)}{2 M^2 M_h} \,  h_{1T}^{\perp} \, H_{1LL}^{\perp}\right] 
\nn \\ & \quad
-\sin 2\theta\, \sin (2 \phi_h +\phi_R - \phi_S) 
	\,{\cal I}\left[ \frac{2(\mb{p}_T \cdot \h)^2
		- \mb{p}_T^2}{2 M^2} \, h_{1T}^{\perp}\, H_{1LT}\right]
\nn \\ & \quad
-\sin^2\theta\, \sin (\phi_h +2\phi_R-\phi_S) 
	\,{\cal I}\left[\frac{2 (\mb{p}_T \cdot \mb{k}_T)(\mb{p}_T \cdot \h)
		- (\mb{k}_T \cdot \h)\mb{p}_T^2}{2 M^2 M_h}  
 \, h_{1T}^{\perp}\,H_{1TT}\right]
%\displaybreak[0]
\nn \\ & \quad
+\sin 2\theta\, \sin (4 \phi_h - \phi_R- \phi_S) 
%\nn \\ & \quad \times
	\,{\cal I}\Biggl[\Biggl(\frac{\mb{k}_T^2\lf[2(\mb{p}_T \cdot \h)^2
		- \mb{p}_T^2\rg]}{4 M^2 M_h^2} - 2 (\mb{k}_T \cdot \h)
\nn \\ & \quad \times
\frac{\lf[
4(\mb{p}_T\cdot\h)^2(\mb{k}_T\cdot\h)
	-2(\mb{p}_T\cdot\h)(\mb{p}_T\cdot\mb{k}_T)
	-\mb{p}_T^2(\mb{k}_T\cdot\h)
\rg]}{4 M^2 M_h^2}\Biggr)  \,
     h_{1T}^{\perp} \, H_{1LT}^{\perp}\Biggr]
\nn \\ & \quad
+\sin^2 \theta\, \sin (5 \phi_h - 2 \phi_R-\phi_S) 
%\nn \\ & \quad \times
	\,{\cal I}\Biggl[\Biggl(\frac{2 \mb{k}_T^2 (\mb{k}_T \cdot \h)
\lf[2(\mb{p}_T \cdot \h)^2
		- \mb{p}_T^2\rg]}{4 M^2 M_h^3}
+ \lf[\mb{k}_T^2 - 4 (\mb{k}_T \cdot \h)^2\rg] 
\nn \\ & \quad \times
\frac{
\lf[
4(\mb{p}_T\cdot\h)^2(\mb{k}_T\cdot\h)
	-2(\mb{p}_T\cdot\h)(\mb{p}_T\cdot\mb{k}_T)
	-\mb{p}_T^2(\mb{k}_T\cdot\h)
\rg]
}{4 M^2 M_h^3}\Biggr)   \,
     h_{1T}^{\perp} \, H_{1TT}^{\perp}\Biggr]
\Biggr\}, \nn
%\end{split} \end{equation} 
\end{align} 

%%%%%%%%%%%%%%%%%%%%%%%%%%%%%%%%%%%%%%%%%%%%%%%%%%%%%%%%%%%%
\subsubsection{Polarized lepton beam and transversely polarized target} 

\begin{equation} 
\begin{split}
\label{e:crosssecLT}
\de^8\!\sigma_{LT}  
&=
\sum_{q} \frac{\alpha^2 e_q^2}{2 \pi s \xbj y^2}\,\lambda_e 
\,\sst\,C(y)\,\Biggl\{ 
\cos(\phi_h-\phi_{S}) \,{\cal I}\left[\frac{\mb{p}_T\cdot\h}{M}\,
     g_{1T} \, D_1 \right]
%\\ & \quad +
+ \frac{1}{3}\lf(3 \cos^2{\theta} -1\rg)
\\ & \quad \times
 \cos(\phi_h - \phi_S)
\,{\cal I}\left[\frac{\mb{p}_T\cdot\h}{M}\,
     g_{1T} \, B_1 \right]
%\\ & \quad
-\sin 2\theta \cos(\phi_R - \phi_S)
	\,{\cal I}\left[\frac{(\mb{p}_T \cdot \mb{k}_T)}{2 M M_h} 
		\,g_{1T} \, D_{1LT}\right]
 \\ & \quad
-\sin 2\theta \cos(2 \phi_h - \phi_R - \phi_S)
	\,{\cal I}\left[\frac{2(\mb{p}_T \cdot \h)(\mb{k}_T
 \cdot \h) - \mb{p}_T\cdot \mb{k}_T}{2 M M_h}
		\,g_{1T} \, D_{1LT}\right]
 \\ & \quad
-\sin^2 \theta \cos(\phi_h - 2 \phi_R + \phi_S)
	\,{\cal I}\left[\frac{2 (\mb{k}_T\cdot\h) (\mb{p}_T\cdot\mb{k}_T)
	- \mb{k}_T^2(\mb{p}_T\cdot\h)}{2 M M_h^2} 
		\,g_{1T} \, D_{1TT}\right]
 \\ & \quad
-\sin^2 \theta \cos(3 \phi_h - 2 \phi_R - \phi_S)
% \\ & \quad \times 
	\,{\cal I}\Biggl[\Biggl(\frac{4(\mb{k}_T\cdot\h)^2(\mb{p}_T\cdot\h)
	-2(\mb{k}_T\cdot\h)(\mb{p}_T\cdot\mb{k}_T)}{2 M M_h^2}
 \\ & \quad 
	-\frac{\mb{k}_T^2(\mb{p}_T\cdot\h)}{2 M M_h^2}\Biggr)
		\,g_{1T} \, D_{1TT}\Biggr]
%\\ & \quad
-\sin 2\theta \cos(\phi_R - \phi_S)
	\,{\cal I}\left[\frac{(\mb{p}_T \cdot \mb{k}_T)}{2 M M_h} 
		\,f_{1T}^{\perp} \, G_{1LT}\right]
 \\ & \quad
+\sin 2\theta \cos(2 \phi_h - \phi_R - \phi_S)
	\,{\cal I}\left[\frac{2(\mb{p}_T \cdot \h)(\mb{k}_T
 \cdot \h) - \mb{p}_T\cdot \mb{k}_T}{2 M M_h}
		\,f_{1T}^{\perp} \, G_{1LT}\right]
 \\ & \quad
-\sin^2 \theta \cos(\phi_h - 2 \phi_R + \phi_S)
	\,{\cal I}\left[\frac{2 (\mb{k}_T\cdot\h) (\mb{p}_T\cdot\mb{k}_T)
	- \mb{k}_T^2(\mb{p}_T\cdot\h)}{2 M M_h^2} 
		\,f_{1T}^{\perp} \, G_{1TT}\right]
 \\ & \quad
+\sin^2 \theta \cos(3 \phi_h - 2 \phi_R - \phi_S)
% \\ & \quad \times 
	\,{\cal I}\Biggl[\Biggl(\frac{4(\mb{k}_T\cdot\h)^2(\mb{p}_T\cdot\h)
	-2(\mb{k}_T\cdot\h)(\mb{p}_T\cdot\mb{k}_T)}{2 M M_h^2}
 \\ & \quad 
	-\frac{\mb{k}_T^2(\mb{p}_T\cdot\h)}{2 M M_h^2}\Biggr)
		\,f_{1T}^{\perp} \, G_{1TT}\Biggr]
\Biggr\}.
\end{split}
\end{equation} 

The previous formulae describe the $pp$ sector of two-hadron
production with transverse momentum and partial-wave expansion, which we
did not compute in Chap.~\ref{c:twohadron}. 

%%%%%%%%%%%%%%%%%%%%%%%%%%%%%%%%%%%%%%%%%%%%%%%%%%%%%%%%%%%%
\section{Summary}

In this chapter, we examined the description of spin-one hadrons, both in the
role of
targets and of final-state fragments.

To describe spin-one targets, we introduced a rank-two 
spin tensor in the decomposition 
of the spin density matrix. 
The presence of the spin tensor made the decomposition of the correlation
function $\Phi$ more complex than the spin-half case. Neglecting partonic 
transverse momentum, we could describe the leading twist part of
the correlation function by means of five distribution functions
[Eq.~\eqref{e:phispin1}].  Besides the
familiar three functions of the spin-half case ($f_1$, $g_1$ and $h_1$), we
needed to introduce the function $b_1$ and the T-odd function $h_{1LT}$. The
latter was never discussed in the literature before and it is the first
leading-twist T-odd distribution function that survives the integration over
 partonic
transverse momentum.
 
We studied the correlation function in the framework of the helicity
formalism and we obtained for the first time positivity bounds on spin-one
distribution functions [Eqs.~\eqref{e:boundspinone1} and \eqref{e:soffer2}].
In particular, we demonstrated
 that the bounds on the helicity and transversity distribution functions are
different from the spin-half case.
We carried out the analysis of the correlation function including also
transverse momentum, leading to the decomposition of the correlation function
shown in Eqs.~\eqref{e:phispin1withtrans}, 
together with the positivity bounds of
Eqs.~\eqref{e:spinoneboundswithtrans}.

We briefly discussed the cross section for inclusive deep inelastic scattering 
on spin-one targets, emphasizing the contribution of the distribution 
function $b_1$ [Eqs.~\eqref{e:spin1cross1}, \eqref{e:spin1cross2} and \eqref{e:spin1cross4}].

We devoted the second half of the chapter to the analysis of spin-one
hadron as current fragments. We explained how it is possible to probe the
polarization of an outgoing hadron if it undergoes a two-particle decay. 
After studying the correlation function $\Delta$, 
in Sec.~\ref{s:spin1cross} 
we listed all the spin asymmetries occurring  at leading
order in $1/Q$ in spin-one deep inelastic leptoproduction, with or without
integration over the outgoing hadron's transverse momentum.
In particular, we highlighted the single transverse spin asymmetry of
Eq.~\eqref{e:utspinone}. Even though this asymmetry was already included in
the discussion of two-hadron fragmentation functions, in this chapter we made
clear that it has the characteristics of a {\em single-particle} fragmentation
function and it is not related to the two-hadron 
$sp$ interference function $H_{1,
UT}^{\newangle \, sp}$. 

In the next chapter, we are going to study how it is possible to generate
T-odd single-particle fragmentation functions in the framework of a field
theoretical approach.
\renewcommand{\quot}{%
\parbox{7cm}{Models are to be used, not believed.}\\[2mm] H. Teil 
}
%\renewcommand{\quot}{%
%\parbox{7cm}{The sciences do not try to explain, they hardly even try to
%interpret, they mainly make models.}\\ J. von Neumann 
%}

%%%%%%%%%%%%%%%%%%%%%%%%%%%%%%%%%%%%%%%%%%%%%%%%%%%%%%%%%%%%%%%%%%%%%%%%%%%%
\chapter[A model estimate of the Collins function]
{A model estimate\\ of the Collins function}
\label{c:model}

In the previous chapters, we extensively discussed how it is possible to
observe the transversity distribution function
in connection with T-odd chiral-odd fragmentation function, in particular 
the Collins function $H_1^{\perp}$, the two-hadron interference function
$H_{1,UT}^{\newangle \,sp}$ or the
spin-one fragmentation function $H_{1LT}$. 
Unfortunately, at the moment we have scarce or no
information about these functions. 
%Yet, experimentalists 
%whish to extract the transversity distribution using T-odd fragmentation
%functions as analyzing powers. 
Therefore, we need to address the important
issue of estimating them and check whether they
could be measurable. 
Beside giving an indication of the magnitude of unknown fuctions, 
model evaluations, however rough or naive, serve some useful
purposes: they show whether a nonzero function can be obtained in the
framework of known theories, they pave the way for future improved estimates,
they shed light on some crucial properties, and they analyze the consequences of
the assumptions of the model.  In this chapter, we focus on the Collins
function, which could be regarded as the prototype of a T-odd
function, and  present an estimate of it. 

In spite of the apparent difficulty in modeling T-odd effects,
a nonvanishing Collins 
function can be obtained
through a consistent one-loop calculation, in a description where massive
constituent quarks and pions are the only 
effective degrees of freedom and interact via a simple pseudoscalar
coupling, as we have shown in Ref.~\cite{Bacchetta:2001di}. We point out that
Collins himself suggested the idea of dressing the quark propagator as a
possible mechanism to produce a nonzero $H_1^{\perp}$~\cite{Collins:1993kk}
and a more specific way to achieve this goal was mentioned by Suzuky in Ref.~\citen{Suzuki:1999pe}.
The model discussed in our early work was admittedly unfit to reproduce
the phenomenology of the Collins function. Here, we show the result of a 
recent calculation~\cite{Bacchetta:2002es}, which gives a more reasonable
estimate.
%In that work~\cite{Bacchetta:2001di} little care 
%has been devoted to the phenomenology of the Collins function.
%In contrast, our interest lies in obtaining a reasonable estimate of 
%this function and the observable effects induced by it.

At present, only one attempt to theoretically estimate the Collins function
for pions exists~\cite{Artru:1997bh}, 
and little phenomenological information is available from experiments.
The HERMES Collaboration reported the first measurements of single spin 
asymmetries in semi-inclusive DIS~\cite{Airapetian:1999tv,Airapetian:2001eg},
giving an indication of a possibly nonzero Collins function. Preliminary
results have been presented by the SMC
collaboration as well~\cite{Bravar:2000ti}. 
The
Collins function has also been invoked to explain large azimuthal
asymmetries in 
$p p^{\uparrow} \rightarrow \pi X$~\cite{Anselmino:1999pw,Boglione:1999dq}. 
However, all these analyses 
are plagued by large uncertainties, 
%due to the
%possible presence of hadronic effects in both the initial and final states, 
and hence do not allow any conclusive statement yet. 
Recently, a phenomenological estimate of the
Collins function has been proposed~\cite{Efremov:2001cz}, 
combining results from the DELPHI, SMC and HERMES experiments.
However, in spite of all the efforts to pin down the Collins function, the
knowledge we have at present is still insufficient.

In this chapter we calculate the Collins 
function for pions in a chiral invariant approach at a low energy scale,
as we have done in Ref.~\citen{Bacchetta:2002es}.
We use the model of Manohar and Georgi~\cite{Manohar:1984md},
which incorporates chiral symmetry and its spontaneous breaking, two
important aspects of QCD at low energies.
The spontaneous breaking of chiral symmetry leads to the existence
of (almost massless) Goldstone bosons, which are included as effective degrees 
of freedom in the model.
Quarks appear as further degrees of freedom as well. However, 
in contrast with the current quarks of the QCD Lagrangian, the model uses
massive constituent quarks -- a concept
that has been proven very successful in many phenomenological models
at hadronic scales.
With the exception of Ref.~\citen{Ji:1993qx}, the implications of a chiral 
invariant interaction for fragmentation functions into Goldstone bosons 
at low energy scales remain essentially unexplored.

Although the applicability of the Manohar-Georgi model
 is restricted to energies below the scale of chiral symmetry 
breaking $\Lambda_\chi \approx 1\, \rm{GeV}$, this might be
 sufficient to calculate
soft objects.
In this kinematical regime, the chiral power counting allows setting up a 
consistent perturbation theory~\cite{Weinberg:1979kz}.
The relevant expansion parameter is given by $l/\Lambda_\chi$, where $l$ is a 
generic external momentum of a particle participating in the fragmentation.
To guarantee the convergence of the perturbation theory, we restrict the
maximum virtuality $\mu^2$ of the decaying quark to a soft value.
We mostly consider the case $\mu^2 =  1 \, \rm{GeV}^2$.

The outline of the chapter is as follows. We first give the details of our 
model and present the analytical results of our calculation. 
Next, we discuss our results and compare them with known observables,
indicating the choice of the parameters of our model. Then, 
we present the features
of our prediction for the Collins function and its moments. Finally, using the
outcome of our model, we estimate the leading order asymmetries containing 
the Collins function in semi-inclusive DIS and in $e^+e^-$
annihilation into two hadrons.

%%%%%%%%%%%%%%%%%%%%%%%%%%%%%%%%%%%%%%%%%%%%%%%%%%%%%%%%%%%%%%%%%%%%%%%%%%%
\section{Calculation of the Collins function}
\label{s:two}

Considering the fragmentation process of a quark into a pion, 
$q^{\ast}(k) \to \pi(p) Y$, we use the expressions of
 the unpolarized fragmentation function 
$D_1$ and the Collins function $H_1^\perp$ 
given in Eq.~\eqref{e:projfrag}. For convenience, we reproduce that definition 
in a more explicit way:\footnote{Note that this definition of $H_1^{\perp}$ slightly differs from the
original one given by Collins \cite{Collins:1993kk}.}
\begin{subequations} \begin{align} 
D_1(z,z^2 \mb{k}^2_{\st}) &= \left. 
 \frac{1}{4z} \int \de k^+ \;
              \tr[ \Delta (k,p)\, \g^-] \right|_{k^-=\frac{p^-}{z}} \,,
\label{e:d1} \\
\frac{\eps_{\st}^{ij} k_{\st j}}{m_{\pi}}
	      \, H_1^{\perp}(z,z^2 \mb{k}^2_{\st}) 
&= \left. \frac{1}{4z} \int \de k^+ \;
              \tr[ \Delta (k,p)\, \ii \sig^{i-}\g_5]
	\right|_{k^-=\frac{p^-}{z}} \,,  
\label{e:col1}
\end{align} \end{subequations} 
with $m_{\pi}$ denoting the pion mass and $\Delta$ being the correlation
function defined in Eq.~\eqref{e:delta}.
% and $\eps_{\st}^{ij} \equiv \eps^{ij-+}$
%(we specify the plus and minus lightcone components of a generic 4-vector 
%$a^{\mu}$ 
%according to $a^{\pm} \equiv (a^0 \pm a^3)/\sqrt{2}$).
%The correlation function $\Delta(k,p)$ in Eqs.~\eqref{e:d1} and Eq.~\eqref{e:col1}, omitting 
%gauge links, takes the form
%\begin{eqnarray}  
%\Delta(k,p)=\sum_X \, \int &&
%        \frac{\de^{4}\xi}{(2\pi)^{4}}\; \e^{+\ii k \cdot \xi}
%       \langle 0|
%%\,{\cal L}[\xi,0;\mbox{path}]
%\,\psi(\xi)|\pi, X\rangle 
%\nn\\
%&&\mbox{}\times
%\langle \pi, X|
%             \bar{\psi}(0)|0\rangle \,.    
%\label{e:delta}
%\end{eqnarray} 

We now use the chiral invariant model of Manohar and Georgi~\cite{Manohar:1984md}
to calculate the matrix elements in the correlation function.
Neglecting the part that describes free Goldstone bosons, 
the Lagrangian of the model reads
\begin{equation}
{\cal L} =  \bar{\psi} \, ( \ii \derslash + \Vslash - m 
 + g_{\sa} \Aslash \g_5 ) \, \psi \,.
\label{e:lagrangian}
\end{equation}
In Eq.~\eqref{e:lagrangian} the pion field enters through the vector and 
axial vector combinations 
\begin{equation} 
V^{\mu}  =  \frac{\ii}{2} \, [u^{\dagger} , \partial^{\mu} u] \,,\qquad
A^{\mu}  =  \frac{\ii}{2} \, \{u^{\dagger} , \partial^{\mu} u\} \,,
\end{equation} 
with $u = \exp(\ii \, \bm{\tau}  \cdot  \bm{\pi} / 2 F_\pi)$,
where the $\tau_i$ are the generators of the SU(2) flavor group and 
$F_\pi = 93 \, \textrm{MeV}$ represents the pion decay constant.
In absence of resonances, the pion decay constant determines the
scale of chiral symmetry breaking via $\Lambda_\chi = 4 \pi F_\pi$.
The quark mass $m$ and the axial coupling constant $g_{\sa}$ are free 
parameters of the model that are not constrained by chiral symmetry.
The values of these parameters will be specified in Sec.~\ref{s:three}.
Although we limit ourselves here to the SU(2) flavor sector of the model, 
the extension to strange quarks 
is straightforward, allowing in 
particular the calculation of kaon fragmentation functions.
For convenience we write down explicitly those terms of the interaction 
part of the Lagrangian \eqref{e:lagrangian} that are relevant for our 
calculation.
To be specific we need both the interaction of a single pion with a 
quark and the two-pion contact interaction, which can easily be obtained by 
expanding the nonlinear representation $u$ in terms of the pion field:
\begin{subequations} \begin{align} 
{\cal L}_{\pi qq} & =  - \frac{g_\sa}{2 F_{\pi}}\, \bar{\psi}\, \gamma_{\mu} 
 \gamma_5\, \bm{\tau} \cdot \partial^{\mu} \bm{\pi}\, \psi \,,
 \label{e:lpi} \\
{\cal L}_{\pi \pi qq} & =  - \frac{1}{4 F_{\pi}^2}\, \bar{\psi}\, \gamma_{\mu} \,
 \bm{\tau} \cdot ( \bm{\pi} \times \partial^{\mu} \bm{\pi} )\, \psi \,.
 \label{e:lpipi}
\end{align} \end{subequations} 
Performing the numerical calculation of the Collins function,
 it turns out that 
the contact interaction \eqref{e:lpipi},
which is a direct consequence of chiral symmetry, plays a dominant role.

\begin{figure}
        \centering
        \includegraphics[width=3.3cm]{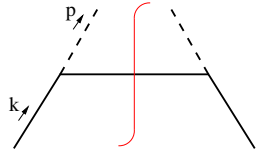}
        \caption{Lowest-order cut diagram describing the fragmentation 
		of a quark into a pion.} 
	\label{f:born}
        \end{figure}

At tree level, the fragmentation of a quark is modeled through the process
$q^{\ast} \to \pi q$, where Fig.~\vref{f:born} represents the corresponding 
unitarity diagram.
Using the Lagrangian in Eq.~\eqref{e:lpi}, the correlation function 
at lowest order reads
\begin{equation} 
\Delta_{(0)}(k,p)  = 
 - \frac{g_\sa^2}{4F_\pi^2}\, \frac{1}{(2\pi)^4}\,
 \frac{(\kslash + m)}{k^2 - m^2} \, \g_5 \, \pslash \, (\kslash - \pslash +m)\,
% \nn \\
%&& \mbox{} \times 
\pslash \, \g_5\frac{(\kslash + m)}{k^2 - m^2}\, 
 2\pi\,\delta\lf((k-p)^2 -m^2\rg) \,. 
\end{equation} 
This correlation function allows to compute the unpolarized fragmentation 
function $D_1$ by means of Eq.~\eqref{e:d1}, leading to
\begin{equation} 
 D_1(z,z^2 k^2_\st)  =  
 \frac{1}{z}\, \frac{g_\sa^2}{4 F_\pi^2}\,
 \frac{1}{16\pi^3}
%\\
%& & \mbox{} \times 
\bigg( 1 - 4\,\frac{1-z}{z^2}
 \,\frac{m^2\, m_\pi^2}{[k_\st^2 +m^2 +(1-z) m_\pi^2 / z^2]^2} \bigg). 
 \label{e:d1tree}
\end{equation} 
Note that the expression in Eq.~\eqref{e:d1tree} is only weakly dependent on the 
transverse momentum of the quark.
In fact, $D_1$ is constant as a function of $k_\st$, if $m_\pi = 0$ and (or) $m = 0$. 
Because our approach is limited to the soft regime, we will 
impose an upper cutoff on the $k_\st$ integration, as will be discussed in 
more detail in Sec.~\ref{s:three}.
This in turn leads to a finite $D_1(z)$ after integration over the 
transverse momentum.

The SU(2) flavor structure of our approach implies the relations
\begin{subequations}\label{e:flavor1flavor2} \begin{gather} 
D_{1}^{u\rightarrow \pi^0} = D_{1}^{\bar{u}\rightarrow \pi^0} 
 = D_{1}^{d\rightarrow \pi^0} = D_{1}^{\bar{d}\rightarrow \pi^0}  =  D_1 \,,
 \label{e:flavor1} \\
D_{1}^{u\rightarrow \pi^+} = D_{1}^{\bar{d}\rightarrow \pi^+} 
 = D_{1}^{\bar{u}\rightarrow \pi^-} = D_{1}^{d\rightarrow \pi^-}  =  2 \, D_1 \,,
 \label{e:flavor2}
\end{gather}
\end{subequations} 
where $D_1$ is the result given in Eq.~\eqref{e:d1tree}.
In the case of unfavored fragmentation processes $D_1$ vanishes at tree 
level, but will be nonzero as soon as one-loop corrections are included. 
According to the chiral power counting, one-loop contributions to $D_1$ are 
suppressed by a factor $l^2/\Lambda_\chi^2$ compared to the tree level result.
The maximum momentum up to which the chiral perturbation expansion 
converges numerically can be determined only by an explicit calculation of the 
one-loop corrections.

\begin{figure}
	\centering
        \begin{tabular}{c}
	\includegraphics[width=3cm]{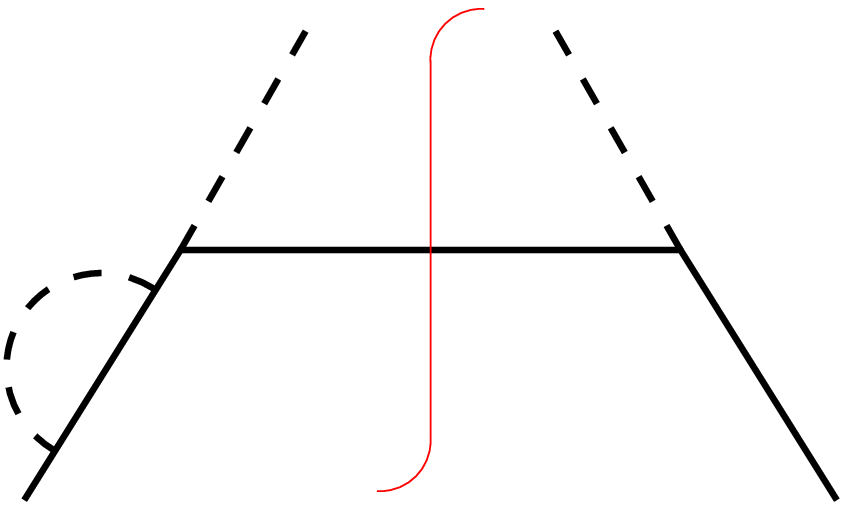}\hspace{1cm}
	\includegraphics[width=3cm]{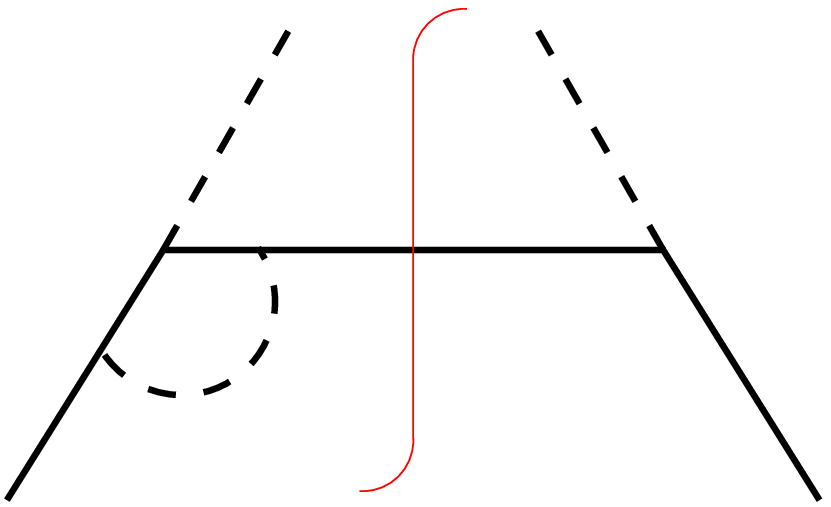}\\
	(a)\hspace{3.7cm}(b)\\ 
	\includegraphics[width=3cm]{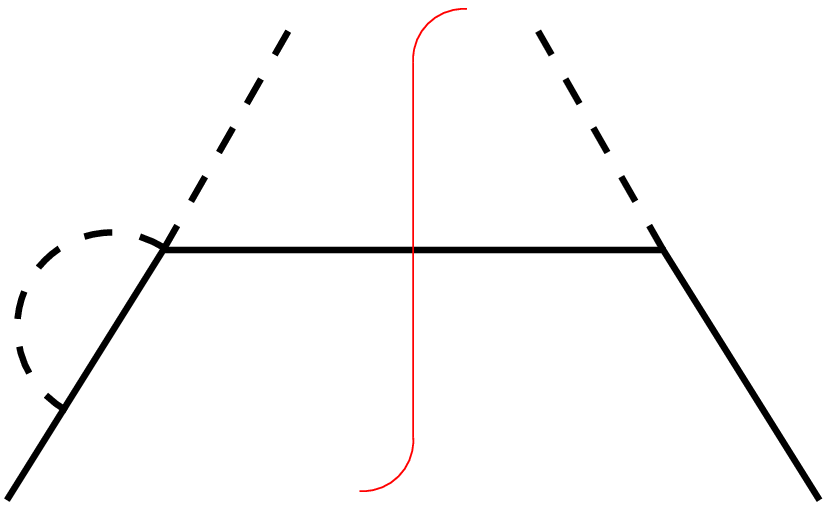}
	\rput(0.8,1){\hspace{1cm} + h.c.}\\
	(c)\\ 
	\end{tabular}
        \caption{One-loop corrections to the fragmentation of a quark
		 into a pion contributing to the Collins function.
                 The Hermitean conjugate diagrams (h.c.) are not shown
                 explicitly.}
        \label{f:1loop}
        \end{figure}

\begin{figure}
	\centering
	\includegraphics[width=9cm]{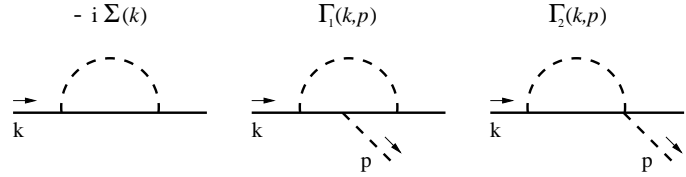}
        \caption{One-loop self-energy, and vertex corrections.}
        \label{f:sigmagammadelta}
        \end{figure}

As in the case of a pseudoscalar quark-pion 
coupling~\cite{Bacchetta:2001di}, the Collins function $H_1^\perp$ turns out to 
be zero in Born approximation.
To obtain a nonzero result, we have to resort to the one-loop level.
In Fig.~\ref{f:1loop} 
the corresponding diagrams are shown, where we have displayed only those 
graphs that contribute to the Collins function.
The explicit calculation of $H_1^\perp$ is similar to our previous 
work~\cite{Bacchetta:2001di}.
The relevant ingredients of the calculation are the self-energy and the vertex 
correction diagrams.
These ingredients are sketched in Fig.~\ref{f:sigmagammadelta} and can 
be expressed analytically as  
\begin{subequations} \begin{align} 
 - \ii \Sigma (k) & = \frac{g_\sa^2}{4 F_\pi^2}
 \int \frac{\de^4 l}{(2 \pi)^4} \, 
 \frac{\lslash \, (\kslash - \lslash - m) \, \lslash}
 {[(k-l)^2 - m^2]\,[l^2 -m_{\pi}^2]} \,, 
 \\
\Gamma_1 (k,p) &= - \ii \frac{g_\sa^3}{8 F_\pi^3}\, \g_5 
 \int \frac{\de^4 l}{(2\pi)^4}\,
% \nn \\ 
%&&\quad \mbox{} \times 
\frac{\lslash \, (\kslash - \pslash - \lslash + m)}
 {[(k - p - l)^2 - m^2]} \,\frac{\pslash \, (\kslash - \lslash - m) \, \lslash}
 {[(k - l)^2 - m^2][l^2 - m_{\pi}^2]} \,,
 \\
\Gamma_2 (k,p) &=  - \ii \frac{g_\sa}{8 F_\pi^3}\, \g_5 
 \int \frac{\de^4 l}{(2\pi)^4}\,
% \nn\\ 
%&&\quad \mbox{} \times 
\frac{(\lslash+\pslash)\,(\lslash-\kslash+m)\,\lslash}
{[(k-l)^2 - m^2)] [l^2-m_\pi^2]} \,,
\end{align} \end{subequations} 
where flavor factors have been suppressed.
For later purpose, we give here the most general parametrization
of the functions $\Sigma$, $\Gamma_1$ and $\Gamma_2$,
\begin{subequations} 
\label{e:siggamma2} \begin{align} 
\Sigma (k) & =  A\,\kslash + B\, m \,, 
 \vphantom{\frac{1}{1}} 
 \label{e:sig} \\
\Gamma_1 (k,p) & =  \frac{g_\sa}{2 F_\pi}\, \g_5 \,
 \Big( C_1 + D_1 \,\pslash + E_1 \,\kslash + F_1\,\pslash \, \kslash \Big) \,,
 \label{e:gamma1} \\
\Gamma_2 (k,p) & =  \frac{g_\sa}{2 F_\pi}\, \g_5\,
 \Big( C_2 + D_2 \,\pslash + E_2 \,\kslash + F_2\,\pslash \, \kslash \Big) \,.
 \label{e:gamma2}
\end{align} \end{subequations} 
The real parts of the functions $A$, $B$, $C_1$, $D_1$ etc. could be 
UV divergent and require in principle a proper renormalization.
Here, we do not need to deal with the question of renormalization at all, 
since only the imaginary parts of the loop diagrams are important 
when calculating the Collins function~\cite{Bacchetta:2001di}.

Taking now flavor factors properly into account, the contributions to the 
correlation function generated by the diagrams (a), (b) and (c) in 
Fig.~\vref{f:1loop} are given by
\begin{subequations} \begin{align}  
\Delta_{(1)}^{(a)}(k,p) &= 
 - 3\,\frac{g_\sa^2}{4 F_\pi^2}\, \frac{1}{(2 \pi)^4} \, 
 \frac{(\kslash + m)}{k^2 -m^2} \, \g_5 \,\pslash\, (\kslash - \pslash + m) \, 
 \nn \\
&\quad \times \pslash \, \g_5\, \frac{(\kslash + m)}{k^2 -m^2} \, 
 \Sigma(k) \, \frac{(\kslash + m)}{k^2 -m^2} \, 
% \nn \\
%& & \qquad \mbox{} \times 
2 \pi \, \delta \lf((k-p)^2 -m^2\rg) \,, 
 \vphantom{\frac{1}{1}} 
 \\
\Delta_{(1)}^{(b)}(k,p) &=  
 \frac{g_\sa}{2F_\pi^2}\, \frac{1}{(2\pi)^4}\,
 \frac{(\kslash + m)}{k^2 - m^2} \, \g_5 \, \pslash \, (\kslash - \pslash +m)
\nn \\
& \quad  \times \Gamma_1(k,p) \,\frac{(\kslash + m)}{k^2 - m^2}\, 
 2 \pi \, \delta \lf((k-p)^2 -m^2\rg) \,,
 \\ 
\Delta_{(1)}^{(c)}(k,p) &=
 - 2 \frac{g_\sa}{2F_\pi^2}\, \frac{1}{(2\pi)^4}\,
 \frac{(\kslash + m)}{k^2 - m^2} \, \g_5 \, \pslash \, (\kslash - \pslash +m)
\nn \\
& \quad  \times \Gamma_2(k,p)\, \frac{(\kslash + m)}{k^2 - m^2}\, 
 2 \pi \, \delta \lf((k-p)^2 -m^2\rg) \,.
\end{align} \end{subequations}  
The correlation functions of the Hermitean conjugate diagrams follow from the
Hermiticity condition
$\Delta_{(1)}^{H.c.}(k,p)=\gamma^0\Delta_{(1)}^{\dagger}(k,p)\,\gamma^0$.

Summing the contributions of all diagrams and inserting the resulting
correlation function in Eq.~\eqref{e:col1}, we eventually obtain the result
\begin{multline}
H_1^{\perp}(z,z^2 \mb{k}^2_{\st}) = 
 \frac{g_\sa^2}{32 \pi^3 F_\pi^2}\, \frac{m_\pi}{1-z}\, \frac{1}{k^2 - m^2} \,
% \nn \\
%&& \quad \mbox{} \times 
\bigg( - 3m \im \big( A+B \big)
% \nn \\
%& & \quad 
- \im \big( C_1  - m E_1 +(k^2  - m^2)F_1 \big) 
% \vphantom{\frac{1}{1}}
 \\
%& \quad 
+ 2 \im \big( C_2  - m E_2 +(k^2  - m^2)F_2 \big)  \bigg) \,
 \bigg|_{k^2 = \frac{z}{1-z}k_{\st}^2 + \frac{m^2}{1-z} + \frac{m_{\pi}^2}{z}} \,.
\label{e:h1}
\end{multline}  
Thus, the Collins function is entirely given by the imaginary parts of 
the coefficients defined in Eqs.~\eqref{e:siggamma2}.  
We can compute these imaginary parts by applying Cutkosky rules to the 
self-energy and vertex diagrams of Fig.~\vref{f:sigmagammadelta}. 
Explicit calculation leads to
\begin{subequations} \label{e:firstlast} \begin{align}  
\im \big( A+B \big) &= \frac{g_\sa^2}{32 \pi^2 F_\pi^2}\,
% \nn \\
%&& \quad \mbox{} \times 
\bigg( 2m_\pi^2 - \frac{k^2-m^2}{2} \,
 \Big(1 - \frac{m^2 - m_{\pi}^2}{k^2}\Big) \bigg)\, I_1 \,,
 \label{e:first} \\
\im \big( C_1 - mE_1 + (k^2-m^2)F_1 \big)& = \frac{g_\sa^2}{32 \pi^2 F_\pi^2} 
% \nn \\
%&& \quad \mbox{} \times 
\, m \, (k^2-m^2) \,
 \bigg( \frac{3k^2+m^2-m_\pi^2}{2 k^2}\, I_1 
 \\
&\quad + 4 m^2 \,\frac{k^2 -m^2 +m_{\pi}^2}{\lambda(k^2,m^2,m_{\pi}^2)}\, 
 \Big( I_1 + (k^2 - m^2 -2m_{\pi}^2)\, I_2 \Big) \bigg) \,,
 \nn \\
\im \big( C_2 - mE_2 + (k^2-m^2)F_2 \big)& = \frac{1}{32 \pi^2 F_\pi^2} 
% \nn \\
%&& \quad \mbox{} \times  
\,m \, (k^2-m^2) \,
 \bigg( 1 - \frac{m^2 - m_\pi^2}{k^2} \bigg)\, I_1 \,,
\end{align} \end{subequations}  
where we have introduced the so-called K\"allen function, 
$\lambda(k^2,m^2,m_{\pi}^2)=[k^2 -(m+m_{\pi})^2][k^2 -(m-m_{\pi})^2]$,
and the factors
\begin{subequations} \begin{align}  
I_1 & =  \int \de^4 l \; \delta \lf(l^2 - m_{\pi}^2\rg)\, \delta
\lf((k-l)^2-m^2\rg)
% \nn \\
% & 
=  \frac{\pi}{2 k^2}\, \sqrt{ \lambda(k^2,m^2,m_{\pi}^2)}
	\;\theta\lf(k^2 -(m+m_{\pi})^2\rg) \,, \\
I_2 & =  \int \de^4 l \; \frac{\delta\lf(l^2 - m_{\pi}^2\rg)\,\delta\lf((k -l)^2-m^2\rg)}
       {(k - p - l)^2 - m^2} \nn \\
 & =  -\frac{\pi}{2 \sqrt{\lambda(k^2,m^2,m_{\pi}^2)}}\,
	\ln{\left| 1+\frac{\lambda(k^2,m^2,m_{\pi}^2)}{k^2m^2 -(m^2-m_{\pi}^2)^2 }
	\right|} 
%\nn \\ 
%&&	\mbox{}\times 
	\;\theta\lf(k^2 -(m+m_{\pi})^2\rg) \,.
% \vphantom{\frac{1}{1}} 
 \label{e:last}
\end{align} \end{subequations}  
These integrals are finite and vanish below the threshold of quark-pion 
production, where the self-energy and vertex diagrams do not possess an 
imaginary part.

Thus, Eq.~\eqref{e:h1} in combination with 
Eqs.~\eqref{e:firstlast} gives the explicit result for the 
Collins function in the Manohar-Georgi model to lowest possible order. 
Because of its chiral-odd nature, the Collins function would vanish 
in this model if we set the mass of the quark to zero.
The same phenomenon has been observed in the calculation
of a chiral-odd twist-3 fragmentation function~\cite{Ji:1993qx}.
The result in Eq.~\eqref{e:h1} corresponds, e.g., to the fragmentation 
$u \to \pi^0$. 
The expressions for the remaining favored transitions are obtained in 
analogy to Eqs.~\eqref{e:flavor1flavor2}.
Unfavored fragmentation processes in the case of the Collins function 
appear only at the two-loop level.

%%%%%%%%%%%%%%%%%%%%%%%%%%%%%%%%%%%%%%%%%%%%%%%%%%%%%%%%%%%%%%%%%%%%%%%%%%%
\section{Estimates and phenomenology}
\label{s:three}

%%%%%%%%%%%%%%%%%%%%%%%%%%%%%%%%%%%%%%%%%%%%%%%%%%%%%%%%%%%%%%%%%%%%%%%%%%%
\subsection{Unpolarized fragmentation function and 
the choice of parameters}

We now present our numerical estimates, where all results for the fragmentation 
functions in this subsection refer to the transition $u \to \pi^+$.
To begin with we calculate the unpolarized fragmentation function $D_1(z)$, 
which is given by
\begin{equation} 
D_1(z)= \pi \int_0^{K^{2}_{\st\,{\rm max}}} \de \mb{K}^2_{\st}\; D_1(z,\mb{K}^2_{\st}),
\end{equation} 
where $\mb{K}_{\st}=- z \mb{k}_{\st}$ denotes the transverse momentum of the 
outgoing hadron with respect to the quark direction. 
The upper limit on the $\mb{K}^2_{\st}$ integration is set by the cutoff on
the fragmenting quark virtuality, $\mu^2$, and corresponds to
\begin{equation} 
K^{2}_{\st\,{\rm max}}=z \, (1-z)\,\mu^2 -z\,m^2-(1-z)\,m_\pi^2 \,.
\end{equation} 
In addition to $m$ and $g_\sa$, the cutoff $\mu^2$ is the third parameter of
our approach that is not fixed {\em a priori}.
However, as will be explained below, the possible values of $\mu^2$ can be
restricted when comparing our results to experimental data. 
Unless otherwise specified, we always use the values
\begin{equation} 
 m = 0.3 \; {\rm GeV}, \qquad g_{\sa}=1, \qquad \mu^2=1\; {\rm GeV}^2 \,.
 \label{e:param1}
\end{equation} 
At the relevant places, the dependence of our results on possible variations 
of these parameters will be discussed.
A few remarks concerning the choice in Eq.~\eqref{e:param1} are in order.
The value of $m$ is a typical mass of a constituent quark.
The choice for the axial coupling can be seen as a kind of average number
of what has been proposed in the literature.
For instance, in a simple SU(6) spin-flavor model for the proton one finds 
$g_\sa \approx 0.75$ in order to obtain the correct value for the axial charge 
of the nucleon~\cite{Manohar:1984md}.
On the other hand, large $N_c$ arguments favor a value of the order 
of 1~\cite{Weinberg:1990xn}, while, according to a recent calculation 
in a relativistic point-form approach~\cite{Boffi:2001zb}, a $g_\sa$ slightly
above 1 seems to be required for describing the axial charge of the nucleon.
Finally, our choice for $\mu^2$ ensures that the momenta of the outgoing pion
and quark, in the rest frame of the fragmenting quark, remain below values
of the order $0.5 \, \rm {GeV}$.
In this region we believe chiral perturbation theory to be applicable, meaning
 that our leading order result can provide a reliable estimate.

\begin{figure}
        \centering
	\rput(-0.5,3.1){\rotatebox{90}
	{{\boldmath $D_1$}} }
	\rput(4.2,-0.3){\boldmath $z$}
	\includegraphics[width=8cm]{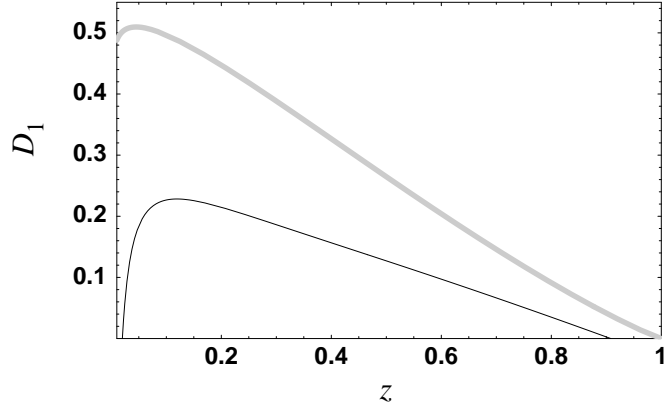}
	\mbox{}\\[3mm]
        \caption{Model result for the unpolarized quark fragmentation 
	function $D_1^{u \rightarrow \pi^+}$ (solid line) and comparison with
	the parametrization of Ref.~{\protect \cite{Kretzer:2000yf}} 
	(gray line).}
	\label{f:d1}
        \end{figure}

In Fig.~\vref{f:d1} we show the result for the unpolarized fragmentation
function $D_1^{u \rightarrow \pi^+}$. 
Notice that in general the fragmentation functions vanish outside the 
kinematical limits, which in our model are given by
\begin{equation} 
z_{{\rm max},{\rm min}}=
	\frac{1}{2}\lf[\left(1-\frac{m^2-m_{\pi}^2}{\mu^2}\right) 
%\nn \\
%	&&\qquad
\pm \sqrt{\left(1-\frac{m^2-m_{\pi}^2}{\mu^2}\right)^2 -4\,
	\frac{m_{\pi}^2}{\mu^2}}\rg],
\end{equation} 
corresponding to the situation when the upper limit of the $K_{\st}^2$ 
integration becomes equal to zero.
We consider our tree level result as a pure valence-type part of 
$D_1^{u \rightarrow \pi^+}$.
The sea-type (unfavored) transition $\bar{u} \to \pi^+$ is strictly zero 
at leading order.
Therefore, we compare the model result to the valence-type quantity 
$D_1^{u\rightarrow \pi^+}-D_1^{\bar{u}\rightarrow \pi^+}$, where the
fragmentation functions have been taken from the parametrization of Kretzer\footnote{Other 
parametrizations~\protect{\cite{Kniehl:2000fe,Bourhis:2000gs,Kretzer:2001pz}} 
use
a starting energy scale $Q^2 \ge 2\, \rm{GeV}^2$, which is too high to allow a
comparison with our results. Moreover, the valence parts of these other
parametrizations would have to be obtained with some extra assumptions and 
seems 
too arbitrary.}
\cite{Kretzer:2000yf} at a scale $Q^2=1 \, \rm{GeV}^2$.
Obviously, the $z$ dependence of both curves is in nice agreement,
which is a nontrivial result.
For example, in the pseudoscalar model that we used in our previous 
work~\cite{Bacchetta:2001di}, $D_1$ behaves quite differently and peaks 
at an intermediate $z$ value. 

On the other hand, we underestimate the parametrization of 
Ref.~\citen{Kretzer:2000yf} by about a factor of 2. 
Some remarks are in order at this point.
Although a part of the discrepancy might be attributed to the uncertainty 
in the value of
$g_\sa$, the most important point is to address the question as to what extent
we can compare our estimate with existing parametrizations.
The parametrization of~\cite{Kretzer:2000yf} serves basically as input
function of the perturbative 
QCD evolution equations, used to describe high-energy
$e^+ e^-$ data, and displays the typical logarithmic dependence on the scale
$Q^2$. A value of $Q^2 = 1$ GeV$^2$ is believed to be already beyond the
limit of applicability of perturbative QCD calculations.
On the other hand, our approach displays, to a first approximation, a linear
dependence on the cutoff $\mu^2$. It is supposed to be valid at low scales 
and it is also stretched to the limit of its applicability for
$\mu^2 = 1$ GeV$^2$.
In this context it should also be investigated to what extent the inclusion
of one-loop corrections, which allow for the additional decay channel
$q^{\ast} \to \pi \pi q$, will increase the result for $D_1$ at 
$\mu^2 = 1 \, \rm{GeV}^2$.
Finally, we want to remark that
to our knowledge there exists no strict one-to-one correspondence
between the quark virtuality $\mu^2$ and the scale used in the evolution 
equation of fragmentation functions, which in semi-inclusive DIS
is typically identified with the photon virtuality $Q^2$.
For all these reasons, 
a smooth matching of our calculation and the parametrization 
of~\cite{Kretzer:2000yf} cannot necessarily be expected.
Despite these caveats, the correct $z$ behavior displayed by our result 
for $D_1$ suggests that 
the calculation can well be used as an
input for evolution equations at a low scale. 
In the next subsection we will elaborate more on this point in connection with
the Collins function. 

	\begin{figure}
        \centering
	\rput(-0.5,3.1){\rotatebox{90}
	{{\boldmath $\langle | K_{T} | \rangle$}{\small(GeV)}} }
	\rput(4.2,-0.3){\boldmath $z$}
        \includegraphics[width=8cm]{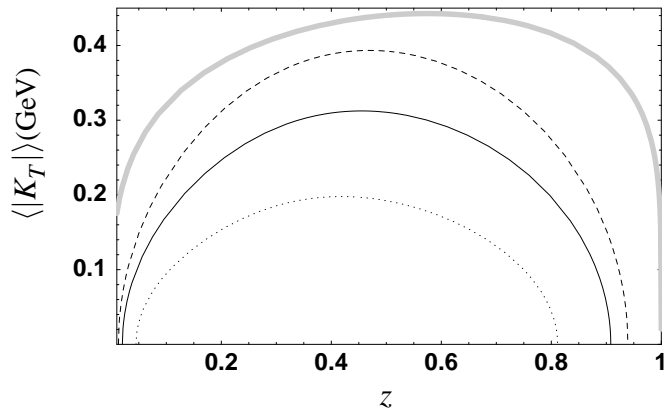}
	\mbox{}\\[3mm]
        \caption{Model result for the average hadron transverse momentum 
	for different choices of the cutoff:  $\mu^2=0.5$ GeV$^2$ 
	(dotted line),
	$\mu^2=1$ GeV$^2$ (solid line), $\mu^2=1.5$ GeV$^2$ (dashed line)
	and comparison with a fit to experimental results from 
        DELPHI~{\protect \cite{Abreu:1996na}} (gray line).} 
	\label{f:trans}
        \end{figure}

The best indication of the appropriate value of the cutoff $\mu^2$ may be 
obtained when
comparing our calculation to experimental data of the average transverse 
momentum of the outgoing hadron with respect to the quark, which we evaluate
according to
\begin{equation} 
\langle |\mb{K}_{\st}|(z)\rangle = \frac{\pi}{D_1(z)} 
	\int_0^{K^{2}_{\st\,{\rm max}}}
	\de \mb{K}^2_{\st}\,|\mb{K}_{\st}|\, D_1(z, \mb{K}^2_{\st}) \,.
\end{equation} 
In Fig.~\vref{f:trans} we show the result of this observable as a function of 
$z$ for three different choices of the parameter $\mu^2$. 
As a comparison, we also show a fit (taken from Ref.~\citen{Anselmino:1999pw}) to 
experimental data obtained by the DELPHI Collaboration~\cite{Abreu:1996na}. 
As in the case of $D_1(z)$, the shape of our result is very similar to the 
experimental one, which we consider as an encouraging result.
For $\mu^2 = 1\, \rm{GeV}^2$ our curve is about $30\%$ below the data.
Such a disagreement is not surprising, keeping in mind that at LEP energies
higher order perturbative QCD effects (e.g. gluon bremsstrahlung, unfavored
fragmentations, etc.) 
play an important role, leading in general to a broadening of the
$\mb{K}_\st$ distribution.
For experiments at lower energies, however, where perturbative QCD contributions 
can be neglected in a first approximation, it may be possible to exhaust 
the experimental value for $\langle |\mb{K}_{\st}| (z)\rangle$ 
with genuine soft 
contributions as described in our model.
This in turn would determine the appropriate value of the cutoff $\mu^2$.
For example, such a method of matching our calculation
with experimental conditions could be applied at HERMES kinematics, 
even though the method is somewhat
hampered since $\mb{K}_\st$ is not directly measured in semi-inclusive DIS.
In this case, one rather observes the transverse momentum of the outgoing hadron 
with respect to the virtual photon, $\mb{P}_{h \perp}$, which 
depends on both $\mb{K}_\st$ and the transverse momentum of the 
partons inside the target $\mb{p}_\st$, as shown in Eq.~\eqref{e:phperpavg}.
%At leading order in the hard scattering cross section one 
%\begin{equation} \begin{split} 
%\langle P^2_{h\perp}\rangle(x,z)& = z^2 \frac{\pi \int
%	\de \mb{p}^2_{\st}\, p_{\st}^2\, f_1(x, \mb{p}^2_{\st})}{f_1(x)}+\frac{\pi \int
%	\de \mb{K}^2_{\st}\, K_{\st}^2\, D_1(z, \mb{K}^2_{\st})}{D_1(z)} \\
%	&= z^2\,\langle \mb{p}^2_{\st}\rangle (x) + \langle \mb{K}^2_{\st}\rangle(z) \,,
%\end{split} \end{equation}  
%where $x$ represents the Bjorken variable.

%%%%%%%%%%%%%%%%%%%%%%%%%%%%%%%%%%%%%%%%%%%%%%%%%%%%%%%%%%%%%%%%%%%%%%%%%%%
\subsection{Collins function}

\begin{figure}
        \centering
	\rput(-0.5,3.1){\rotatebox{90}
	{{\boldmath $H_1^{\perp}$}} }
	\rput(4.4,-0.3){\boldmath $z$}  
        \includegraphics[width=8cm]{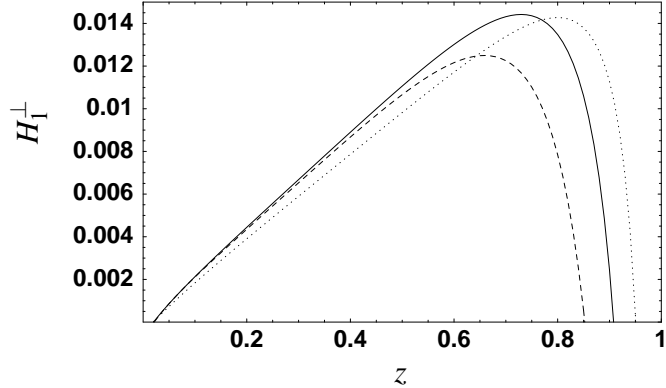}
       	\mbox{}\\[3mm]
	\caption{Model result for the Collins function for different values
	of the constituent quark mass: $m=0.2$ GeV (dotted line),
	$m=0.3$ GeV (solid line), $m=0.4$ GeV (dashed line).} 
	\label{f:collwithm}
        \end{figure}

\begin{figure}
        \centering
	\rput(-0.5,3.1){\rotatebox{90}
	{{\boldmath $H_1^{\perp} / D_1$ }} }
	\rput(4.2,-0.3){\boldmath $z$}
        \includegraphics[width=8cm]{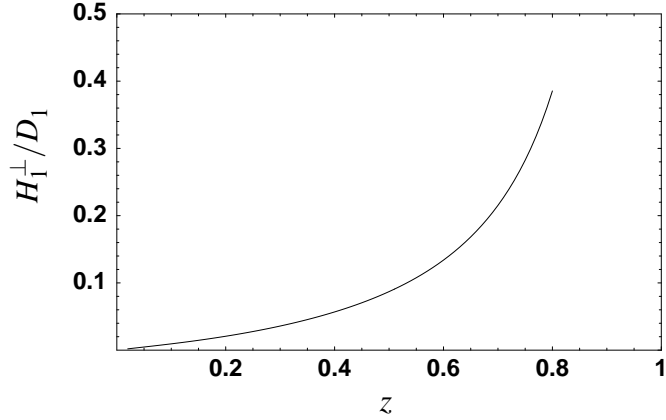}
	\mbox{}\\[3mm]
        \caption{Model result for $H_1^{\perp}/D_1$.}
	\label{f:ratio}
        \end{figure}

We now turn to the description of our model result for the Collins function.
In Fig.~\vref{f:collwithm}, $H_1^\perp$ is plotted for three different values of
the constituent quark mass, $m= 0.2, \, 0.3, \, 0.4$ GeV\@. 
In a large $z$ range, the function does not depend strongly on the precise value 
of the quark mass, if we choose it within reasonable limits. 
That is why we can confidently fix $m = 0.3 \, \rm{GeV}$ for our numerical studies. 
It is very interesting to observe that the behavior of the unpolarized
fragmentation function $D_1$ is quite distinct from that of 
the Collins function:
while the former is decreasing as $z$ increases, the latter is growing.
The different behavior of the two functions becomes even more evident
when looking at their ratio, shown in Fig.~\ref{f:ratio}. At present, there
exists some evidence of $z$ behavior of the Collins function and it is in
agreement with our results. We briefly discuss this subject in Sec.~\vref{s:comparison}.

The ratios of the Collins function or any of its moments 
with $D_1$ are almost independent of the coupling constant $g_{\sa}$. 
The reason is that the one-loop correction containing the 
contact interaction is only proportional to $g_{\sa}^2$, as $D_1$ is, 
and is dominating over the 
others.
Furthermore, the ratio $H_1^{\perp}/D_1$ is nearly independent of the
cutoff $\mu^2$.
In conclusion, the prediction shown in Fig.~\vref{f:ratio} is almost
independent of the choice of parameters in our approach.

At this point we would like to add some general remarks concerning the
$z$ behavior of our results.
It turns out that the shape of all the results does not vary much when
changing the parameters within reasonable limits.
In particular, variations of $g_\sa$ and of the cutoff $\mu^2$ only change
the normalization of the curves but not their shape.
In this sense our calculation of fragmentation functions has a good
predictive power for the $z$ behavior of the function.
This has a direct practical consequence if one uses, for instance, our
result of the Collins function as input in an evolution equation:
the $z$ dependence of the input function can be adjusted to the shape of 
our $H_1^\perp$, while its normalization can be kept free in order to account
for uncertainties in the values of $g_\sa$ and $\mu^2$.

\begin{figure}
        \centering
	\rput(-0.5,3.1){\rotatebox{90}
	{{\boldmath $H_1^{\perp (1/2)} / D_1$ }} }
	\rput(4.2,-0.3){\boldmath $z$}
        \includegraphics[width=8cm]{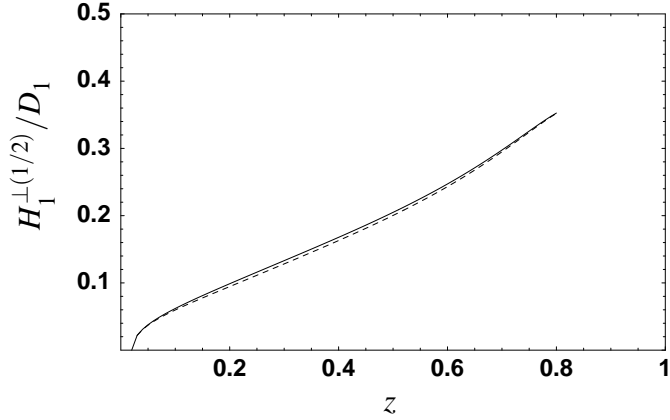}
	\mbox{}\\[3mm]
        \caption{Model result for $H_1^{\perp (1/2)}/D_1$
	(solid line) and comparison with the product
	$(\langle |\mb{K}_{\st}|\rangle/2 z m_{\pi})\, (H_1^{\perp}/
	 D_1)$ (dashed line). 
	Note that the positivity bound requires the ratio to
	be smaller than $0.5$.} 
	\label{f:ratiomom0}
        \end{figure}

In Fig.~\vref{f:ratiomom0} we plot the ratio
\begin{equation}	
 \frac{H_1^{\perp
(1/2)}(z)}{D_1(z)} \equiv
\frac{\pi}{D_1(z)} \int \de K_{\st}^2\, \frac{|\mb{K}_{\st}|}{2 z m_{\pi}}\,
H_1^{\perp}(z,K_{\st}^2) \,,
\label{e:ratiomom0}
\end{equation} 
which enters the transverse single spin asymmetry of Eq.~\eqref{e:asym1}.
This quantity rises roughly linearly within a large $z$ range, leading to a 
similar $z$ behavior of the transverse spin asymmetry.
$H_1^{\perp(1/2)}/D_1$ is no longer independent of the cutoff $\mu^2$, but rather 
the same dependence as in the case of $\langle |\mb{K}_{\st}|\rangle$ 
(Fig.~\vref{f:trans}) can be assumed.
In Fig.~\vref{f:ratiomom0}, this ratio  is compared to the expression
\begin{equation}	
 \frac{\langle |\mb{K}_{\st}|(z)\rangle}{2 z m_{\pi}}\,
	 \frac{H_1^{\perp}(z)}{D_1(z)} = 
\pi \,\frac{H_1^{\perp}(z)}{D_1^2(z)} \int \de K_{\st}^2\, \frac{|\mb{K}_{\st}|}
{2 z m_{\pi}}\, 
D_1(z,K_{\st}^2) \,.
\end{equation}
A very close agreement between the two different curves can be observed,
indicating that the model predicts a
quite similar transverse momentum dependence of both the Collins function and $D_1$.
In the literature, this feature is sometimes assumed in phenomenological 
parametrizations of $H_1^\perp$.
Note, however, that in our approach deviations from this simple behavior can be 
expected, if $D_1$ is also calculated consistently to the one-loop 
order.
	
The Collins function has to fulfill the positivity 
bound of Eq.~\eqref{e:boundcollins}, which can be rewritten as
\begin{equation}
\frac{|\mb{K}_{\st}|}{2 z m_{\pi}}\,
H_1^{\perp}(z,K_{\st}^2) \leq \frac{1}{2}\,D_1(z,K_{\st}^2).
\end{equation} 
Integration over $K_{\st}^2$ gives the simplified expression
\begin{equation}
\frac{H_1^{\perp (1/2)}(z)}{D_1(z)}\leq \frac{1}{2} \, ,
\label{e:boundint}
\end{equation}	
which is satisfied by our model calculation. 
It is clear, however, that increasing the value of $\mu^2$ will eventually 
result in a violation of the positivity condition.
To avoid such a violation, we should calculate $D_1$ and $H_1^{\perp}$ 
consistently at the same order, i.e., the one-loop corrections to $D_1$ 
should be included.
By doing so, the positivity bound will be fulfilled even at larger values 
of $\mu^2$, for which our numerical results are no longer trustworthy.

\begin{figure}
        \centering
	\rput(-0.5,3.1){\rotatebox{90}
	{{\boldmath $H_1^{\perp (1/2)} / D_1$ }} }
	\rput(4.2,-0.3){\boldmath $z$}
        \includegraphics[width=8cm]{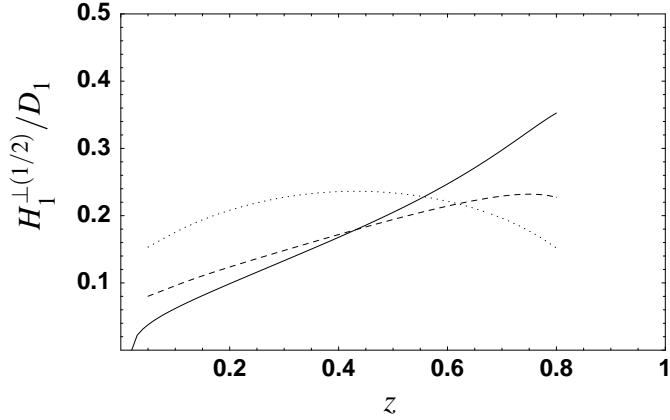}
	\mbox{}\\[3mm]
        \caption{Model result for $H_1^{\perp (1/2)}/D_1$
	(solid line) and comparison with the same ratio, where 
        $H_1^{\perp(1/2)}$ is calculated according to 
        Eq.~\eqref{e:collansatz} 
	with $M_C=0.3$ GeV (dashed line) and $M_C=0.7$ GeV (dotted line).}
	\label{f:collansatz}
        \end{figure}

From our results, we expect an increasing
behavior of the azimuthal asymmetry in $p^\uparrow p\to\pi\,X$ as
function of $x_F$, qualitatively similar to what has been predicted 
in Ref.~\citen{Artru:1997bh} in the context of the Lund fragmentation model. 
At this point, it is also interesting to discuss the comparison of our results 
with the ones obtained using the so-called ``Collins guess''. 
On the basis of very general assumptions, Collins suggested a possible
behavior for the transverse spin asymmetry containing 
$H_1^\perp$~\cite{Collins:1993kk}. 
This suggestion has been used in the literature 
(see, e.g.,
Refs.~\citen{Oganessyan:1998ma,Kotzinian:1999dy,DeSanctis:2000fh,Ma:2001ie})  
to propose the following shape for the Collins function:
\begin{equation}
 H_1^{\perp(1/2)}(z) \approx \pi \int \de K_{\st}^2 \, 
\frac{|\mb{K}_\st|}{2 z} \,
\frac{M_C} {M_C^2 + {K_{\st}^2/z^2}}\, D_1(z,K_{\st}^2),
\label{e:collansatz}
\end{equation}	
with the parameter $M_C$ ranging between 0.3 and $0.7 \,
\rm{GeV}$.\footnote{Note that even this particular form does not 
correspond precisely to what proposed in
Ref.~\citen{Collins:1993kk}, even if it is often referred
to as ``Collins ansatz''.} 
Using our model outcome for the unpolarized fragmentation function, we apply
Eq.~\eqref{e:collansatz} to estimate  $H_1^{\perp(1/2)}$, and in
Fig.~\ref{f:collansatz} we show how this compares to Eq.~\eqref{e:ratiomom0}.  
There is a rough qualitative agreement with the Collins ansatz for the lowest 
value of the parameter $M_C$, although it is not growing fast enough 
compared to Eq.~\eqref{e:ratiomom0}. 
On the other hand, in the Manohar-Georgi model there is no agreement with 
the Collins ansatz  for high values of the parameter $M_C$, which might indicate
that the relation suggested in Eq.~\eqref{e:collansatz} should be 
handled with care.

\begin{figure}
        \centering
	\rput(-0.5,3.1){\rotatebox{90}
	{{\boldmath $H_1^{\perp (1)} / D_1$ }} }
	\rput(4.2,-0.3){\boldmath $z$}
        \includegraphics[width=8cm]{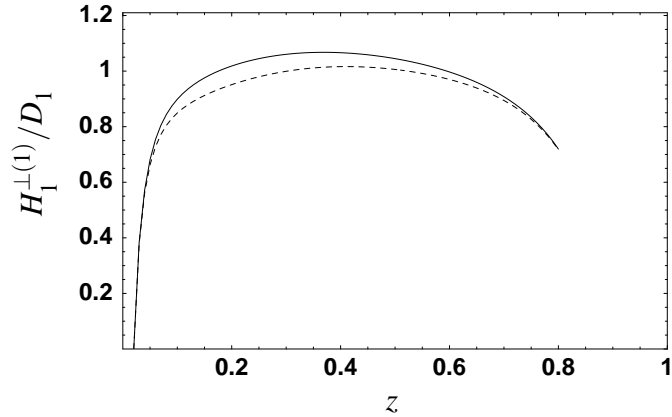}
	\mbox{}\\[3mm]
        \caption{Model result $H_1^{\perp (1)}/D_1$
	(solid line) and comparison with the product
	$(\langle K_{\st}^2\rangle/2 z^2 m_{\pi}^2) (H_1^{\perp}/D_1^{\perp})$
	(dashed line).} 
	\label{f:ratiomom}
        \end{figure}

Finally, in Fig.~\vref{f:ratiomom} we display the quantity
\begin{equation}	
 \frac{H_1^{\perp(1)}(z)}{D_1(z)} \equiv
\frac{\pi}{D_1(z)} \int \de K_{\st}^2\, \frac{K_{\st}^2}{2 z^2 m_{\pi}^2}\,
H_1^{\perp}(z,K_{\st}^2) \,,
\end{equation} 
because this ratio appears in the weighted asymmetries of Eq.~\eqref{e:asym2}.
In Fig.~\vref{f:ratiomom}, the expression
\begin{equation}	
 \frac{\langle K_{\st}^2\rangle (z)}{2 z^2 m_{\pi}^2}
	 \frac{H_1^{\perp}(z)}{D_1(z)} = 
\pi \,\frac{H_1^{\perp}(z)}{D_1^2(z)} \int \de K_{\st}^2 \frac{K_{\st}^2}
{2 z^2 m_{\pi}^2} 
D_1(z,K_{\st}^2) \,
\end{equation}
is also shown for comparison.
Once again, there is a remarkable agreement between the two different
expressions, confirming the quite similar $K_\st$ behavior of $H_1^\perp$
and $D_1$.

%%%%%%%%%%%%%%%%%%%%%%%%%%%%%%%%%%%%%%%%%%%%%%%%%%%%%%%%%%%%%%%%%%%%%%%%%%%
\subsection{Asymmetries in semi-inclusive DIS and $e^+e^-$ annihilation}

We turn now to estimates of possible observables containing the Collins
function. We will take into consideration one-particle inclusive DIS,
 where the Collins function appears in connection with the transversity
distribution of the nucleon, 
and  $e^+e^-$ annihilation into two hadrons belonging to two different jets.

In the first case, 
we show predictions for both
transverse spin asymmetries defined in
Eqs.~\eqref{e:asym1} and \eqref{e:asym2}. Different calculations can be found
in the literature, e.g., 
in Refs.~\citen{Anselmino:2000mb,Korotkov:1999jx,Ma:2001ie}.
To estimate the magnitude of the asymmetries, 
we need inputs for the
distribution functions, in particular for the transversity distribution. 
Several model calculations of this function
are available at present (see \cite{Barone:2001sp} for a comprehensive
review). We refrain from considering many
different examples and rather restrict the analysis to two 
limiting situations. In the first case we adopt the
``nonrelativistic'' assumption $h_1 = g_1$, while in the second case we 
exhaust the upper bound on the transversity distribution, i.e., 
$h_1 = \frac{1}{2}(f_1 + g_1)$~\cite{Soffer:1995ww}. 
We use the simple parametrization
of $g_1$ and $f_1$ suggested in Ref.~\citen{Brodsky:1995kg}.
 At the moment, more sophisticated
parametrizations are available, taking scale
evolution into account also. 
However, all these parametrizations are compatible with each other to the 
extent of our purpose here, which is to give an estimate of the asymmetries
for a low scale. 
We focus on the production of $\pi^+$, where the contribution of
down quarks is negligible, not only because of the presence of unfavored
fragmentation functions, but also because the transversity distribution 
for down
quarks appears to be much smaller than for up quarks in model calculations.

\begin{figure}
        \centering
	\rput(-0.5,3.1){\rotatebox{90}
	{{\boldmath $\langle \sin{\phi} \rangle_{UT}$ }} }
	\rput(4.2,-0.3){\boldmath $\xbj$}
        \includegraphics[width=8cm]{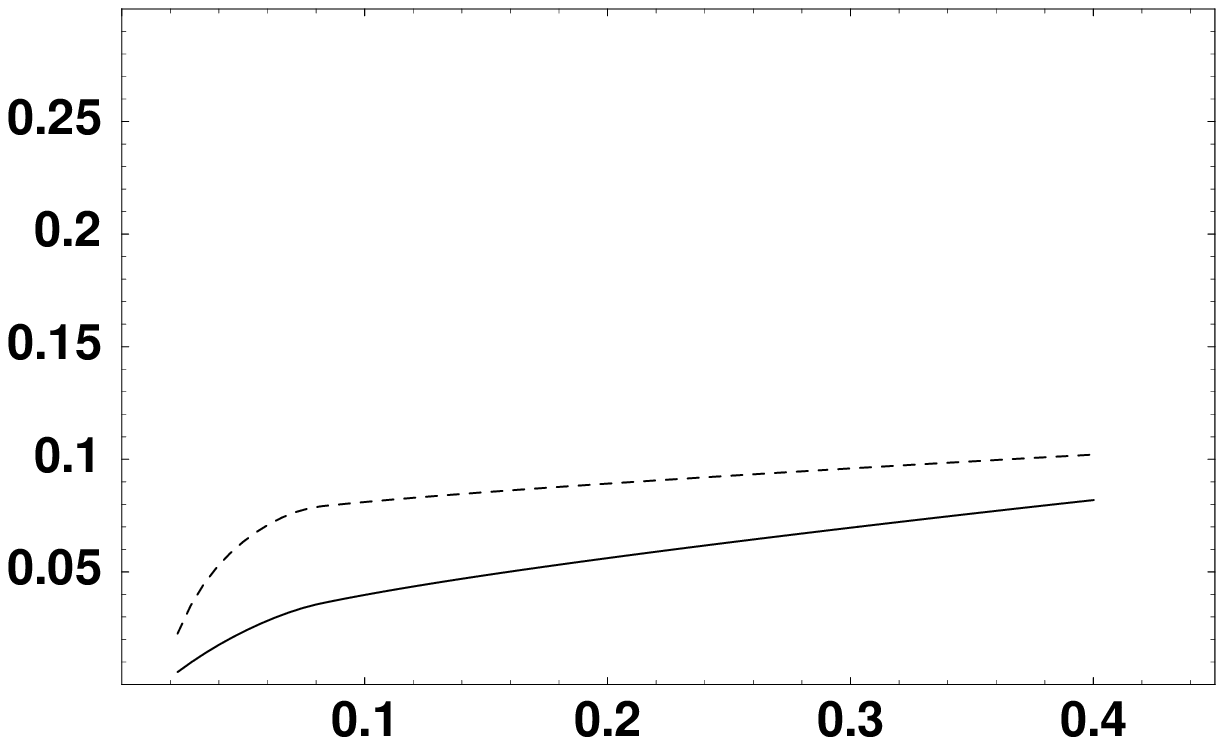}
	\mbox{}\\[3mm]
        \caption{Azimuthal transverse spin asymmetry 
	$\langle\sin{\phi}\rangle_{UT}$ as a function of $\xbj$.
	Solid line: assuming $h_1 = g_1$. Dashed line: assuming
	$h_1 =\frac{1}{2} (f_1+ g_1)$. The functions $f_1$ and
	$g_1$ are taken from {\protect\cite{Brodsky:1995kg}}.}
	\label{f:axnw}
        \end{figure}

\begin{figure}
        \centering
	\rput(-0.5,3.1){\rotatebox{90}
	{{\boldmath $\langle \sin{\phi} \rangle_{UT}$ }} }
	\rput(4.2,-0.3){\boldmath $z_h$}
        \includegraphics[width=8cm]{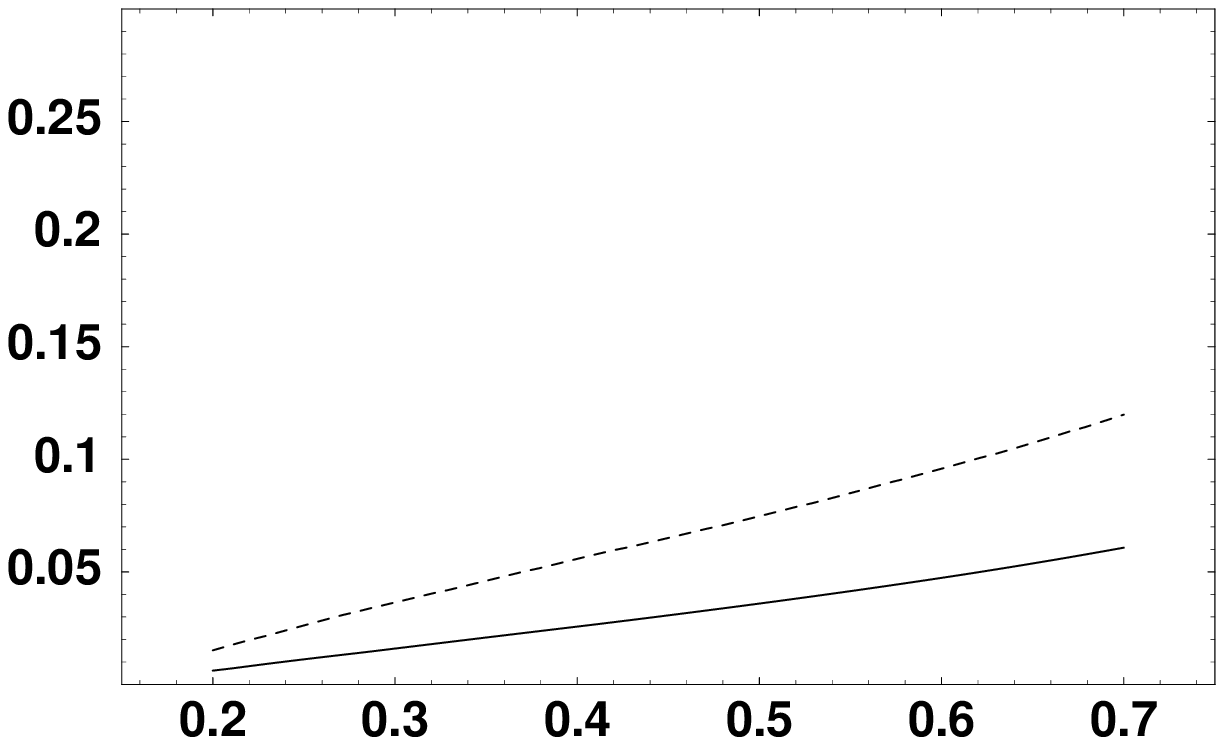}
	\mbox{}\\[3mm]
        \caption{Azimuthal transverse spin asymmetry 
	$\langle\sin{\phi}\rangle_{UT}$ as a function of $z_h$.
	Solid line: assuming $h_1 = g_1$. Dashed line: assuming
	$h_1 =\frac{1}{2} (f_1+ g_1)$. The functions $f_1$ and
	$g_1$ are taken from {\protect\cite{Brodsky:1995kg}}.} 
	\label{f:aznw}
        \end{figure}

In Fig.~\ref{f:axnw} we present the azimuthal asymmetry
defined in Eq.~\eqref{e:asym1} as a function of $\xbj$, 
after integrating numerator and denominator over
the variables $y$ and $z_h$, for the two cases described above. 
In Fig.~\ref{f:aznw}, we present the same asymmetry as a function of $z_h$,
after integrating over $y$ and $\xbj$. 
As already mentioned before, our prediction is supposed to be valid at a low
energy scale of about $1\, \rm{GeV}^2$. Neglecting evolution effects, 
it could be utilized for comparison with
experiments at a scale of a few GeV$^2$.
We assume the value of the transverse
polarization to be $|\mb{S}_T| = 0.75$.
In performing the integrations, we apply
the kinematical cuts typical of the HERMES experiment, as described
in~\cite{Airapetian:1999tv}. Therefore, our prediction is particularly
significant for HERMES, which is expected to be the first experiment 
to measure this asymmetry.
In principle, the simultaneous study of the $\xbj$ and $z_h$ dependence of the
asymmetry yields separate information on the distribution and fragmentation
parts and allows one to
extract both up to a normalization factor~\cite{Korotkov:1999jx}. Note,
however, that this procedure relies on the assumption of up-quark dominance 
and is valid only if the $\xbj$ dependence of the asymmetry can be ascribed
entirely to the distribution functions and the $z_h$ dependence entirely to the
fragmentation functions. Kinematical cuts could partially spoil this
situation.  
We would like to stress that our calculation predicts an asymmetry up to the
order of 10\%, which
should be within experimental reach, and suggests the possibility of
distinguishing between different assumptions on the transversity distribution.

\begin{figure}
        \centering
	\rput(-0.5,3.1){\rotatebox{90}
{{\boldmath $\langle\frac{|P_{h \perp}|}{m_{\pi}}\sin{\phi}\rangle_{UT}$}} }
	\rput(4.2,-0.3){\boldmath $\xbj$}
        \includegraphics[width=8cm]{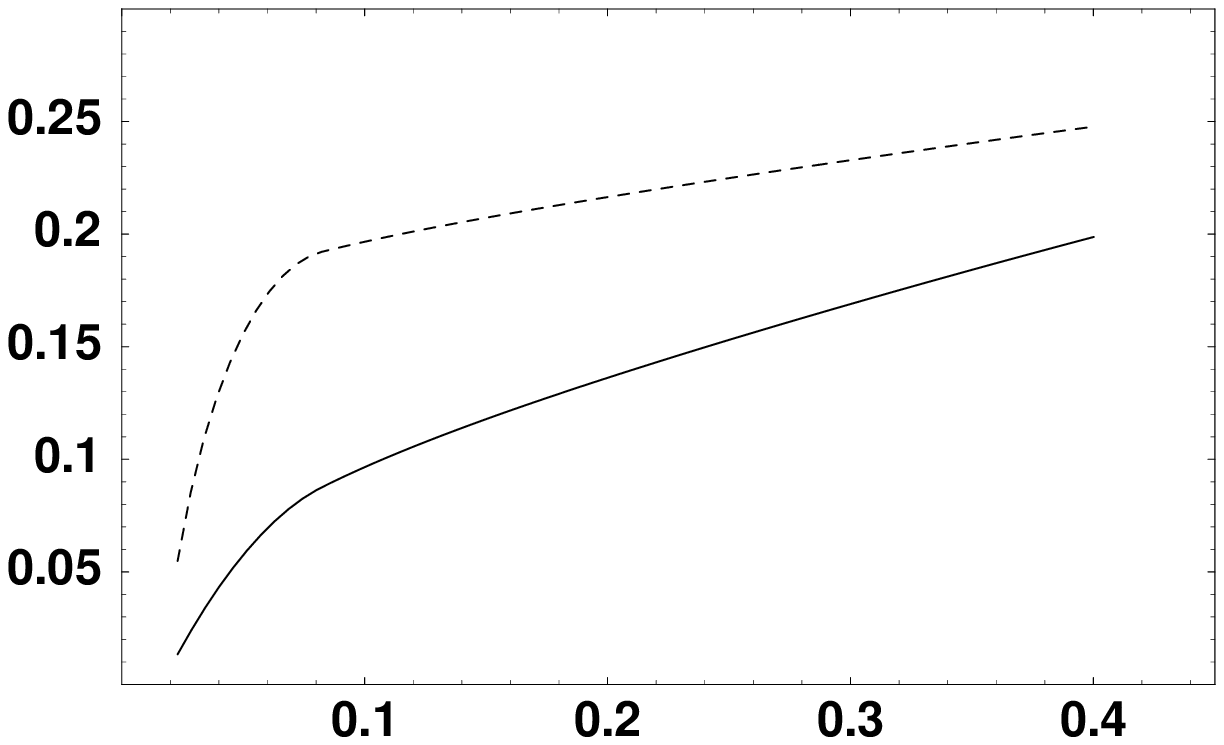}
	\mbox{}\\[3mm]
        \caption{Azimuthal spin asymmetry 
	$\langle |\mb{P}_{h \perp}|/m_{\pi}\,\sin{\phi}\rangle_{UT}$ 
	as a function of $\xbj$.
	Solid line: assuming $h_1 = g_1$. Dashed line: assuming
	$h_1 =\frac{1}{2} (f_1+ g_1)$. The functions $f_1$ and
	$g_1$ are taken from {\protect\cite{Brodsky:1995kg}}.}         
	\label{f:axw}
        \end{figure}

\begin{figure}
        \centering
	\rput(-0.5,3.1){\rotatebox{90}
{{\boldmath $\langle\frac{|P_{h \perp}|}{m_{\pi}}\sin{\phi}\rangle_{UT}$}} }
	\rput(4.2,-0.3){\boldmath $z_h$}
        \includegraphics[width=8cm]{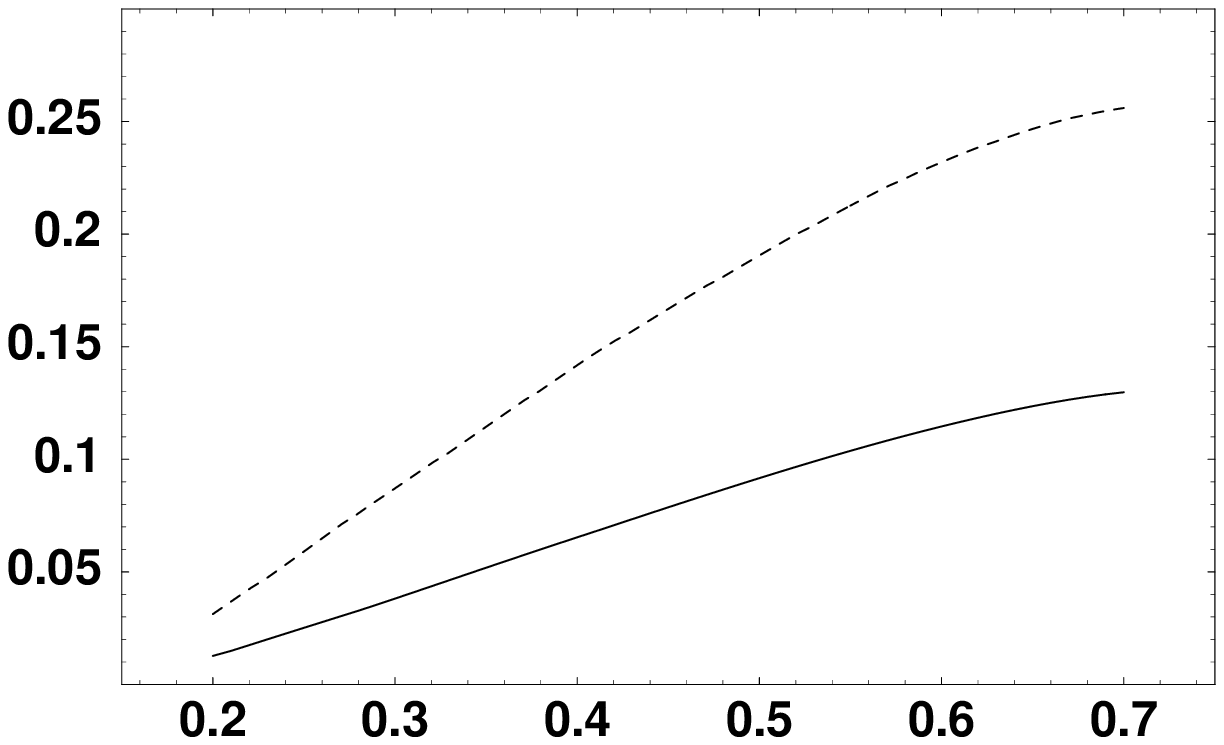}
	\mbox{}\\[3mm]
        \caption{Azimuthal spin asymmetry 
	$\langle |\mb{P}_{h \perp}|/m_{\pi}\,\sin{\phi}\rangle_{UT}$
	as a function of $z_h$.
	Solid line: assuming $h_1 = g_1$. Dashed line: assuming
	$h_1 =\frac{1}{2} (f_1+ g_1)$. The functions $f_1$ and
	$g_1$ are taken from {\protect\cite{Brodsky:1995kg}}.} 
	\label{f:azw}
        \end{figure}

Using the same procedure as before, we have estimated the asymmetry defined in
Eq.~\eqref{e:asym2}, containing the weighting with 
$|\mb{P}_{h \perp}|/m_{\pi}$. 
The results are shown in Fig.~\ref{f:axw} as a function of $\xbj$  and  
in Fig.~\ref{f:azw} as a function of $z_h$. The magnitude of this asymmetry is
higher than in the unweighted case, which is partially due to the
fact that the weighting  enhances the asymmetry by about a factor of
2.  

In addition to  appearing in semi-inclusive DIS in connection with the transversity
distribution of the nucleon, the Collins function can be independently 
extracted from another
process, i.e.\
electron-positron 
annihilation into two hadrons belonging to two
back-to-back jets~\cite{Boer:1997mf,Boer:1998qn}. We restrict ourselves to the
case of photon exchange only.
In this process, one of the two hadrons
(say hadron 2) defines the scattering plane together with the leptons and 
determines the direction with respect to which the azimuthal 
angles must be
measured. The cross section is differential in five variables, e.g.\
$z_1, z_2, y, |\mb{P}_{h \perp}|, \phi$. The variables $z_1$ and
$z_2$ are the longitudinal fractional momenta of the two hadrons. In the
center of mass frame $y=(1+\cos{\theta})/2$, where $\theta$ is the angle of
hadron 2 with respect to the momentum of the incoming leptons. The vector
$\mb{P}_{h \perp}$ denotes the transverse component of the momentum of hadron
1 and $\phi$ is its azimuthal angle with respect to the scattering plane. For a
more detailed description of the kinematical variables we refer to 
\cite{Boer:1997mf,Boer:1998qn}.

We define the azimuthal asymmetry  
\begin{equation} \begin{split} 
\left\langle P_{h \perp}^2 \cos{2 \phi}
\right\rangle_{e^+ e^-} (\theta,z_1,z_2) 
% \nn \\
&=  \frac{\int  \de^2 \mb{P}_{h\perp}
\,\lvert \mb{P}_{h\perp}\rvert^2\,\cos{2 \phi}\;
\de^5\sigma_{e^+ e^-}}
{\int \de^2 \mb{P}_{h\perp} \,\lvert \mb{P}_{h\perp}\rvert^2\,
\de^5\sigma_{e^+ e^-}}  \\
&= \frac{2 \, \sin^2 {\theta}}{1+\cos^2{\theta}}\,
\frac{H_1^{\perp (1)} (z_1)\,\bar{H}_1^{\perp (1)}(z_2)}
{\left( D_1 (z_1)\,\bar{D}_1^{(1)}(z_2)
	+ D_1^{(1)} (z_1)\,\bar{D}_1(z_2)\right)},  
\end{split} \end{equation}  
where
summations over quark flavors are understood.	
The weighting with a second power of $\lvert \mb{P}_{h\perp}\rvert$ 
in the numerator is 
necessary to obtain a deconvoluted expression. We prefer to use the same
weighting in the denominator as well, to avoid a modification of the 
asymmetry just caused by the weighting. 

\begin{figure}
        \centering
	\rput(-0.5,3.1){\rotatebox{90}
	{{\boldmath $\langle P_{h \perp}^2\,\cos{2 \phi} \rangle_{e^+e^-}$ }} }
	\rput(4.2,-0.3){\boldmath $z_1$}
        \includegraphics[width=8cm]{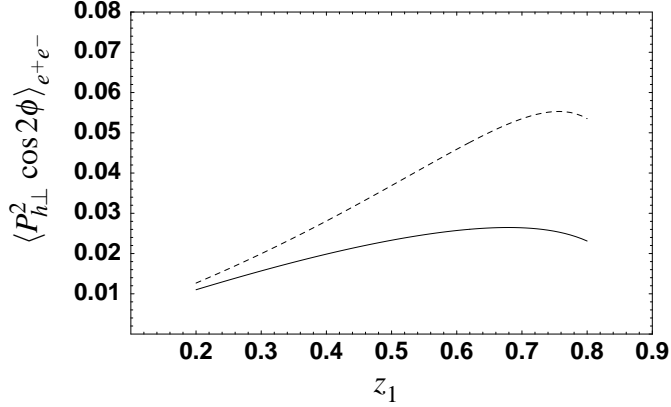}
	\mbox{}\\[3mm]
       \caption{Azimuthal asymmetry 
	$\langle P_{h \perp}^2 \cos{2 \phi}\rangle_{e^+ e^-}$ 
	for $e^+ e^-$ annihilation into two hadrons, 
	integrated over the range $0.2 \leq z_2 \leq 0.8$ (solid line), and
	over the range $0.5 \leq z_2 \leq 0.8$ (dashed line).} 
	\label{f:aee}
        \end{figure}

In Fig.~\vref{f:aee} we present the estimate of the asymmetry defined above,
entirely based on our model. The asymmetry has been integrated over $z_2$ and
$\theta$, leaving the dependence on $z_1$ alone. 
We have extended the $\theta$ integration interval all the way to $[0,\pi]$, to
obtain a conservative estimate. In fact, limiting the interval to
$[{\pi/4},{3 \pi/4}]$ will enhance the asymmetry by a factor of 2,
approximately.
Because the Collins function increases with
increasing $z$, we also get a larger asymmetry by restricting
the integration range for $z_2$. 
As an illustration of this feature, in Fig.~\vref{f:aee} we present
two results, obtained from two different integration ranges.
Our prediction is supposed to
be valid only at low energy scales and should be evolved for comparison with
higher energy experiments. It is important to note that we estimate the 
asymmetry to be of the order of about 5\%, 
and thus it should observable in experiments.

%%%%%%%%%%%%%%%%%%%%%%%%%%%%%%%%%%%%%%%%%%%%%%%%%%%%%%%%%%%%
\section{Comparison with existing data}
\label{s:comparison}

The HERMES collaboration recently measured the single {\em longitudinal} spin
asymmetry
\begin{equation} 
\langle \sin \phi_h \rangle_{UL}  \propto 
 \frac{1}{Q} \Big[ \Big( c_1 \, h_{L}(x) + c_2 \, h_1(x) \Big) 
                  H_1^{\perp (1/2)}(z) \,
                  + \, \text{other terms} \Big]. 
 \label{e:ul}
\end{equation} 
On purpose, we avoid entering the details of the formula. It is difficult to
extract information on the Collins function and on the transversity
distribution from this asymmetry~\cite{Boglione:2000jk}. 
In a recent analysis of this 
asymmetry, Efremov et al.~\cite{Efremov:2001cz} extracted a behavior
$H_1^\perp / D_1 \propto z$ for $z \le 0.7$, although some questionable
assumptions were used to obtain this result.

\begin{figure}
	\centering
	\rput(-0.5,3.1){\rotatebox{90}
	{{\boldmath $H_1^{\perp (1/2)} / D_1$ }} }
	\rput(4.2,-0.3){\boldmath $z$}
	\includegraphics[width=8cm]{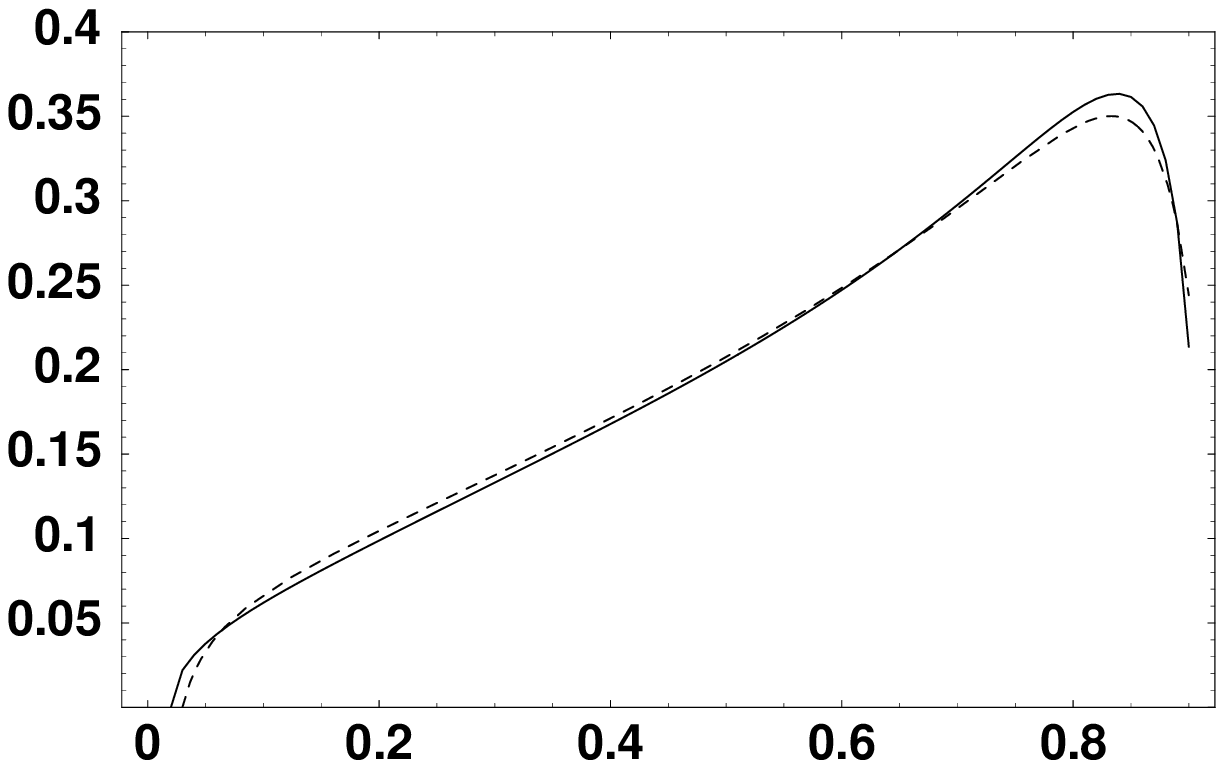}
	\mbox{}\\[3mm]
	\caption{Model result for the ratio $H_1^{\perp (1/2)} / D_1$ as a
	function of $z$ and the simple analytic parametrization 
	of \protect{Eq.~\eqref{e:param}}. 
	\label{f:ratioparam}}
	\end{figure}

If we assume that the {\it other terms} in (\ref{e:ul}) are small, then the
$z_h$ dependence of the asymmetry should be almost entirely due to the Collins
function. To compare this behavior with our model estimate, as a first step we 
parametrize our result for the ratio $H_1^{\perp (1/2)} / D_1$ with a simple
analytic form 
\begin{equation} 
\frac{H_1^{\perp (1/2)}(z)}{D_1(z)} \approx 0.316\, z + 0.0345\,\frac{1}{1-z}-
0.00359\, \frac{1}{(1 - z)^2}.
\label{e:param}
\end{equation} 
Fig.~\vref{f:ratioparam} shows the result of our model 
together with this parametrization (note that we extended the plot to higher
values of $z$ compared to Fig.~\vref{f:ratiomom0}).

\begin{figure}
	\centering
        \includegraphics[width=8cm]{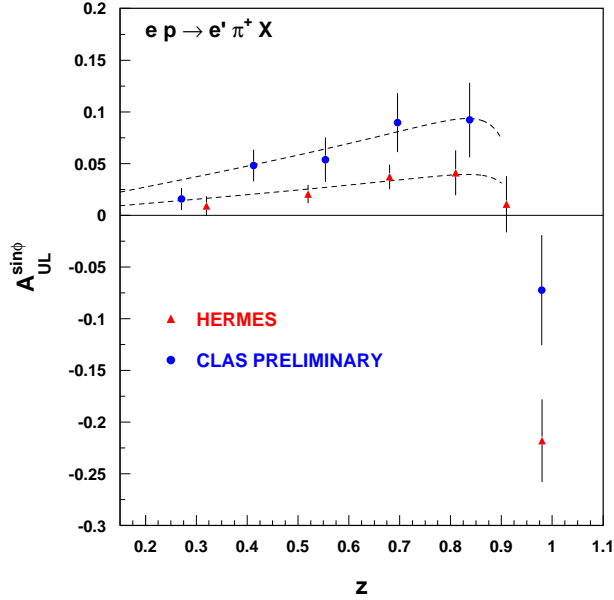}
	\caption{Comparison between 
the results of our model and data from the HERMES and 
CLAS experiments.} 
\label{f:fit}
\end{figure}

In Fig.~\ref{f:fit} we compare our parametrization 
with
HERMES data on $\langle \sin \phi_h \rangle_{UL}$~\cite{Airapetian:2001eg} and
preliminary data on the same asymmetry from the CLAS collaboration at JLAB~\cite{Avakian:2002}. Note that we
arbitrarily normalized our curve to take into account the unknown distribution 
functions and prefactors. The
agreement of the $z_h$ shape is remarkable. 

%%%%%%%%%%%%%%%%%%%%%%%%%%%%%%%%%%%%%%%%%%%%%%%%%%%%%%%%%%%%%%%%%%%%%%%%%%%
\section{Summary}
\label{s:four}

We have estimated the Collins fragmentation function for pions at a low
energy
scale by means of the Manohar-Georgi model.
This model 
contains three essential features of nonperturbative QCD:
massive quark degrees of freedom, chiral symmetry and 
its spontaneous breaking (with pions as Goldstone bosons). 
Because of the chiral invariant interaction between pions
and
quarks, the fragmentation process can be evaluated in a perturbative expansion.
The constituent quark
mass,
the axial pion-quark coupling $g_\sa$ and the maximum virtuality $\mu^2$
of the
fragmenting quark are free parameters of our approach.
The quark mass and $g_\sa$ 
are constrained within reasonable limits.
To ensure the convergence of the chiral perturbation expansion, $\mu^2$
cannot exceed a typical hadronic scale.
We have mostly considered the value $\mu^2 = 1\,\rm{GeV}^2$, which
guarantees that
the momenta of the particles produced in the fragmentation process
stay well below the scale of chiral symmetry breaking, $\Lambda_\chi \approx 1$
GeV\@.
To determine the appropriate value of $\mu^2$, 
the average transverse momentum of a data set could
be used.
In any case, we observed that variations of the free parameters within 
reasonable limits have only a minor
influence on the shape of the results, implying that
our approach has a good 
predictive power for
the $z$ behavior of the various functions.

We have found that the Manohar-Georgi model reproduces reasonably well the
unpolarized
pion fragmentation function and the average transverse momentum of a
produced hadron
as a function of $z$, supporting the idea of describing the fragmentation
process
by such a chiral invariant approach.

Compared to the unpolarized fragmentation function, modeling the
Collins function
is considerably more difficult, mainly because of its chiral-odd and
T-odd nature.
In our approach, the helicity flip required to generate a chiral-odd
object
is caused by the mass of the constituent quark, while
the T-odd behavior is produced via one-loop corrections.
The Collins function exhibits a quite distinct behavior from the
unpolarized fragmentation
function.
In particular, the ratio $H_1^\perp / D_1$ is strongly increasing with
increasing $z$.

On the basis of our results, we have calculated
the transverse single-spin asymmetry in semi-inclusive DIS 
where the Collins function appears in combination with the
transversity
of the nucleon.
This observable will be measured in the near future at HERMES and could
also be
investigated at COMPASS, Jlab (upgraded) and EIC.
For typical HERMES kinematics the asymmetry is of the order of $10\%$, giving
support to the
intention of extracting the nucleon transversity in this way.
We believe that our estimate of the Collins
function, despite its uncertainties,
can be very useful for this extraction. Finally, we found an encouraging
agreement between the shape of our estimate and the trend of the single spin
asymmetry measured by the HERMES
collaboration~\cite{Airapetian:1999tv,Airapetian:2001eg}, although it is not
clear if this asymmetry is originated solely by the Collins function.

More information on the Collins function from the experimental side is
required.
In this respect, the most promising experiment seems to be $e^+ e^-$
annihilation into two hadrons, where
$H_1^\perp$ appears squared in an azimuthal $\cos{2\phi}$ asymmetry.
According to our calculation, an asymmetry of the order of $5\%$ can be
expected, which should be measurable at high luminosity
accelerators, such as BABAR and BELLE~\cite{Ogawa:2001}.
Dedicated measurements of the Collins function would be extremely
important for
the extraction of the transversity distribution.
Moreover, they could answer the question whether a chiral invariant
Lagrangian can be
used to model the Collins function.

\renewcommand{\quot}{%
\parbox{5cm}{I was born not knowing and have had only a little time to change
that here and there.}\\
[2mm]
R. Feynman
}

%%%%%%%%%%%%%%%%%%%%%%%%%%%%%%%%%%%%%%%%%%%%%%%%%%%%%%%%%%%%%%%%%%%%%%%%%%
\chapter{Conclusions and outlook}

The main focus of this thesis was to present
some ways of accessing the transversity
distribution of quarks inside hadrons, $h_1$. This was the main motivation 
to perform a thorough analysis of one-particle and two-particle inclusive 
deep inelastic scattering at
leading order in $1/Q$, with a particular attention to T-odd fragmentation
functions. 
This thesis is certainly not the first work on this important subject, however
it complements the existing literature in many respects. 

%%%%%%%%%%%%%%%%%%%%%%%%%%%%%%%%%%%%%%%%%%%%%%%%%%%%%%%%%%%%
\section{Conclusions}

Several observables containing the
transversity distribution appear all along the thesis, but three of them are particularly promising and
have been highlighted.
The first method presented to access transversity 
is the measurement of the azimuthal single spin asymmetry described in
Eq.~\eqref{e:asym2}, or the similar asymmetry described in 
Eq.~\eqref{e:asym1}. In
these asymmetries, the transversity distribution appears in combination with
the Collins fragmentation function, which provides the chirality
flip necessary to compensate the chiral-odd nature of $h_1$. 

The Collins
function was introduced in Ref.~\citen{Collins:1993kk} as early as
1993. However, it was looked at with some skepticism because it is a 
single-particle T-odd
fragmentation function. It was conjectured to be small or to vanish
altogether~\cite{Jaffe:1998hf}. 
In this thesis, we tackled the question by calculating directly
the 
Collins function in a consistent and time-reversal invariant field theoretical 
approach. Our model calculation dispels the doubts about the vanishing of  
the Collins function and, more in general, of T-odd fragmentation functions.
Moreover, our calculation aspires to describe the Collins function in a
qualitative and quantitative way. At the moment, there is no unquestionable
experimental information on the Collins function. However, 
the single longitudinal spin 
asymmetry measured by the HERMES collaboration~\cite{Airapetian:1999tv,Airapetian:2001eg} can be
originated by the Collins function. We tested our model
on this experimental results and we found an encouraging agreement between the
shape of our estimate and the trend of the data.

It would be of remarkable importance to collect extra information on the  
Collins function, not only to allow the extraction of the transversity
distribution, but also to open a new window on polarization in
hadronization processes. 
New measurements could consider different asymmetries in deep
inelastic scattering (HERMES, COMPASS, CLAS, EIC), $pp$ scattering (RHIC)
and $e^+e^-$ annihilation into
two hadrons belonging to two jets (BABAR, BELLE). 

The second observable containing the transversity presented in this thesis 
is the asymmetry of Eq.~\eqref{e:asym3}, involving a chiral-odd T-odd
fragmentation function generated by the interference between the production of 
two hadrons in the $s$ and $p$ waves. This asymmetry was indicated for the
first time by Jaffe, Jin and Tang in 1998~\cite{Jaffe:1998hf}. However, their 
analysis suffered two limitations: it did not present a complete treatment
of two-hadron fragmentation functions and it relied just on one specific
mechanism to give some phenomenological 
indications. 
An improved analysis of two-hadron fragmentation was presented by Bianconi,
Boffi, Jakob and Radici~\cite{Bianconi:1999cd}, together with different models 
of the interference fragmentation
function~\cite{Bianconi:1999uc,Radici:2001na}. In this thesis, we complemented 
their formalism by performing a partial-wave decomposition of two-hadron
fragmentation functions, with and without including partonic transverse
momentum.\footnote{The results of the partial-wave expansion are presented
here for the first time and still await publication.}  We were able to separate
the contribution 
of the $s$ and $p$ waves and we recovered the results of Jaffe, Jin and Tang in
the transverse momentum integrated $sp$ interference sector. 

Besides rediscussing the interference fragmentation function in a more
exhaustive framework, another interesting asymmetry came out of the analysis,
as shown in Eq.~\eqref{e:asym4}. This asymmetry contains a chiral-odd T-odd
function pertaining to the pure $p$-wave sector of two-hadron production. Such 
a contribution was already mentioned by Collins, Heppelmann and Ladinsky in
1994~\cite{Collins:1994kq} and identified more formally by 
Ji in the same year~\cite{Ji:1994vw}. However, our work is the first one to
clearly distinguish the $p$-wave from the $sp$-interference part in
the single transverse spin asymmetry. 
It is useful to stress once more that the $p$-wave
contribution should not be overlooked.  It
vanishes when integrating over the full range of the polar angle $\theta$,
unlike the $sp$ interference term. On the other hand, it might be
larger than the latter -- especially in the presence of a spin-one resonance --thus offering an excellent way to access the transversity
distribution.

We looked more carefully at $p$-wave two-hadron fragmentation functions
from a different point of view, i.e.\ comparing them with
 spin-one fragmentation functions, introduced for the first time by
Ji~\cite{Ji:1994vw}. 
In the thesis, we 
carried out the analysis of spin-one functions in a 
more detailed way compared to the original work of Ji. 
We showed that they overlap almost
exactly with pure $p$-wave two-hadron fragmentation functions, 
except for the lack of a dependence on the invariant mass squared.
We included partonic
transverse momentum  and we presented the complete cross section of spin-one
leptoproduction at leading order in $1/Q$. 
As in the case of the Collins function, spin-one functions could provide us
with fresh information on spin effects in hadronization.

An important achievement of the thesis  is the connection between the
formalism of correlation
functions and of helicity matrices. The two approaches were known to be
equivalent at leading order in $1/Q$, but the connection had to be worked out
more explicitly. We expressed in the
helicity matrix formalism all the partonic functions we considered -- spin-half and spin-one distribution functions, without and with transverse momentum,
and one-hadron, two-hadron and spin-one
fragmentation functions, without and with transverse momentum. A significant
outcome of this examination was the derivation of positivity bounds for all of 
them. Positivity bounds can be valuable tools, as they provide guidance 
to estimate
unknown functions and they test the consistency of model calculations.

%%%%%%%%%%%%%%%%%%%%%%%%%%%%%%%%%%%%%%%%%%%%%%%%%%%%%%%%%%%%
\section{Outlook}

The formalism of two-hadron and spin-one fragmentation
functions has to be completed by analyzing all the azimuthal asymmetries
containing the transversity distribution function, in case partonic
transverse momentum is included. The only step in
this direction was presented in Ref.~\citen{Bacchetta:2001bt} and was limited
to the spin-one contributions.
Subleading twist contributions have been neglected in the thesis and they 
deserve further survey.

Another very useful extension would be to calculate the cross section of 
$e^+ e^-$ annihilation into two couples of hadrons belonging to two different
jets, including in the analysis 
the distinction of the $s$- and $p$-wave contributions. 
A measurement of two-hadron fragmentation functions in this process would be
extremely interesting first of all because -- as already mentioned -- 
they contain valuable
information on the role of polarization in fragmentation processes, and
secondly because --
as in the case of the Collins function -- an independent measurement of
these functions would be very important for a clear extraction of the
transversity distribution.

For what concerns our model calculation of T-odd functions,
an immediate application would be to use our results to estimate other
asymmetries containing the Collins function. Furthermore, 
the shape of our results could be 
used to choose a specific analytical form to parametrize the Collins
function and fit future experimental data.

Another possible development is to use a similar approach 
to estimate T-odd interference
fragmentation functions and T-odd spin-one fragmentation functions. The
analysis is complicated by the presence of a vector meson, which is 
 beyond the reach of our model as formulated at present. In principle, it is
possible to extend the model. In any case, even without any extension
it would be already
possible to explore the effect of one-loop corrections on ``incoherent''
two-pion production, and compare the relative
importance of $sp$ interference and pure $p$-wave contributions, even in the
absence of a resonance (e.g.\ in $\pi^+ \pi^+$ production).

From a more formal point of view, the use of a chiral invariant approach for
describing fragmentation functions at low energy scales should be investigated 
further. First of all, we should check if the chiral perturbation expansion 
is reliable. Secondly, we should try to describe fragmentation
functions entirely, and not their ``valence'' part only. Finally, it would be 
interesting to explore the connection between our one-loop corrections and 
the recent models of T-odd
effects~\cite{Brodsky:2002cx}, which are also based on one-loop effects.

\renewcommand{\quot}{}
\bibliography{mybiblio}

%%%%%%%%%%%%%%%%%%%%%%%%%%%%%%%%%%%%%%%%%%%%%%%%%%%%%%%%%%%%%%%%%%%%%%%%%%
\chapter*{Summary}
\markboth{Summary}{Summary}
\addcontentsline{toc}{chapter}{Summary}

This thesis discusses three different ways to observe the transverse spin of
quarks inside the nucleons. This problem 
relates to one
of the unanswered questions of present day physics: what is the
internal structure of protons and neutrons?

One of the ways to investigate such a question is to study the process of
{\em deep inelastic scattering}, in which 
a nucleon is bombarded with a focused 
beam of electrons (or other leptons) at very high energies. The electrons
penetrate inside the nucleon and the way they are scattered yields information 
on its inner structure. A down-to-earth analogy would be
shooting at a car with a machine gun, penetrating its hood and recording the
 way bullets bounce on its internal components -- engine, gear box and all. 

Deep inelastic scattering provides us with snapshots of the
nucleon at high resolution, a few percents of the proton size. 
Extracting the relevant information on the structure of the nucleon 
is a complex task and requires a lot of theoretical effort. Due to the high
energies and high momenta involved, we resort to the techniques of
{\em quantum field theory}, which incorporates relativity and quantum
mechanics.
At low resolution -- in low energy experiments -- nucleons appear to be
unbreakable particles with a certain mass, electric charge and spin.
At high resolution -- in deep inelastic scattering -- nucleons display
an extremely intricate structure, 
involving a
very high number of smaller particles, {\em quarks} and {\em gluons}.
At present, we are not able to match these two pictures, in other words
we are not able 
to explain how quarks and gluons interact to give origin to the
characteristics of the nucleons.

A key issue in understanding the nucleon is to explain the composition of
its spin
in terms of the underlying quark and gluon structure. 
So far, we acquired a good 
knowledge of the {\em helicity} distribution of quarks inside targets with a 
{\em longitudinal} spin. 
The exploration of the {\em transverse} spin distribution (or {\em transversity}) 
of quarks
in targets with a {\em transverse} spin
would give us a new perspective on the dynamics inside the nucleon. To a 
certain extent, the helicity and the transverse spin distributions
represent a front and a side view of the nucleon spin.

Observing the transversity of quarks is an arduous task. Several 
approaches have been suggested in theory, but none of them has been
put into practice yet.
In this thesis, three ways to probe the transversity
are highlighted and analyzed. 
They are far from exhausting all possible methods, but they seem to be 
promising and could be adopted in experiments in the near future. All of
them require that in the deep inelastic process we keep track not only of 
the scattered electron, but also of one or two of the fragments coming from
the disintegration of the nucleon. The detection of the fragments offers the
opportunity to  unravel more details of the
structure of the target, but the price to pay is that we have to deal not
only with the way the quarks are arranged in the nucleon, but also with the way
they produce the final fragments, i.e.\ with {\em fragmentation functions}.

From a more technical point of view, this thesis starts with reviewing the
formalism to deal with deep inelastic scattering. The {\em parton distribution 
functions} are introduced, describing the way quarks and gluons are
disposed inside the nucleon. Totally inclusive deep inelastic scattering,
where only the scattered electron is detected, is discussed as the first and
simplest way to observe distribution functions.
To identify some way to measure the quark transversity, 
this thesis analyzes
one-particle and
two-particle inclusive deep inelastic scattering, where one and two 
of the outgoing hadrons are detected in coincidence with the electron. These
processes require the introduction of {\em parton fragmentation 
functions}, describing the way a quark evolves into final state hadrons.
This thesis shows that transversity 
can be measured 
in the above processes, in connection with three
different fragmentation functions: the first one requires the observation  
of an unpolarized final state hadron with transverse
momentum, the second requires the observation of the interference between the
$s$- and $p$-wave production of two hadrons, the third requires the 
observation of pure $p$-wave two-hadron production, or equivalently 
of a spin-one resonance.
All these fragmentation functions fall in the category of {\em T-odd}
fragmentation functions: they require the presence of final state
interactions, or else they are forbidden by time-reversal invariance. The last 
part of the thesis looks at the possibility of modeling this kind of
fragmentation functions and investigates 
whether they can be large enough to allow
the extraction of transversity.

%\include{thanks}

%%%%%%%%%%%%%%%%%%%%%%%%%%%%%%%%%%%%%%%%%%%%%%%%%%%%%%%%%%%%%%%%%%%%%%%%%%%%%
\end{document}